%% file: main.tex
\documentclass[prb,twocolumn,notitlepage,superscriptaddress,nofootinbib]{revtex4-2}
\usepackage{bbm}
\usepackage{amsmath,amssymb}
\usepackage[hidelinks,colorlinks,linkcolor=blue,
citecolor=blue,urlcolor=blue]{hyperref}
\usepackage{graphicx,siunitx}
\usepackage[dvipsnames]{xcolor}
\usepackage{flafter}
\usepackage{simpler-wick}
\usepackage{pifont}

\usepackage[normalem]{ulem}

\usepackage{xcolor}
\DeclareUnicodeCharacter{3000}{\textcolor{red}{BAD!!}}

\newcommand{\be}{\begin{equation}}
\newcommand{\ee}{\end{equation}}
\newcommand{\bea}{\begin{equation} \begin{aligned}}
\newcommand{\eea}{\end{aligned} \end{equation} }
\newcommand{\bi}{\begin{itemize}}
\newcommand{\ei}{\end{itemize}}

\renewcommand{\be}{\beta}

\newcommand{\bpm}{\begin{pmatrix}}
\newcommand{\epm}{\end{pmatrix}}

\usepackage{amsmath,amsfonts,amssymb,amsthm,epsfig,array}
\usepackage{dsfont}
\usepackage{slashed}
\usepackage{graphics}

\usepackage{float}
\usepackage{verbatim}

\usepackage{color}
\usepackage{tabularx}
\usepackage[mathscr]{euscript}
\usepackage{mathtools}

\newcommand{\bsl}[1]{\boldsymbol{#1}} 
\newcommand{\mbf}[1]{\boldsymbol{#1}}

\renewcommand{\mod}{\,\mathrm{mod}\,}

\newcommand{\bra}[1]{\langle #1|}
\newcommand{\ket}[1]{|#1 \rangle}
\newcommand{\braket}[2]{\left\langle #1 | #2  \right\rangle}

\newcommand{\ii}{\mathrm{i}}

\renewcommand{\Re}{\mathop{\mathrm{Re}}}
\renewcommand{\Im}{\mathop{\mathrm{Im}}}

\usepackage{bm}
\usepackage{environ}
\NewEnviron{eqs}{%
\begin{equation}\begin{split}
    \BODY
\end{split}\end{equation}
}

\let\oldAA\AA
\renewcommand{\AA}{\text{\normalfont\oldAA}}

\newcommand{\C}{\mathcal{C}}

\newcommand{\K}{\text{K}}

\begin{document}

\title{``Berry Trashcan'' Model of Interacting Electrons in Rhombohedral Graphene}

\author{B. Andrei Bernevig}
\affiliation{Department of Physics, Princeton University, Princeton, New Jersey 08544, USA}
\affiliation{Donostia International Physics Center, P. Manuel de Lardizabal 4, 20018 Donostia-San Sebastian, Spain}
\affiliation{IKERBASQUE, Basque Foundation for Science, Bilbao, Spain}
\author{Yves H. Kwan}
\affiliation{Princeton Center for Theoretical Science, Princeton University, Princeton NJ 08544, USA}

\date{\today}

\begin{abstract}
    We present a model for interacting  electrons in a continuum band structure that resembles a trashcan, with a flat bottom of radius $k_b$ beyond which the dispersion increases rapidly with velocity $v$. 
    The form factors of the Bloch wavefunctions can be well-approximated by the Girvin-MacDonald-Platzman algebra, which encodes the uniform Berry curvature. We demonstrate how this model captures the salient features of the low-energy Hamiltonian for electron-doped pristine $n$-layer rhombohedral graphene (R$n$G) for appropriate values of the displacement field, and provide corresponding expressions for $k_b$.
     In the regime where the Fermi wavevector is close to $k_b$,  we analytically solve the Hartree-Fock equations for a gapped Wigner crystal in several limits of the model. We introduce a new method, the sliver-patch approximation, which extends the previous few-patch approaches and is crucial in both determining the full Chern number (beyond \mbox{mod 3}) of the ground state and gapping the Hartree-Fock solution. A key parameter is the Berry flux $\varphi_{\text{BZ}}$ enclosed by the (flat) bottom of the band. We analytically show that there is a ferromagnetic coupling between the signs of $\varphi_{\text{BZ}}$ and the Chern number $C$ of the putative Wigner crystal. We also study the competition between the $C=0$ and $1$ solutions as a function of the interaction potential for parameters relevant to R$n$G. 
    By exhaustive comparison to numerical Hartree-Fock calculations, we demonstrate how the analytic results capture qualitative trends of the phase diagram, as well as quantitative details such as the enhancement of the effective velocity.
    Our analysis paves the way for an analytic and numerical examination of the stability and competition beyond mean-field theory of the Wigner crystals in this model.
\end{abstract}

\maketitle

\section{Introduction}

 \begin{figure}
 \centering
\includegraphics[width=0.65\columnwidth]{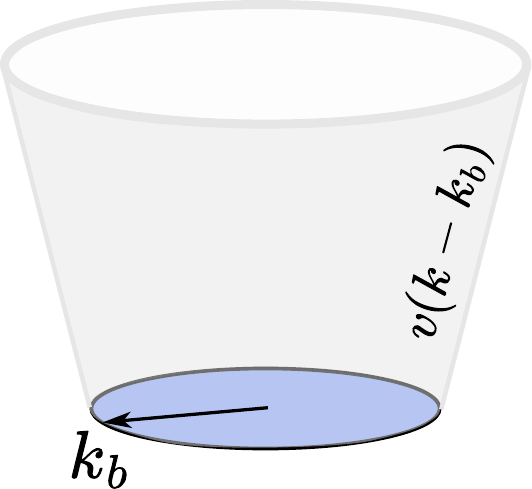} 
\caption{Schematic of the Berry Trashcan model. The dispersion consists of a flat bottom with radius $k_b$ (shaded blue), with sharply dispersing walls of velocity $v$. The form factors are described by the Girvin-MacDonald-Platzman algebra for all momenta. The Berry curvature enclosed by the flat bottom is controlled by $\varphi_\text{BZ}$.}
\label{fig:main_schematic}
\end{figure}

Rhombohedral-stacked multilayer graphene (R$n$G) has emerged as a fertile platform for realizing a panoply of correlated phenomena~\cite{zhou2021half,zhou2021superconductivity,shi2020electronic,myhro2018large,han2023orbital,han2024correlated,han2024large,sha2024observation,han2024signatureschiralsuperconductivityrhombohedral,patterson2024superconductivityspincantingspinorbit,yang2024diverseimpactsspinorbitcoupling,liu2024spontaneous,zhang2024layer,zhang2024graphite}, such as flavor symmetry-breaking and  superconductivity. The presence of an interlayer potential significantly flattens the band structure at low energies, enabling interactions to drive interesting behaviors. 
The recent discovery of Chern insulators (CIs) and fractional Chern insulators (FCIs)~\cite{neupert2011fci,regnault2011fci,sheng2011fractional,sun2011nearly,tang2011high} in pentalayer graphene twisted and aligned on one side with the hBN substrate at $\theta=0.77^\circ$~\cite{Lu2024fractional} has also invigorated experimental investigations into such phenomena in the family of R$n$G/hBN superlattices~\cite{chen2020tunable,zhou2021half,xie2024even,choi2024electricfieldcontrolsuperconductivity,waters2024interplayelectroniccrystalsinteger,lu2025extended,aronson2024displacementfieldcontrolledfractionalchern,zhou2024layer,han2024engineering,ding2024electricalswitchingchiralityrhombohedral,zheng2024switchablecherninsulatorisospin,xiang2025continuouslytunableanomaloushall,wang2025electricalswitchingcherninsulators}, and further emphasized the topology inherent in such platforms. The combination of the lattice mismatch and the twist angle $\theta$ generates a moir\'e pattern and enables the notion of a filling factor $\nu$ relative to the moir\'e unit cell. Owing to the single-sided alignment to hBN, the direction of an externally applied displacement field can be used to tune the system into the moir\'e-proximate (moir\'e-distant) regime if the active charge carriers are localized towards (away from) the aligned hBN, leading to distinct phenomenology. On the moir\'e-distant side, the interplay~\cite{Lu2024fractional,lu2025extended} of CIs and FCIs at commensurate $\nu$ with other phenomena that are extended along the density axis has sparked theoretical investigations~\cite{dong2024AHC1,zhou2024fractional,dong2024theorypentalayer,guo2024fractional,kwan2023mfci3,dong2024stability,soejima2024AHC2,tan2024parent,zeng2024sublattice,xie2024integerfractional,crepel2024efficientpredictionsuperlatticeanomalous,kudo2024quantumanomalousquantumspin,sarma2024thermal,huang2024selfconsistent,yu2024MFCI4,shavit2024entropy,huang2025displacement,huang2024impurityinducedthermalcrossoverfractional,wei2025edge,patri2024extended,tan2024wavefunctionapproachfractionalanomalous,zhou2024newclassesquantumanomalous,zeng2024berryphasedynamicssliding} and prompted debates regarding the precise role played by the moir\'e pattern. In particular, it has been suggested that if the effective moir\'e coupling is sufficiently weak, the observed physics may be connected to Wigner crystallization\footnote{This is also referred to as an anomalous Hall crystal if the Chern number $C\neq 0$.} rooted in the pristine R$n$G limit~\cite{kwan2023mfci3,dong2024AHC1,zhou2024fractional,dong2024theorypentalayer}. Such Wigner crystals have been obtained in numerical Hartree-Fock (HF) calculations of R$n$G.

It is desirable to have simplified models that can yield better analytical and physical insight into the array of strongly-interacting and correlated phenomena observed in R$n$G and related platforms. In this work, we introduce an idealized interacting continuum Hamiltonian (see Fig.~\ref{fig:main_schematic}) that captures the salient features of the low-energy conduction electrons in pristine R$n$G under certain conditions. The kinetic energy of the model resembles a trashcan, with a flat bottom encircled by steeply dispersing walls. The form factors of the Bloch wavefunctions encode the nearly uniform Berry curvature, and can be well approximated by the Girvin-MacDonald-Platzman (GMP) algebra~\cite{girvin1986magnetoroton}. For this reason, we call this the ``Berry Trashcan'' model. The usefulness of the model is illustrated with an analytical mean-field study of Wigner crystals whose length scale coincides with that set by flat bottom of the trashcan. In these HF studies, we introduce the sliver-patch approximation, which generalizes the few-patch constructions that have been used to study orders such as superconductivity and density waves in other systems~\cite{le2009superconductivity,nandkishore2012chiral}. In R$n$G, patch methods applied to the $K_M,K'_M$ corners of the Wigner crystal Brillouin zone have been used to constrain the Chern number $C$ modulo 3~\cite{crepel2024efficientpredictionsuperlatticeanomalous, soejima2024AHC2,dong2024stability}. Being  $C\mod 3$ analyses, they do not distinguish between, say $C=-2,1,4$, and, with the exception of Ref.~\cite{dong2024stability}, neglect the large gapless regions of R$n$G around the $K_M$-$M_M$-$K'_M$ lines.
The sliver-patch HF analysis of our Berry trashcan model is analytically tractable in certain limits, and we obtain the mean-field energy and the full Chern number $C$ of the Wigner crystal, show the presence of a ferromagnetic coupling between $C$ and the Berry curvature of the underlying band, and compare with exhaustive numerical HF calculations. 

Our simple Berry Trashcan model can be addressed with other techniques such as exact diagonalization. The analytical tractability in certain limits may also enable inclusion of fluctuations to investigate the (in)stability of the Wigner crystals~\cite{collective_unpub}. However, we emphasize that our results do not claim that the insulating ground states in R$n$G/hBN are directly connected to a moir\'e-less Wigner crystal. Experiments currently show different behavior in R$n$G samples with and without alignment to hBN, suggesting that the moir\'e could be crucial in understanding the physics of the system~\cite{kwan2023mfci3}. We also note that the FCIs observed in Ref.~\cite{Lu2024fractional} have so far not been obtained in unbiased multiband calculations~\cite{yu2024MFCI4}.

\section{The Berry Trashcan Model}

Near charge neutrality, the low-energy band structure of R$n$G is located near the two valleys of graphene. The continuum model~\cite{bistritzer2011moire} for small momenta $\bm{k}$ around the valley $K$ Dirac point is $H_K(\bm{k})+H_D$, explicitly given in App.~\ref{secapp:Ham_symmetries_RnGhBN}~\cite{herzog2024MFCI2}. $H_D$ models the externally applied displacement field as an onsite potential $V(l-\frac{n-1}{2})$ that depends linearly on the layer index $l=0,\ldots,n-1$. This acts to separate the lowest valence and conduction bands, and flatten the dispersion at the band edges. The single-particle Hamiltonian obeys continuous translation invariance, $C_3$ rotation symmetry, and an antiunitary intravalley symmetry $M_1\mathcal{T}$ which takes $(k_x,k_y)\rightarrow (k_x,-k_y)$. 

The essential features of R$n$G can be captured in the limit where we retain just the vertical interlayer hopping $t_1=355.16\,$meV and the intralayer Dirac cones of $H_K(\bm{k})$
\begin{equation}
    [h(\bm{k})]_{ll'}=v_F\delta_{ll'}\bm{k}\cdot\bm{\sigma}+t_1\delta_{l,l'+1}\sigma^++t_1\delta_{l,l'-1}\sigma^-,
\end{equation}
where $v_F=542.1\,$meVnm, $\bm{\sigma}=(\sigma_x,\sigma_y)$ are Pauli matrices in sublattice space, and $\sigma^\pm=\frac{1}{2}({\sigma_x\pm i\sigma_y})$. We refer to $h(\bm{k})$ as the chiral Hamiltonian, which possesses an enhanced set of symmetries, including a chiral symmetry $\Sigma=\sigma_z$ and full intravalley $SO(2)$ rotation. For small $v_Fk/t_1$, where $k=|\bm{k}|$, the spectrum of $h(\bm{k})$ decomposes into two low-energy eigenstates localized near the $(l,\sigma)=(0,A)$ and $(n-1,B)$ sites, with the other states at significantly higher energies $|E|\gtrsim t_1$ due to the interlayer dimerization. We therefore focus on the former, which are encoded with the chiral basis wavefunctions~\cite{koshino2010interlayer,herzog2024MFCI2,soejima2024AHC2,tan2024parent, crepel2024efficientpredictionsuperlatticeanomalous} 
\begin{gather}
    [\psi_A(\bm{k})]_{l\sigma}=\frac{(-v_Fk_+/t_1)^l}{N(\bm{k})}\delta_{\sigma,A}\\ [\psi_B(\bm{k})]_{l\sigma}=\frac{(-v_Fk_-/t_1)^{n-l-1}}{N(\bm{k})}\delta_{\sigma,B},
\end{gather}
where $k_\pm=k_x\pm ik_y$, and $N(\bm{k})=\sqrt{\frac{1-(v_Fk/t_1)^{2n}}{1-(v_Fk/t_1)^2}}$ is a normalization factor. For $v_Fk/t_1<1$, $\psi_A(\bm{k})$, which only has weight on sublattice $A$, has maximal amplitude on the bottom $l=0$ layer, and exponentially decays into the higher layers, while the opposite occurs for $\psi_B(\bm{k})$ whose maximal amplitude is on the top $l=n-1$ layer. Projecting $h(\bm{k})$ and the interlayer potential into the chiral basis leads to
\begin{equation}
h(\bm{k})+H_D\rightarrow \begin{pmatrix}\label{eq:hk_HD}
        V(k) & \frac{-t_1}{N(\bm{k})^2}(-v_Fk_-/t_1)^n \\
        \frac{-t_1}{N(\bm{k})^2}(-v_Fk_+/t_1)^n & -V(k)
    \end{pmatrix},
\end{equation}
where $V(k)$ is given in Eq.~\ref{eqapp:Vk}. For positive $V$ and small $v_Fk/t_1$, $V(k)$ is negative so that the lowest conduction band, which is the focus of this paper, is primarily built out of $\psi_B(\bm{k})$ and localized near the top layer $l=n-1$. For our purposes, $V$ should be sufficiently large to generate a sizable gap with the valence band, but not enough to distort the dispersion into a ``Mexican hat'' shape with a prominent local maximum at $k=0$ as considered in Ref.~\cite{soejima2024AHC2}.

To define the Berry Trashcan model, we need to specify its dispersion and form factors. 
For the former, we note that as long as $n$ is not too small, the conduction band of Eq.~\ref{eq:hk_HD} consists of a relatively flat bottom that rapidly disperses upwards above some momentum. As illustrated in Fig.~\ref{fig:main_schematic}, the trashcan model parametrizes this with an exactly flat bottom up to a boundary radius $k_b$, beyond which the dispersion increases linearly with velocity $v$, i.e.~$E(k)=v(k-k_b)\theta(k-k_b)$. As shown in App.~\ref{subsecapp:trashcan_parameterization}, $k_b$ and $v$ can be extracted analytically, and for $n=5$ we find $k_b\simeq 0.51\frac{t_1}{v_F}$ and $v\simeq 0.45v_F$. The corresponding values for other $n$ are listed in Tab.~\ref{tab:trashcan_params}. In the limit $n\rightarrow \infty$, we note that $k_b\rightarrow \frac{t_1}{v_F}$, where the use of the chiral basis is no longer fully justified.

For large $V$, the conduction band eigenfunction for small $k$ is dominated by $\psi_B(\bm{k})$. In App.~\ref{subsecapp:form_factor_approximations}, we show that the form factors of $\psi_B(\bm{k})$ can be approximated by
\begin{equation}\label{eq:Mkq_GMP}
    M_{\bm{k},\bm{q}}=\braket{\psi_B(\bm{k}+\bm{q})}{\psi_B(\bm{k})}\approx e^{-\frac{v_F^2}{2t_1^2}(q^2+2i\bm{q}\times\bm{k})},
\end{equation}
which becomes exact as $v_Fk/t_1\rightarrow0$, and remains a good approximation for a range of $k$ and intermediate values of $n$. Eq.~\ref{eq:Mkq_GMP} takes the same form as that of the lowest Landau level, from which we can read off the uniform Berry curvature $\Omega(\bm{k})=2\beta$, where $\beta\equiv v_F^2/t_1^2$. The density operator thus obeys the Girvin-MacDonald-Platzman (GMP) algebra~\cite{girvin1986magnetoroton}. Based on our above extraction of the trashcan radius $k_b$, the Berry flux enclosed by the flat bottom is $\simeq 0.52\pi$ for $n=5$\footnote{The GMP approximation underestimates the Berry curvature of R$n$G for finite $k$, see App.~\ref{subsecapp:form_factor_approximations}.}, and reaches a maximum of $2\pi$ for $n\rightarrow \infty$.
We note that the overlaps in the GMP approximation obey a triangle area law 
\begin{equation}
\langle \psi_B(\bm{k}) \ket{\psi_B(\bm{k}')}=|\langle \psi_B(\bm{k}) \ket{\psi_B(\bm{k}')}|e^{- i\varphi_{\bm{k}\rightarrow \bm{k}'}},
\end{equation}
where $\varphi_{\bm{k}\rightarrow \bm{k}'}$ is the signed Berry curvature enclosed by the loop $\bm{0}\rightarrow \bm{k}\rightarrow \bm{k}
'\rightarrow \bm{0}$.

When the chemical potential lies near the flat bottom of the band structure, it is reasonable to consider a fully valley- and spin-polarized system~\cite{antebi2024stoner,bernevig2021TBGIII,lian2021TBGIV,bernevig2021TBGV}. Including density-density interactions with interaction potential $V_{\bm{q}}$, the full Hamiltonian projected to the chiral conduction band is\footnote{This corresponds to the charge neutrality interaction scheme, i.e.~the quartic term is normal-ordered with respect to the occupied valence bands. Alternative interaction schemes can trivially be recast into Eq.~\ref{eq:trashcan_ham} by adjusting $E(\bm{k})$~\cite{kwan2023mfci3}.}
\begin{align}\label{eq:trashcan_ham}
    H=& \sum_k E(\mbf{k}) \gamma_{\mbf{k}}^\dagger \gamma_{\mbf{k}} \\
    +& \frac{1}{2\Omega_{tot}}  \sum_{\bm{q}, \bm{k}, \bm{k'}} V_{\mbf{q}} M_{\mbf{k},\mbf{q}} M_{\mbf{k}',-\mbf{q}} \gamma_{\mbf{k}+ \mbf{q}}^\dagger \gamma_{\mbf{k}'- \mbf{q}}^\dagger \gamma_{\mbf{k}'} \gamma_{\mbf{k}}
\end{align}
where $\Omega_{tot}=L_xL_y$ is the system area, and $\gamma^\dagger_{\bm{k}}$ is the creation operator for the chiral basis state $\psi_B(\bm{k})$. Inserting the GMP form factors leads to the simple form $V_{\mbf{q}} M_{\mbf{k},\mbf{q}} M_{\mbf{k}',-\mbf{q}} = V_{\mbf{q}} e^{-\beta q^2} e^{-i\beta\bm{q}\times (\bm{k}- \bm{k'})}$. For the analytical mean-field calculations described later, it will often be useful to specialize to an exponential interaction $V_{\bm{q}}=V_0 e^{-\alpha q^2}$, which will appear in conjunction with the magnitude of Bloch overlaps as $V_{|\mbf{k-k'}|}| \langle \psi_B(\mbf{k})|\psi_B(\mbf{k'}) \rangle|^2=V_0e^{-\phi  |\mbf{k-k'}|^2}$, with $\phi\equiv \alpha+\beta$. The exponential interaction can be fitted well to the gate-screened Coulomb interaction for short gate distances (see App.~\ref{subsubsecapp:f_GMP_exp}).

In App.~\ref{secapp:classical}, we show that the model for $\phi=0$ admits an interesting `phases-only' limit with an enhanced set of symmetries. 
However, we emphasize that the model can be defined for any $V_{\bm{q}}$.
While we focus on fully spin- and valley-polarized states, the Hamiltonian can be straightforwardly generalized to include the other valley and spin flavors which are related by time-reversal and spin-$SU(2)$ symmetries.

\section{Mean-field Wigner Crystal and The Sliver-Patch Approximation }

 \begin{figure}
 \centering
\includegraphics[width=0.8\columnwidth]{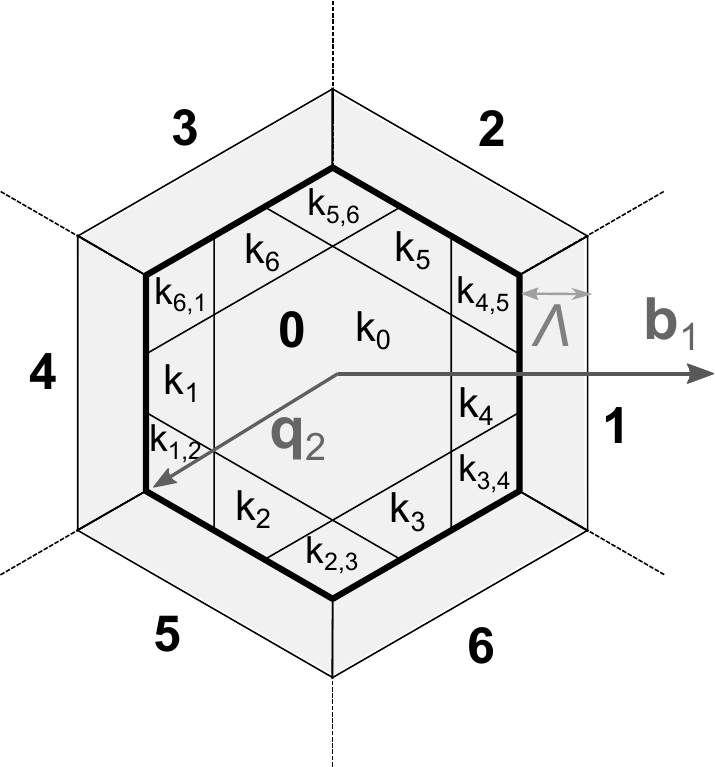} 
\caption{
Mean-field treatment of Wigner crystals in the Berry Trashcan model for electron densities roughly coincident with the flat bottom of radius $k_b$. The model contains continuum states (shaded) within seven BZ's $0,\ldots 6$. BZ 0, which coincides with the reduced BZ of the Wigner crystal, is bordered by the the thick hexagon. $\bm{b}_1$ is a primitive reciprocal lattice vector, while $\bm{q}_2$ connects the high-symmetry points $\Gamma_M$ and $K_M$. Due to the sharp dispersion outside the first BZ, we only retain single-particle states within a momentum cutoff that is spaced by $\Lambda$ from BZ 0. The reduced BZ is partitioned into regions depending on the allowed hybridizations at finite momenta (see main text). 
}
\label{fig:BZzones}
\end{figure} 

We now present an analytical HF study of insulating Wigner crystals with one electron per Wigner unit cell and various Chern numbers $C$. We consider a hexagonal reconstructed Brillouin zone (BZ) where the wavevector $\bm{q}_2$ of its $K_M$ corner is close to the trashcan radius $k_b$, which implicitly sets the electronic density. This setting allows for a tractable mean-field calculation, because the translation symmetry-breaking only involves momenta near the boundary of the fundamental BZ (referred to as BZ 0) owing to the sharp dispersion beyond $k_b$. In particular, we decompose the momentum space into regions as shown in Fig.~\ref{fig:BZzones}, where states outside the outer hexagonal cutoff, at distance $\Lambda$ from the boundary of BZ 0, are removed from the Hilbert space due to the large kinetic penalty (see App.~\ref{secapp:1dtrashcansetup} and \ref{secapp:1dtrashcan_HFanalysis} for an analysis of a simpler toy 1D problem). $\Lambda<q_2$ should be chosen so that the kinetic scale $v\Lambda$ is of order the typical inter-particle interaction energy.
Any reduced momentum, i.e.~any $\bm{k}$ taking values in BZ 0, can be assigned to various non-overlapping regions. The region $\bm{k}_0$, which lies within the flat part of the dispersion, corresponds to states which do not have partners $\bm{k}_0+\bm{G}$ separated by a reciprocal lattice vector (RLV) that lie within the cutoff, while the gapless `sliver' $\bm{k}_j$ corresponds to states that have one partner in BZ $j$ that lies within the cutoff. Finally, the gapless `patch' region $\bm{k}_{j,j+1}$ refers to states have partners in both BZ $j$ and $j+1$.

The HF solution is encoded in the density matrix $O_{ij}(\bm{k})\equiv \langle \gamma^\dagger_{\bm{k}+C^{i-1}_6\bm{b}_1}\gamma_{\bm{k}+C^{j-1}_6\bm{b}_1}\rangle$, where $\bm{b}_1$ is one of the primitive RLVs (see Fig.~\ref{fig:BZzones}). Consistent with the numerical HF calculations (see App.~\ref{secapp:HFphasediagrams}), we impose the symmetries\footnote{The $C_6$ we refer to here is an intravalley $C_6$ symmetry that is a subgroup of the emergent $SO(2)$ symmetry, which is therefore distinct from the microscopic $C_6$ rotation that would interchange the graphene valleys.} generated by $C_6 \gamma_{\mbf{k}}^\dagger C_6^{-1} = e^{i\frac{\pi n}{6} } \gamma_{C_6 \mbf{k}}^\dagger$ and $M_1\mathcal{T}\gamma_{\mbf{k}}^\dagger (M_1 \mathcal{T})^{-1} = \gamma_{M_1\mathcal{T}\mbf{k}}^\dagger$ on the solution (see App.~\ref{subsecapp:symmetries_general_model}), so that the independent order parameters can be chosen as those with $\bm{k}$ in the $\bm{k}_0,\bm{k}_1$ or $\bm{k}_{12}$ regions. The interacting part of the HF Hamiltonian involves components at momentum transfers $\bm{q}=0$, characterized by a band renormalization field $f_{\bm{k}}$, and $|\bm{q}|=b_1$, characterized by a hybridization field $g_{j,\bm{k}}$
\begin{align}
    H^{\text{HF,int}}=&\sum_{\bm{k}}f_{\bm{k}}\gamma^\dagger_{\bm{k}}\gamma_{\bm{k}}+\sum_{j=1}^6\sum_{\bm{k}}g_{j,\bm{k}}\gamma_{\mbf{k}+ C_6^{j-1} \mbf{b_{1}}}^\dagger \gamma_{\mbf{k}}.
\end{align}
Note that the mean fields $f_{\bm{k}}$ and $g_{j,\bm{k}}$, whose explicit expressions are provided in App.~\ref{secapp:2dtrashcan_genparam}, are functions of $O_{ij}(\bm{k})$. We discuss the conditions for an insulating HF solution in App.~\ref{secapp:2dmodel_HF}, which we will assume are met in the discussion below.

\subsection{Chern number and high-symmetry points}

We first focus on the high-symmetry points, which together constrain the Chern number of the HF solution modulo 6 because of $C_6$ symmetry~\cite{fang2012bulk} (see App.~\ref{subsecapp:symChern} for more details). 
At the $\Gamma_M$-point for an insulating state, we have no choice but to fill the state at $\bm{k}=0$, leading to a $C_6$ eigenvalue $\eta_{\Gamma}=e^{i\frac{\pi n}{6}}$. For the $M_M$-point at $\bm{k}=-\bm{b}_1/2$, up to an overall energy shift, we have the $2\times 2$ Hamiltonian
\begin{equation}
    H^\text{HF}_{-\frac{\bm{b}_1}{2}}=\begin{pmatrix}
        \gamma^\dagger_{0,-\frac{\bm{b}_1}{2}}&\gamma^\dagger_{1,-\frac{\bm{b}_1}{2}}
    \end{pmatrix}
    \begin{pmatrix}
        0 & g_{1,-\frac{\bm{b}_1}{2}}^*\\
        g_{1,-\frac{\bm{b}_1}{2}} & 0
    \end{pmatrix}
    \begin{pmatrix}
        \gamma_{0,-\frac{\bm{b}_1}{2}}\\\gamma_{1,-\frac{\bm{b}_1}{2}}
    \end{pmatrix},
\end{equation}
where $\gamma^\dagger_{j,\bm{k}}$ is the creation operator in BZ $j$ with reduced momentum $\bm{k}$, e.g.~$\gamma^\dagger_{1,\bm{k}}\equiv \gamma^\dagger_{\bm{k}+\bm{b}_1}$.
Filling the lower-energy solution leads to the order parameter $O_{10,-\frac{\bm{b}_1}{2}}=-\frac{1}{2}\text{sgn}\,g_{1, -\frac{\mbf{b_{1}}}{2}}$ and a $C_2$ eigenvalue $\xi_M=e^{i\frac{n\pi}{2}}\text{sgn}\,O_{10,-\frac{\bm{b}_1}{2}}$. Note that the parity of the Chern number satisfies $(-1)^C=-\text{sgn}\,g_{1, -\frac{\mbf{b_{1}}}{2}}$.

Finally, considering the $K_M$ point at $\bm{k}=\bm{q}_2$ leads to 
\begin{equation}\label{eq:Hq2}
    H^{\text{HF}}_{\bm{q}_2}=\begin{pmatrix}
        \gamma^\dagger_{0,\bm{q}_2}&\gamma^\dagger_{1,\bm{q}_2}&\gamma^\dagger_{2,\bm{q}_2}
    \end{pmatrix}
    \begin{pmatrix}
        0 & g_{12}^* & g_{12} \\
        g_{12} & 0 & g_{12}^* \\
        g_{12}^* & g_{12} & 0
    \end{pmatrix}
    \begin{pmatrix}
        \gamma_{0,\bm{q}_2}\\\gamma_{1,\bm{q}_2}\\\gamma_{2,\bm{q}_2}
    \end{pmatrix},
\end{equation}
where $g_{12}=g_{1,\bm{q}_2}$. This has three eigenvectors $\phi_\alpha=\frac{1}{\sqrt{3}}(1,e^{i\frac{2\pi m}{3}},e^{i\frac{4\pi m}{3}})$ with $C_3$ eigenvalues $\theta_K=e^{i\frac{n\pi}{3}}e^{-i\frac{2\pi m}{3}}$, for $m=0,1,2$. The lowest energy solution corresponds to the value of $m$ satisfying $\text{arg}(g_{12}e^{-i\frac{2(m+1)\pi}{3}})\in[0,\frac{2\pi}{3}]$, which constrains $C=m\mod 3$~\cite{soejima2024AHC2,dong2024stability,crepel2024efficientpredictionsuperlatticeanomalous} and leads to $O_{01,\bm{q}_2}+O_{20,\bm{q}_2}+O_{12,\bm{q}_2}=e^{i\frac{2\pi C}{3}}$. By combining this with the $M_M$-points, we can now further constrain the Chern number $C$ mod $6$ using 
$e^{i \frac{\pi}{3} C} =(-1)^F \eta_{\Gamma} \theta_{K} \xi_{M} $, where $F=n$ is determined via $C_6^6= (-1)^F$~\cite{fang2012bulk}. Later, we will use information away from the high-symmetry points to determine the full Chern number $C$ (see App.~\ref{subsecapp:full_Chern}).

\subsection{General solution for $\mathbf{k}_1$}

We now parameterize the general solution for the sliver region $\bm{k}_1$ (the other slivers are related by $C_6$).
The Hamiltonian for the $\bm{k}_1$ region is
\begin{gather}\label{eq:Hk1_Ham}
    H_{\bm{k}_1}=\sum_{\bm{k}_1}\begin{pmatrix}
        \gamma^\dagger_{0,\bm{k}_1} & \gamma^\dagger_{1,\bm{k}_1}
    \end{pmatrix}[d_{0,\bm{k}_1}+\bm{d}_{\bm{k}_1}\cdot\bm{\sigma}]\begin{pmatrix}
        \gamma_{0,\bm{k}_1} \\ \gamma_{1,\bm{k}_1}
    \end{pmatrix}\\
    d_{0,\bm{k}_1}=\frac{1}{2}\left(E_{\mbf{k_1}} +f_{\mbf{k_1}}+E_{\mbf{k_1}+ \mbf{b_{1}}} +f_{\mbf{k_1+ b_{1}}}\right)\\
    d_{z,\bm{k}_1}=\frac{1}{2}\left(E_{\mbf{k_1}} +f_{\mbf{k_1}}-E_{\mbf{k_1}+ \mbf{b_{1}}} -f_{\mbf{k_1+ b_{1}}}\right)\\
    d_{x,\bm{k}_1}=\text{Re}\,g_{1,\bm{k}_1},\quad d_{y,\bm{k}_1}=\text{Im}\,g_{1,\bm{k}_1},
\end{gather}
where occupation of the lower HF band leads to $O_{01,\mbf{k_1}}=- \frac{1}{2} \frac{g_{1\mbf{k_1}}}{|\bm{d}_{k_1}|}$. For later computation of the total energy, it will be useful to integrate $O_{01,\mbf{k_1}}$ along $k_{1x}$. 
We will invoke the `thin sliver' approximation where $\Lambda$ is small, such that we can neglect the  weak dependence of the hybridization field $g_{1\mbf{k_1}}$ on $k_{1x}$, as justified in App.~\ref{subsecapp:k1_k12_limit} and the numerical results of App.~\ref{secapp:HFwavefunctions}.
Note that $O_{01,\bm{k}_1}$ still depends strongly on $k_{1x}$ owing to the sharp dispersion perpendicular to the BZ boundary. Considering just the component of the interaction-renormalized dispersion perpendicular to the $\bm{k}_1$ sliver, we have $d_{z,\bm{k}_1}=-v'\delta {k_{1x}}$ where $\bm{\delta k}_1=\bm{k}_1+\frac{\bm{b}_1}{2}$, leading to
\begin{equation}\label{eq:O01_k1y}
    O_{01,k_{1y}}\equiv \sum_{\delta k_{1x}}O_{01,\bm{k}_1}\approx-\frac{g_{1,k_{1y}}L_x}{4\pi v'}\ln \left(\frac{2v'\Lambda}{|g_{1, k_{1y}}|}\right).
\end{equation}
The summation above is exact for $\frac{|g_{1,k_{1y}}|}{v'\Lambda}\rightarrow 0$. The parametrization $d_{z,\bm{k}_1}=-v'\delta {k_{1x}}$ means that the flat bottom of the trashcan is effectively treated as hexagonal, which is expected to only quantitatively influence the results compared to using a circular flat bottom for sufficiently strong interactions. Increasing $v'$ reduces the low-energy density of states for hybridization, therefore suppressing the magnitude of $O_{01,k_{1y}}$.

The effective dispersion $v'$ is obtained by considering the band renormalization $f_{\bm{k}}$. Analytical progress can be made using the exponential interaction in the limit of small $\phi$, where we find $v'=\frac{1}{2}(v+2b_1\mathcal{O})$
with $\mathcal{O}=n_e V_0\phi>0$ proportional to the average electronic density $n_e$. At energies above the hybridization scale $|g_{1,\bm{k}_1}|$, the upper and lower eigenvalues of Eq.~\ref{eq:Hk1_Ham} are approximately $ E_{+,\bm{k}_1}\simeq (v+b_{1}\mathcal{O})\delta k_{1x}$ and $E_{-,\bm{k}_1}\simeq -b_{1}\mathcal{O}\delta k_{1x}$, i.e.~the effective velocity is amplified by interactions to  $(v+b_{1}\mathcal{O})$. This enhancement of the dispersion due to the Fock self-energy is reflected in the numerical HF calculations in App.~\ref{secapp:HFbandstructure}.

\subsection{Energy competition including just $\mathbf{k}_{12}$ patches}

We now turn to a computation of the total mean-field energy $E^{\text{HF}}$ in order to study the competition between HF solutions with different $C$. Of particular interest is the dependence of the competition on quantities such as $\varphi_{\text{BZ}}=3\sqrt{3}\beta q_2^2$, which measures the integrated Berry curvature over BZ 0. For this subsection, we consider the limit where we only retain contributions from the $\bm{k}_{12}$ region and its symmetry-related counterparts in both the mean fields and the total energy (see App.~\ref{subsecapp:Etot_k12_only_limit}). Similar approaches have already been pursued in Refs.~\cite{soejima2024AHC2,dong2024stability,crepel2024efficientpredictionsuperlatticeanomalous}. Concentrating only on these patches neglects the gapless regions around the boundary of the BZ (i.e.~the $\mbf{k}_1$ momenta), which need to be included for a proper mean-field treatment of Wigner crystallization, as done in the next subsection.

In the thin sliver approximation where the $\bm{k}_{12}$ patches are each very small with momentum area $A_{\bm{k}_{12}}$, we neglect the variation of quantities within the patches and set $\bm{k}_{12}=\bm{q}_2$. In this limit, the symmetry-breaking order parameter is characterized by $O_{01,\bm{q}_2}+O_{20,\bm{q}_2}+O_{12,\bm{q}_2}=e^{i\frac{2\pi C}{3}}$, which is fully determined by $C\mod 3$, while the hybridization field is captured by $g_{12}$, whose full expression is provided in App.~\ref{subsecapp:Etot_k12_only_limit}. Importantly, $g_{12}$ depends on both $C\mod 3$ and $\varphi_{\text{BZ}}$. We find
\begin{eqnarray}
    E^\text{HF}
    \propto &(e^{-\phi b_1^2}-e^{-\phi q_2^2})\cos\frac{\varphi_\text{BZ}}{3}\cos \frac{2\pi C}{3}\\
    -&(e^{-\phi b_1^2}+e^{-\phi q_2^2})\sin\frac{\varphi_\text{BZ}}{3}\sin \frac{2\pi C}{3},
\end{eqnarray}
where we have used the exponential interaction introduced below Eq.~\ref{eq:trashcan_ham} for concreteness\footnote{This does not lead to any loss of generality, since the results only depend on the interaction potential at momentum transfers $q=q_2$ and $q=b_1$.}. The first (second) term on each line above corresponds to the Hartree (Fock) contribution. In the limit $\phi=0$, we find $E^{\text{HF}}\propto -\sin\frac{\varphi_\text{BZ}}{3}\sin \frac{2\pi C}{3}$ which indicates a ferromagnetic coupling between the enclosed Berry flux and the Chern number of the Wigner crystal. In other words, $\varphi_{\text{BZ}}>0$ leads to $C=1\mod 3$ being the ground state. The $C=-1\mod 3$ solution becomes lower energy when the Berry flux reaches $\varphi_{\text{BZ}}=3\pi$, but our estimates of the Berry trashcan parameters extracted from R$n$G imply that such large Berry fluxes are not possible in our regime of interest.  Note that $\phi=0$ translates to the condition $V_{b_1}=V_{q_2}e^{\beta(b_1^2-q_2^2)}$, which is unphysical because a realistic interaction potential, such as gate-screened Coulomb, is expected to decay with momentum transfer. This therefore amplifies the Hartree energy controlled by $V_{b_1}$, which for small $\varphi_{\text{BZ}}$ penalizes $C=0\mod 3$ relative to the other solutions, and prevents it from ever becoming the ground state. 

 \begin{figure}
 \centering
\includegraphics[width=0.8\columnwidth]{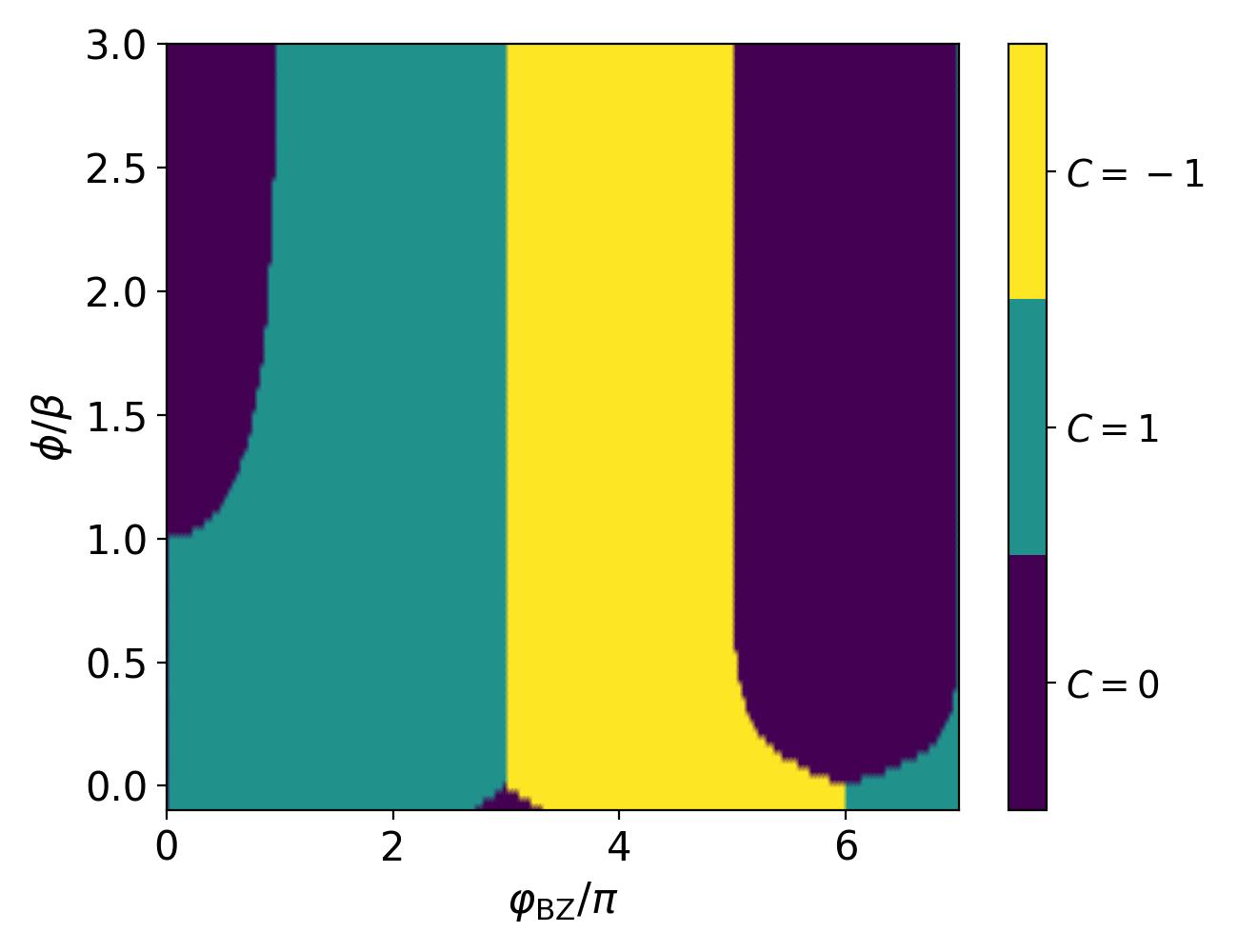} 
\caption{Chern number $C\mod 3$ of the HF ground state when considering just the $\bm{k}_{12}$ patches, as a function of $\phi=\alpha+\beta$  and the Berry flux $\varphi_{\text{BZ}}=3\sqrt{3}\beta q_2^2$ in the BZ. We consider the GMP form factors with Berry curvature $2\beta$ and an exponential interaction with exponent $\alpha$.}
\label{fig:k12only_Eorder}
\end{figure}

In Fig.~\ref{fig:k12only_Eorder}, we show the energy competition as a function of $\varphi_{\text{BZ}}$ and $\phi$. For $\phi>\beta$ (i.e.~$\alpha>0$), the $C=0\mod 3$ state is the lowest energy from $\varphi_{\text{BZ}}=0$ up to some threshold value where a transition to the $C=1\mod 3$ state occurs. When $\phi\geq 0$, the lowest energy solution is also always self-consistent, in the sense that it corresponds to the lowest eigenvector of Eq.~\ref{eq:Hq2}. For $\phi\rightarrow \infty$ where only the Fock term survives, the $C=0\mod 3$ state is the ground state up to $\varphi_{\text{BZ}}=\pi$. Taking $q_2$ corresponding to the moir\'e BZ of $\theta=0.77^\circ$ R$n$G/hBN, the Coulomb interaction with $V_{q_2}/V_{b_1}=\sqrt{3}$ can be matched to $\phi/\beta\approx 1.89$, where the first transition between $C=0\mod3$ and $C=1\mod3$ occurs at $\varphi_\text{BZ}\approx 0.9\pi$.

\subsection{Energy competition including $\mathbf{k}_1$ and $\mathbf{k}_{12}$}

The reintroduction of the $\bm{k}_1$ region to the mean-field analysis is important not only because it refines the classification of the Chern number to $C\mod 6$ due to the $M_M$-points (enabled by the emergent $C_6$ symmetry of our model), but it also accounts for all remaining parts of the BZ where translation symmetry-breaking will be non-negligible owing to degeneracies in the folded band dispersion. In the thin sliver approximation, the problem after using $C_6$ symmetry therefore consists of the small $\bm{k}_{12}$ patch with momentum area $A_{\bm{k}_{12}}$ and the $\bm{k}_1$ sliver with momentum area $A_{\bm{k}_1}\simeq q_2 \Lambda$. 
The $\bm{k}_{12}$ patch region is described by the hybridization $g_{12}=g_{1,k_{1y}=-\frac{q_2}{2}}$ and the order parameter $\mathcal{O}_{012}=\sum_{\bm{k}_{12}}(O_{01,\bm{k}_{12}}+O_{20,\bm{k}_{12}}+O_{12,\bm{k}_{12}})\simeq \sum_{\bm{k}_{12}}(O_{01,\bm{q}_{2}}+O_{20,\bm{q}_{2}}+O_{12,\bm{q}_{2}})$. The $\bm{k}_1$ region is characterized by $k_{1y}$-dependent quantities $g_{1,k_{1y}}$ and $O_{01,k_{1y}}$ (which is determined by $g_{1,k_{1y}}$, see Eq.~\ref{eq:O01_k1y}). 
As we present in detail in App.~\ref{secapp:2dmodel_energy}, the mean-field equations reduce to a self-consistent equation for $g_{1,k_{1y}}$, from which we can obtain the total energy $E^\text{HF}$ and verify the self-consistency between $C\mod 6$ and the hybridization field.
The only possible effect of the $\bm{k}_0$ region, which is fully occupied for an insulating state\footnote{Hence, the precise details of the dispersion in the $\bm{k}_0$ region does not affect the mean-field analysis.}, that we retain is in the $C$-independent renormalized velocity parameter $v'$ that enters Eq.~\ref{eq:O01_k1y}. In App.~\ref{subsecapp:full_Chern}, we further show how the Chern number (without modding by an integer) can be extracted from the properties of the mean-field Hamiltonian around the entire BZ boundary.

We first consider the limit $\phi=0$, where we find that $g_{1,k_{1y}}$ can be expressed as
\begin{eqnarray}
    & g_{1,k_{1y}} = 2\text{Re} G \cos(\frac{\varphi_{\text{BZ}}}{6}\frac{2k_{1y}}{q_2})+2i\text{Im} G \sin(\frac{\varphi_{\text{BZ}}}{6}\frac{2k_{1y}}{q_2})
\end{eqnarray}
for some complex $G$ whose implicit expression is provided in App.~\ref{subsecapp:k1_k12_HF_phi0}. As shown there, we find that for reasonable values of $\varphi_{\text{BZ}}>0$ that are not too large, the ground state has $C=1\mod3$, in agreement with the above analysis with just the $\bm{k}_{12}$ patches. The Chern number can be refined to $C=1\mod 6$ by considering the $M_M$ points, and to $C=1$ by analyzing the full $k_{1y}$-dependence of $g_{1,k_{1y}}$.
In App.~\ref{subsecapp:k1_k12_linearphi_HF}, we find that a finite $\phi>0$ leads to the ground state being $C=0$ for sufficiently small values of $\varphi_\text{BZ}$. Increasing $\phi$ suppresses the Hartree energy cost, which enlarges the $C=0$ phase.

We also repeat the calculation in the absence of Hartree terms (App.~\ref{subsecapp:k1_k12_Fockonly_phi0}). This is equivalent to setting $V_{b_1}=0$, though we maintain the exponential interaction form $V_{\bm{q}}=V_0 e^{-\alpha q^2}$ for $q<b_1$. First considering the $\phi=0$ limit,
we find that the hybridization field winds around the edge of the BZ with constant magnitude
\begin{eqnarray}\label{eq:Fockonly_phi0_g1}
g_{1,k_{1y}}=2Ge^{i\frac{\varphi_{\text{BZ}}}{6}\frac{2k_{1y}}{q_2}}.
\end{eqnarray} 
In terms of dimensionless variables $z=\frac{G}{v\Lambda}$, $\kappa=\frac{A_{\bm{k}_{12}}}{A_{\bm{k}_1}}\cos(\frac{\varphi_{\text{BZ}}}{6}+\frac{2\pi C}{3})$ and $\gamma=\frac{(2\pi)^2v'}{q_2V_0}$, we have 
\begin{eqnarray}
    z=\frac{-\kappa}{W_0\left(|\kappa|e^\gamma\right)},
\end{eqnarray}
where $W_0$ is the principal branch of the Lambert $W$ function, and the total mean-field energy up to a constant is
\begin{eqnarray}
E^\text{HF}\propto -\frac{\kappa^2\gamma}{W_0(|\kappa|e^\gamma)}\left(2+\frac{1}{W_0(|\kappa|e^\gamma)}\right).
\end{eqnarray}
The ground state is always self-consistent, and corresponds to $C=m$ for $\varphi_\text{BZ}\in[2\pi(m-\frac{1}{2}),2\pi(m+\frac{1}{2})]$ with $m$ integer. In other words, the ground state chooses the Chern number whose integrated Berry curvature is closest to $\varphi_\text{BZ}$\footnote{Ref.~\cite{dong2024stability} has previously obtained a similar `Berry curvature rounding' due the Fock term in a treatment that uses a small-$q$ expansion and considers the BZ boundary.}.
For small values of $\varphi_\text{BZ}$, the $C=0$ solution has a larger charge density modulation than that of $C=1$ (App.~\ref{subsecapp:k1_k12_Fockonly_phi0}), and hence will be relatively disfavored if the Hartree penalty is re-introduced.  
In App.~\ref{secapp:fockonly_linearphi}, we show that the positions of the phase boundaries are unchanged\footnote{In fact, for $\varphi_\text{BZ}=m\pi$ and arbitrary interaction potentials $V_{\bm{q}}$ (but still neglecting the Hartree contribution), we find a degeneracy between the solutions for $C=(-m+1)\mod 3$ and $C=(-m+2)\mod 3$.} for finite $\phi>0$ (with the Hartree term still switched off), but the winding of the order parameter is corrected from Eq.~\ref{eq:Fockonly_phi0_g1}. In particular for $0<\varphi_{\text{BZ}}<2\pi$, the phase of $g_{1,k_{1y}}$ rotates less (more) rapidly for $C=0$ ($C=1$).

\section{Discussion}

In this work, we introduced the Berry Trashcan model, and performed an analytical mean-field investigation into the possible Wigner crystalline phases.
Beyond providing insight into the competition between Wigner crystals of different Chern numbers, the results here will be useful for the analytic understanding of other aspects such as the collective excitations~\cite{kwan2023mfci3,zeng2024berryphasedynamicssliding} and stability properties beyond mean-field theory~\cite{guo2024beyondmeanfieldstudieswignercrystal}, in particular those arising from the gapless phonon fluctuations~\cite{collective_unpub}.
The effects of a superlattice potential~\cite{crepel2024efficientpredictionsuperlatticeanomalous}, for example that induced by (near)-alignment to hBN, will also be incorporated into the analysis.

The mean-field treatment above is sensible when the momentum scale $q_2$ of the putative Wigner crystal is close to $k_b$, and interactions are not strong enough to significantly hybridize the $\bm{k}_0$ region around $\Gamma_M$ with the higher folded bands. If $q_2$ is significantly smaller than $k_b$, then the thin sliver approximation invoked in the analytic solutions becomes less justified, and the problem is less suited to a mean-field description. 
On the other hand, full spin-valley polarization, which is expected within the flat-band regime~\cite{antebi2024stoner,bernevig2021TBGIII,lian2021TBGIV,bernevig2021TBGV}, is unlikely for significantly larger $q_2$ where the Fermi level lies in the steep part of the dispersion\footnote{If the system does remain spin-valley polarized though, then the Hartree-Fock approach to Wigner crystallization would be more justified, since beyond mean-field fluctuations involving the flat bottom region would be suppressed by the kinetic energy penalty.}. 
In addition, Eq.~\ref{eq:Mkq_GMP} does not quantitatively describe the form factors of R$n$G for momenta $k$ approaching $t_1/v_F$. 

Regardless of the regime of validity of the present mean-field analysis, the Berry Trashcan model represents a concrete Hamiltonian for any electronic density, and can be readily addressed with various techniques. Crucially, it is a simple idealized Hamiltonian that captures key features of the lowest conduction band of R$n$G for appropriately tuned values of the interlayer potential $V$, including a flat region with a momentum scale $k_b$ beyond which the dispersion rapidly increases with slope $v$, and a GMP density algebra associated with a constant Berry curvature which is accurate for small momenta.

\acknowledgments
We thank Miguel Gonçalves, Jonah Herzog-Arbeitman, Felipe Mendez-Valderrama, Nicolas Regnault, Minxuan Wang and Jiabin Yu for helpful discussions.
B.A.B. was supported by the
Gordon and Betty Moore Foundation through Grant No.~GBMF8685 towards the Princeton theory program, the Gordon and Betty Moore Foundation’s EPiQS Initiative (Grant No.~GBMF11070), the Office of Naval Research (ONR Grant No.~N00014-20-1-2303), the Global Collaborative Network Grant at Princeton University, the Simons Investigator Grant No.~404513, the BSF Israel US foundation No.~2018226, the NSF-MERSEC (Grant No.~MERSEC DMR 2011750), Simons Collaboration on New Frontiers in Superconductivity (SFI-MPS-NFS-00006741-01), and the Schmidt Foundation at the Princeton University, European Research Council (ERC) under the European Union’s Horizon 2020 research and innovation program (Grant Agreement No. 101020833).

\clearpage

\input{main.bbl}

\clearpage

\newpage

\appendix

\onecolumngrid

\tableofcontents

\clearpage
\newpage

For the convenience of the reader, we provide a summary of the contents of the Appendices:
\begin{itemize}
\item In App.~\ref{secapp:Ham_symmetries_RnGhBN}, we introduce the interacting Hamiltonian for R$n$G/hBN. 
\item In App.~\ref{secapp:pristine}, we discuss in more detail the limit of pristine R$n$G. We introduce various approximations, including the chiral/holomorphic wavefunctions, and the GMP limit for the form factors. We also discuss an analytical estimate of the Berry Trashcan model parameters from the R$n$G model. 
\item In App.~\ref{secapp:classical}, we introduce a variant of the Berry Trashcan model which has an enhanced set of symmetries. 
\item In App.~\ref{secapp:HFphasediagrams}, we perform numerical self-consistent Hartree-Fock 
(HF) calculations on various limits of the R5G model. Our calculations consider competition between the $C=0,1$ Wigner crystals and their properties.
\item In App.~\ref{secapp:1dtrashcansetup}, as a warm-up to the Berry Trashcan model, we introduce a toy 1D model, which is treated in App.~\ref{secapp:1dtrashcan_HFanalysis} with analytical HF. 
\item In App.~\ref{secapp:2dmodel_setup}, we introduce the Berry Trashcan model, which is designed to capture the salient features of R$n$G. 
\item In App.~\ref{secapp:2dmodel_HF} and \ref{secapp:2dmodel_energy}, we discuss the Berry Trashcan model in certain regimes with analytical HF.
\end{itemize}

\section{Hamiltonian and symmetries for R$n$G/hBN superlattice}\label{secapp:Ham_symmetries_RnGhBN}

In this appendix section, we review the Hamiltonian and symmetries of the continuum model for R$n$G/hBN. We focus on the properties for a single spin and valley (in particular valley $K$). The presentation of the model follows Ref.~\cite{herzog2024MFCI2}. 

\subsection{Single-particle continuum model for R$n$G/hBN}

At low energies near charge neutrality, the low-energy band structure of R$n$G is localized near the two valleys $\eta K$ of graphene, where $\eta=\pm 1$ is a valley index. Expanding in momenta $\bm{p}$ about the Dirac momentum $\bm{K}_G=(\frac{4\pi}{3a_G})$, where $a_G=2.46\,$\r{A} is the graphene lattice constant, the matrix Hamiltonian for R$n$G in valley $K$ reads~\cite{moon2014electronic,park2023topologicalflat,jung2014abinitio,jung2014accurate,herzog2024MFCI2}
\begin{equation}\label{eq:H_K}
    H_{\K}(\bsl{p}) = \bpm
v_F\mbf{p} \cdot \pmb{\sigma}  & t^\dag(\mbf{p}) & t'^\dagger &   &\\
t(\mbf{p}) & \ddots & \ddots & t'^\dagger \\
t' & \ddots & v_F\mbf{p} \cdot \pmb{\sigma} & t^\dagger(\mbf{p})\\
& t' & t(\mbf{p})  & v_F\mbf{p} \cdot \pmb{\sigma}
\epm + H_{ISP}
\end{equation}
where $\bsl{p}=-\ii \nabla$, $v_F$ is the graphene Fermi velocity, and $\bsl{\sigma}=(\sigma_x,\sigma_y)$ are Pauli matrices in sublattice subspace. Note that $H_K(\bm{p})$ is a $2n\times 2n$ matrix in layer $(l=0,\dots,n-1)$ and sublattice $(\sigma=A,B)$ space, and is ordered according to $(0,A),(0,B),(1,A),\ldots,(n-1,B)$, where $(l,\sigma)$ indexes the layer $l$ and sublattice $\sigma $ degree of freedom. $t(\mathbf{p})$ and $t'$ are sublattice matrices describing the interlayer tunneling processes
\begin{equation}
    t(\mbf{p}) = -\bpm v_4 p_+ & -t_1 \\ v_3 p_- &  v_4 p_+ \epm, \qquad  \qquad t' = \bpm 0 & 0 \\ t_2 & 0 \epm
\end{equation}
where $p_\pm = p_x \pm \ii p_y$,  $v_F$ is the Fermi velocity, $t_1,v_3,v_4$ are parameters describing hopping between consecutive layers, and $t_2$ describes hopping between next-nearest layers. Note that we take $v_3=v_4$ in this work. $H_{ISP}$ is the inversion symmetric polarization
\begin{equation}
    [H_{ISP}]_{l\sigma,l'\sigma'}=V_{ISP}\left|l-\frac{n-1}{2}\right|\delta_{l,l'}\delta_{\sigma,\sigma'}.
\end{equation}
We use the parameter values $v_F=542.1\,\text{meVnm}, t_1=355.16\,\text{meV},t_2=-7\,\text{meV},v_3=v_4=34\,\text{meVnm},V_{ISP}=16.65\,\text{meV}$ from Ref.~\cite{herzog2024MFCI2}.

We also add a single-particle term $H_D$ that models an externally applied displacement field, and is implemented as a linearly varying layer potential of amplitude $V$
\begin{equation}\label{eqapp:displacement}
    [H_{D}]_{l\sigma,l'\sigma'}=V\left(l-\frac{n-1}{2}\right)\delta_{l,l'}\delta_{\sigma,\sigma'}.
\end{equation}

In this work, we will primarily focus on pristine R$n$G, but we describe the Hamiltonian of R$n$G/hBN for completeness. We continue to follow the presentation of Ref.~\cite{herzog2024MFCI2} --- other treatments of the moir\'e coupling can be found in Ref.~\cite{moon2014electronic,jung2014abinitio,zhang2019nearly,park2023topologicalflat}. Consider the case where R$n$G is aligned with twist angle $\theta$ to the hBN substrate adjacent to layer $l=0$. The combination of twist angle and lattice mismatch leads to a moir\'e pattern which can be characterized by the difference between the corresponding valley $K$ Dirac momenta
\begin{equation}
    \bm{q}_1=\bm{K}_G-\bm{K}_{hBN}=\frac{4\pi}{3a_G}\left(1-\frac{1}{1+\epsilon_\text{lat}}R(-\theta)\right)\hat{x},
\end{equation}
where $R(\theta)$ is a counter-clockwise rotation by $\theta$, and $\epsilon_{\text{lat}}=(a_\text{hBN}-a_\text{G})/a_{\text{G}}\simeq 0.017$ parameterizes the lattice mismatch. We also define $\bm{q}_{j+1}=R\left(\frac{2\pi}{3}\right)\bm{q}_j$. The shortest moir\'e reciprocal lattice vectors (RLVs) are
\begin{equation}
    \bm{b}_1=\bm{q}_2-\bm{q}_3,\quad \bm{b}_2=\bm{q}_3-\bm{q}_1,\quad \bm{b}_3=\bm{q}_1-\bm{q}_2.
\end{equation}
The effect of the aligned hBN can captured by integrating out the hBN degrees of freedom, leading to an effective moir\'e potential (with $2\times2$ sublattice structure) acting only on the bottom graphene layer $l=0$
\begin{equation}
    V_\xi(\mbf{r}) = V_0 + \left[V_1 e^{i\psi_\xi}\sum_{j=1}^3 e^{i \mbf{b}_j\cdot\mbf{r}}\bpm 1& \omega^{-j} \\ \omega^{j+1} &\omega \epm + h.c.\right]
\end{equation}
where $\omega=\exp\left(\frac{2\pi i}{3}\right)$, and $V_0,V_1,\psi_\xi$ are moir\'e coupling parameters whose values can be found in Ref.~\cite{herzog2024MFCI2}. Note that $\xi=0,1$ represents two inequivalent orientations of hBN related by $180^\circ$ rotation. The moir\'e part of the Hamiltonian is then 
\begin{equation}
    [H_{\text{moir\'e},\xi}(\bsl{r})]_{l \sigma,l' \sigma'} = \left[ V_\xi(\mbf{r}) \right]_{\sigma\sigma} \delta_{l0}\delta_{ll'}.
\end{equation}
The total non-interacting Hamiltonian for valley $K$ in real space $\bm{r}$ reads
\begin{equation}
    H_{\K, \xi}(\mbf{r}) = H_{\K}(- i \pmb{\nabla})+H_D + H_{\text{moir\'e},\xi}(\bsl{r}).
\end{equation}

It will be convenient to also express the Hamiltonian in momentum space. To this end, we introduce the real-space continuum creation operator $c^\dagger_{\bm{r},l\sigma}$, where the valley and spin indices are suppressed (unless otherwise stated, we consider valley $K$ and spin $\uparrow$). The creation operator in momentum space is  
\begin{equation}
c^\dagger_{\bsl{k},\bsl{G},l\sigma} = \frac{1}{\Omega_\text{tot}} \int d^2 r e^{\ii (\bsl{k}+\bsl{G})\cdot\bsl{r}} 
c^\dagger_{\bsl{r},l\sigma},
\end{equation}
where $\Omega_\text{tot}$ is the area of the whole system, $\bsl{k}$ is in the first moir\'e Brillouin zone (BZ), and $\bsl{G}$ is a RLV.

In momentum space, the total matrix Hamiltonian is
\begin{equation}
\left[ \tilde{H}_{\K,\xi}(\bsl{k}) \right]_{\bsl{G}\bsl{G}'} = H_{\K}(\bsl{k}+\bsl{G})\delta_{\bsl{G}\bsl{G}'} + H_D\delta_{\bm{G}\bm{G}'}+\left[ H_{\text{moir\'e},\xi} \right]_{\bsl{G}\bsl{G}'},
\end{equation}
where
\begin{equation}
\left[H_{\text{moir\'e},\xi} \right]_{\bsl{G}l\sigma,\bsl{G}'l'\sigma'} = \left[  V_0 \delta_{\bsl{G}\bsl{G}'} + V_1 e^{\ii\psi_\xi}\sum_{j=1}^3 \delta_{\bsl{G},\bsl{G}'+\bsl{b}_j} \bpm 1& \omega^{-j} \\ \omega^{j+1} &\omega \epm + V_1 e^{-\ii\psi_\xi}\sum_{j=1}^3 \delta_{\bsl{G},\bsl{G}'-\bsl{b}_j} \bpm 1& \omega^{-j-1} \\ \omega^{j} &\omega^* \epm  \right]_{\sigma\sigma'} \delta_{l0}\delta_{ll'} .
\end{equation}

The full single-particle model for R$n$G/hBN can be diagonalized to obtain Bloch eigenvectors $U^\eta_{\bm{G}\alpha,m}(\bm{k})$ and eigenvalues $E_n^\eta(\bm{k})$ satisfying
\begin{equation}\label{eq:Hsp_eigen}
     \sum_{\mbf{G}\alpha, \mbf{G}'\beta} U^{\eta *}_{\mbf{G}\alpha, m} (\mbf{k}) \left[\tilde{H}_{K,\xi}(\bm{k})\right]_{\mbf{G}\alpha, \mbf{G}'\beta}(\mbf{k}) U^\eta_{\mbf{G}'\beta,n}(\mbf{k}) = \delta_{mn} E_n^\eta(\mbf{k}),
\end{equation}
where $m,n$ are band labels, and $\alpha,\beta$ are composite indices for sublattice and layer. The Bloch eigenvectors are chosen to obey the embedding relation $U_{\mbf{G}-\mbf{g}_i,\alpha,n}(\mbf{k}+\mbf{g}_i)=U_{\mbf{G},\alpha,n}(\mbf{k})$.

\subsection{Symmetries of R$n$G/hBN model}\label{subsecapp:symmetryRnG/hBN}

The continuum Hamiltonian $H_{\K,\xi}$ obeys a set of single-valley symmetries. For general parameters, we have moir\'e translation $T_{\bm{R}_0}$ and $C_3$ rotation, which transform the creation operators as
\begin{gather}
    C_3 c^\dagger_{\bsl{r},l\sigma  } C_3^{-1} = \sum_{\sigma'}c^\dagger_{C_3\bsl{r},l\sigma} e^{i\frac{2\pi}{3}(l-\lfloor\frac{n}{2}\rfloor)} \left[ e^{-i\frac{\pi}{3}\sigma_3} \right]_{\sigma'\sigma}\\
    T_{\bsl{R}_0} c^\dagger_{\bsl{r},l\sigma } T_{\bsl{R}_0}^{-1} =  c^\dagger_{\bsl{r}+\bsl{R}_0,l\sigma },
\end{gather}
where $\bm{R}_0$ is a moir\'e lattice vector. 

If the coupling to the hBN is switched off ($V_0=V_1=0$), the pristine R$n$G model in a single valley additionally has continuous translation symmetry $T_{\bm{r}_0}$, where $\bm{r}_0$ is any vector, and $(M_1\mathcal{T})$, which is an antiunitary symmetry that is combination of a mirror that flips $x$ and time-reversal
\begin{gather}
    T_{\bsl{r}_0} c^\dagger_{\bsl{r},l\sigma } T_{\bsl{r}_0}^{-1} =  c^\dagger_{\bsl{r}+\bsl{r}_0,l\sigma }\\
    (M_1\mathcal{T})c^\dagger_{\bsl{r},l\sigma }(M_1\mathcal{T})^{-1}=c^\dagger_{M_1\bsl{r},l\sigma }. 
\end{gather}
Note that $(M_1\mathcal{T})$ takes $(p_x,p_y)$ to $(p_x,-p_y)$, where $\bm{p}$ is measured from the Dirac momentum. The presence of $(M_1\mathcal{T})$ in the R$n$G model follows from the fact that $H_K(\bm{p})$ (Eq.~\ref{eq:H_K}) only depends on real combinations of $p_x$ and $ip_y$.

\subsection{Interactions}

For the interacting term $\hat{H}_\text{int}$, we consider a density-density interaction with interaction potential $V(\bm{q})$, which will be specified where necessary. In terms of the single-particle eigenstates of the non-interacting Hamiltonian (see Eq.~\ref{eq:Hsp_eigen}), we define the density operator 
\begin{equation}
\rho_{\mbf{q}+\mbf{G}} = \sum_{\mbf{k} mn \eta s} M_{mn}^{\eta}(\mbf{k},\mbf{q}+\mbf{G}) c^\dag_{\eta,\mbf{k}+\mbf{q}, m,s} c_{\eta,\mbf{k}, n,s}
\end{equation}
where $c^\dag_{\eta,\mbf{k}, m,s}$ is a creation operator for momentum $\bm{k}$, band $m$, valley $\eta$, and spin $s$, and we have introduced the form factor
\begin{equation}
    M_{mn}^\eta(\mbf{k},\mbf{q}+\mbf{G}) = \sum_{\mbf{G}'\alpha}U^{\eta *}_{\mbf{G}+\mbf{G}',\alpha,m}(\mbf{k}+\mbf{q}) U^\eta_{\mbf{G}',\alpha,n}(\mbf{k}).
\end{equation}
In the so-called CN interaction scheme~\cite{kwan2023mfci3}, we consider a density-density interaction normal-ordered to the state corresponding to fully occupied valence bands of the single-particle Hamiltonian
\begin{equation}
    \hat{H}_{\text{int,CN}}=\frac{1}{2\Omega_{\text{tot}}}\sum_{\mbf{q},\mbf{G}} V(\mbf{q}+\mbf{G}) : \rho_{\mbf{q}+\mbf{G}}\rho_{-\mbf{q}-\mbf{G}}:,
\end{equation}
where the normal-ordering symbol $:\hat{O}:$ places all annihilation (creation) operators on the right for conduction (valence) electrons in $\hat{O}$, keeping track of minus signs. Other choices of interaction schemes, such as the average (AVE) scheme, differ from the above by an effective one-body term~\cite{kwan2023mfci3}. For certain parameter regimes, the different schemes incorporate extrinsic moir\'e effects in qualitatively distinct ways~\cite{kwan2023mfci3,yu2024MFCI4}. As we will primarily focus on the moir\'e-less limit ($V_0=V_1=0$) in this work, the choice of interaction scheme is not critical, and we will use the CN scheme for simplicity. 

\clearpage

\section{Pristine R$n$G model and chiral wavefunctions}\label{secapp:pristine}

In this appendix section, we consider the pristine R$n$G model (i.e.~no coupling to the hBN so that $V_0=V_1=0$) in more detail, and introduce its description in terms of the chiral/holomorphic wavefunctions. We also discuss the interacting Hamiltonian.

\subsection{Holomorphic wavefunctions}\label{subsecapp:holomorphicwfns}
The essential physics of the R$n$G model (Eq.~\ref{eq:H_K}) can be extracted by first considering the limit where $v_3=v_4=t_2=V_{ISP}=0$ and vanishing displacement field $V=0$, leading to the so-called chiral Hamiltonian
\begin{equation}\label{eqapp:chiralham}
    [h_n(\bm{k})]_{ll'}=v_F\delta_{ll'}\bm{k}\cdot\bm{\sigma}+t_1\delta_{l,l'+1}\sigma^++t_1\delta_{l,l'-1}\sigma^-
\end{equation}
where $\sigma^\pm=\frac{1}{2}({\sigma_x\pm i\sigma_y})$. 

$h_n(\bm{k})$ has additional symmetries, namely chiral symmetry $\Sigma h_n(\bm{k})\Sigma^\dagger=-h_n(\bm{k})$ with $\Sigma=\sigma_z$ acting as the identity in layer space, spacetime inversion $D_{ll'}[\mathcal{IT}]=\delta_{l,n-1-l'}\sigma_x\mathcal{K}$, and $SO(2)$ rotation~\cite{herzog2024MFCI2}. The latter takes the form
\begin{equation}
    h_n(R_\theta\bm{k})=D_\theta h_n(\bm{k})D^\dagger_\theta,\quad\quad [D_\theta]_{ll'}=e^{i\theta(l-\lfloor{\frac{n}{2}}\rfloor)}e^{-i\theta\sigma_z/2}.
\end{equation}
Note that $C_3$ is represented as $D[C_3]=-D_{\frac{2\pi}{3}}$, where the minus sign arises from imposing $C_3^3=1$.

By considering the characteristic polynomial for small $v_Fk/t_1$, one can show that the eigenvalues are approximately
\begin{equation}\label{eqapp:Ekchiral}
    E(\bm{k})=\pm (v_Fk)^{n}/t_1^{n-1},\quad \pm t_1+\ldots
\end{equation}
where there are $2(n-1)$ states with $|E|\sim t_1$ corresponding to the dimerized interlayer states. At low energies, the physics is dominated by the degree-$n$ Dirac node at the Dirac momentum which primarily consists of $(0,A)$ and $(n-1,B)$ orbitals. 

We now define the (anti-)holomorphic normalized states~\cite{herzog2024MFCI2}
\begin{gather}\label{eqapp:chiralstates}
    [\psi_A(\bm{k})]_{l\sigma}=\frac{(-v_Fk_+/t_1)^l}{N(\bm{k})}\delta_{\sigma,A}\\\label{eqapp:chiralstates_B}
    [\psi_B(\bm{k})]_{l\sigma}=\frac{(-v_Fk_-/t_1)^{n-l-1}}{N(\bm{k})}\delta_{\sigma,B}\\
    N(\bm{k})=\sqrt{\frac{1-(v_Fk/t_1)^{2n}}{1-(v_Fk/t_1)^2}}
\end{gather}
where $k=|\bm{k}|$. Only the norm $ N(\bm{k})=  N(k)$ has both holomorphic and anti-holomorphic parts of $\bm{k}$ --- the rest of the wavefunction for the $A$ ($B$) basis is holomorphic (anti-holomorphic). These are also called chiral states, since they diagonalize the chiral operator $\Sigma$. For $v_Fk/t_1<1$, $\psi_A(\bm{k})$, which is localized on sublattice $A$, has predominant weight on the bottom $l=0$ layer, and exponentially decays into the higher layers, while the opposite occurs for $\psi_B(\bm{k})$. It can be shown that low-energy states of Eq.~\ref{eqapp:Ekchiral} are built out of the chiral states. The chiral Hamiltonian (Eq.~\ref{eqapp:chiralham}) projected onto the chiral states $\Psi(\bm{k})=[\psi_A(\bm{k}),\psi_B(\bm{k})]$ is
\begin{equation}
\Psi^\dagger(\bm{k})h_n(\bm{k})\Psi(\bm{k})=\frac{-t_1}{N(\bm{k})^2}\begin{pmatrix}
        0 & (-v_Fk_-/t_1)^n \\
        (-v_Fk_+/t_1)^n & 0
    \end{pmatrix}.
\end{equation}
We can also project the displacement field (Eq.~\ref{eqapp:displacement}) onto the chiral basis, yielding
\begin{gather}
\Psi^\dagger(\bm{k})H_D(\bm{k})\Psi(\bm{k})=
\begin{pmatrix}
    V(k)&0\\
    0&-V(k)
\end{pmatrix}\\
V(k)=V\left[-\frac{n-1}{2}+\frac{(n-1)\left(\frac{v_Fk}{t_1}\right)^{2n+2}+\left(\frac{v_Fk}{t_1}\right)^2-n\left(\frac{v_Fk}{t_1}\right)^{2n}}{\left(1-\left(\frac{v_Fk}{t_1}\right)^2\right)\left(1-\left(\frac{v_Fk}{t_1}\right)^{2n}\right)}\right].\label{eqapp:Vk}
\end{gather}
For positive $V$ and small $v_Fk/t_1$, $V(k)$ is negative such that the lowest conduction band near $k=0$ is primarily built out of $\psi_B(\bm{k})$, which is mostly on layer $l=n-1$. For large positive $V$, we can estimate the correction to the dispersion from the $v_3,v_4$ terms in Eq.~\ref{eq:H_K} by evaluating its expectation value with $\psi_B(\bm{k})$. For the $n=5$ pentalayer case of most interest here, we have
\begin{equation}\label{eqapp:v3v4_dispersion}
    \bra{\psi_B(\bm{k})}  H^{n=5}_{v_3,v_4}\ket{\psi_B(\bm{k})}=\frac{2k^2v_3v_F(t_1^2+v_F^2k^2)(t_1^4+v_F^4k^4)}{N(\bm{k})^2t_1^{7}}.
\end{equation}
The first two panels of Fig.~\ref{fig:N18_t0.77_Hrad1.701_U0.024_pstrnone+0+0+none+none_epsr5.00_singleplotC1} show a representative band structure for the full R5G model, while the first two panels of Fig.~\ref{fig:N18_t0.77_Hrad1.701_U0.024_pstrt1_v0_only+1+1+exp+none_epsr5.00_singleplotC1} show the corresponding dispersion for the chiral Hamiltonian with the above $v_3,v_4$ correction.

\subsection{Form factor approximations}\label{subsecapp:form_factor_approximations}

In this appendix subsection, we restrict to a finite displacement field $V>0$ such that the lowest conduction band wavefunction at low energies is well-approximated by $\psi_B(\bm{k})$. The exact overlap of the chiral basis wavefunction is 
\begin{equation}
    \braket{\psi_B(\bm{k_1})}{\psi_B(\bm{k_2})}=\frac{1}{N(\bm{k_1})N(\bm{k_2})}\frac{1-\left(\frac{v_F^2k_{1+}k_{2-}}{t_1^2}\right)^n}{1-\frac{v_F^2k_{1+}k_{2-}}{t_1^2}}.
\end{equation}
For moderate $n$ (such as $n=5$ for pentalayer) and small $v_Fk/t_1$, we can neglect terms that go as $(v_Fk/t_1)^{2n}$. We Taylor expand in $({v_F^2k_{1+}k_{2-}}/{t_1^2})$ and re-exponentiate the result, leading to
\begin{equation}
    \braket{\psi_B(\bm{k_1})}{\psi_B(\bm{k_2})}\approx \frac{\sqrt{\left(1-\left(\frac{v_Fk_{1}}{t_1}\right)^2\right)\left(1-\left(\frac{v_Fk_{2}}{t_1}\right)^2\right)}}{1-\left(\frac{v_F^2k_{1+}k_{2-}}{t_1^2}\right)}\simeq e^{-\frac{v_F^2}{2t_1^2}(k_1^2+k_2^2-2k_{1+}k_{2-})}.
\end{equation}
The resulting form factor $M_{\bm{k},\bm{q}}$ becomes
\begin{equation}\label{eqapp:Mkq_exp}
    M_{\bm{k},\bm{q}}=\braket{\psi_B(\bm{k}+\bm{q})}{\psi_B(\bm{k})}\approx e^{-\frac{v_F^2}{2t_1^2}(q^2+2i\bm{q}\times\bm{k})}.
\end{equation}
We call Eq.~\ref{eqapp:Mkq_exp} the exponential form factor approximation. Note that this has exactly the same form as that of the lowest Landau level, where the density operator obeys the Girvin-MacDonald-Platzman (GMP) algebra~\cite{girvin1986magnetoroton}. Hence, we will also refer to Eq.~\ref{eqapp:Mkq_exp} as the GMP limit.

We now discuss the Berry curvature of the basis $\psi_B(\bm{k})$ in Eq.~\ref{eqapp:chiralstates_B} (the properties of $\psi_A(\bm{k})$ are closely related, e.g.~its Berry curvature has the opposite sign). We first compute the Berry connection $A_\mu(\bm{k})=i\langle\psi(\bm{k}) |\partial_\mu|\psi(\bm{k})\rangle$, where we have dropped to chiral basis label $B$ for simplicity. To this end, we decompose the chiral Bloch functions as
\begin{equation}
    \psi(\bm{k})=\frac{\phi(k_-)}{N(k)},\quad \phi_l(k_-)=(-v_Fk_-/t_1)^{n-l-1},\quad N(k)=\sqrt{\frac{1-(v_Fk/t_1)^{2n}}{1-(v_Fk/t_1)^2}}=\sqrt{\sum_l |\phi_l(k_-)|^2}
\end{equation}
where we have omitted the sublattice index $\sigma$ since the wavefunctions are fully localized on sublattice $B$. The anti-holomorphic part $\phi(k_-)$ satisfies
\begin{equation}
    A^\phi_-=i\langle \phi|\partial_-|\phi\rangle=i\partial_- [N(k)^2],
\end{equation}
leading to 
\begin{gather}
    A_+=i\langle \psi|\partial_+|\psi\rangle=iN(k)\partial_+[1/N(k)]=-i\partial_+\ln N(k)\\
    A_-=i\langle \psi|\partial_-|\psi\rangle=-i\partial_-\ln N(k)+\frac{1}{N(k)^2}A_-^\phi=i\partial_-\ln N(k).
\end{gather}
Using $\partial_{x} = \partial_{+} + \partial_{-}$ and $  \partial_{y} =i(  \partial_{+} - \partial_{-})$ yields
\begin{equation}
    A_x=-\partial_y \ln N(k),\quad A_y=\partial_x\ln N(k)
\end{equation}
leading to the Berry curvature
\begin{equation}
    \Omega(\bm{k})=\partial_x A_y-\partial_y A_x=(\partial_x^2+\partial_y^2)\ln N(k)=\frac{1}{k}\frac{\partial}{\partial k}\left({k}\frac{\partial}{\partial k}\ln N(k)\right).
\end{equation}
Note that for moderate/large $n$ and small $v_Fk/t_1$, we can neglect the term $(v_Fk/t_1)^{2n}$ in the normalization $N(k)$. If we assume that $v_Fk/t_1$ is small, then we may approximate $\ln N(k)\approx \frac{1}{2}(v_F k/t_1)^2$. In this limit, we find a uniform Berry curvature
\begin{equation}
    \Omega(\bm{k})\approx \frac{2v_F^2}{t_1^2}
\end{equation}
which is equal to that of the GMP limit in Eq.~\ref{eqapp:Mkq_exp}.

\begin{figure}
    \centering
    \includegraphics[width = 0.8\linewidth]{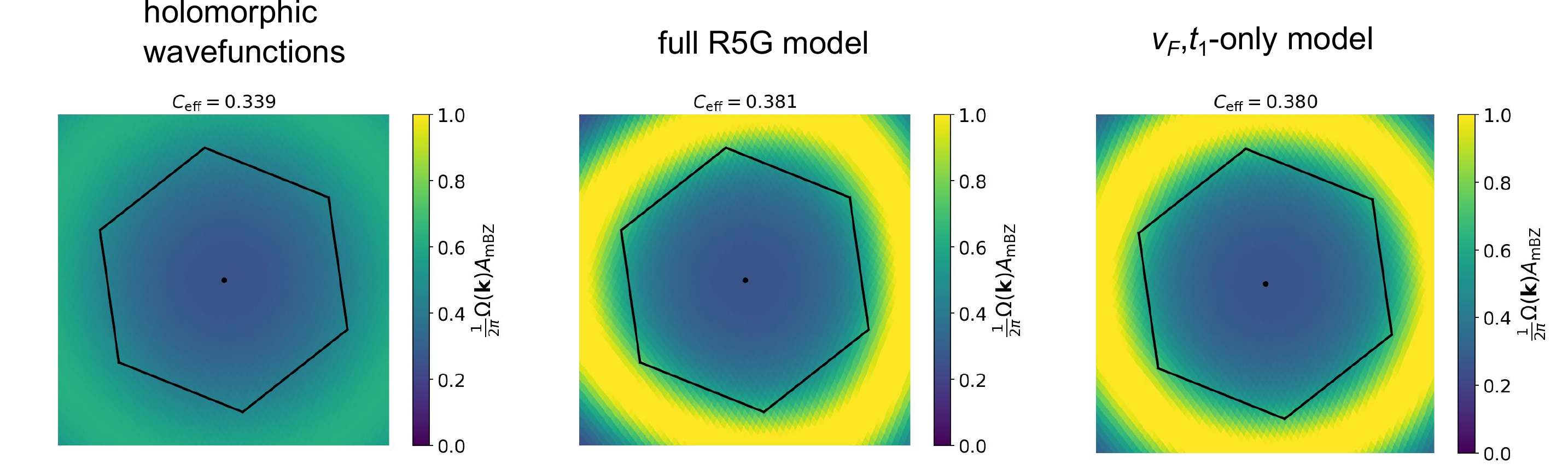}
    \caption{Berry curvature distribution of the lowest conduction band of R5G in valley $K$ with $V=30\,\text{meV}$ and using different approximations. The hexagon indicates the first BZ at $\theta=0.77^\circ$ centered at the graphene Dirac point $\bm{k}=0$ (black dot). We also indicate the integrated Berry curvature $C_\text{eff}$ over the BZ. The color scale is clamped at 1.  Left: Berry curvature computed using the holomorphic wavefunctions of App.~\ref{subsecapp:holomorphicwfns}. The Berry curvature at $\bm{k}=0$ is equal to the uniform Berry curvature of the GMP limit where $\frac{1}{2\pi}\Omega(\bm{k})A_{\text{BZ}}\simeq 0.26$. Middle: Berry curvature computed using the Bloch functions of the full R5G model. Right: Berry curvature computed using the Bloch functions computed in the case where only the $v_F$ and $t_1$ of the R5G model are kept. }
    \label{figapp:RLG_Berry_curvature}
\end{figure}

In Fig.~\ref{figapp:RLG_Berry_curvature}, we show the Berry curvature $\Omega(\bm{k})$ of the lowest conduction band of R5G in the holomorphic limit, of the full R5G model, and of the R5G model with just $v_F,t_1$ graphene parameters ($V$ is still included). We set the interlayer potential $V=30\,\text{meV}$.  Note that $\Omega(\bm{k})$ computed with the holomorphic wavefunctions is independent of $V$. We observe from Fig.~\ref{figapp:RLG_Berry_curvature} that the Berry curvature distribution within the first BZ appropriate for $\theta=0.77^\circ$ is roughly equal across the three calculations, though the holomorphic wavefunctions do not capture the large peak in Berry curvature in a ring that lies outside the first BZ. Note that the value of the uniform Berry curvature in the GMP limit is equal to that at $\bm{k}=0$ computed with the holomorphic wavefunctions. $\Omega(\bm{k})$ for the full R5G model as well as the $v_F,t_1$-only model is peaked somewhat outside the first BZ. The main difference with keeping just $v_F$ and $t_1$ coefficients in the R$5$G parameters is that the trigonal warping is removed, though the integrated Berry curvature within the first BZ is roughly unchanged. 
In all cases in Fig.~\ref{figapp:RLG_Berry_curvature}, the effective Chern number $C_\text{eff}$, i.e.~the integrated Berry curvature over the BZ, at $\theta=0.77^\circ$ is in the range $C_\text{eff}\simeq 0.3-0.4$. In the GMP limit, we find $C_\text{eff}\simeq 0.26$, which underestimates the realistic value in R$n$G somewhat.

\subsection{Interacting Hamiltonian}

We consider the interacting Hamiltonian projected to the first conduction band of R$n$G restricted to a single spin and valley $K$
\begin{equation}
    \gamma_{\mbf{k}}^\dagger = \sum_{l\sigma} [\psi(\bm{k})]_{l\sigma} c_{\mbf{k},l\sigma}^\dagger
\end{equation}
\begin{align}\label{eqapp:Hintproject}
    H=& \sum_k E(\mbf{k}) \gamma_{\mbf{k}}^\dagger \gamma_{\mbf{k}} + \frac{1}{2\Omega_{tot}}  \sum_{\bm{q}, \bm{k}, \bm{k'}} V_{\mbf{q}} M_{\mbf{k},\mbf{q}} M_{\mbf{k}',-\mbf{q}} \gamma_{\mbf{k}+ \mbf{q}}^\dagger \gamma_{\mbf{k}'- \mbf{q}}^\dagger \gamma_{\mbf{k}'} \gamma_{\mbf{k}}\\
    =&\sum_k E(\mbf{k}) \gamma_{\mbf{k}}^\dagger \gamma_{\mbf{k}} + \frac{1}{2\Omega_{tot}}  \sum_{\bm{q}}V_{\mbf{q}}:\tilde{\rho}_{\mbf{q}}\tilde{\rho}_{-\mbf{q}}:
    \\=& \sum_k \left(E(\mbf{k}) -\frac{1}{2\Omega_{tot}}\sum_{\bm{q}\text{ s.t. }\bm{k}+\bm{q}\in \mathscr{H}}V_{\bm{q}}|M_{\bm{k},\bm{q}}|^2\right)\gamma_{\mbf{k}}^\dagger \gamma_{\mbf{k}} + \frac{1}{2\Omega_{tot}}  \sum_{\bm{q}, \bm{k}, \bm{k'}} V_{\mbf{q}} M_{\mbf{k},\mbf{q}} M_{\mbf{k}',-\mbf{q}} \gamma_{\mbf{k}+ \mbf{q}}^\dagger \gamma_{\mbf{k}} \gamma_{\mbf{k}'- \mbf{q}}^\dagger \gamma_{\mbf{k}'}\label{eqapp:Hintprojectrewrite} 
\end{align}
where $\gamma^\dagger_{\bm{k}}$ is the creation operator for the first conduction band (for a single spin in valley $K$), $\psi(\bm{k})$ and $E(\bm{k})$ are the corresponding Bloch eigenvector and energy, $\Omega_{tot}$ is the system area, $V_{\bm{q}}$ is the interaction potential, and $\tilde{\rho}_{\bm{q}}=\sum_{\bm{k}}M_{\bm{k},\bm{q}}\gamma^\dagger_{\bm{k}+\bm{q}}\gamma_{\bm{k}}$ is the density operator projected to the first conduction band. The interacting part above uses the charge neutrality (CN) interaction scheme, where the interaction is normal-ordered relative to the state corresponding to filled valence bands. 

In the last line, the four-fermion term has been reshuffled so that the combination $\tilde{\rho}_{\bm{q}}\tilde{\rho}_{-\bm{q}}$ appears explicitly, at the cost of altering the two-fermion term. This extra piece at $\bm{k}$ corresponds to one-half of the Fock self-energy contribution arising from an occupied state at $\bm{k}+\bm{q}$, summed over all $\bm{q}$. The summation is thus constrained to $\bm{k}+\bm{q}$ belonging to the allowed set of single-particle states $\mathscr{H}$ (no restriction is explicitly indicated in the summation of the four-fermion term, since the Bloch operators are defined to vanish if the momentum argument lies outside of $\mathscr{H})$. Since having the interaction in the form $\tilde{\rho}_{\bm{q}}\tilde{\rho}_{-\bm{q}}$ is often useful for analytic results, we sometimes consider the so-called \emph{`density-density' approximation} where the extra contribution to the two-fermion term is neglected
\begin{equation}\label{eqapp:Hdensityapprox}
    H'=\sum_k E(\mbf{k}) \gamma_{\mbf{k}}^\dagger \gamma_{\mbf{k}} + \frac{1}{2\Omega_{tot}}  \sum_{\bm{q}, \bm{k}, \bm{k'}} V_{\mbf{q}} M_{\mbf{k},\mbf{q}} M_{\mbf{k}',-\mbf{q}} \gamma_{\mbf{k}+ \mbf{q}}^\dagger  \gamma_{\mbf{k}}\gamma_{\mbf{k}'- \mbf{q}}^\dagger \gamma_{\mbf{k}'}.
\end{equation}

Finally, we explicitly write the form of the normal-ordered four-fermion interaction term using the GMP form factors of Eq.~\ref{eqapp:Mkq_exp}
\begin{eqnarray}\label{eqapp:Hint_GMP}
H_\text{int}=\frac{1}{2\Omega_{tot}}  \sum_{\bm{q}, \bm{k}, \bm{k'}} V_{\mbf{q}} e^{-\frac{v_F^2}{t_1^2}q^2} e^{-i\frac{v_F^2}{t_1^2}\bm{q}\times (\bm{k}- \bm{k'})} \gamma_{\mbf{k}+ \mbf{q}}^\dagger \gamma_{\mbf{k}'- \mbf{q}}^\dagger \gamma_{\mbf{k}'} \gamma_{\mbf{k}}.
\end{eqnarray}

\subsection{Berry Trashcan parameterization}\label{subsecapp:trashcan_parameterization}

In this appendix subsection, we discuss how the parameters of the Berry Trashcan model introduced in App.~\ref{secapp:2dmodel_setup} can be extracted from the R$n$G Hamiltonian. Our goal is to obtain the radius $k_b$ of the flat bottom, as well as the velocity $v$ of the dispersive region. 

We consider the $V=0$ dispersion  of the conduction band of the chiral Hamiltonian
    \begin{eqnarray} 
& E(k) = t \frac{1-\left(\frac{v_F k}{t_1} \right)^{2  } }{1-\left(\frac{v_F k}{t_1} \right)^{2n} } {\left(\frac{v_Fk}{t_1}\right)^{n} } .
   \end{eqnarray} 
Note that if the band structure disperses rapidly beyond the flat bottom, a finite $V$ does not significantly affect the extraction of $v$ and $k_b$ below, and primarily serves to generate a finite gap with the valence band. For small $\frac{v_Fk}{t_1}$, the energy is nearly zero due to the numerator. Note that this expression is not valid for $\frac{v_Fk}{t_1}>1$, where $E(k)$ eventually reaches 0 due to the diverging denominator, but such values of $k$ will not explicitly enter the results of our analysis. The maximum of $E(k)$ is approximately at $\frac{v_Fk}{t_1}=1$. We expand to fourth order in wavevector around this point
\begin{eqnarray}
    &E(k) \approx t \frac{-\frac{1}{360} (\frac{v_F k}{t_1}-1)^4 \left(1-n^2\right) \left(7 n^2+2\right)+\frac{1}{6} (\frac{v_F k}{t_1}-1)^2 \left(1-n^2\right)+(\frac{v_F k}{t_1}-1)+1}{n}.
\end{eqnarray}
As $\frac{v_Fk}{t_1}$ is reduced from 1, $E(k)$ begins to decrease, then reach a saddle point where the gradient is maximal (which persists over a broad range), before flattening out to approach $E(k)=0$ for small $k$. This saddle point is at $k_1$:
\begin{eqnarray}
    & \frac{\partial^2 E(k)}{\partial k ^2}=\frac{v_F^2}{t_1^2}( \frac{1-n^2}{3 n}-\frac{(\frac{v_F k}{t_1}-1)^2 \left(1-n^2\right) \left(7 n^2+2\right)}{30 n})=0  \nonumber \\ & 
    \implies k_1 =\frac{t_1}{v_F}(1-\frac{\sqrt{10}}{\sqrt{7 n^2+2}}),
\end{eqnarray}
where we choose the solution with $\frac{v_Fk}{t_1}<1$.  The slope at the saddle point is 
\begin{eqnarray}
    &v=\frac{\partial E(k)}{\partial k}|_{k=k_1}= v_F \left(\frac{2 \sqrt{10} \left(n^2-1\right) \sqrt{\frac{1}{7 n^2+2}}}{9 n}+\frac{1}{n}\right)
\end{eqnarray}
which sets the velocity parameter $v$.

At this momentum, the energy is
\begin{eqnarray}
    &E(k_1) = t_1 \frac{-\frac{225 n^2}{7 n^2+2}-36 \sqrt{10} \sqrt{\frac{1}{7 n^2+2}}+61}{36 n}.
\end{eqnarray}
We extract the radius $k_b$ as the momentum at which the linear approximation of $E(k)$ derived at $k=k_1$ reaches zero energy. This leads to
\begin{eqnarray}
    &k_b= -\frac{E(k_1)-vk_1}{v} = \frac{t_1}{v_F} \frac{\left(n^2-1\right) \left(4 \sqrt{10 \left(7 n^2+2\right)}-15\right)}{18 \left(7 n^2+\frac{2}{9} \left(n^2-1\right) \sqrt{10 \left(7 n^2+2\right)}+2\right)}.
\end{eqnarray}

Finally, we can compute the Berry flux $\varphi_b$ enclosed by the flat bottom of the dispersion (assuming the GMP limit with constant Berry curvature $2v_F^2/t_1^2$) as $\varphi_b=2\frac{v_F^2}{t_1^2}\pi k_b^2$.

\begin{table}\label{tab:trashcan_params}
\begin{tabular}{|c|c|c|c|c|c|c|c|}
\hline
$n$                       & 4        & 5        & 6        & 7        & 8        & 9        & 10       \\ \hline
$\frac{v}{v_F}$           & 0.496812 & 0.453537 & 0.423876 & 0.402288 & 0.385875 & 0.372977 & 0.362575 \\ \hline
$\frac{v_Fk_b}{t_1}$      & 0.441615 & 0.509194 & 0.561653 & 0.603716 & 0.638268 & 0.667193 & 0.691782 \\ \hline
$\frac{\varphi_{b}}{\pi}$ & 0.390048 & 0.518557 & 0.630908 & 0.728945 & 0.814771 & 0.890293 & 0.957126 \\ \hline
\end{tabular}
\caption{Extraction of Berry Trashcan model parameters from R$n$G. $v$ is the velocity, $k_b$ is the radius of the flat bottom, and $\varphi_b$ is the Berry flux enclosed by the flat bottom. We use the R$n$G parameters  $v_F=542.1\,\text{meVnm}, t_1=355.16\,\text{meV}$.}
\end{table}

\begin{figure}
    \centering
    \includegraphics[width = 0.6\linewidth]{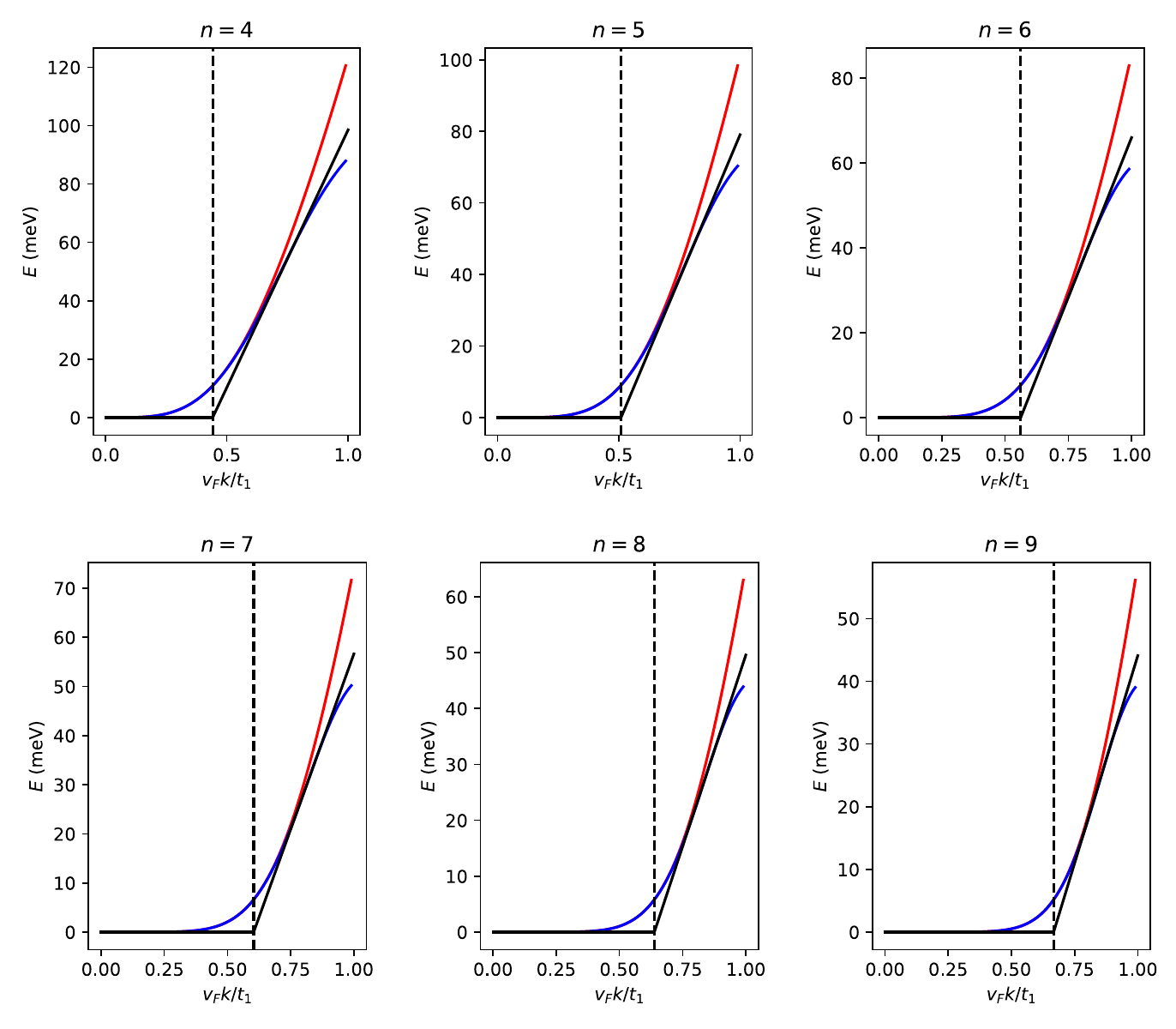}
    \caption{Berry trashcan parametrization of the dispersion of R$n$G for $n=4,\ldots,9$. We keep only $v_F=542.1\,\text{meVnm}, t_1=355.16\,\text{meV}$ in the Hamiltonian, and consider the lowest conduction band (red) at $V=0$. Blue lines show the conduction band dispersion of the model projected onto the chiral wavefunctions. Black lines show the corresponding trashcan parametrization.}
    \label{figapp:trashcan_Ek_param}
\end{figure}

The extracted parameters for various values of $n$ are tabulated in Tab.~\ref{tab:trashcan_params}. In Fig.~\ref{figapp:trashcan_Ek_param}, we show the analytically extracted dispersion (black) for $n=4,\ldots,9$, and compare it with the conduction band dispersion of the chiral Hamiltonian (blue) and the full Hamiltonian with just the $v_F$ and $t_1$ terms (red).

\section{Classical limit of the Interacting Model}\label{secapp:classical}

In this appendix section, we consider a `classical' limit of the interacting Hamiltonian Eq.~\ref{eqapp:Hintproject} with the GMP form factors of Eq.~\ref{eqapp:Mkq_exp}. We call the resulting model classical because it commutes with an extensive number of commuting operators in the absence of a momentum cutoff.

\subsection{Derivation of classical Hamiltonian}

Recall the interacting part of the Hamiltonian in Eq.~\ref{eqapp:Hintproject} and the GMP form factors of Eq.~\ref{eqapp:Mkq_exp}
\begin{gather}
    H^\text{int}=\frac{1}{2\Omega_{tot}}  \sum_{\bm{q}, \bm{k}, \bm{k'}} V_{\mbf{q}} M_{\mbf{k},\mbf{q}} M_{\mbf{k}',-\mbf{q}} \gamma_{\mbf{k}+ \mbf{q}}^\dagger \gamma_{\mbf{k}'- \mbf{q}}^\dagger \gamma_{\mbf{k}'} \gamma_{\mbf{k}}\\
    M_{\bm{k},\bm{q}}=e^{-\frac{v_F^2}{2t_1^2}(q^2+2i\bm{q}\times\bm{k})}\\
    \rightarrow H^\text{int}=\frac{1}{2\Omega_{tot}}  \sum_{\bm{k},\bm{k'},\bm{q}}^{\{\bm{k}, \bm{k'},\bm{k}+\bm{q},\bm{k}'-\bm{q}\}\in\mathscr{H}} V_{\mbf{q}}e^{-\frac{v_F^2}{ t^2} q^2 } e^{-i \frac{v_F^2}{ t_1^2}  \mbf{q}\times (\mbf{k}- \mbf{k'}) }\gamma_{\mbf{k}+ \mbf{q}}^\dagger \gamma_{\mbf{k}'- \mbf{q}}^\dagger \gamma_{\mbf{k}'} \gamma_{\mbf{k}},
\end{gather}
where in the last line, we have been explicit about the range of momentum summation in the case that a finite momentum cutoff $\mathscr{H}$ is used (in the context of the Berry Trashcan model, this cutoff would be comparable to the radius $k_b$ of the flat bottom). In particular, $\sum_{\bm{k},\bm{k'},\bm{q}}^{\{\bm{k}, \bm{k'},\bm{k}+\bm{q},\bm{k}'-\bm{q}\}\in\mathscr{H}}$ implies a summation over $\bm{k},\bm{k}',\bm{q}$ with the condition that $\{\bm{k}, \bm{k'},\bm{k}+\bm{q},\bm{k}'-\bm{q}\}$ all lie within $\mathscr{H}$.

We now consider a interaction potential $V_{q}= V_0 e^{\frac{ v_F^2}{t_1^2} q^2}$ that cancels the Gaussian part of the GMP form factors. Note that such an interaction potential is not physical if it is defined for all $q$, since it diverges for $q\rightarrow \infty$. We rewrite
\begin{align}
    H^\text{int}=&\frac{V_0}{2\Omega_{tot}}  \sum_{\bm{k},\bm{k'},\bm{q}}^{\{\bm{k}, \bm{k'},\bm{k}+\bm{q},\bm{k}'-\bm{q}\}\in\mathscr{H}} e^{-i \frac{v_F^2}{ t_1^2}  \mbf{q}\times (\mbf{k}- \mbf{k'}) }\gamma_{\mbf{k}+ \mbf{q}}^\dagger \gamma_{\mbf{k}'- \mbf{q}}^\dagger \gamma_{\mbf{k}'} \gamma_{\mbf{k}}\\
    =&\frac{V_0}{4\Omega_{tot}}  \sum_{\bm{k},\bm{k'},\bm{q}}^{\{\bm{k}, \bm{k'},\bm{k}+\bm{q},\bm{k}'-\bm{q}\}\in\mathscr{H}}\left(  e^{- i \frac{v_F^2}{t^2}  \mbf{q} \times (\mbf{k-k'})}-  e^{ i \frac{v_F^2}{t^2}  \mbf{q} \times (\mbf{k-k'})}\right) \gamma_{\mbf{k}+ \mbf{q}}^\dagger \gamma_{\mbf{k}'- \mbf{q}}^\dagger \gamma_{\mbf{k}'} \gamma_{\mbf{k}}\\
    =&\frac{V_0}{2\Omega_{tot}}  \sum_{\bm{k},\bm{k'},\bm{q}}^{\{\bm{k}, \bm{k'},\bm{k}+\bm{q},\bm{k}'-\bm{q}\}\in\mathscr{H}}i\sin  \left(\frac{v_F^2}{t^2}  \mbf{q} \times (\mbf{k'-k})\right) \gamma_{\mbf{k}+ \mbf{q}}^\dagger \gamma_{\mbf{k}'- \mbf{q}}^\dagger \gamma_{\mbf{k}'} \gamma_{\mbf{k}}\\
    =&\frac{V_0}{2\Omega_{tot}}  \sum_{\bm{k},\bm{k'},\bm{q}}^{\{\bm{k}, \bm{k'},\bm{k}+\bm{q},\bm{k}'-\bm{q}\}\in\mathscr{H}}i\sin  \left(\frac{v_F^2}{t^2}  \mbf{q} \times (\mbf{k'-k})\right) \gamma_{\mbf{k}+ \mbf{q}}^\dagger \gamma_{\mbf{k}}\gamma_{\mbf{k}'- \mbf{q}}^\dagger \gamma_{\mbf{k}'}. 
\end{align}
Note that in the second line, we first antisymmetrize the form factor part. Only after that, do we rewrite the interaction in the final line into the form $\gamma^\dagger \gamma \gamma^\dagger \gamma$ without any one-body terms. For general form factors and interaction potentials, we would have obtained an extra one-body contribution $- \sum_{\mbf{k,k'}} (V_{\mbf{k'-k}} |\langle \mbf{k'} |\mbf{k}\rangle |^2 - V_0) \gamma_{\mbf{k'}}^\dagger \gamma_{\mbf{k'}}$. We introduce the \emph{band} particle-hole (ph) operator
\begin{gather}
    \rho_{\mbf{k}, \mbf{q} }\equiv \gamma_{\mbf{k}+\mbf{q}}^\dagger \gamma_{\mbf{k}} \nonumber \\ 
    [\rho_{\mbf{k}, \mbf{q} }, \rho_{\mbf{k'}, \mbf{q'} }] =\delta_{\mbf{k}, \mbf{k'}+ \mbf{q'} } \gamma_{\mbf{k}+\mbf{q}}^\dagger \gamma_{\mbf{k' }}- \delta_{\mbf{k'}, \mbf{k}+ {\mbf{q}} } \gamma_{\mbf{k}'+\mbf{q'}}^\dagger \gamma_{\mbf{k }}=
    \delta_{\mbf{k}, \mbf{k'}+ \mbf{q'} } \rho_{\bm{k}',\bm{q}+\bm{q}'}- \delta_{\mbf{k'}, \mbf{k}+ {\mbf{q}} } \rho_{\bm{k},\bm{q}+\bm{q}'}
\end{gather}
which differs from the \emph{projected} ph operator $\tilde{\rho}_{\mbf{k},\mbf{q}}={M}_{\mbf{k},\mbf{q}}{\rho}_{\mbf{k},\mbf{q}}$ by the form factor. We now perform further algebraic manipulations on $H^{\text{int}}$
\begin{align}
    H^{\text{int}}=&\frac{V_0}{2\Omega_{tot}}  \sum_{\bm{k},\bm{k'},\bm{q}}^{\{\bm{k}, \bm{k'},\bm{k}+\bm{q},\bm{k}'-\bm{q}\}\in\mathscr{H}} i\sin  \left(\frac{v_F^2}{t^2}  \mbf{q} \times (\mbf{k'-k})\right) {\rho}_{\mbf{k},\mbf{q}}{\rho}_{\mbf{k}',-\mbf{q}}\\
    =&\frac{iV_0}{2\Omega_{tot}}  \sum_{\bm{k},\bm{k'},\bm{q}}^{\{\bm{k}, \bm{k'},\bm{k}+\bm{q},\bm{k}'-\bm{q}\}\in\mathscr{H}} \left[\sin \left( \frac{v_F^2}{t^2}  \mbf{q} \times \mbf{k}'\right)\cos  \left( \frac{v_F^2}{t^2}  \mbf{q} \times \mbf{k}\right)-\sin\left(   \frac{v_F^2}{t^2}  \mbf{q} \times \mbf{k}\right)\cos\left(   \frac{v_F^2}{t^2}  \mbf{q} \times \mbf{k}'\right)
    \right]{\rho}_{\mbf{k},\mbf{q}}{\rho}_{\mbf{k}',-\mbf{q}}\\
     =&-\frac{iV_0}{2\Omega_{tot}}  \sum_{\bm{k},\bm{k'},\bm{q}}^{\{\bm{k}, \bm{k'},\bm{k}+\bm{q},\bm{k}'-\bm{q}\}\in\mathscr{H}}\sin\left(   \frac{v_F^2}{t^2}  \mbf{q} \times \mbf{k}\right)\cos\left(   \frac{v_F^2}{t^2}  \mbf{q} \times \mbf{k}'\right)
    \left({\rho}_{\mbf{k}',-\mbf{q}}{\rho}_{\mbf{k},\mbf{q}}+{\rho}_{\mbf{k},\mbf{q}}{\rho}_{\mbf{k}',-\mbf{q}}\right)\\
    =&-\frac{iV_0}{2\Omega_{tot}}  \sum_{\bm{k},\bm{k'},\bm{q}}^{\{\bm{k}, \bm{k'},\bm{k}+\bm{q},\bm{k}'-\bm{q}\}\in\mathscr{H}}\sin\left(   \frac{v_F^2}{t^2}  \mbf{q} \times \mbf{k}\right)\cos\left(   \frac{v_F^2}{t^2}  \mbf{q} \times \mbf{k}'\right)
    \left(2\rho_{\mbf{k}, \mbf{q} }  \rho_{\mbf{k'},- \mbf{q} }  -\delta_{\mbf{k, k'-q}} \gamma_{\mbf{k'}}^\dagger \gamma_{\mbf{k'}}+ \delta_{\mbf{k', k+q}} \gamma_{\mbf{k}}^\dagger \gamma_{\mbf{k}}\right)\\
    =&-\frac{iV_0}{\Omega_{tot}}  \sum_{\bm{k},\bm{k'},\bm{q}}^{\{\bm{k}, \bm{k'},\bm{k}+\bm{q},\bm{k}'-\bm{q}\}\in\mathscr{H}}\sin\left(   \frac{v_F^2}{t^2}  \mbf{q} \times \mbf{k}\right)\cos\left(   \frac{v_F^2}{t^2}  \mbf{q} \times \mbf{k}'\right)
    \rho_{\mbf{k}, \mbf{q} }  \rho_{\mbf{k'},- \mbf{q} },
\end{align}
where to obtain the last line, we considered the momentum cutoff to be $C_2$-symmetric so that terms like $\sum_{ \{\mbf{k}, \mbf{k}'\}\in\mathscr{H}}  \sin{( \frac{v_F^2}{t^2}  \mbf{k'} \times \mbf{k} )}  \cos{( \frac{v_F^2}{t^2}  \mbf{k'} \times \mbf{ k}))}  \gamma_{\mbf{k'}}^\dagger \gamma_{\mbf{k'}} $ vanish. In the presence of a dispersion that sharply increases beyond some momentum radius that is much smaller than $t_1/v_F$, it is natural that the important terms are those for small $\mbf{k},\mbf{k}',\mbf{q}$, leading to the small-angle approximation to the trigonometric functions in the interacting part of the Hamiltonian
\begin{align}
    H^{\text{int}}&\approx -\frac{iV_0}{\Omega_{tot}}  \frac{v_F^2}{t^2} \sum_{\bm{k},\bm{k'},\bm{q}}^{\{\bm{k}, \bm{k'},\bm{k}+\bm{q},\bm{k}'-\bm{q}\}\in\mathscr{H}}\left(\mbf{q} \times \mbf{k}\right)
    \rho_{\mbf{k}, \mbf{q} }  \rho_{\mbf{k'},- \mbf{q} }\\
    &=-\frac{iV_0}{\Omega_{tot}}  \frac{v_F^2}{t^2} \sum_{\bm{k},\bm{q}}^{\{\bm{k},\bm{k}+\bm{q}\}\in\mathscr{H}}\left(\mbf{q} \times \mbf{k}\right)
    \rho_{\mbf{k}, \mbf{q} }  \rho_{- \mbf{q}}\label{eqapp:clasHam}
\end{align}
where we have introduced the \emph{band} density operator 
\begin{equation}
    \rho_{\mbf{q}}=\sum_{\bm{k}}^{\{\bm{k},\bm{k}+\bm{q}\}\in\mathscr{H}}\rho_{\mbf{k},\mbf{q}}=\sum_{\bm{k}}^{\{\bm{k},\bm{k}+\bm{q}\}\in\mathscr{H}}\gamma^\dagger_{\mbf{k}+\mbf{q}}\gamma_{\mbf{k}}
\end{equation}
which obeys the commutator
\begin{align}
    [\rho_{\mbf{k}, \mbf{q} }, \rho_{\mbf{q'} }] &=\sum_{\bm{k}'}^{\{\bm{k}',\bm{k}'+\bm{q}'\}\in\mathscr{H}}[\rho_{\mbf{k}, \mbf{q} }, \rho_{\mbf{k}',\mbf{q}'}]\\
    &=\sum_{\bm{k}'}^{\{\bm{k}',\bm{k}'+\bm{q}'\}\in\mathscr{H}}\left(\delta_{\mbf{k}, \mbf{k'}+ \mbf{q'} } \gamma_{\mbf{k}+\mbf{q}}^\dagger \gamma_{\mbf{k' }}- \delta_{\mbf{k'}, \mbf{k}+ {\mbf{q}} } \gamma_{\mbf{k}'+\mbf{q'}}^\dagger \gamma_{\mbf{k }}\right)\\
    &=\delta_{\{\bm{k}-\bm{q}',\bm{k}\}\in\mathscr{H}}\gamma_{\mbf{k}+\mbf{q}}^\dagger \gamma_{\mbf{k}-\mbf{q}'}-
    \delta_{\{\bm{k}+\bm{q},\bm{k}+\bm{q}+\bm{q}'\}\in\mathscr{H}}\gamma_{\mbf{k}+\mbf{q}+\mbf{q}'}^\dagger \gamma_{\mbf{k}}\\
    &=\delta_{\{\bm{k}-\bm{q}',\bm{k}\}\in\mathscr{H}}\rho_{\bm{k}-\bm{q}',\bm{q}+\bm{q}'}-
    \delta_{\{\bm{k}+\bm{q},\bm{k}+\bm{q}+\bm{q}'\}\in\mathscr{H}}\rho_{\bm{k},\bm{q}+\bm{q}'}
\end{align}
where $\delta_{\{\bm{k}-\bm{q}',\bm{k}\}\in\mathscr{H}}$ vanishes unless $\{\bm{k}-\bm{q}',\bm{k}\}\in\mathscr{H}$, etc. We also have
\begin{align}
    [\rho_{\mbf{q} }, \rho_{\mbf{q'} }] &=\sum_{\bm{k},\bm{k}'}^{\{\bm{k},\bm{k}+\bm{q},\bm{k}',\bm{k}'+\bm{q}'\}\in\mathscr{H}}\left(\delta_{\mbf{k}, \mbf{k'}+ \mbf{q'} } \gamma_{\mbf{k}+\mbf{q}}^\dagger \gamma_{\mbf{k' }}- \delta_{\mbf{k'}, \mbf{k}+ {\mbf{q}} } \gamma_{\mbf{k}'+\mbf{q'}}^\dagger \gamma_{\mbf{k }}\right)\\
    &=\sum_{\bm{k}}^{\{\bm{k},\bm{k}+\bm{q},\bm{k}-\bm{q}'\}\in\mathscr{H}}\gamma_{\mbf{k}+\mbf{q}}^\dagger \gamma_{\mbf{k}-\bm{q}'}- \sum_{\bm{k}}^{\{\bm{k},\bm{k}+\bm{q},\bm{k}+\bm{q}+\bm{q}'\}\in\mathscr{H}}\gamma_{\mbf{k}+\bm{q}+\mbf{q'}}^\dagger \gamma_{\mbf{k }}\\
    &=\sum_{\bm{k}}^{\{\bm{k},\bm{k}+\bm{q},\bm{k}-\bm{q}'\}\in\mathscr{H}}\gamma_{\mbf{k}+\mbf{q}}^\dagger \gamma_{\mbf{k}-\bm{q}'}- \sum_{\bm{k}}^{\{\bm{k}+\bm{q},\bm{k}+\bm{q}-\bm{q}',\bm{k}-\bm{q}'\}\in\mathscr{H}}\gamma_{\mbf{k}+\bm{q}}^\dagger \gamma_{\mbf{k }-\bm{q}'}\\
    &=\sum_{\bm{k}}^{\{\bm{k}+\bm{q},\bm{k}-\bm{q}'\}\in\mathscr{H}}\left(\delta_{\bm{k}\in\mathscr{H}}-\delta_{\bm{k}+\bm{q}-\bm{q}'\in\mathscr{H}}\right)\gamma_{\mbf{k}+\bm{q}}^\dagger \gamma_{\mbf{k }-\bm{q}'}\\
    &=\sum_{\bm{k}}^{\{\bm{k}+\bm{q}+\bm{q}',\bm{k}\}\in\mathscr{H}}\left(\delta_{\bm{k}+\bm{q}'\in\mathscr{H}}-\delta_{\bm{k}+\bm{q}\in\mathscr{H}}\right)\rho_{\mbf{k },\bm{q}+\bm{q}'}.
\end{align}
The above vanishes if the momentum cutoff $\mathscr{H}$ includes all momenta. 

We now take the commutator of the band density operator with the approximated interacting part of the Hamiltonian
\begin{align}
    \sum_{\bm{k},\bm{q}}^{\{\bm{k},\bm{k}+\bm{q}\}\in\mathscr{H}}\left(\mbf{q} \times \mbf{k}\right)\left[
    \rho_{\mbf{k}, \mbf{q} }  \rho_{- \mbf{q}},\rho_{\mbf{q'} }\right]&=\sum_{\bm{k},\bm{q}}^{\{\bm{k},\bm{k}+\bm{q}\}\in\mathscr{H}}\left(\mbf{q} \times \mbf{k}\right)\left(\rho_{\mbf{k}, \mbf{q} }\left[
      \rho_{- \mbf{q}},\rho_{\mbf{q'} }\right]+\left[
    \rho_{\mbf{k}, \mbf{q} }  ,\rho_{\mbf{q'} }\right]\rho_{- \mbf{q}}\right)\\
    &=\sum_{\bm{k},\bm{q}}^{\{\bm{k},\bm{k}+\bm{q}\}\in\mathscr{H}}\left(\mbf{q} \times \mbf{k}\right)\rho_{\mbf{k}, \mbf{q} }\sum_{\bm{k}'}^{\{\bm{k}'-\bm{q},\bm{k}'-\bm{q}'\}\in\mathscr{H}}\left(\delta_{\bm{k}'\in\mathscr{H}}-\delta_{\bm{k}'-\bm{q}-\bm{q}'\in\mathscr{H}}\right)\gamma_{\mbf{k}'-\bm{q}}^\dagger \gamma_{\mbf{k }'-\bm{q}'}\\
    &+\sum_{\bm{k},\bm{q}}^{\{\bm{k},\bm{k}+\bm{q}\}\in\mathscr{H}}\left(\mbf{q} \times \mbf{k}\right)\left(\delta_{\bm{k}-\bm{q}'\in\mathscr{H}}\gamma_{\mbf{k}+\mbf{q}}^\dagger \gamma_{\mbf{k}-\mbf{q}'}-
    \delta_{\bm{k}+\bm{q}+\bm{q}'\in\mathscr{H}}\gamma_{\mbf{k}+\mbf{q}+\mbf{q}'}^\dagger \gamma_{\mbf{k}}\right)\rho_{- \mbf{q}}.
\end{align}
In the case that $\mathscr{H}$ includes all momenta (i.e.~no momentum cutoff), then the above vanishes since
\begin{align}
    \sum_{\bm{k},\bm{q}}\left(\mbf{q} \times \mbf{k}\right)\left[
    \rho_{\mbf{k}, \mbf{q} }  \rho_{- \mbf{q}},\rho_{\mbf{q'} }\right]&=\sum_{\bm{k},\bm{q}}\left(\mbf{q} \times \mbf{k}\right)\left(\gamma_{\mbf{k}+\mbf{q}}^\dagger \gamma_{\mbf{k}-\mbf{q}'}-
   \gamma_{\mbf{k}+\mbf{q}+\mbf{q}'}^\dagger \gamma_{\mbf{k}}\right)\rho_{- \mbf{q}}\\
   &=\sum_{\bm{k},\bm{q}}\left(\mbf{q} \times \mbf{q}'\right)
   \gamma_{\mbf{k}+\mbf{q}+\mbf{q}'}^\dagger \gamma_{\mbf{k}}\rho_{- \mbf{q}}\\
   &=\sum_{\bm{q}}\left(\mbf{q} \times \mbf{q}'\right)
   \rho_{ \mbf{q}+\bm{q}'}\rho_{- \mbf{q}}=0,
\end{align}
where we used the fact that the band density operators commute in the absence of a momentum cutoff. Hence when $\mathscr{H}$ includes all momenta, the approximate interaction Hamiltonian in Eq.~\ref{eqapp:clasHam} commutes with all the band density operators $\rho_{\bm{q}}$ for any $\bm{q}$, which also commute amongst themselves.

\clearpage

\section{Self-consistent Hartree-Fock calculations of R$5$G with no moir\'e potential}\label{secapp:HFphasediagrams}

In this appendix section, we compute numerical self-consistent HF phase diagrams of pristine pentalayer graphene (R5G). Throughout, the moir\'e potential is switched off $(V_0=V_1=0)$, such that any gapped state implies the formation of a Wigner crystal that breaks the continuous translation symmetry. 

\subsection{Phase diagrams and non-interacting model approximations}\label{secapp:nonintapproxcutoff}
Here, we perform self-consistent HF calculations at `$\nu=1$' for different twist angles $\theta$ and interlayer potentials $V$. Note that due to the absence of a moir\'e potential, the `twist angle' refers to the choice of BZ (i.e.~that of R$n$G/hBN at twist angle $\theta$) which fixes the lattice periodicity of any Wigner crystal. We choose the electron density to correspond to one electron per Wigner unit cell. The goal is to study the properties of the gapped $C=0,1$ phases within different approximations on the model, to be detailed below. We use the charge neutrality (CN) interaction scheme and the dual-gate screened interaction potential $V_{\bm{q}}=\frac{e^2}{2\epsilon_0\epsilon q}\tanh \frac{q\xi}{2}$  with $\xi=20\,\text{nm}$ and relative dielectric constant $\epsilon=5$ unless otherwise stated. {Note that there is a factor of $4\pi \epsilon_0$ difference in the definition of $\epsilon$ compared to Ref.~\cite{bernevig2021TBGIII}.

We first outline the type of information presented in each of the phase diagrams in Figs.~\ref{fig:Hrad20.001_nvalact-2_epsr5.00_HFtypenone+0+0+0+none+none+0} to \ref{fig:Hrad1.701_theta0.77_V0.024_gateexpq2_HFtypet1_v0_only+1+1+exp+none}. $\text{max}[n(\bm{k})]-\text{min}[n(\bm{k})]$ indicates the maximum difference in occupation numbers $n(\bm{k})$ across the BZ for the lowest energy HF state. A gapped state requires this to be zero. 
The subplot labelled `Chern $C$' indicates the Chern number, computed using the method of Ref.~\cite{fukui2005chern}, of the lowest energy HF state if it is gapped. `$C=0$ exists' and `$C=1$ exists' indicate whether such a Chern state can be converged to in HF (regardless of whether it is the global minimum). 
Note that our HF calculations are restricted so that only valley $K$ and spin $\uparrow$ are active. Hence for any solution with Chern number $C$ polarized in valley $K$, there will also be a degenerate solution with Chern number $-C$ polarized in valley $K'$ by time-reversal symmetry.
The rest of the subplots in each phase diagram show more detailed information of the $C=0$ and $C=1$ HF state, including the the energy width of the occupied HF bands, and the HF gap. We also compare the relative energy (in units of meV per moir\'e unit cell) between the $C=0$ and $C=1$ HF states if they can both be obtained, including a breakdown into kinetic, Hartree, and Fock contributions.

For sufficiently large $\bm{k}$, the kinetic energy $E(\bm{k})$ of a Bloch state is large enough relative to the interaction strength that such Bloch states only affect the phases quantitatively, and can be neglected from the calculation. (Note that we only ever consider the lowest conduction and highest valence band, since the other bands are split off from charge neutrality by $\sim|t_1|$.) We now describe the two different ways that the continuum band structure is truncated for HF calculations:
\begin{enumerate}
\item The first method is called `band truncation', which is the standard method used for moir\'e continuum models. Here, the pentalayer bands are folded into the specified BZ. Keeping $m+n$ active bands means that the highest $m$ folded bands below charge neutrality and the lowest $n$ folded bands above charge neutrality are kept in the calculation. Note that because the BZ is smaller for smaller $\theta$, a larger number of folded conduction bands need to be retained if a specified level of quantitative accuracy is desired. As we are interested in broad trends in the phase diagrams, we will keep $m,n$ fixed across different twist angles for a given phase diagram.  If valence bands are retained $(m\neq0)$, care needs to be taken to implement the CN interaction scheme correctly.
\item Since the moir\'e potential is switched off in this appendix subsection, there is another method that we can employ, which is called `circular cutoff truncation'. Here, we pick a cutoff circle in momentum space centered at the Dirac momentum, whose radius is quoted in units of $q_1$. Only states with momenta lying within this cutoff are kept in the calculation.  
\end{enumerate}

We first discuss the results of calculations which use the `band truncation' method:
\begin{itemize}
\item Fig.~\ref{fig:Hrad20.001_nvalact-2_epsr5.00_HFtypenone+0+0+0+none+none+0} shows the phase diagram using the least amount of approximations in this appendix section. We use the full pentalayer Hamiltonian (Eq.~\ref{eq:H_K}) and use the band truncation method with two (folded) valence bands and four (folded) conduction bands. For most of the $V-\theta$ phase diagram, the ground state is the gapped $C=1$ insulator. Larger twist angles favor the $C=1$ state, which consistently has a better Hartree energy than the $C=0$ state. This is consistent with the fact that the $C=0$ Wigner crystal has a more inhomogeneous charge density~\cite{kwan2023mfci3}.
Note that the `occupied width' denotes the total bandwidth including the occupied valence bands, which explains its large value. 
\item In Fig.~\ref{fig:Hrad20.001_nvalact-2_epsr5.00_HFtypenoVISP+0+0+0+none+none+0}, the only difference from Fig.~\ref{fig:Hrad20.001_nvalact-2_epsr5.00_HFtypenone+0+0+0+none+none+0} is  that $V_{ISP}$ has been switched off. The general trends of the phase diagram remain the same. 
\item In Fig.~\ref{fig:Hrad20.001_nvalact0_epsr5.00_HFtypenone+0+0+0+none+none+0}, the only difference from Fig.~\ref{fig:Hrad20.001_nvalact-2_epsr5.00_HFtypenone+0+0+0+none+none+0} is that only four (folded) conduction bands are kept, and none of the valence bands are included. While the general trends of the phase diagram remain the same, it can be seen that the removal of the valence bands slightly favors the $C=0$ phase. For instance at $\theta=0.77^\circ$, the $C=1$ state has lower energy than the $C=0$ state by around $0.3\,\text{meV}$ per unit cell when the valence bands are kept, but without the valence bands, the energetic competition is significantly closer. Note that for larger twist angles, the region where the occupied HF bandwidth is smallest and the HF gap is largest moves to larger $V$. This is consistent with the fact that a larger $\theta$ corresponds to a larger BZ, and hence requires a larger interlayer potential to flatten the band structure.
\end{itemize}

The rest of the figures in this subsection employ the circular Hilbert space cutoff for the conduction band:
\begin{itemize}
\item 
In Figs.~\ref{fig:Hrad1.301_epsr5.00_HFtypenone+0+0+none+none} to \ref{fig:Hrad1.701_epsr5.00_HFtypenone+0+0+none+none}, we show phase diagrams for a momentum cutoff of radius $1.3q_1,1.5q_1,1.7q_1$ respectively using the full pentalayer Hamiltonian (Eq.~\ref{eq:H_K}). The results using the largest cutoff $1.7q_1$ (Fig.~\ref{fig:Hrad1.701_epsr5.00_HFtypenone+0+0+none+none}) are quantitatively similar to those using $0+4$ band truncation (Fig.~\ref{fig:Hrad20.001_nvalact-2_epsr5.00_HFtypenone+0+0+0+none+none+0}). The agreement worsens as the cutoff radius is reduced. For example, at cutoff radius $1.3q_1$ (Fig.~\ref{fig:Hrad1.301_epsr5.00_HFtypenone+0+0+none+none}), the competition at $\theta=0.77^\circ$ is more clearly in favor of $C=0$. However, the properties of the states themselves, such as the HF gap, remain qualitatively well captured (see also the detailed HF band structures in App.~\ref{secapp:HFbandstructure}).

\item In Figs.~\ref{fig:Hrad1.301_epsr5.00_HFtypet1_v0_v34_only+0+0+none+none} to \ref{fig:Hrad1.701_epsr5.00_HFtypet1_v0_v34_only+0+0+none+none}, we repeat the analysis of Figs.~\ref{fig:Hrad1.301_epsr5.00_HFtypenone+0+0+none+none} to \ref{fig:Hrad1.701_epsr5.00_HFtypenone+0+0+none+none}, except that $t_2$ and $V_{ISP}$ are neglected in the pentalayer Hamiltonian. The results are qualitatively similar.

\item In Figs.~\ref{fig:Hrad1.301_epsr5.00_HFtypet1_v0_only+1+1+exp+none} to \ref{fig:Hrad1.701_epsr5.00_HFtypet1_v0_only+1+1+exp+none}, we repeat the analysis of Figs.~\ref{fig:Hrad1.301_epsr5.00_HFtypenone+0+0+none+none} to \ref{fig:Hrad1.701_epsr5.00_HFtypenone+0+0+none+none}, except that:
\begin{itemize}
    \item Only $v_F$, $t_1$ and $V$ are kept in the pentalayer Hamiltonian (Eq.~\ref{eqapp:chiralham}).
    \item A $SO(2)$-symmetric $v_3,v_4$ correction to the dispersion is included in perturbation theory (Eq.~\ref{eqapp:v3v4_dispersion}).
    \item The exponential form factor approximation is used (Eq.~\ref{eqapp:Mkq_exp}).
\end{itemize}
The key finding in Figs.~\ref{fig:Hrad1.301_epsr5.00_HFtypet1_v0_only+1+1+exp+none} to \ref{fig:Hrad1.701_epsr5.00_HFtypet1_v0_only+1+1+exp+none} is that the combination of the above approximations energetically favor the $C=0$ state. However, the $C=1$ state can still be converged to in HF for a large range of parameters, and its properties remain qualitatively similar (see also the detailed HF band structures in App.~\ref{secapp:HFbandstructure}). Trends such as larger $\theta$ relatively favoring the $C=1$ solution still hold. Crucially, the enhancement from $C_3$ to an effective (intralayer) $C_6$ symmetry (which would be broken by trigonal warping) in the HF solutions does not qualitatively change their competition or identification.

\item In Figs.~\ref{fig:Hrad1.301_epsr5.00_HFtypet1_v0_only+1+1+exp+cutoff} to \ref{fig:Hrad1.701_epsr5.00_HFtypet1_v0_only+1+1+exp+cutoff}, we repeat the analysis of Figs.~\ref{fig:Hrad1.301_epsr5.00_HFtypet1_v0_only+1+1+exp+none} to \ref{fig:Hrad1.701_epsr5.00_HFtypet1_v0_only+1+1+exp+none}, except we use the `density-density' approximation for the interaction term (Eq.~\ref{eqapp:Hdensityapprox}). The results of the density-density approximation at some interlayer potential $V'$ appear to be qualitatively similar to those without the density-density approximation at some higher interlayer potential $V>V'$, which can be rationalized as follows. The density-density approximation effectively amounts to a shift of the one-body dispersion, which corresponds to the negative of the Fock self-energy corresponding to fully filling the states within the momentum cutoff (compare Eq.~\ref{eqapp:Hintprojectrewrite} and Eq.~\ref{eqapp:Hdensityapprox}). This shift is positive and peaked at the Dirac momentum, and therefore mimics an enhanced interlayer potential. 
\end{itemize}

We summarize the main takeaways from the HF phase diagrams presented in this subsection. The $C=0$ vs $C=1$ competition obeys the following general trends. The $C=1$ state consistently has a lower Hartree (higher Fock) energy than the $C=0$ state, though this difference decreases for larger twist angles, at least for the range $0.6^\circ\leq \theta \leq 1.1^\circ$ studied here. Overall, the $C=1$ state is relatively favored for larger twist angles. This is consistent (see later discussion in Sec.~\ref{secapp:2dmodel_energy}) with the fact that the first BZ encloses greater integrated Berry curvature for larger $\theta$. The region where the HF bandwidth is smallest and the HF gap is largest moves to larger $V$ for increasing $\theta$. The $C=1$ state is relatively favored when the number of bands or the momentum cutoff radius is increased. Many of the approximations, such as only keeping the dominant $v_F$ and $t_1$ terms in the continuum Hamiltonian and using the exponential form factor approximation (Eq.~\ref{eqapp:Mkq_exp}), tend to favor the $C=0$ phase. The latter finding is consistent with the fact that the exponential form factor approximation underestimates the actual Berry curvature of the R$n$G Hamiltonian (see e.g.~Fig.~\ref{figapp:RLG_Berry_curvature}). However, the $C=1$ can still be obtained as a metastable state for a wide range of parameters, and overall the trends of the $C=0$ vs $C=1$ competition remain unchanged. Importantly, restoring an emergent (intravalley) $C_6$ symmetry in the Wigner crystals does not fundamentally change the physics.

\begin{figure}
    \centering
    \includegraphics[width = 0.9\linewidth]{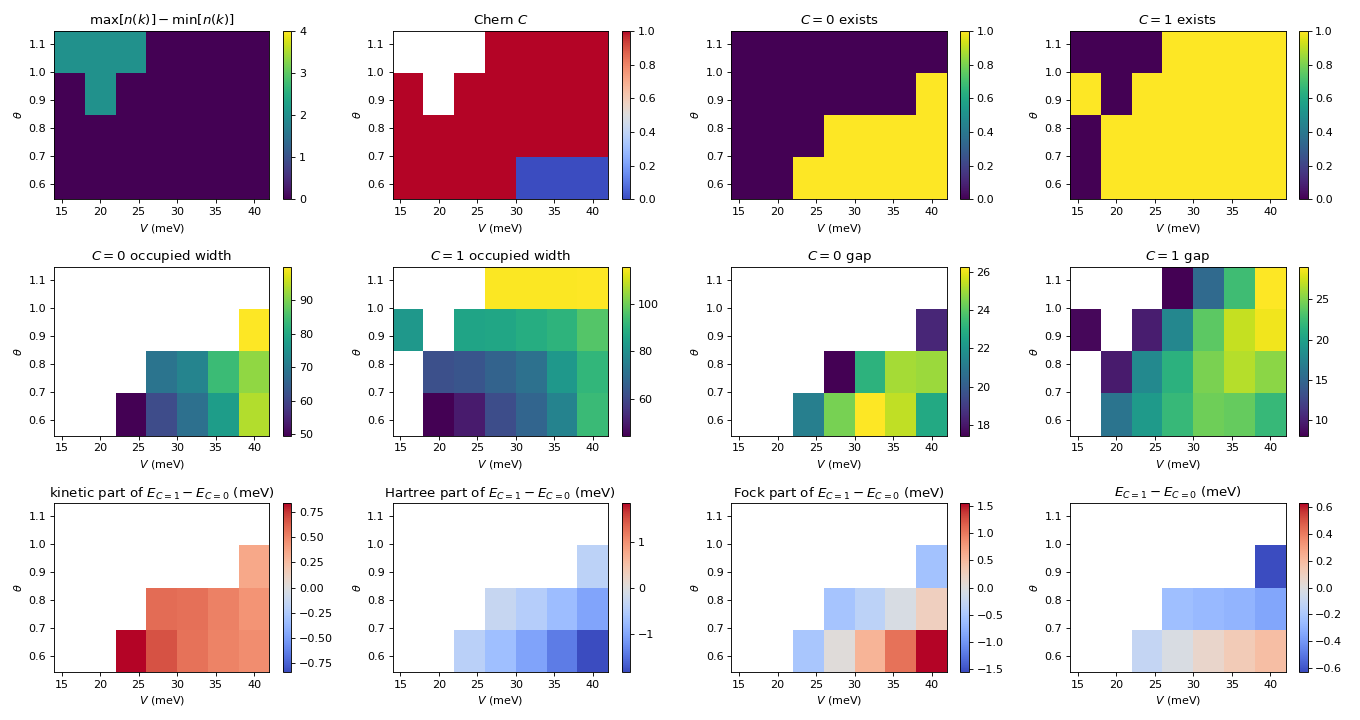}
    \caption{HF phase diagram for R$5$G. The full R$5$G Hamiltonian (Eq.~\ref{eq:H_K}) without approximations is used. No approximations on the form factors are made. Only valley $K$ and spin $\uparrow$ is included. The Hilbert space is truncated based on band index.\\\textbf{System parameters}: $N_1=N_2=12$; $2+4$ active bands; CN interaction scheme; $\epsilon=5$; $\xi=20\,\text{nm}$}
    \label{fig:Hrad20.001_nvalact-2_epsr5.00_HFtypenone+0+0+0+none+none+0}
\end{figure}

\begin{figure}
    \centering
    \includegraphics[width = 0.9\linewidth]{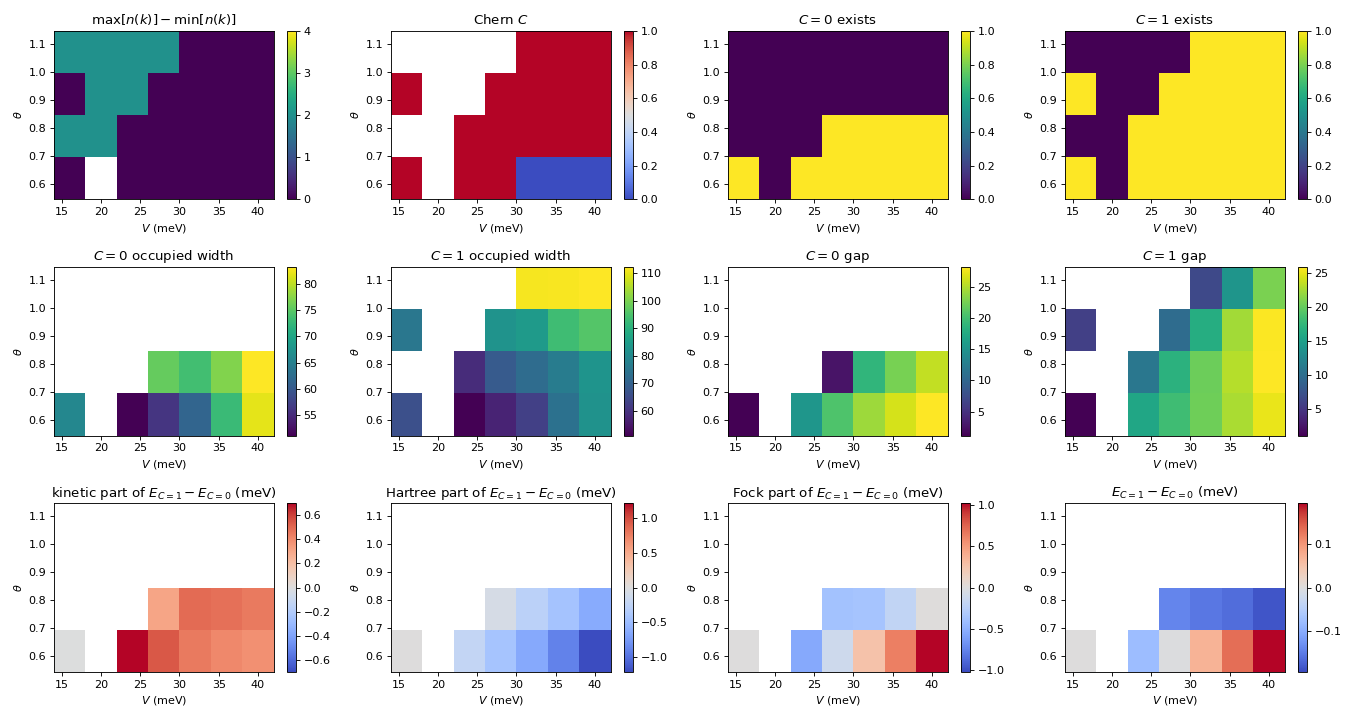}
    \caption{HF phase diagram for R$5$G. The full R$5$G Hamiltonian (Eq.~\ref{eq:H_K}) without approximations is used, except that $V_{ISP}=0$. No approximations on the form factors are made. Only valley $K$ and spin $\uparrow$ is included. The Hilbert space is truncated based on band index.\\\textbf{System parameters}: $N_1=N_2=12$; $2+4$ active bands; CN interaction scheme; $\epsilon=5$; $\xi=20\,\text{nm}$}
    \label{fig:Hrad20.001_nvalact-2_epsr5.00_HFtypenoVISP+0+0+0+none+none+0}
\end{figure}

\begin{figure}
    \centering
    \includegraphics[width = 0.9\linewidth]{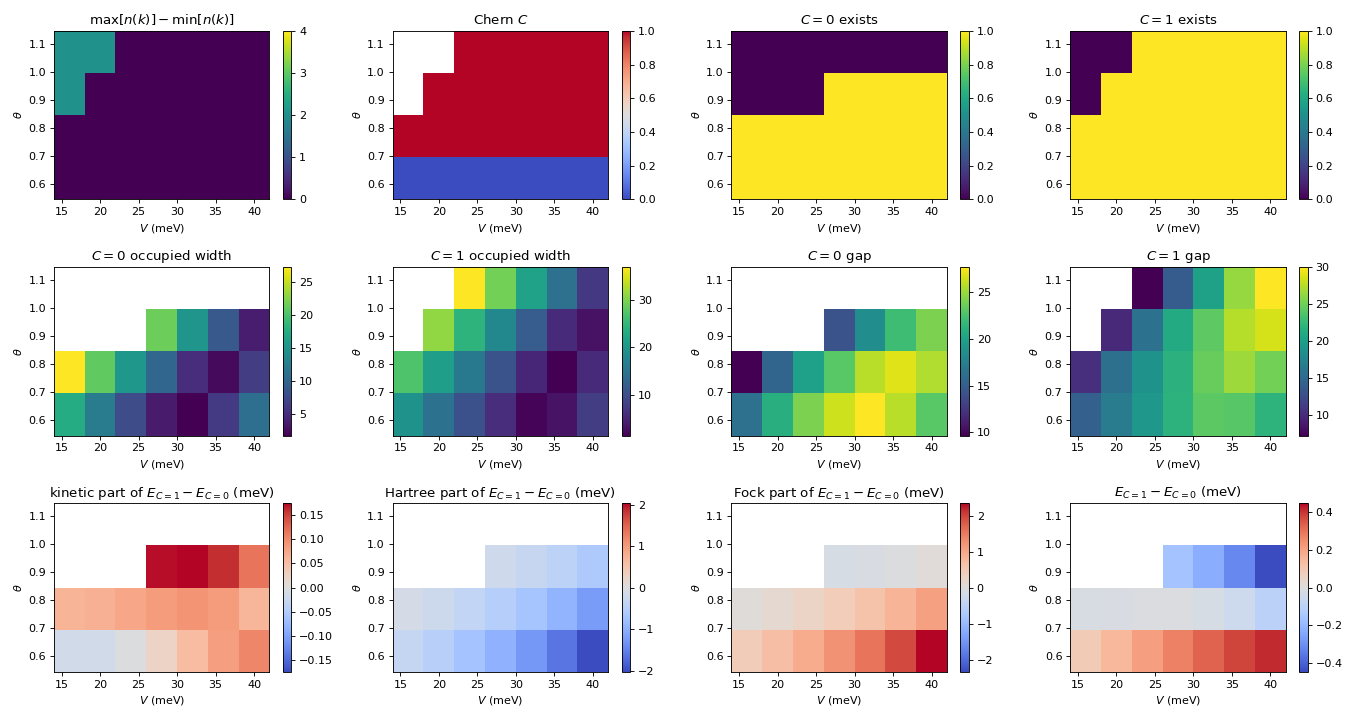}
    \caption{HF phase diagram for R$5$G. The full R$5$G Hamiltonian (Eq.~\ref{eq:H_K}) without approximations is used. No approximations on the form factors are made. Only valley $K$ and spin $\uparrow$ is included. The Hilbert space is truncated based on band index.\\\textbf{System parameters}: $N_1=N_2=12$; $0+4$ active bands; CN interaction scheme; $\epsilon=5$; $\xi=20\,\text{nm}$}
    \label{fig:Hrad20.001_nvalact0_epsr5.00_HFtypenone+0+0+0+none+none+0}
\end{figure}

\begin{figure}
    \centering
    \includegraphics[width = 0.9\linewidth]{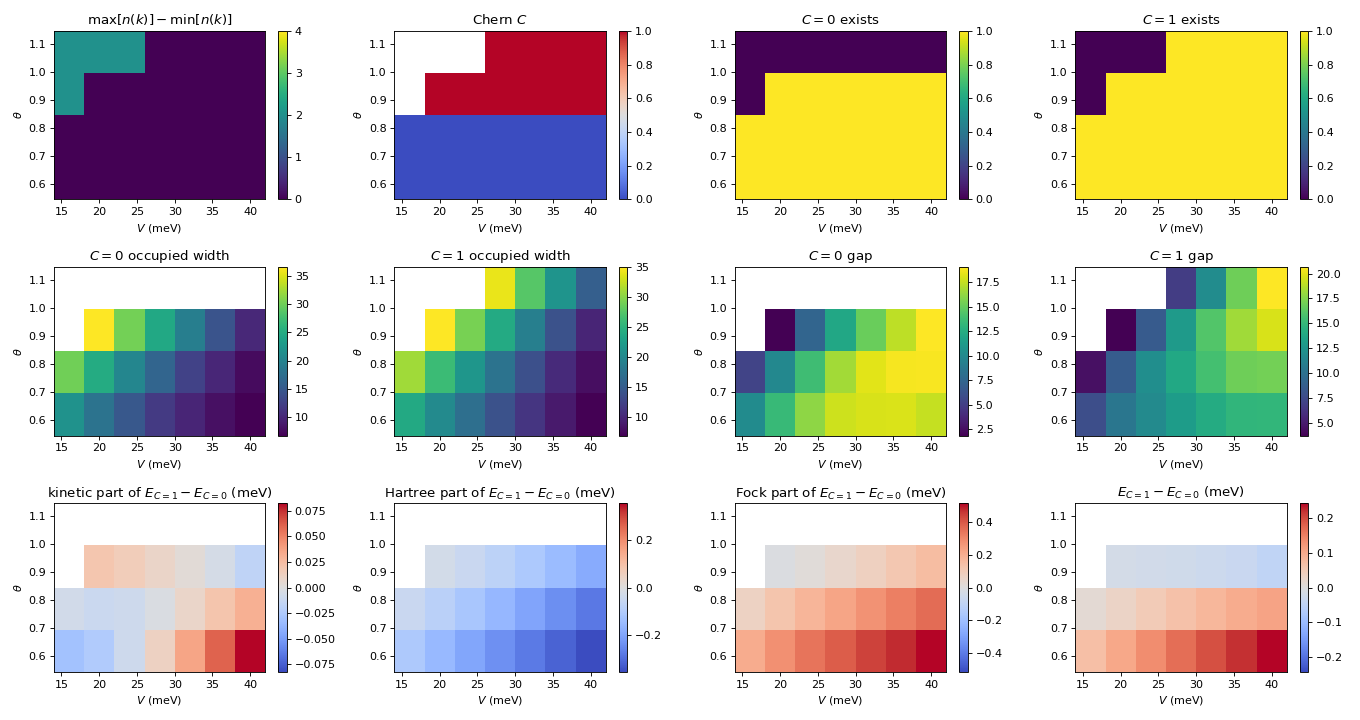}
    \caption{HF phase diagram for R$5$G. The full R$5$G Hamiltonian (Eq.~\ref{eq:H_K}) without approximations is used. No approximations on the form factors are made. Only valley $K$ and spin $\uparrow$ is included. The Hilbert space is made of conduction band states truncated based on a circular momentum cutoff. \\\textbf{System parameters}: $N_1=N_2=18$; Hilbert space cutoff radius $1.3q_1$; CN interaction scheme; $\epsilon=5$; $\xi=20\,\text{nm}$}
    \label{fig:Hrad1.301_epsr5.00_HFtypenone+0+0+none+none}
\end{figure}

\begin{figure}
    \centering
    \includegraphics[width = 0.9\linewidth]{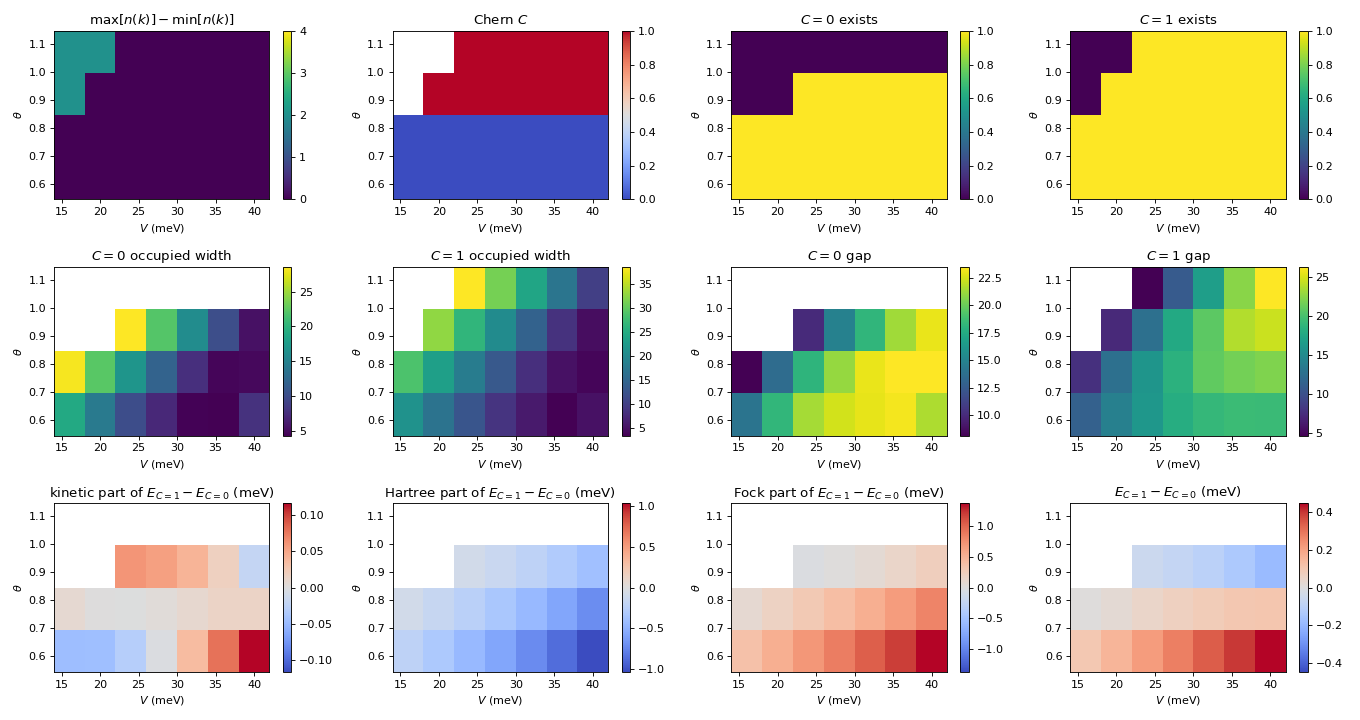}
    \caption{HF phase diagram for R$5$G. The full R$5$G Hamiltonian (Eq.~\ref{eq:H_K}) without approximations is used. No approximations on the form factors are made. Only valley $K$ and spin $\uparrow$ is included. The Hilbert space is made of conduction band states truncated based on a circular momentum cutoff. \\\textbf{System parameters}: $N_1=N_2=18$; Hilbert space cutoff radius $1.5q_1$; CN interaction scheme; $\epsilon=5$; $\xi=20\,\text{nm}$}
    \label{fig:Hrad1.501_epsr5.00_HFtypenone+0+0+none+none}
\end{figure}

\begin{figure}
    \centering
    \includegraphics[width = 0.9\linewidth]{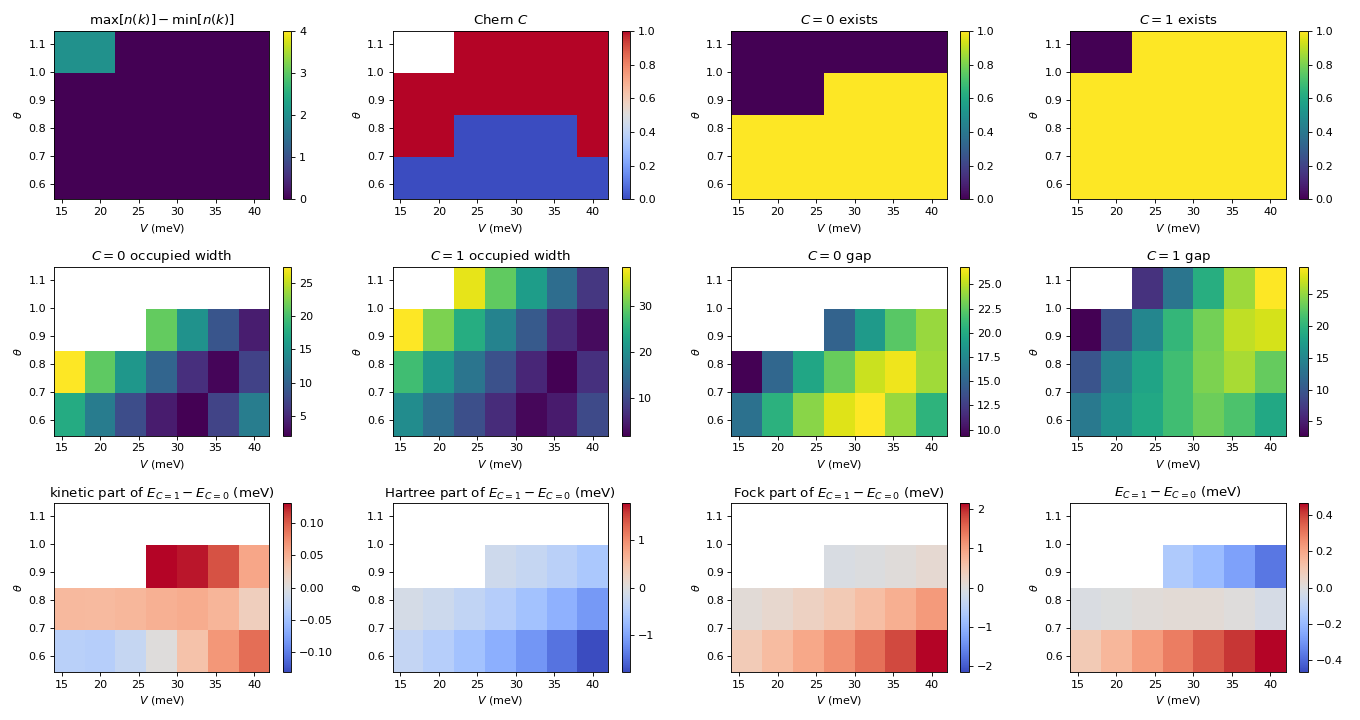}
    \caption{HF phase diagram for R$5$G. The full R$5$G Hamiltonian (Eq.~\ref{eq:H_K}) without approximations is used. No approximations on the form factors are made. Only valley $K$ and spin $\uparrow$ is included. The Hilbert space is made of conduction band states truncated based on a circular momentum cutoff. \\\textbf{System parameters}: $N_1=N_2=18$; Hilbert space cutoff radius $1.7q_1$; CN interaction scheme; $\epsilon=5$; $\xi=20\,\text{nm}$}
    \label{fig:Hrad1.701_epsr5.00_HFtypenone+0+0+none+none}
\end{figure}

\begin{figure}
    \centering
    \includegraphics[width = 0.9\linewidth]{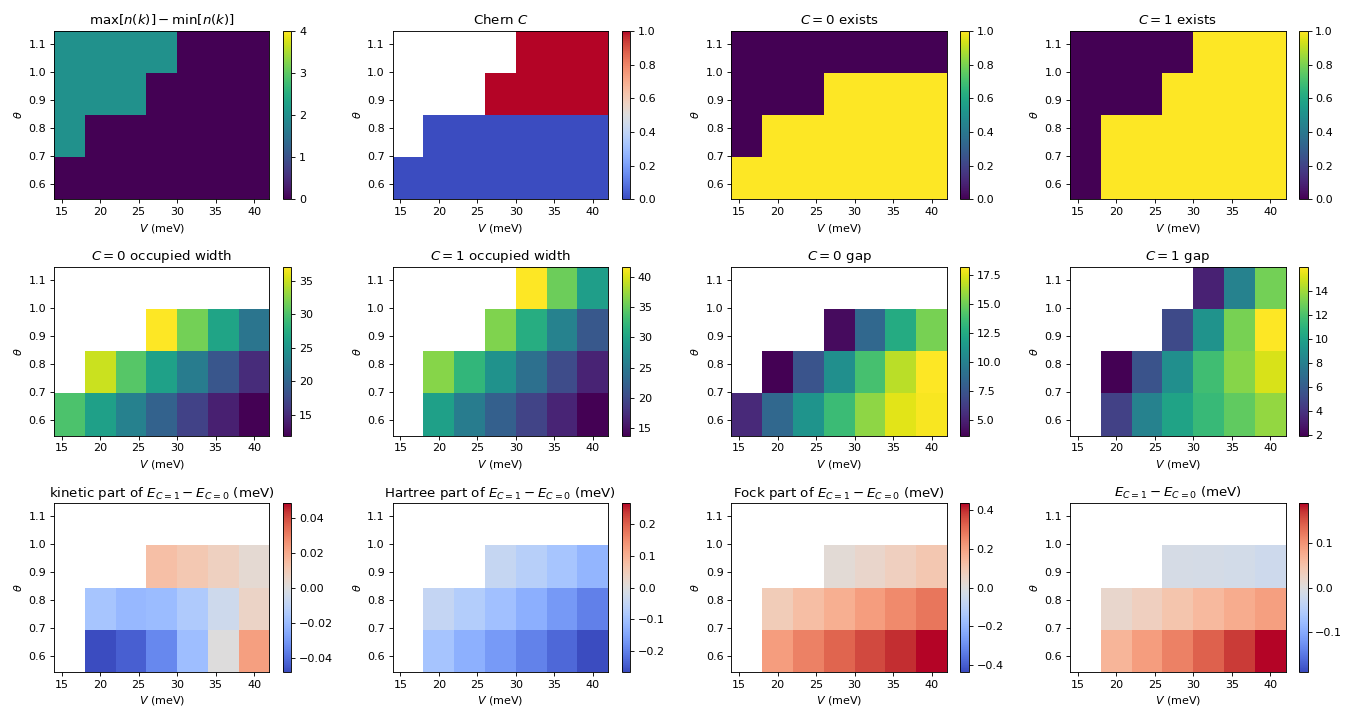}
    \caption{HF phase diagram for R$5$G. In the R$5$G Hamiltonian (Eq.~\ref{eq:H_K}), only the $v_F,t_1,v_3,v_4$ terms are kept. No approximations on the form factors are made. Only valley $K$ and spin $\uparrow$ is included. The Hilbert space is made of conduction band states truncated based on a circular momentum cutoff. \\\textbf{System parameters}: $N_1=N_2=18$; Hilbert space cutoff radius $1.3q_1$; CN interaction scheme; $\epsilon=5$; $\xi=20\,\text{nm}$}
    \label{fig:Hrad1.301_epsr5.00_HFtypet1_v0_v34_only+0+0+none+none}
\end{figure}

\begin{figure}
    \centering
    \includegraphics[width = 0.9\linewidth]{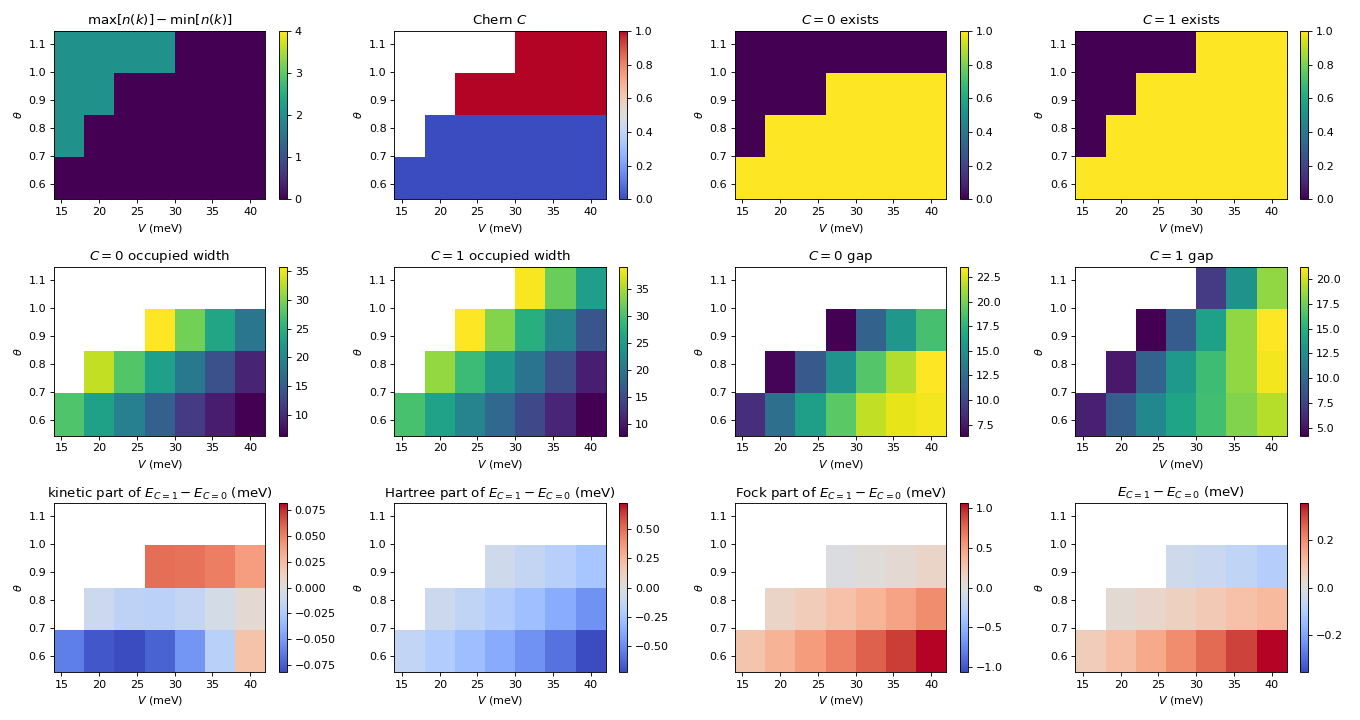}
    \caption{HF phase diagram for R$5$G. In the R$5$G Hamiltonian (Eq.~\ref{eq:H_K}), only the $v_F,t_1,v_3,v_4$ terms are kept. No approximations on the form factors are made. Only valley $K$ and spin $\uparrow$ is included. The Hilbert space is made of conduction band states truncated based on a circular momentum cutoff. \\\textbf{System parameters}: $N_1=N_2=18$; Hilbert space cutoff radius $1.5q_1$; CN interaction scheme; $\epsilon=5$; $\xi=20\,\text{nm}$}
    \label{fig:Hrad1.501_epsr5.00_HFtypet1_v0_v34_only+0+0+none+none}
\end{figure}

\begin{figure}
    \centering
    \includegraphics[width = 0.9\linewidth]{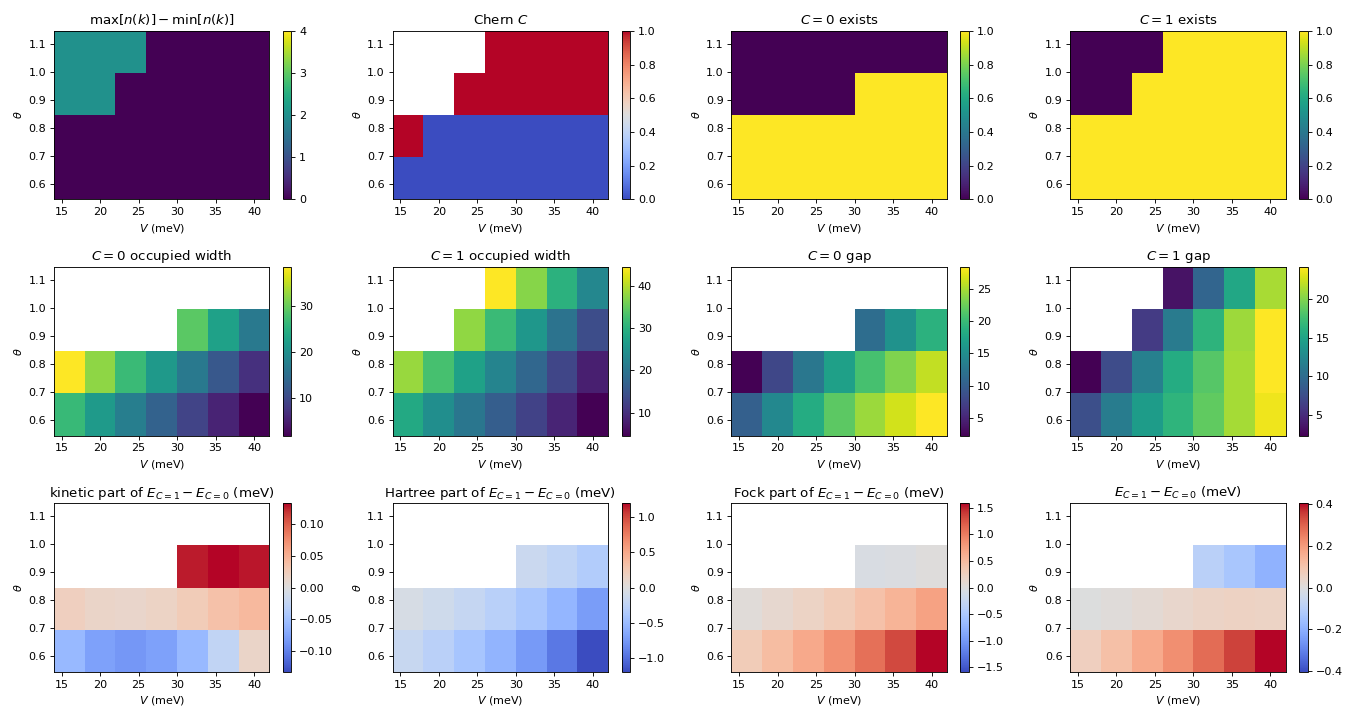}
    \caption{HF phase diagram for R$5$G. In the R$5$G Hamiltonian (Eq.~\ref{eq:H_K}), only the $v_F,t_1,v_3,v_4$ terms are kept. No approximations on the form factors are made. Only valley $K$ and spin $\uparrow$ is included. The Hilbert space is made of conduction band states truncated based on a circular momentum cutoff. \\\textbf{System parameters}: $N_1=N_2=18$; Hilbert space cutoff radius $1.7q_1$; CN interaction scheme; $\epsilon=5$; $\xi=20\,\text{nm}$}
    \label{fig:Hrad1.701_epsr5.00_HFtypet1_v0_v34_only+0+0+none+none}
\end{figure}

\begin{figure}
    \centering
    \includegraphics[width = 0.9\linewidth]{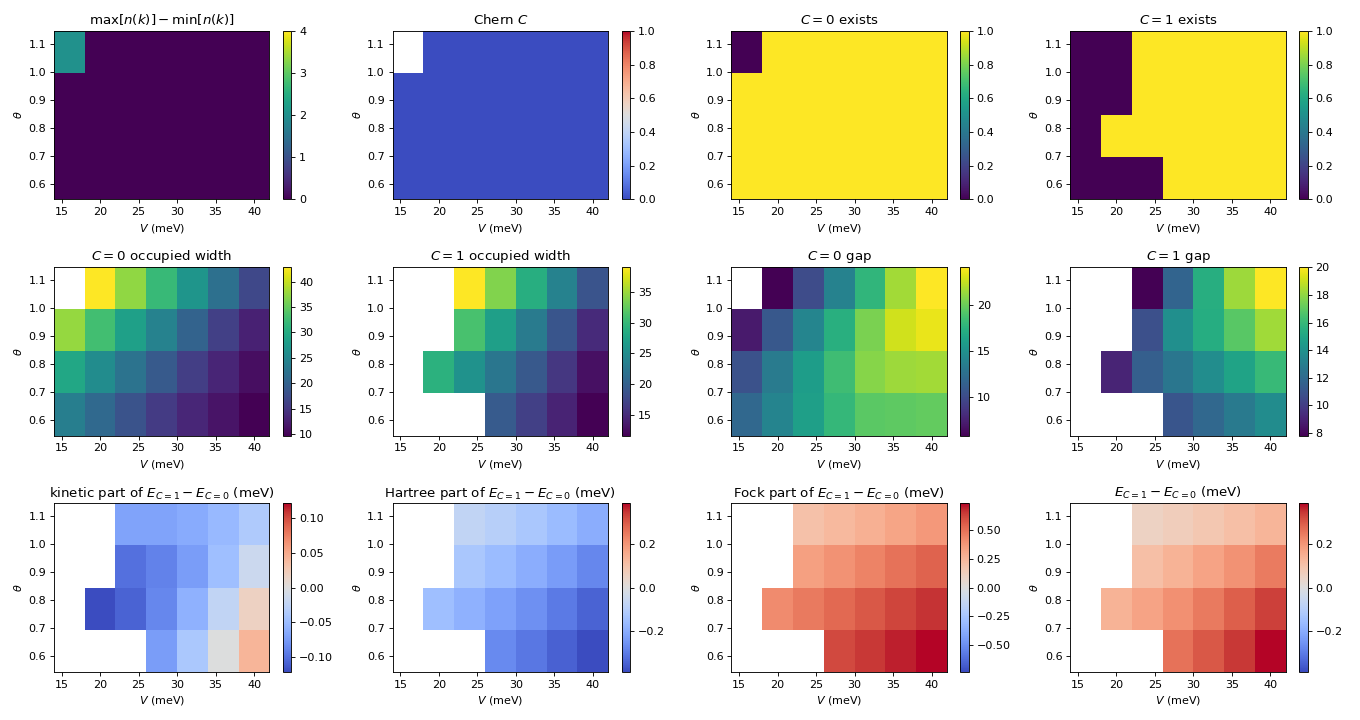}
    \caption{HF phase diagram for R$5$G. In the R$5$G Hamiltonian (Eq.~\ref{eq:H_K}), only the $v_F,t_1$ terms are kept. A $SO(2)$-symmetric dispersion correction from $v_3,v_4$ (Eq.~\ref{eqapp:v3v4_dispersion}) is included. The exponential form factor approximation (Eq.~\ref{eqapp:Mkq_exp}) is used. Only valley $K$ and spin $\uparrow$ is included. The Hilbert space is made of conduction band states truncated based on a circular momentum cutoff. \\\textbf{System parameters}: $N_1=N_2=18$; Hilbert space cutoff radius $1.3q_1$; CN interaction scheme; $\epsilon=5$; $\xi=20\,\text{nm}$}
    \label{fig:Hrad1.301_epsr5.00_HFtypet1_v0_only+1+1+exp+none}
\end{figure}

\begin{figure}
    \centering
    \includegraphics[width = 0.9\linewidth]{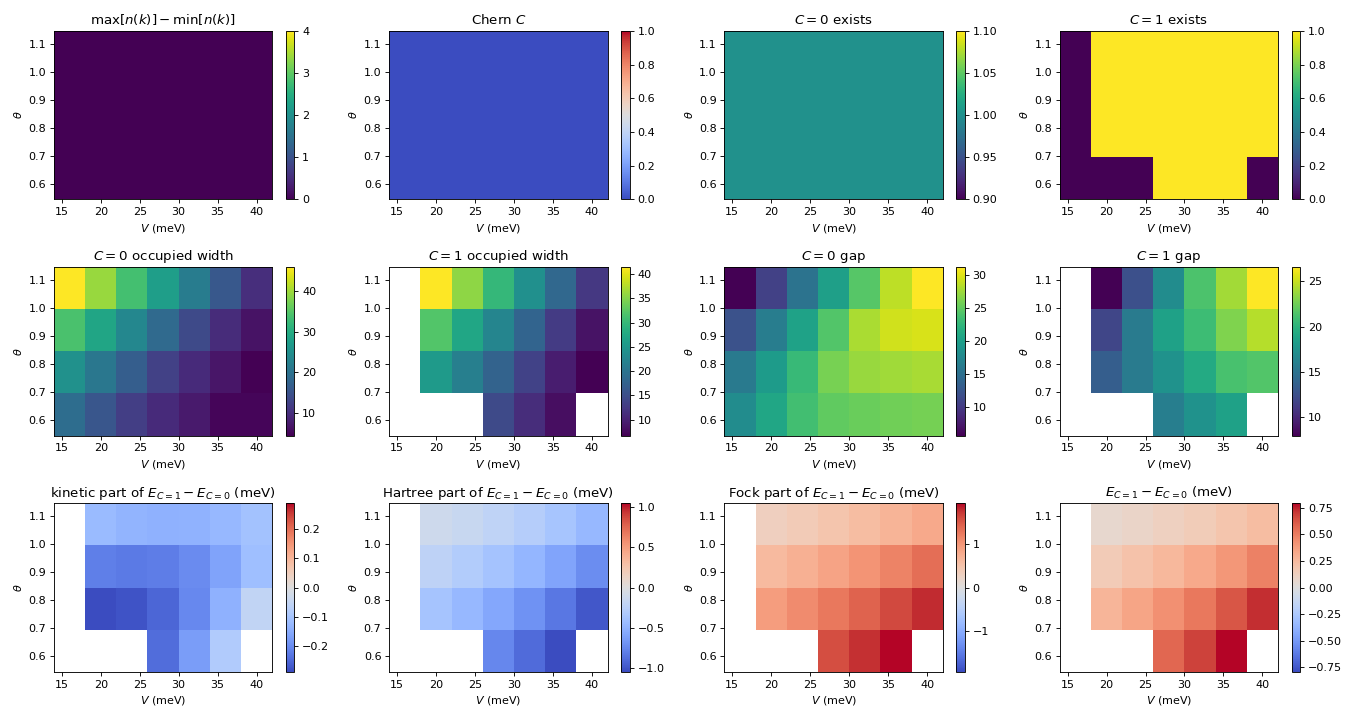}
    \caption{HF phase diagram for R$5$G. In the R$5$G Hamiltonian (Eq.~\ref{eq:H_K}), only the $v_F,t_1$ terms are kept. A $SO(2)$-symmetric dispersion correction from $v_3,v_4$ (Eq.~\ref{eqapp:v3v4_dispersion}) is included. The exponential form factor approximation (Eq.~\ref{eqapp:Mkq_exp}) is used. Only valley $K$ and spin $\uparrow$ is included. The Hilbert space is made of conduction band states truncated based on a circular momentum cutoff. \\\textbf{System parameters}: $N_1=N_2=18$; Hilbert space cutoff radius $1.5q_1$; CN interaction scheme; $\epsilon=5$; $\xi=20\,\text{nm}$}
    \label{fig:Hrad1.501_epsr5.00_HFtypet1_v0_only+1+1+exp+none}
\end{figure}

\begin{figure}
    \centering
    \includegraphics[width = 0.9\linewidth]{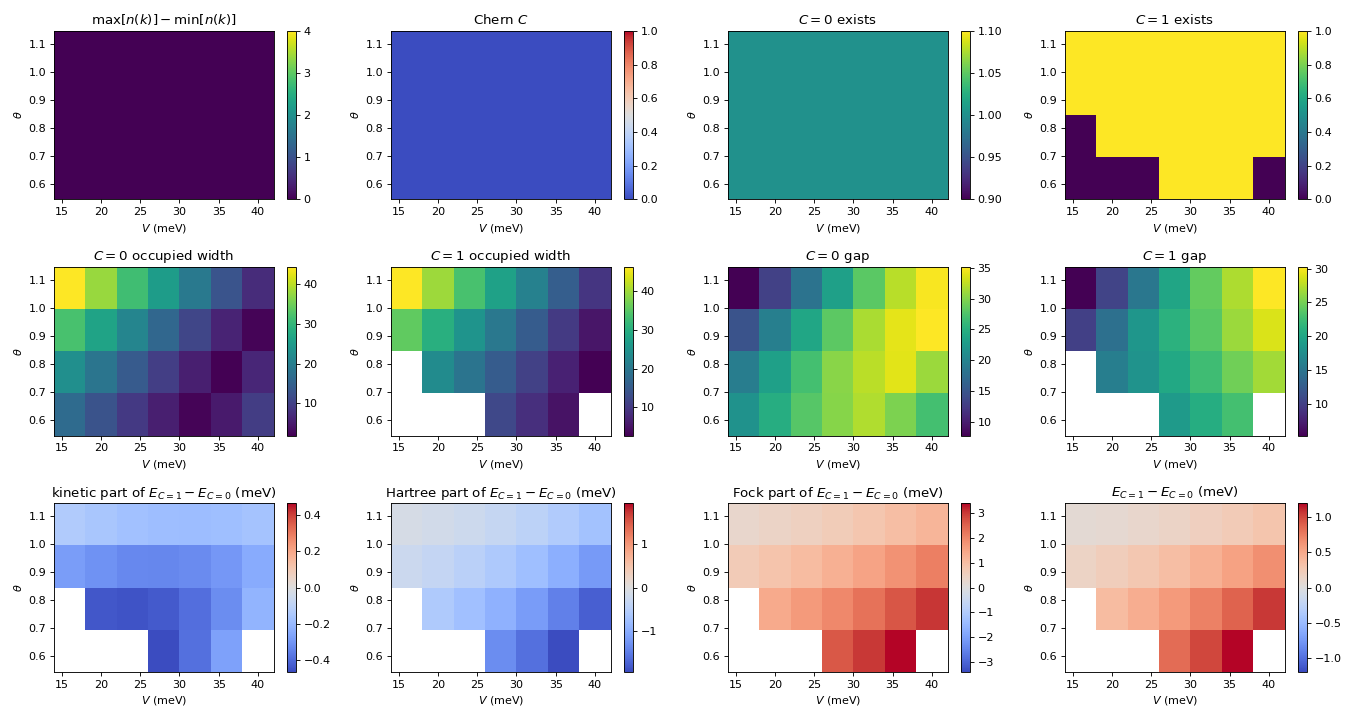}
    \caption{HF phase diagram for R$5$G. In the R$5$G Hamiltonian (Eq.~\ref{eq:H_K}), only the $v_F,t_1$ terms are kept. A $SO(2)$-symmetric dispersion correction from $v_3,v_4$ (Eq.~\ref{eqapp:v3v4_dispersion}) is included. The exponential form factor approximation (Eq.~\ref{eqapp:Mkq_exp}) is used. Only valley $K$ and spin $\uparrow$ is included. The Hilbert space is made of conduction band states truncated based on a circular momentum cutoff. \\\textbf{System parameters}: $N_1=N_2=18$; Hilbert space cutoff radius $1.7q_1$; CN interaction scheme; $\epsilon=5$; $\xi=20\,\text{nm}$}
    \label{fig:Hrad1.701_epsr5.00_HFtypet1_v0_only+1+1+exp+none}
\end{figure}

\begin{figure}
    \centering
    \includegraphics[width = 0.9\linewidth]{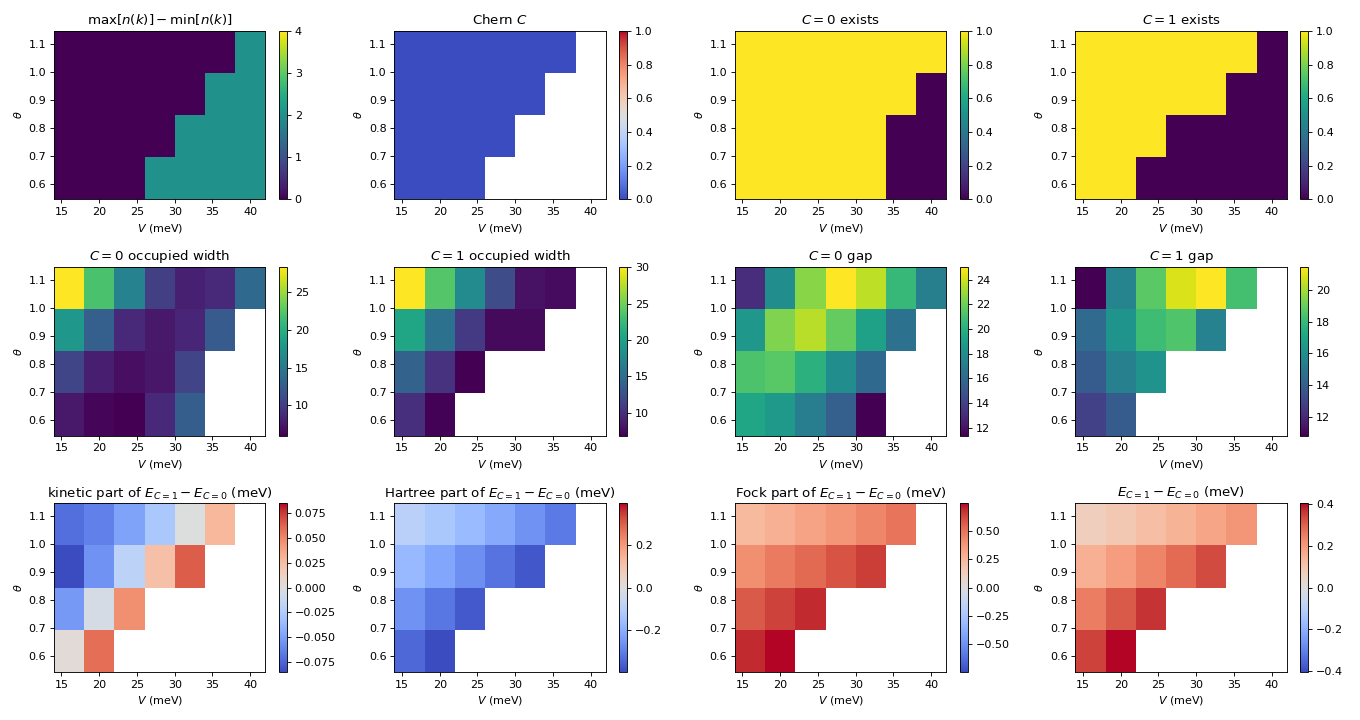}
    \caption{HF phase diagram for R$5$G. In the R$5$G Hamiltonian (Eq.~\ref{eq:H_K}), only the $v_F,t_1$ terms are kept. A $SO(2)$-symmetric dispersion correction from $v_3,v_4$ (Eq.~\ref{eqapp:v3v4_dispersion}) is included. The exponential form factor approximation (Eq.~\ref{eqapp:Mkq_exp}) is used. The density-density approximation to the interaction Hamiltonian (Eq.~\ref{eqapp:Hdensityapprox}) is used. Only valley $K$ and spin $\uparrow$ is included. The Hilbert space is made of conduction band states truncated based on a circular momentum cutoff. \\\textbf{System parameters}: $N_1=N_2=18$; Hilbert space cutoff radius $1.3q_1$; CN interaction scheme; $\epsilon=5$; $\xi=20\,\text{nm}$}
    \label{fig:Hrad1.301_epsr5.00_HFtypet1_v0_only+1+1+exp+cutoff}
\end{figure}

\begin{figure}
    \centering
    \includegraphics[width = 0.9\linewidth]{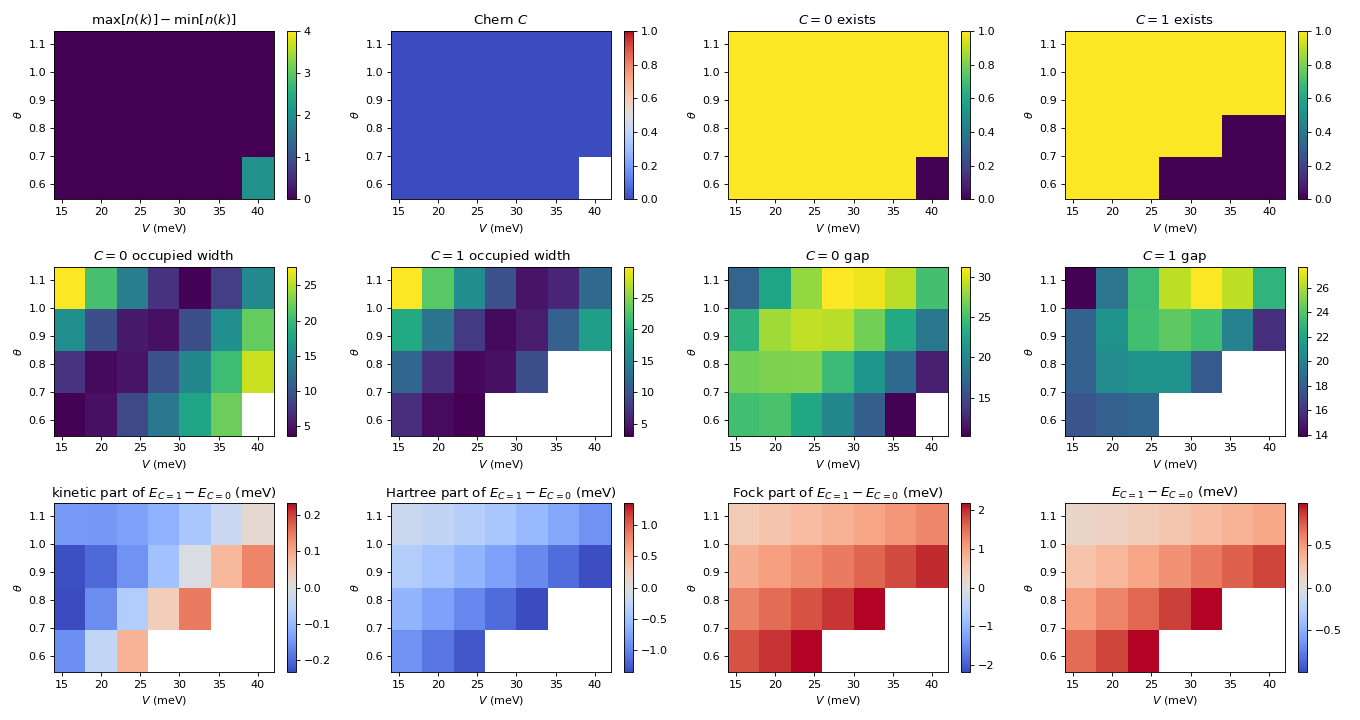}
    \caption{HF phase diagram for R$5$G. In the R$5$G Hamiltonian (Eq.~\ref{eq:H_K}), only the $v_F,t_1$ terms are kept. A $SO(2)$-symmetric dispersion correction from $v_3,v_4$ (Eq.~\ref{eqapp:v3v4_dispersion}) is included. The exponential form factor approximation (Eq.~\ref{eqapp:Mkq_exp}) is used. The density-density approximation to the interaction Hamiltonian (Eq.~\ref{eqapp:Hdensityapprox}) is used. Only valley $K$ and spin $\uparrow$ is included. The Hilbert space is made of conduction band states truncated based on a circular momentum cutoff. \\\textbf{System parameters}: $N_1=N_2=18$; Hilbert space cutoff radius $1.5q_1$; CN interaction scheme; $\epsilon=5$; $\xi=20\,\text{nm}$}
    \label{fig:Hrad1.501_epsr5.00_HFtypet1_v0_only+1+1+exp+cutoff}
\end{figure}

\begin{figure}
    \centering
    \includegraphics[width = 0.9\linewidth]{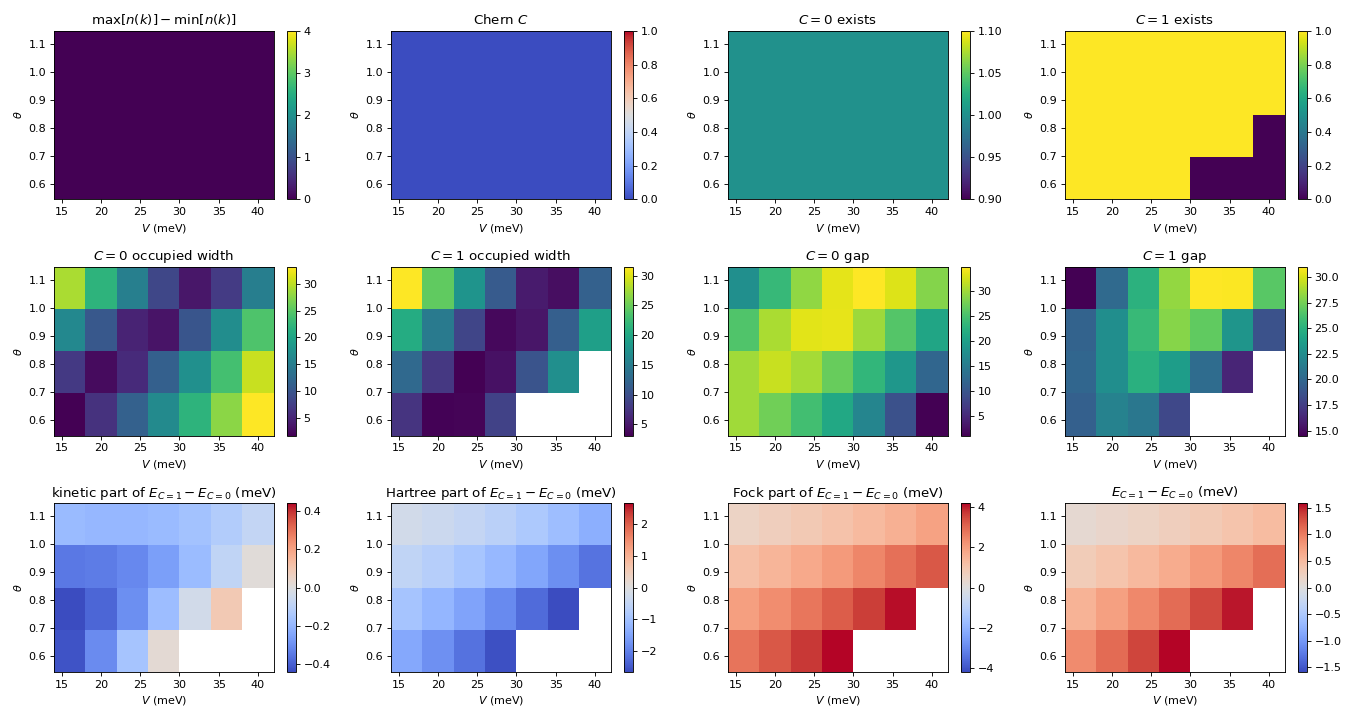}
    \caption{HF phase diagram for R$5$G. In the R$5$G Hamiltonian (Eq.~\ref{eq:H_K}), only the $v_F,t_1$ terms are kept. A $SO(2)$-symmetric dispersion correction from $v_3,v_4$ (Eq.~\ref{eqapp:v3v4_dispersion}) is included. The exponential form factor approximation (Eq.~\ref{eqapp:Mkq_exp}) is used. The density-density approximation to the interaction Hamiltonian (Eq.~\ref{eqapp:Hdensityapprox}) is used. Only valley $K$ and spin $\uparrow$ is included. The Hilbert space is made of conduction band states truncated based on a circular momentum cutoff. \\\textbf{System parameters}: $N_1=N_2=18$; Hilbert space cutoff radius $1.7q_1$; CN interaction scheme; $\epsilon=5$; $\xi=20\,\text{nm}$}
    \label{fig:Hrad1.701_epsr5.00_HFtypet1_v0_only+1+1+exp+cutoff}
\end{figure}

\clearpage

\subsection{Interaction potentials}
In this appendix subsection, we perform self-consistent HF calculations at $\nu=1$ for different choices of the interaction potential $V_{\bm{q}}$ for fixed $\theta=0.77^\circ$ and $V=24\,\text{meV}$. For each figure, we consider different values of the overall interaction strength $5/\epsilon$ and a length scale $\xi$ that parameterizes the interaction. The interaction potentials are normalized so that they take the same value at $q=0$ (since the interaction potentials we consider can vary greatly in their functional form, the phase diagrams can have significantly different ranges of $5/\epsilon$). All calculations employ the circular Hilbert space cutoff in momentum space with radius $1.7q_1$. We summarize the results of the calculations:
\begin{itemize}
\item
Figs.~\ref{fig:Hrad1.701_theta0.77_V0.024_gatedual_HFtypenone+0+0+none+none} to \ref{fig:Hrad1.701_theta0.77_V0.024_gateexpq2_HFtypenone+0+0+none+none} show phase diagrams using the full pentalayer Hamiltonian (Eq.~\ref{eq:H_K}). In Fig.~\ref{fig:Hrad1.701_theta0.77_V0.024_gatedual_HFtypenone+0+0+none+none}, we use the dual-gate screened interaction $V_{\bm{q}}=\frac{e^2}{2\epsilon q}\tanh\frac{q\xi}{2}$. Lowering the gate distance $\xi$ relatively favors the $C=1$ phase. One reason for this is that the ratio $V(|\bm{b_1}|)/V(q)$ for small $q$ gets larger for smaller $\xi$, relatively increasing the importance of the Hartree term, which benefits $C=1$ over $C=0$. 
\item
In Fig.~\ref{fig:Hrad1.701_theta0.77_V0.024_gatequadratic_HFtypenone+0+0+none+none}, we consider $V_{\bm{q}}=\frac{e^2}{2\epsilon q}\left(\frac{q\xi}{2}-\frac{q^3\xi^3}{12}\right)$, which is an expansion of the dual-gate screened interaction for small $q$. For the range of momentum transfers within the cutoff, this expansion is only justified for small $\xi$. Indeed, the results agree with the dual-gate screened interaction only for $\xi\lesssim 5\,\text{nm}$. 
\item
In Fig.~\ref{fig:Hrad1.701_theta0.77_V0.024_gateexpmq2_HFtypenone+0+0+none+none}, we consider $V_{\bm{q}}=\frac{\xi e^2}{2\epsilon }e^{-\xi^2q^2}$. The phase diagram shows that the $C=1$ state can be favored for smaller $\xi$, which is consistent with the fact that the Hartree term is relatively strengthened.
\item
In Fig.~\ref{fig:Hrad1.701_theta0.77_V0.024_gateexpq2_HFtypenone+0+0+none+none}, we consider $V_{\bm{q}}=\frac{\xi e^2}{2\epsilon }e^{\xi^2q^2}$, which is not a physical interaction potential for all momenta since it blows up for large enough $q$. The motivation for considering this is that for $\xi=v_F/t_1\simeq 1.53\,\text{nm}$, $V_{\bm{q}}$ cancels the Gaussian part of the form factor $M_{\bm{k},\bm{q}}$ in the exponential form factor approximation (Eq.~\ref{eqapp:Mkq_exp}). The phase diagram shows that the $C=1$ state can be stabilized, but the $C=0$ state is not obtained in HF. In this limit, the coefficients of the four-fermion term in the Hamiltonian of Eq.~\ref{eqapp:Hintproject} are just pure phases. If the phases are further neglected, the interaction term identically vanishes by Pauli antisymmetry, and the Hamiltonian is purely non-interacting and cannot sustain a Wigner crystal. This underscores the importance of the phases, which encodes the Berry curvature of the underlying single-particle Bloch functions. These phases, which are a purely 2D property, can still stabilize a gapped HF solution (though not necessarily as the lowest energy solution).
\item
Figs.~\ref{fig:Hrad1.701_theta0.77_V0.024_gatedual_HFtypet1_v0_only+1+1+exp+none} to \ref{fig:Hrad1.701_theta0.77_V0.024_gateexpq2_HFtypet1_v0_only+1+1+exp+none} show corresponding results for the case that only $v_F$ and $t_1$ are kept in the pentalayer Hamiltonian (Eq.~\ref{eqapp:chiralham}), a $SO(2)$-symmetric $v_3,v_4$ correction to the dispersion is included in perturbation theory, and the exponential form factor approximation is used (Eq.~\ref{eqapp:Mkq_exp}).
The results show qualitatively similar trends as Figs.~\ref{fig:Hrad1.701_theta0.77_V0.024_gatedual_HFtypenone+0+0+none+none} to \ref{fig:Hrad1.701_theta0.77_V0.024_gateexpq2_HFtypenone+0+0+none+none}, though as in App.~\ref{secapp:nonintapproxcutoff}, these approximations on the pentalayer model and the form factors tend to favor $C=0$. 
\end{itemize}

\begin{figure}
    \centering
    \includegraphics[width = 0.9\linewidth]{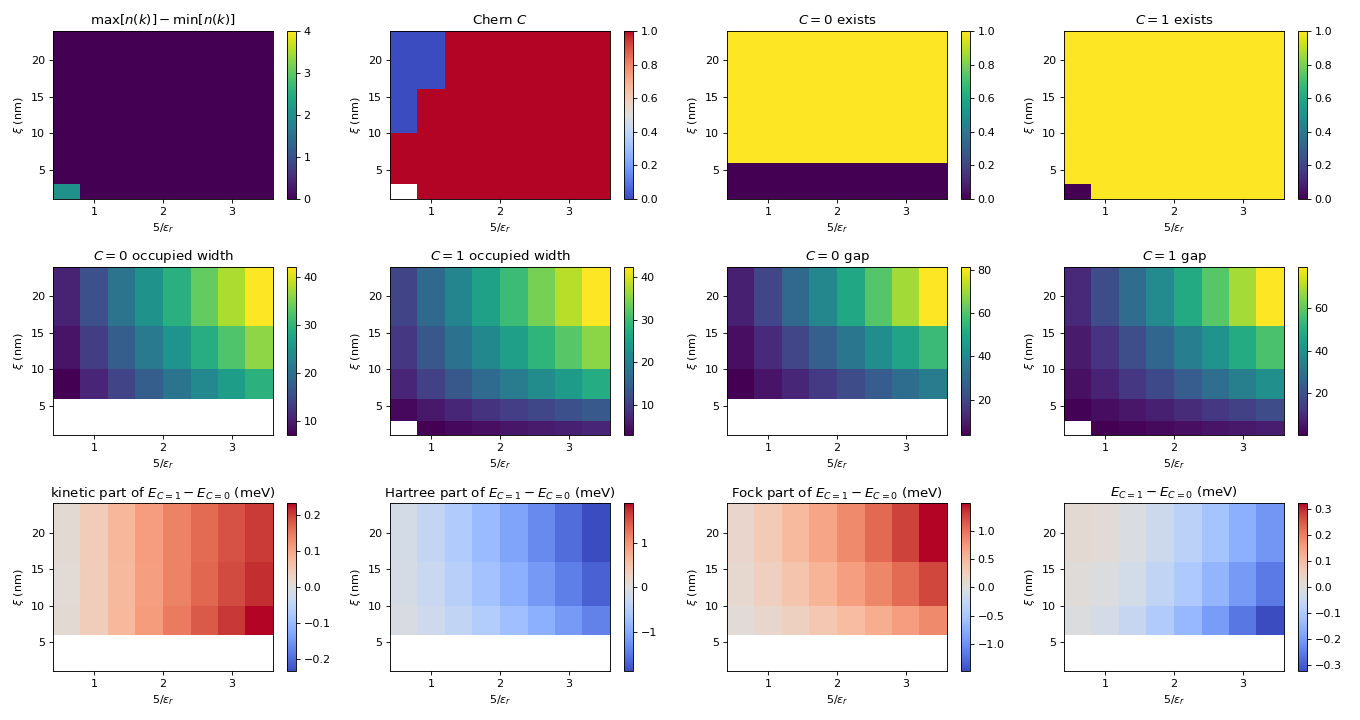}
    \caption{HF phase diagram for R$5$G. The full R$5$G Hamiltonian (Eq.~\ref{eq:H_K}) without approximations is used. No approximations on the form factors are made. Only valley $K$ and spin $\uparrow$ is included. The Hilbert space is made of conduction band states truncated based on a circular momentum cutoff. The interaction potential takes the dual-gate screened form $V_{\bm{q}}=\frac{e^2}{2\epsilon q}\tanh\frac{q\xi}{2}$. \\\textbf{System parameters}: $N_1=N_2=18$; Hilbert space cutoff radius $1.7q_1$; CN interaction scheme.}
    \label{fig:Hrad1.701_theta0.77_V0.024_gatedual_HFtypenone+0+0+none+none}
\end{figure}

\begin{figure}
    \centering
    \includegraphics[width = 0.9\linewidth]{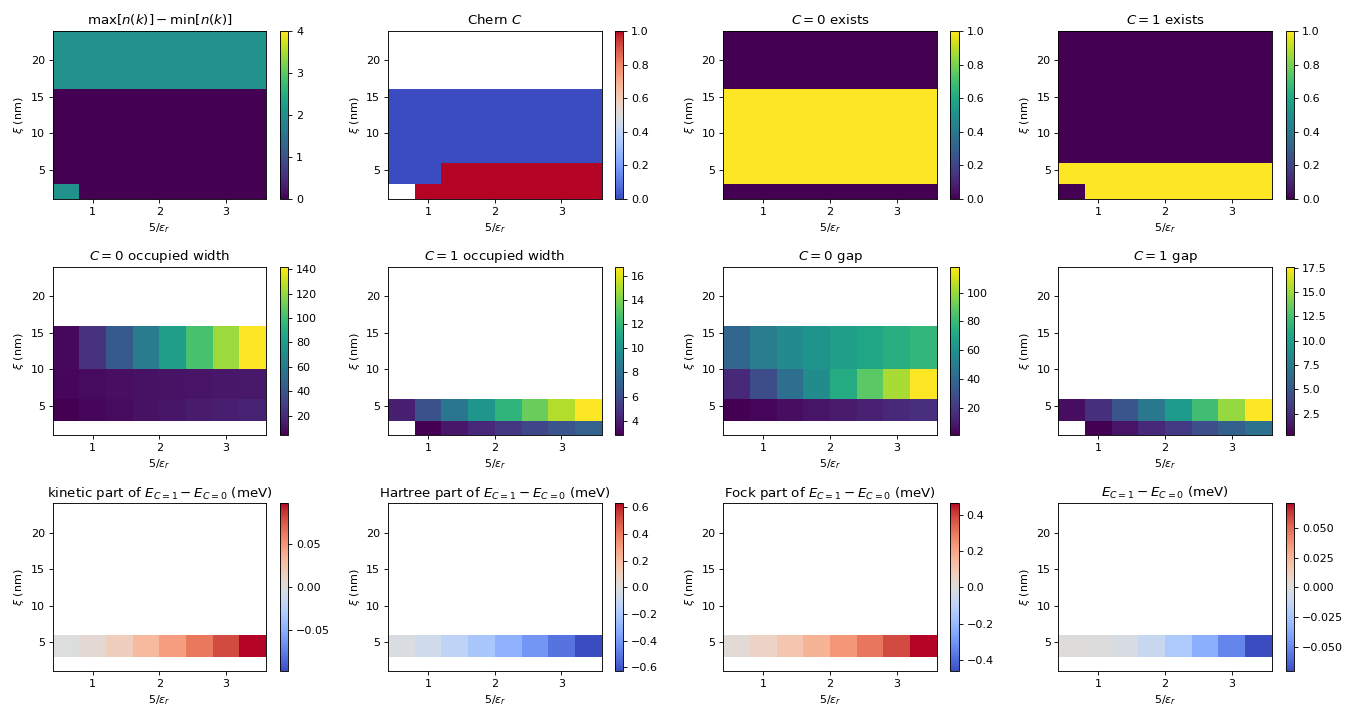}
    \caption{HF phase diagram for R$5$G. The full R$5$G Hamiltonian (Eq.~\ref{eq:H_K}) without approximations is used. No approximations on the form factors are made. Only valley $K$ and spin $\uparrow$ is included. The Hilbert space is made of conduction band states truncated based on a circular momentum cutoff. The interaction potential takes the form $V_{\bm{q}}=\frac{e^2}{2\epsilon q}\left(\frac{q\xi}{2}-\frac{q^3\xi^3}{12}\right)$, which is an expansion of the dual-gate screened interaction for small $q$. \\\textbf{System parameters}: $N_1=N_2=18$; Hilbert space cutoff radius $1.7q_1$; CN interaction scheme.}
    \label{fig:Hrad1.701_theta0.77_V0.024_gatequadratic_HFtypenone+0+0+none+none}
\end{figure}

\begin{figure}
    \centering
    \includegraphics[width = 0.9\linewidth]{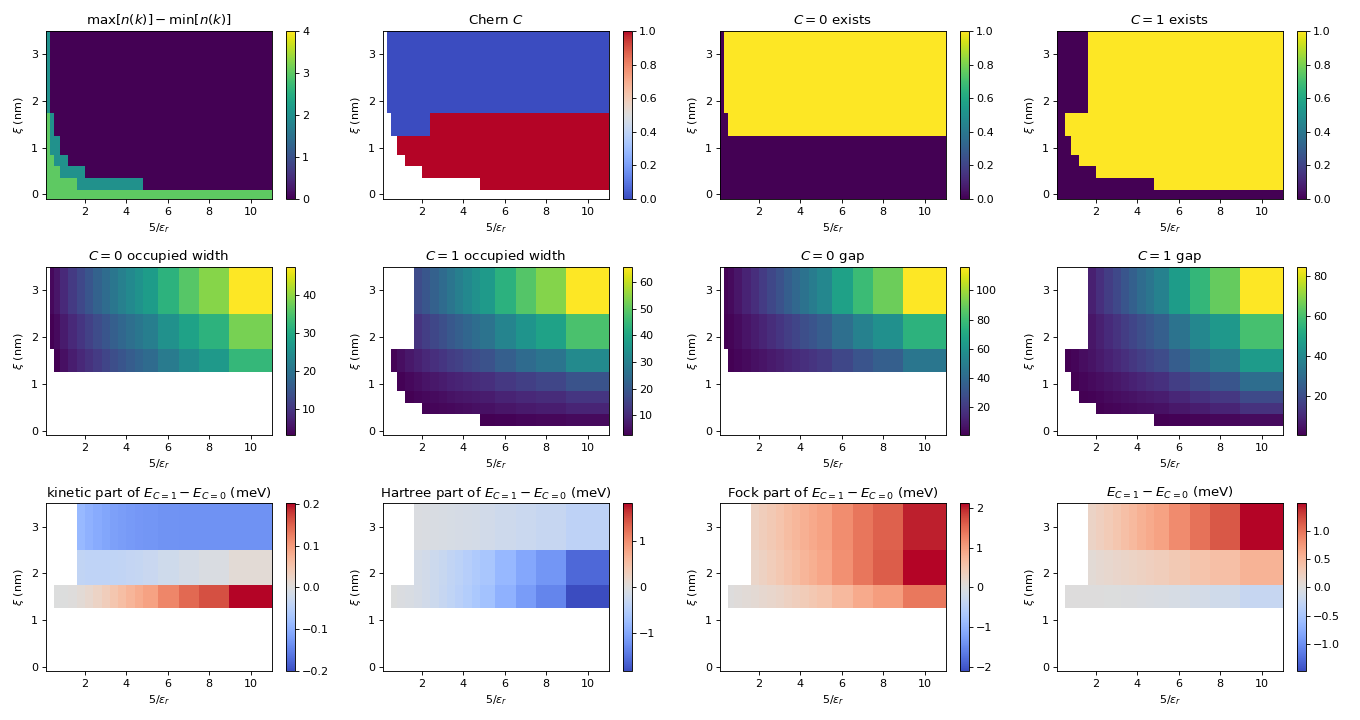}
    \caption{HF phase diagram for R$5$G. The full R$5$G Hamiltonian (Eq.~\ref{eq:H_K}) without approximations is used. No approximations on the form factors are made. Only valley $K$ and spin $\uparrow$ is included. The Hilbert space is made of conduction band states truncated based on a circular momentum cutoff. The interaction potential takes the form $V_{\bm{q}}=\frac{\xi e^2}{2\epsilon }e^{-\xi^2q^2}$. \\\textbf{System parameters}: $N_1=N_2=18$; Hilbert space cutoff radius $1.7q_1$; CN interaction scheme.}
    \label{fig:Hrad1.701_theta0.77_V0.024_gateexpmq2_HFtypenone+0+0+none+none}
\end{figure}

\begin{figure}
    \centering
    \includegraphics[width = 0.9\linewidth]{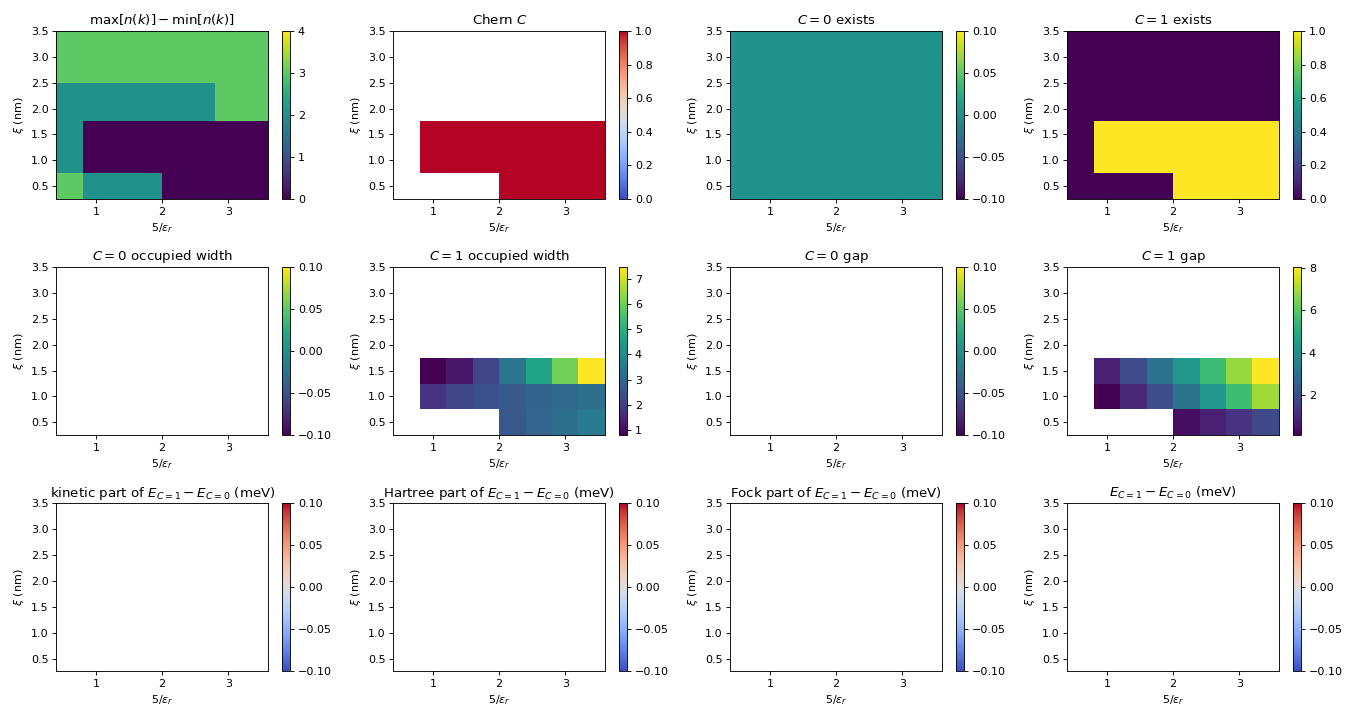}
    \caption{HF phase diagram for R$5$G. The full R$5$G Hamiltonian (Eq.~\ref{eq:H_K}) without approximations is used. No approximations on the form factors are made. Only valley $K$ and spin $\uparrow$ is included. The Hilbert space is made of conduction band states truncated based on a circular momentum cutoff. The interaction potential takes the form $V_{\bm{q}}=\frac{\xi e^2}{2\epsilon }e^{\xi^2q^2}$. \\\textbf{System parameters}: $N_1=N_2=18$; Hilbert space cutoff radius $1.7q_1$; CN interaction scheme.}
    \label{fig:Hrad1.701_theta0.77_V0.024_gateexpq2_HFtypenone+0+0+none+none}
\end{figure}

\begin{figure}
    \centering
    \includegraphics[width = 0.9\linewidth]{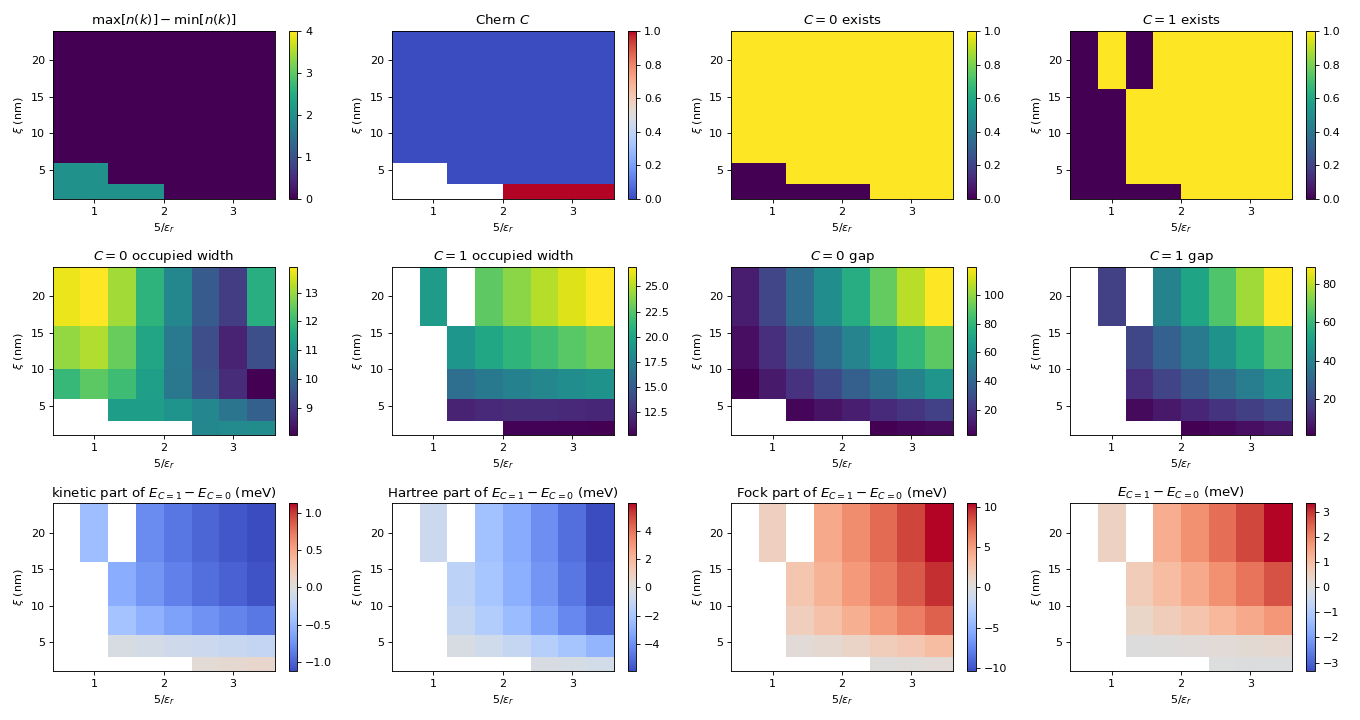}
    \caption{HF phase diagram for R$5$G. In the R$5$G Hamiltonian (Eq.~\ref{eq:H_K}), only the $v_F,t_1$ terms are kept. A $SO(2)$-symmetric dispersion correction from $v_3,v_4$ (Eq.~\ref{eqapp:v3v4_dispersion}) is included. The exponential form factor approximation (Eq.~\ref{eqapp:Mkq_exp}) is used. Only valley $K$ and spin $\uparrow$ is included. The Hilbert space is made of conduction band states truncated based on a circular momentum cutoff. The interaction potential takes the dual-gate screened form $V_{\bm{q}}=\frac{e^2}{2\epsilon q}\tanh\frac{q\xi}{2}$. \\\textbf{System parameters}: $N_1=N_2=18$; Hilbert space cutoff radius $1.7q_1$; CN interaction scheme.}
    \label{fig:Hrad1.701_theta0.77_V0.024_gatedual_HFtypet1_v0_only+1+1+exp+none}
\end{figure}

\begin{figure}
    \centering
    \includegraphics[width = 0.9\linewidth]{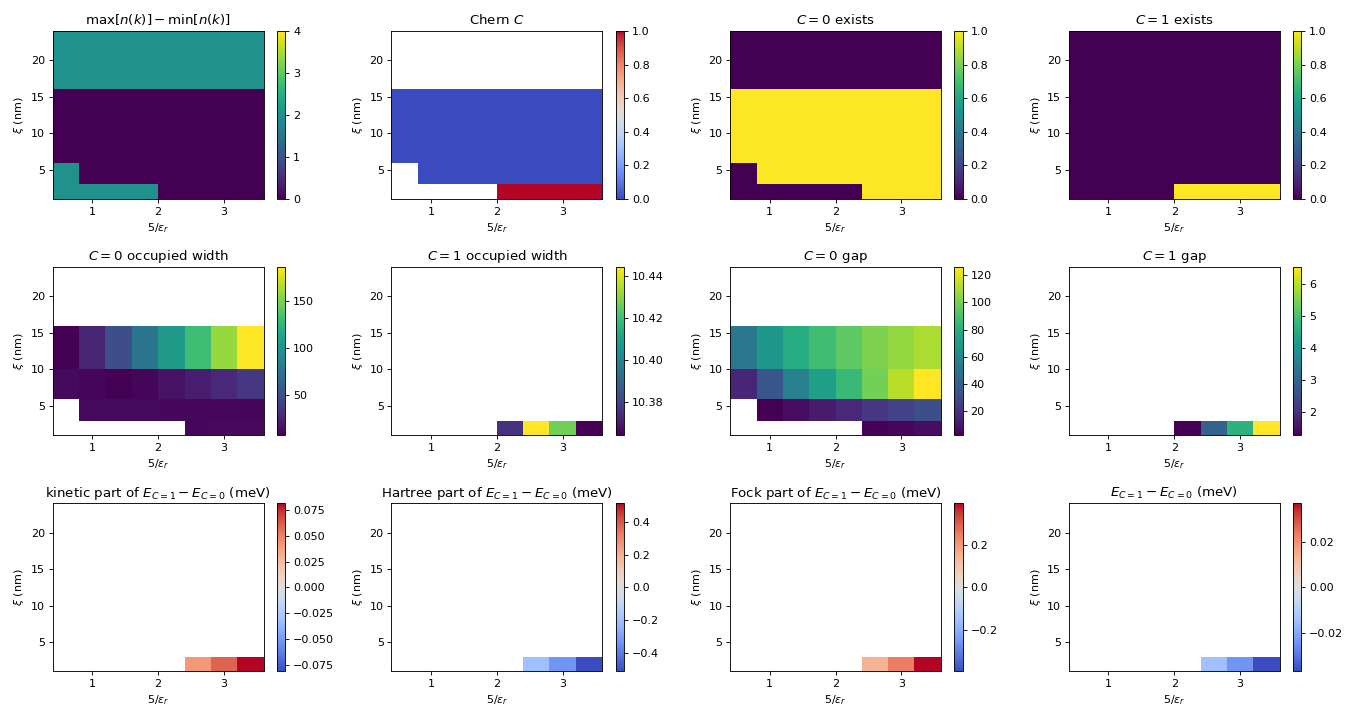}
    \caption{HF phase diagram for R$5$G. In the R$5$G Hamiltonian (Eq.~\ref{eq:H_K}), only the $v_F,t_1$ terms are kept. A $SO(2)$-symmetric dispersion correction from $v_3,v_4$ (Eq.~\ref{eqapp:v3v4_dispersion}) is included. The exponential form factor approximation (Eq.~\ref{eqapp:Mkq_exp}) is used. Only valley $K$ and spin $\uparrow$ is included. The Hilbert space is made of conduction band states truncated based on a circular momentum cutoff. The interaction potential takes the form $V_{\bm{q}}=\frac{e^2}{2\epsilon q}\left(\frac{q\xi}{2}-\frac{q^3\xi^3}{12}\right)$, which is an expansion of the dual-gate screened interaction for small $q$. \\\textbf{System parameters}: $N_1=N_2=18$; Hilbert space cutoff radius $1.7q_1$; CN interaction scheme.}
    \label{fig:Hrad1.701_theta0.77_V0.024_gatequadratic_HFtypet1_v0_only+1+1+exp+none}
\end{figure}
\begin{figure}
    \centering
    \includegraphics[width = 0.9\linewidth]{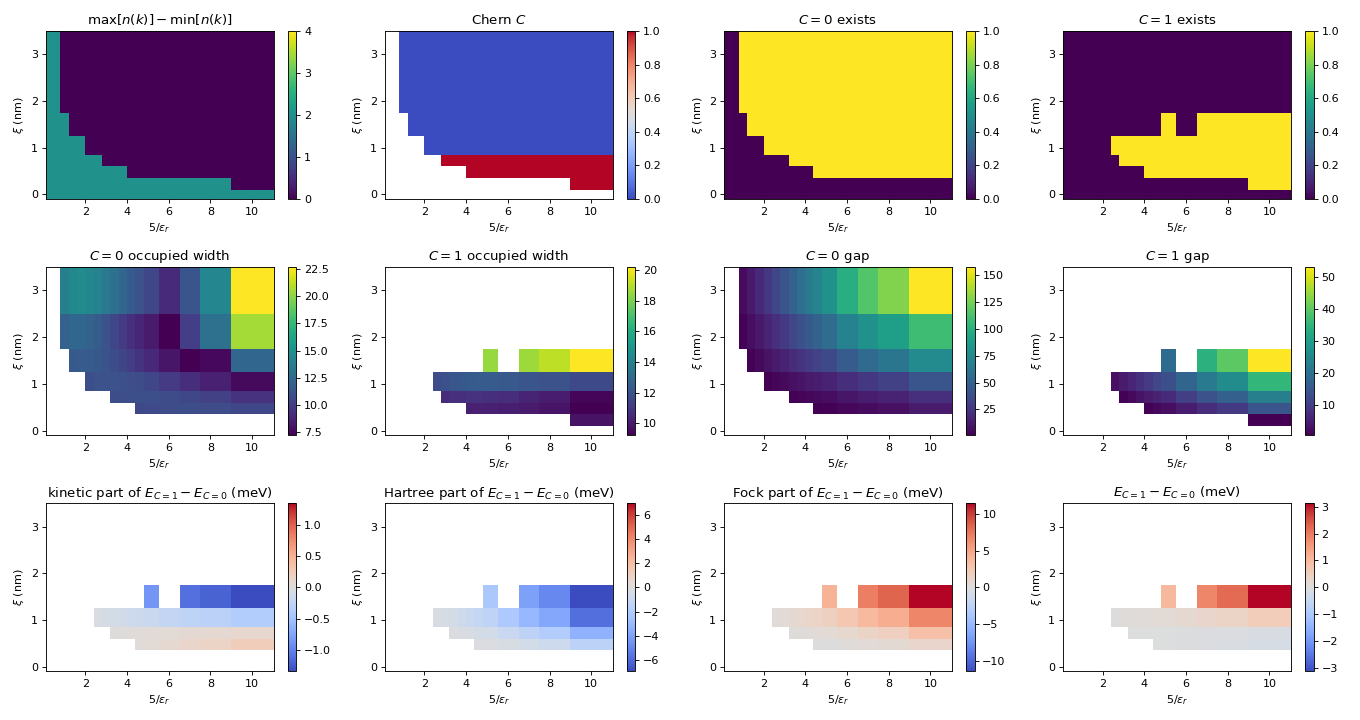}
    \caption{HF phase diagram for R$5$G. In the R$5$G Hamiltonian (Eq.~\ref{eq:H_K}), only the $v_F,t_1$ terms are kept. A $SO(2)$-symmetric dispersion correction from $v_3,v_4$ (Eq.~\ref{eqapp:v3v4_dispersion}) is included. The exponential form factor approximation (Eq.~\ref{eqapp:Mkq_exp}) is used. Only valley $K$ and spin $\uparrow$ is included. The Hilbert space is made of conduction band states truncated based on a circular momentum cutoff. The interaction potential takes the form $V_{\bm{q}}=\frac{\xi e^2}{2\epsilon }e^{-\xi^2q^2}$. \\\textbf{System parameters}: $N_1=N_2=18$; Hilbert space cutoff radius $1.7q_1$; CN interaction scheme.}
    \label{fig:Hrad1.701_theta0.77_V0.024_gateexpmq2_HFtypet1_v0_only+1+1+exp+none}
\end{figure}

\begin{figure}
    \centering
    \includegraphics[width = 0.9\linewidth]{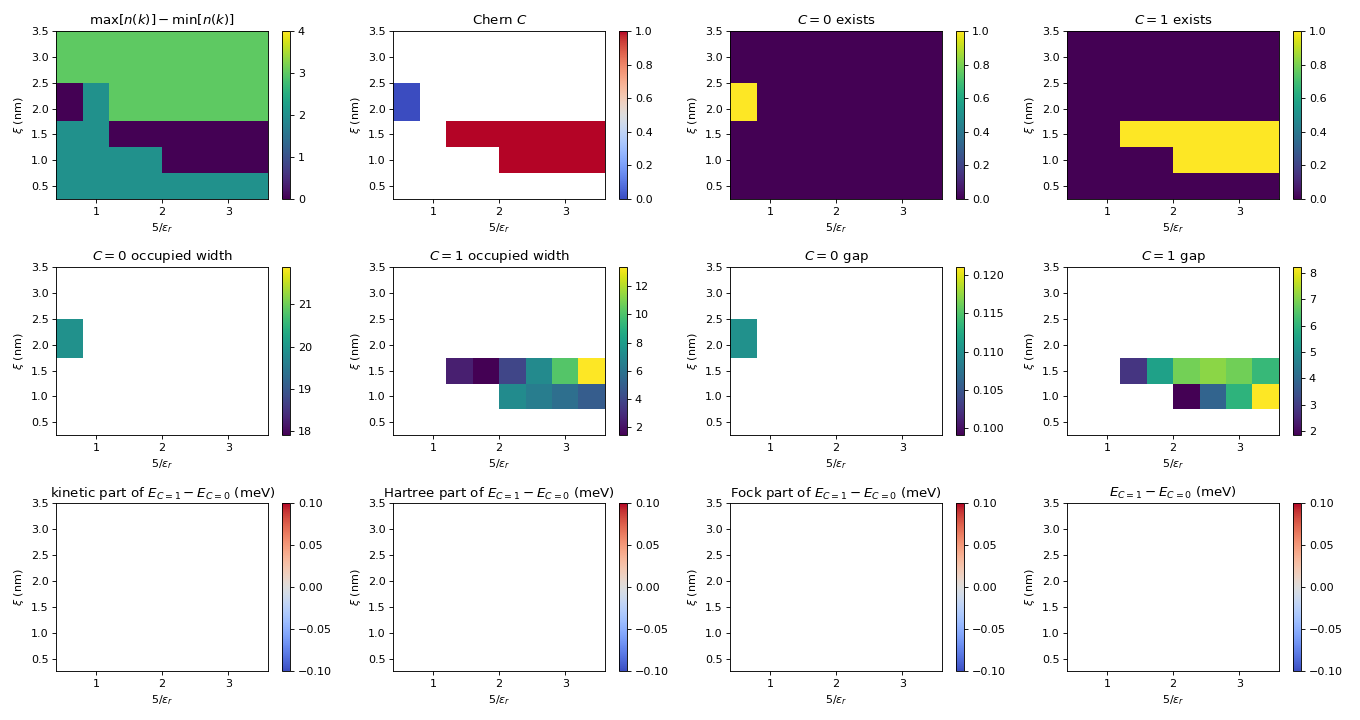}
    \caption{HF phase diagram for R$5$G. In the R$5$G Hamiltonian (Eq.~\ref{eq:H_K}), only the $v_F,t_1$ terms are kept. A $SO(2)$-symmetric dispersion correction from $v_3,v_4$ (Eq.~\ref{eqapp:v3v4_dispersion}) is included. The exponential form factor approximation (Eq.~\ref{eqapp:Mkq_exp}) is used. Only valley $K$ and spin $\uparrow$ is included. The Hilbert space is made of conduction band states truncated based on a circular momentum cutoff. The interaction potential takes the form $V_{\bm{q}}=\frac{\xi e^2}{2\epsilon }e^{\xi^2q^2}$. \\\textbf{System parameters}: $N_1=N_2=18$; Hilbert space cutoff radius $1.701q_1$; CN interaction scheme.}
    \label{fig:Hrad1.701_theta0.77_V0.024_gateexpq2_HFtypet1_v0_only+1+1+exp+none}
\end{figure}

\subsection{Self-consistent Hartree-Fock band structures}\label{secapp:HFbandstructure}

In this appendix subsection, we study the self-consistent numerical HF band structures of the Wigner crystals obtained in the phase diagrams of App.~\ref{secapp:HFphasediagrams}. We employ a circular Hilbert space cutoff in this subsection. We first outline the information presented in each of Figs.~\ref{fig:N18_t0.77_Hrad1.301_U0.024_pstrnone+0+0+none+none_epsr5.00_singleplotC1} to \ref{fig:N18_t0.77_Hrad1.701_U0.024_pstrt1_v0_only+1+1+exp+none_epsr5.00_singleplotC0}. The kinetic energy is plotted in the BZ. Since the Hamiltonians studied here have no moir\'e potential, the kinetic energy can also be plotted as a function of the absolute value of the unfolded momentum $k$ for all $\bm{k}$ within the momentum cutoff. If there is no $SO(2)$ symmetry, then generally this plot will show some spread for a fixed magnitude $k$. The HF band structure is plotted in the BZ. Finally, we show a color plot of the (logarithm of the) expectation value of the occupation $n(\bm{k})$ in the HF ground state as a function of $\bm{k}$. This color plot also shows visually the scale of the Hilbert space cutoff relative to the BZ. 

\subsubsection{$C=1$ band structure and properties}\label{subsecapp:C1bandstructure}

Here, we discuss results for the $C=1$ phase: 
\begin{itemize}
    \item In Fig.~\ref{fig:N18_t0.77_Hrad1.301_U0.024_pstrnone+0+0+none+none_epsr5.00_singleplotC1}, we use the full pentalayer Hamiltonian (Eq.~\ref{eq:H_K}) and a small cutoff radius $1.3q_1$ (note that the $K_M$ point is at radius $q_1$). While the lowest non-interacting band is flat, the Fock self-energy of the filled HF band leads to a large dip around $\Gamma_M$ (see e.g.~later discussion around Eq.~\ref{eqapp:H'k0} for a derivation of this feature within an analytical calculation).  In Fig.~\ref{fig:N18_t0.77_Hrad1.701_U0.024_pstrnone+0+0+none+none_epsr5.00_singleplotC1}, we increase the cutoff radius to $1.7q_1$, which more fully resolves the dramatic steepening of the kinetic energy for $k/q_1\gtrsim 1$. This is clear from comparing the kinetic and HF dispersion of the lowest two bands when following the path $\Gamma_M\rightarrow M_M\rightarrow \Gamma_M$. The HF band structure is qualitatively similar to that using the smaller cutoff, but the $n(\bm{k})$ plot shows clearly that the occupation is quickly suppressed outside of the first BZ. 
\item
In Figs.~\ref{fig:N18_t0.77_Hrad1.301_U0.024_pstrt1_v0_v34_only+0+0+none+none_epsr5.00_singleplotC1} and \ref{fig:N18_t0.77_Hrad1.701_U0.024_pstrt1_v0_v34_only+0+0+none+none_epsr5.00_singleplotC1}, we repeat the calculations except we neglect $t_2$ and $V_{ISP}$ in the pentalayer Hamiltonian. This leads to a small increase in the velocity of the band for $k/q_1 < 1$. The HF band structure and momentum occupations remain similar.
\item
In Figs.~\ref{fig:N18_t0.77_Hrad1.301_U0.024_pstrt1_v0_only+1+1+exp+none_epsr5.00_singleplotC1} and \ref{fig:N18_t0.77_Hrad1.701_U0.024_pstrt1_v0_only+1+1+exp+none_epsr5.00_singleplotC1}, we show corresponding results for the case that only $v_F$ and $t_1$ are kept in the pentalayer Hamiltonian (Eq.~\ref{eqapp:chiralham}), a $SO(2)$-symmetric $v_3,v_4$ correction to the dispersion is included in perturbation theory, and the exponential form factor approximation is used (Eq.~\ref{eqapp:Mkq_exp}). Because of the $SO(2)$ symmetry, the kinetic energy is only a function of the magnitude $k$. We note that despite all the approximations to the single-particle model and the form factors, the HF band structure and momentum occupations remain similar. 
\end{itemize}

\begin{figure}
    \centering
    \includegraphics[width = 1.0\linewidth]{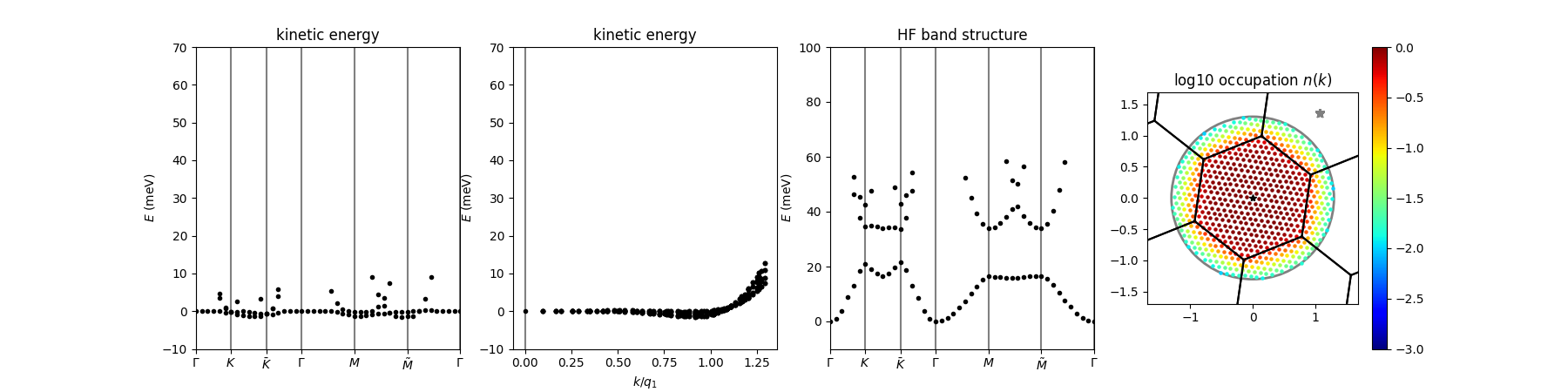}
    \caption{HF band structure and properties of the $C=1$ insulator. The full R$5$G Hamiltonian (Eq.~\ref{eq:H_K}) without approximations is used. No approximations on the form factors are made. Only valley $K$ and spin $\uparrow$ is included. The Hilbert space is made of conduction band states truncated based on a circular momentum cutoff. \\\textbf{System parameters}: $N_1=N_2=18$; Hilbert space cutoff radius $1.3q_1$; CN interaction scheme; $\epsilon=5$; $\xi=20\,\text{nm}$}
    \label{fig:N18_t0.77_Hrad1.301_U0.024_pstrnone+0+0+none+none_epsr5.00_singleplotC1}
\end{figure}

\begin{figure}
    \centering
    \includegraphics[width = 1.0\linewidth]{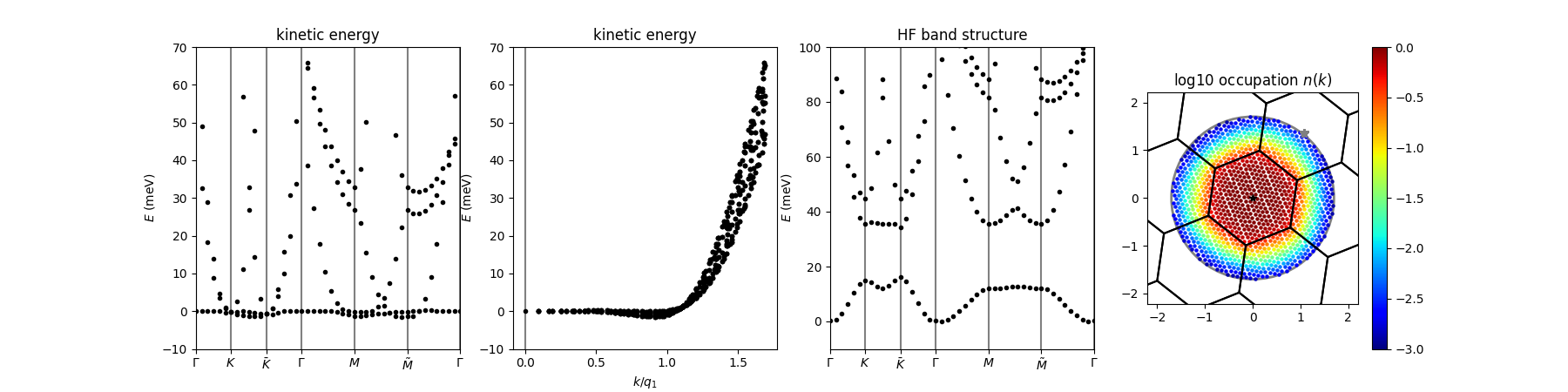}
    \caption{HF band structure and properties of the $C=1$ insulator. The full R$5$G Hamiltonian (Eq.~\ref{eq:H_K}) without approximations is used. No approximations on the form factors are made. Only valley $K$ and spin $\uparrow$ is included. The Hilbert space is made of conduction band states truncated based on a circular momentum cutoff. \\\textbf{System parameters}: $N_1=N_2=18$; Hilbert space cutoff radius $1.7q_1$; CN interaction scheme; $\epsilon=5$; $\xi=20\,\text{nm}$; $\theta=0.77^\circ$; $V=24\,\text{meV}$}
    \label{fig:N18_t0.77_Hrad1.701_U0.024_pstrnone+0+0+none+none_epsr5.00_singleplotC1}
\end{figure}

\begin{figure}
    \centering
    \includegraphics[width = 1.0\linewidth]{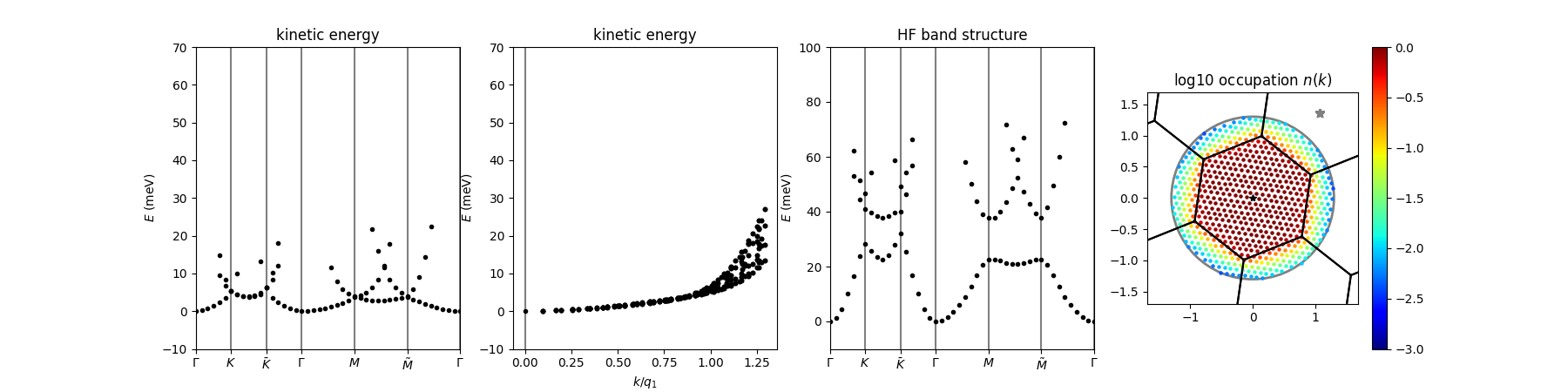}
    \caption{HF band structure and properties of the $C=1$ insulator. In the R$5$G Hamiltonian (Eq.~\ref{eq:H_K}), only the $v_F,t_1,v_3,v_4$ terms are kept. No approximations on the form factors are made. Only valley $K$ and spin $\uparrow$ is included. The Hilbert space is made of conduction band states truncated based on a circular momentum cutoff. \\\textbf{System parameters}: $N_1=N_2=18$; Hilbert space cutoff radius $1.3q_1$; CN interaction scheme; $\epsilon=5$; $\xi=20\,\text{nm}$; $\theta=0.77^\circ$; $V=24\,\text{meV}$}
    \label{fig:N18_t0.77_Hrad1.301_U0.024_pstrt1_v0_v34_only+0+0+none+none_epsr5.00_singleplotC1}
\end{figure}

\begin{figure}
    \centering
    \includegraphics[width = 1.0\linewidth]{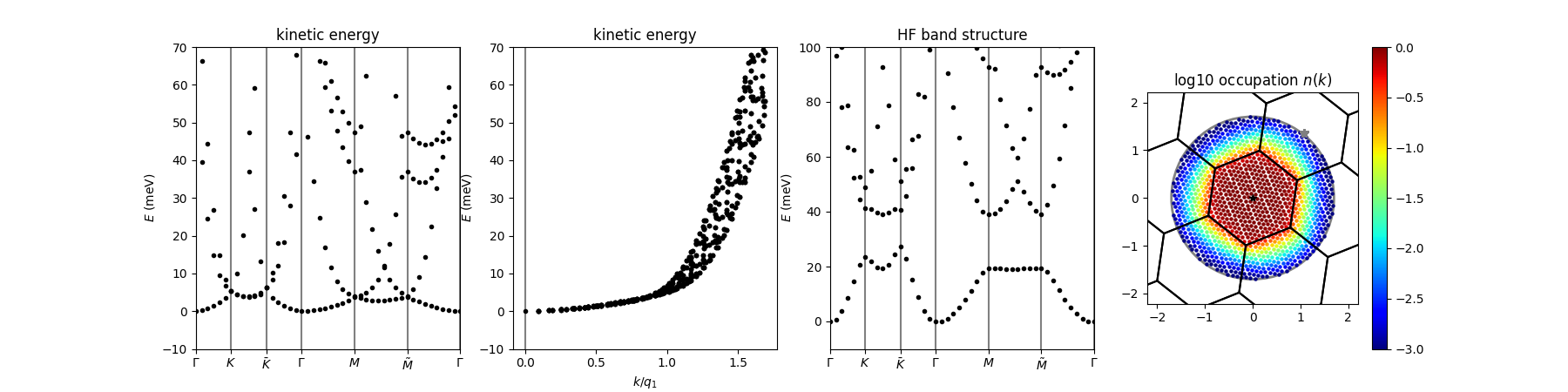}
    \caption{HF band structure and properties of the $C=1$ insulator. In the R$5$G Hamiltonian (Eq.~\ref{eq:H_K}), only the $v_F,t_1,v_3,v_4$ terms are kept. No approximations on the form factors are made. Only valley $K$ and spin $\uparrow$ is included. The Hilbert space is made of conduction band states truncated based on a circular momentum cutoff. \\\textbf{System parameters}: $N_1=N_2=18$; Hilbert space cutoff radius $1.7q_1$; CN interaction scheme; $\epsilon=5$; $\xi=20\,\text{nm}$; $\theta=0.77^\circ$; $V=24\,\text{meV}$}
    \label{fig:N18_t0.77_Hrad1.701_U0.024_pstrt1_v0_v34_only+0+0+none+none_epsr5.00_singleplotC1}
\end{figure}

\begin{figure}
    \centering
    \includegraphics[width = 1.0\linewidth]{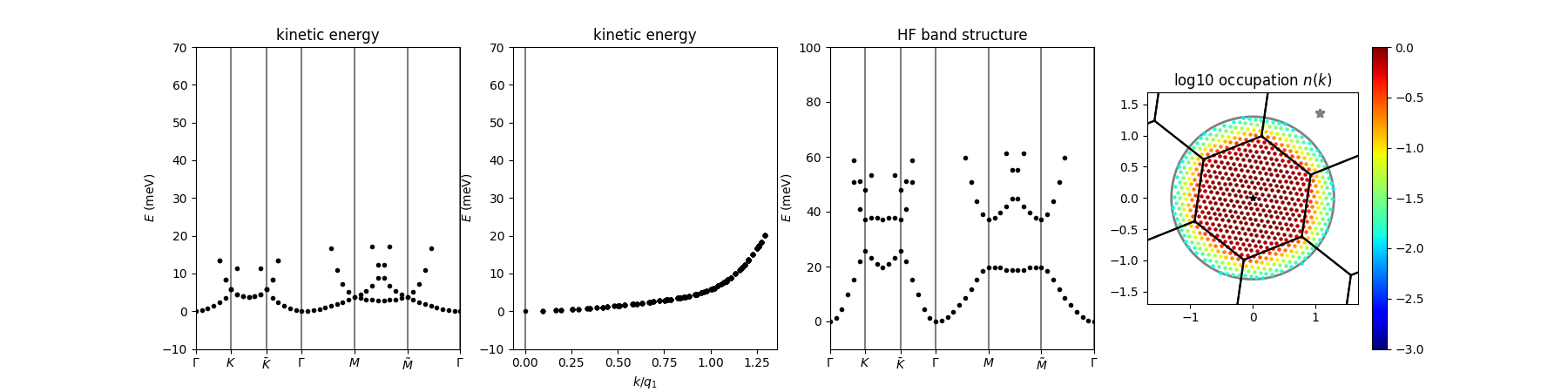}
    \caption{HF band structure and properties of the $C=1$ insulator. In the R$5$G Hamiltonian (Eq.~\ref{eq:H_K}), only the $v_F,t_1$ terms are kept. A $SO(2)$-symmetric dispersion correction from $v_3,v_4$ (Eq.~\ref{eqapp:v3v4_dispersion}) is included. The exponential form factor approximation (Eq.~\ref{eqapp:Mkq_exp}) is used. Only valley $K$ and spin $\uparrow$ is included. The Hilbert space is made of conduction band states truncated based on a circular momentum cutoff. \\\textbf{System parameters}: $N_1=N_2=18$; Hilbert space cutoff radius $1.3q_1$; CN interaction scheme; $\epsilon=5$; $\xi=20\,\text{nm}$; $\theta=0.77^\circ$; $V=24\,\text{meV}$}
    \label{fig:N18_t0.77_Hrad1.301_U0.024_pstrt1_v0_only+1+1+exp+none_epsr5.00_singleplotC1}
\end{figure}

\begin{figure}
    \centering
    \includegraphics[width = 1.0\linewidth]{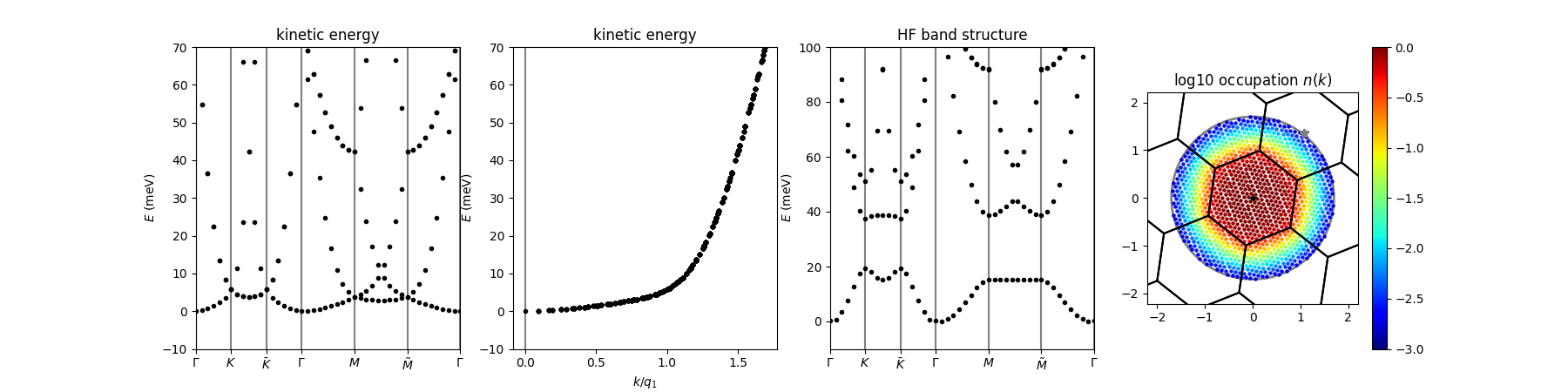}
    \caption{HF band structure and properties of the $C=1$ insulator. In the R$5$G Hamiltonian (Eq.~\ref{eq:H_K}), only the $v_F,t_1$ terms are kept. A $SO(2)$-symmetric dispersion correction from $v_3,v_4$ (Eq.~\ref{eqapp:v3v4_dispersion}) is included. The exponential form factor approximation (Eq.~\ref{eqapp:Mkq_exp}) is used. Only valley $K$ and spin $\uparrow$ is included. The Hilbert space is made of conduction band states truncated based on a circular momentum cutoff. \\\textbf{System parameters}: $N_1=N_2=18$; Hilbert space cutoff radius $1.7q_1$; CN interaction scheme; $\epsilon=5$; $\xi=20\,\text{nm}$; $\theta=0.77^\circ$; $V=24\,\text{meV}$}
    \label{fig:N18_t0.77_Hrad1.701_U0.024_pstrt1_v0_only+1+1+exp+none_epsr5.00_singleplotC1}
\end{figure}

\clearpage

\subsubsection{$C=0$ band structure and properties}\label{subsecapp:C0bandstructure}

Here, we  repeat the calculations of App.~\ref{subsecapp:C1bandstructure} except that we show results for the $C=0$ phase. In particular
\begin{itemize}
    \item Figs.~\ref{fig:N18_t0.77_Hrad1.301_U0.024_pstrnone+0+0+none+none_epsr5.00_singleplotC0} and \ref{fig:N18_t0.77_Hrad1.701_U0.024_pstrnone+0+0+none+none_epsr5.00_singleplotC0} are the analogs of Figs.~\ref{fig:N18_t0.77_Hrad1.301_U0.024_pstrnone+0+0+none+none_epsr5.00_singleplotC1} and \ref{fig:N18_t0.77_Hrad1.701_U0.024_pstrnone+0+0+none+none_epsr5.00_singleplotC1} for the case of the full pentalayer Hamiltonian.  
    \item Figs.~\ref{fig:N18_t0.77_Hrad1.301_U0.024_pstrt1_v0_v34_only+0+0+none+none_epsr5.00_singleplotC0} and \ref{fig:N18_t0.77_Hrad1.701_U0.024_pstrt1_v0_v34_only+0+0+none+none_epsr5.00_singleplotC0} are the analogs of Figs.~\ref{fig:N18_t0.77_Hrad1.301_U0.024_pstrt1_v0_v34_only+0+0+none+none_epsr5.00_singleplotC1} and \ref{fig:N18_t0.77_Hrad1.701_U0.024_pstrt1_v0_v34_only+0+0+none+none_epsr5.00_singleplotC1} for the case where $t_2$ and $V_{ISP}$ are neglected.
    \item Figs.~\ref{fig:N18_t0.77_Hrad1.301_U0.024_pstrt1_v0_only+1+1+exp+none_epsr5.00_singleplotC0} and \ref{fig:N18_t0.77_Hrad1.701_U0.024_pstrt1_v0_only+1+1+exp+none_epsr5.00_singleplotC0} are the analogs of Figs.~\ref{fig:N18_t0.77_Hrad1.301_U0.024_pstrt1_v0_only+1+1+exp+none_epsr5.00_singleplotC1} and \ref{fig:N18_t0.77_Hrad1.701_U0.024_pstrt1_v0_only+1+1+exp+none_epsr5.00_singleplotC1} for the case that only $v_F$ and $t_1$ are kept in the pentalayer Hamiltonian (Eq.~\ref{eqapp:chiralham}), a $SO(2)$-symmetric $v_3,v_4$ correction to the dispersion is included in perturbation theory, and the exponential form factor approximation is used (Eq.~\ref{eqapp:Mkq_exp}).
\end{itemize}
In all cases, the HF band structure and occupation number $n(\bm{k})$ are quantitatively similar between the $C=0$ and $C=1$ states. This highlights the fact that the two states closely resemble each other from the perspective of the mean-field dispersion, especially for parameters where they are in close energy competition, and their distinction lies in the phase structure of the order parameter and the gap opening (see App.~\ref{secapp:HFwavefunctions}).

\begin{figure}
    \centering
    \includegraphics[width = 1.0\linewidth]{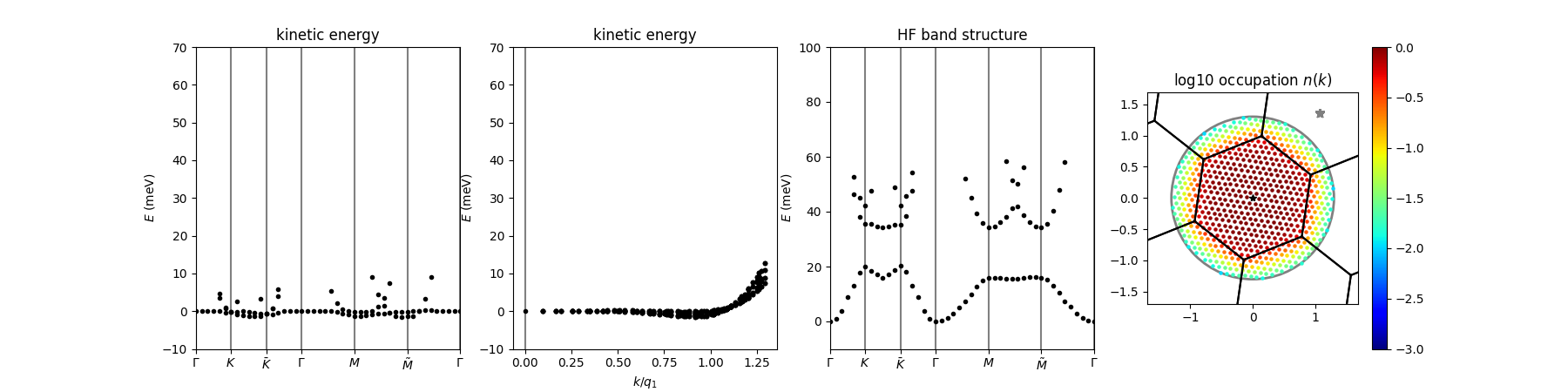}
    \caption{HF band structure and properties of the $C=0$ insulator. The full R$5$G Hamiltonian (Eq.~\ref{eq:H_K}) without approximations is used. No approximations on the form factors are made. Only valley $K$ and spin $\uparrow$ is included. The Hilbert space is made of conduction band states truncated based on a circular momentum cutoff. \\\textbf{System parameters}: $N_1=N_2=18$; Hilbert space cutoff radius $1.3q_1$; CN interaction scheme; $\epsilon=5$; $\xi=20\,\text{nm}$; $\theta=0.77^\circ$; $V=24\,\text{meV}$}
    \label{fig:N18_t0.77_Hrad1.301_U0.024_pstrnone+0+0+none+none_epsr5.00_singleplotC0}
\end{figure}

\begin{figure}
    \centering
    \includegraphics[width = 1.0\linewidth]{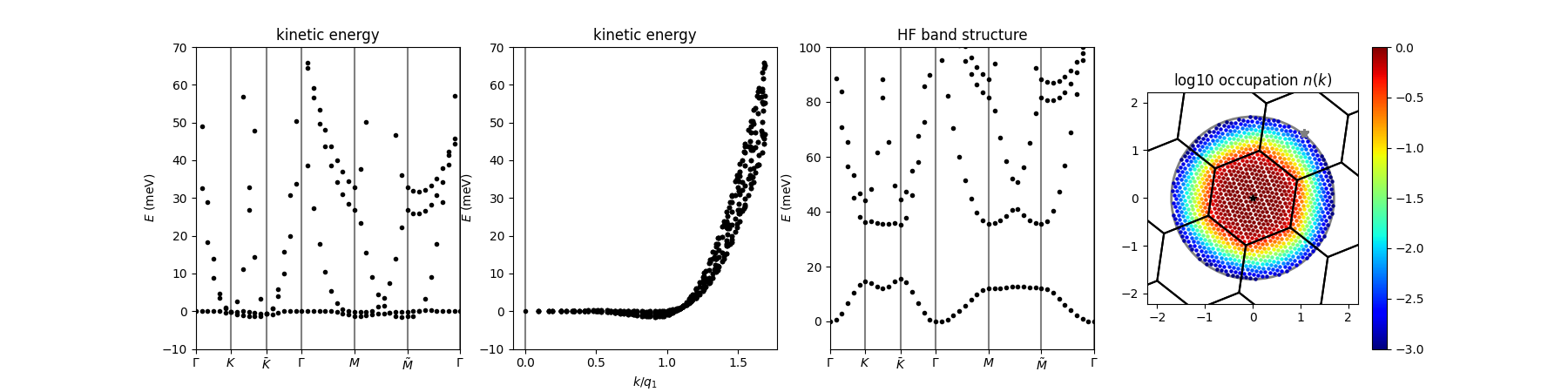}
    \caption{HF band structure and properties of the $C=0$ insulator. The full R$5$G Hamiltonian (Eq.~\ref{eq:H_K}) without approximations is used. No approximations on the form factors are made. Only valley $K$ and spin $\uparrow$ is included. The Hilbert space is made of conduction band states truncated based on a circular momentum cutoff. \\\textbf{System parameters}: $N_1=N_2=18$; Hilbert space cutoff radius $1.7q_1$; CN interaction scheme; $\epsilon=5$; $\xi=20\,\text{nm}$; $\theta=0.77^\circ$; $V=24\,\text{meV}$}
    \label{fig:N18_t0.77_Hrad1.701_U0.024_pstrnone+0+0+none+none_epsr5.00_singleplotC0}
\end{figure}

\begin{figure}
    \centering
    \includegraphics[width = 1.0\linewidth]{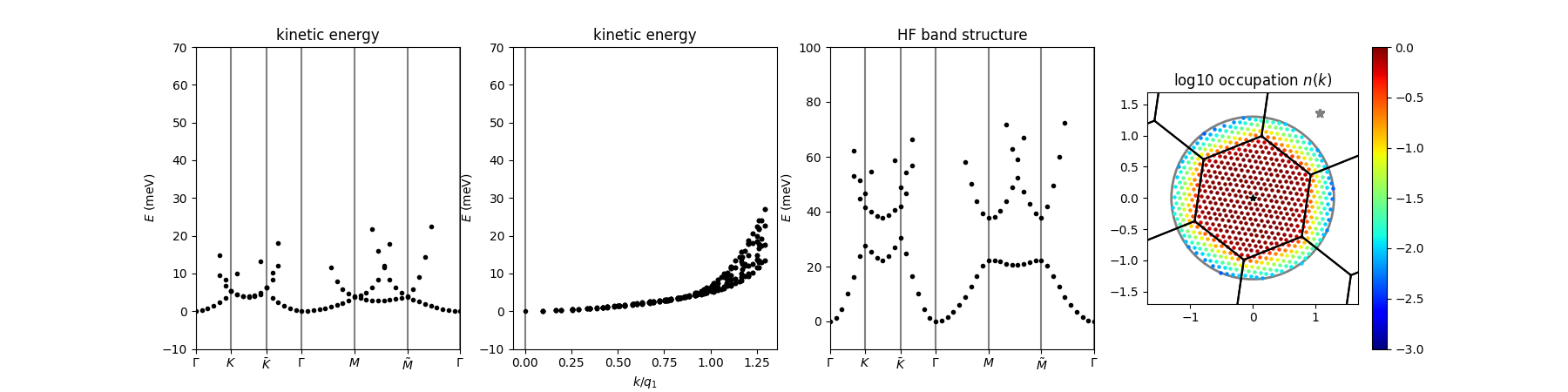}
    \caption{HF band structure and properties of the $C=0$ insulator. In the R$5$G Hamiltonian (Eq.~\ref{eq:H_K}), only the $v_F,t_1,v_3,v_4$ terms are kept. No approximations on the form factors are made. Only valley $K$ and spin $\uparrow$ is included. The Hilbert space is made of conduction band states truncated based on a circular momentum cutoff. \\\textbf{System parameters}: $N_1=N_2=18$; Hilbert space cutoff radius $1.3q_1$; CN interaction scheme; $\epsilon=5$; $\xi=20\,\text{nm}$; $\theta=0.77^\circ$; $V=24\,\text{meV}$}
    \label{fig:N18_t0.77_Hrad1.301_U0.024_pstrt1_v0_v34_only+0+0+none+none_epsr5.00_singleplotC0}
\end{figure}

\begin{figure}
    \centering
    \includegraphics[width = 1.0\linewidth]{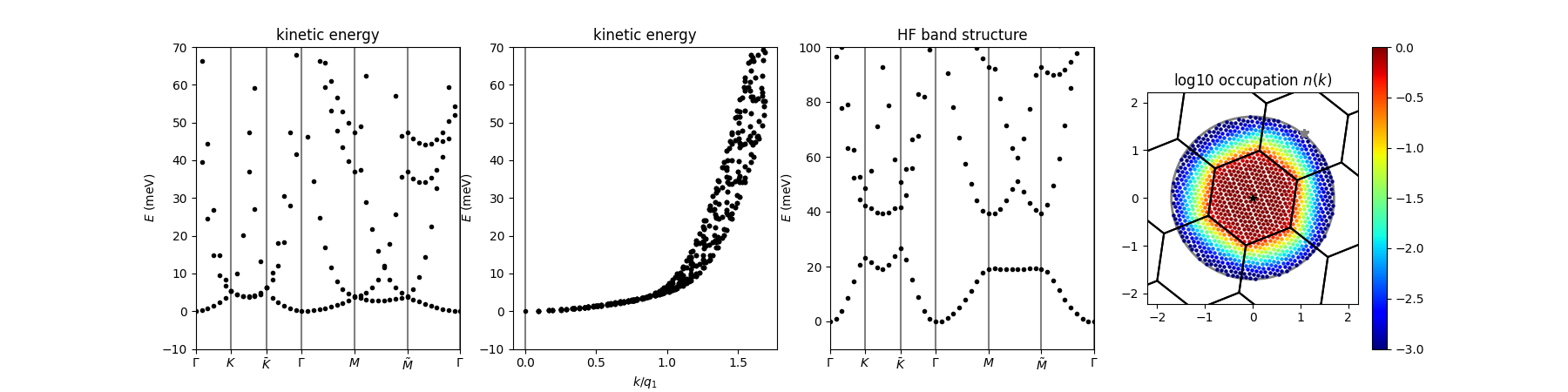}
    \caption{HF band structure and properties of the $C=0$ insulator. In the R$5$G Hamiltonian (Eq.~\ref{eq:H_K}), only the $v_F,t_1,v_3,v_4$ terms are kept. No approximations on the form factors are made. Only valley $K$ and spin $\uparrow$ is included. The Hilbert space is made of conduction band states truncated based on a circular momentum cutoff. \\\textbf{System parameters}: $N_1=N_2=18$; Hilbert space cutoff radius $1.7q_1$; CN interaction scheme; $\epsilon=5$; $\xi=20\,\text{nm}$; $\theta=0.77^\circ$; $V=24\,\text{meV}$}
    \label{fig:N18_t0.77_Hrad1.701_U0.024_pstrt1_v0_v34_only+0+0+none+none_epsr5.00_singleplotC0}
\end{figure}

\begin{figure}
    \centering
    \includegraphics[width = 1.0\linewidth]{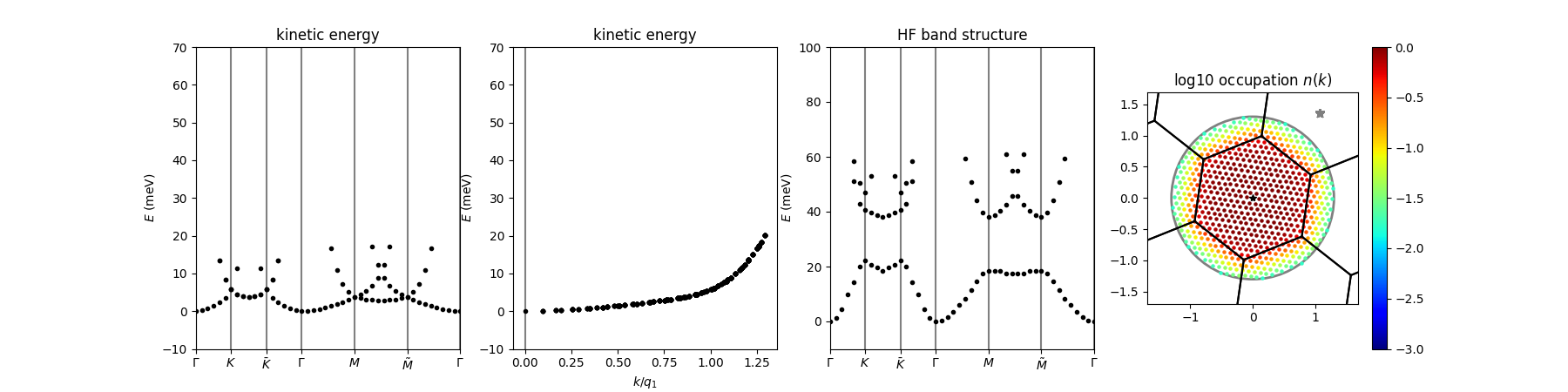}
    \caption{HF band structure and properties of the $C=0$ insulator. In the R$5$G Hamiltonian (Eq.~\ref{eq:H_K}), only the $v_F,t_1$ terms are kept. A $SO(2)$-symmetric dispersion correction from $v_3,v_4$ (Eq.~\ref{eqapp:v3v4_dispersion}) is included. The exponential form factor approximation (Eq.~\ref{eqapp:Mkq_exp}) is used. Only valley $K$ and spin $\uparrow$ is included. The Hilbert space is made of conduction band states truncated based on a circular momentum cutoff. \\\textbf{System parameters}: $N_1=N_2=18$; Hilbert space cutoff radius $1.3q_1$; CN interaction scheme; $\epsilon=5$; $\xi=20\,\text{nm}$; $\theta=0.77^\circ$; $V=24\,\text{meV}$}
    \label{fig:N18_t0.77_Hrad1.301_U0.024_pstrt1_v0_only+1+1+exp+none_epsr5.00_singleplotC0}
\end{figure}

\begin{figure}
    \centering
    \includegraphics[width = 1.0\linewidth]{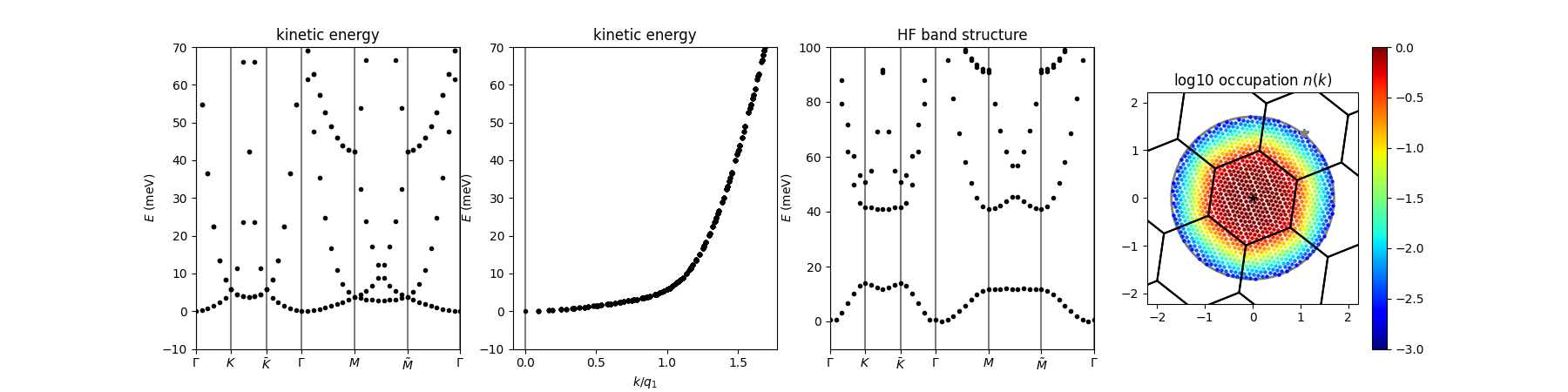}
    \caption{HF band structure and properties of the $C=0$ insulator. In the R$5$G Hamiltonian (Eq.~\ref{eq:H_K}), only the $v_F,t_1$ terms are kept. A $SO(2)$-symmetric dispersion correction from $v_3,v_4$ (Eq.~\ref{eqapp:v3v4_dispersion}) is included. The exponential form factor approximation (Eq.~\ref{eqapp:Mkq_exp}) is used. Only valley $K$ and spin $\uparrow$ is included. The Hilbert space is made of conduction band states truncated based on a circular momentum cutoff. \\\textbf{System parameters}: $N_1=N_2=18$; Hilbert space cutoff radius $1.7q_1$; CN interaction scheme; $\epsilon=5$; $\xi=20\,\text{nm}$; $\theta=0.77^\circ$; $V=24\,\text{meV}$}
    \label{fig:N18_t0.77_Hrad1.701_U0.024_pstrt1_v0_only+1+1+exp+none_epsr5.00_singleplotC0}
\end{figure}

\clearpage

\subsection{Self-consistent Hartree-Fock wavefunctions}\label{secapp:HFwavefunctions}

In this appendix subsection, we show in more detail properties of the HF wavefunctions of the Wigner crystals, as well as the mean-field hybridization field $g_1(\bm{k})$ that couples Bloch states with momenta $\bm{k}$ and $\bm{k}+\bm{b}_1$. The calculations in this subsection employ a circular Hilbert space cutoff. Compared to the previous calculations in this section, we rotate the hexagonal BZ to align with the $k_x$ and $k_y$ axes for convenience. In particular, one pair of edges of the BZ is oriented parallel the to $k_x$ axis. (This orientation of the BZ is similar to that of $\theta\approx0.77^\circ$ R$n$G/hBN.) This convention is also used in the analytical study of App.~\ref{secapp:2dmodel_setup}, \ref{secapp:2dmodel_HF} and \ref{secapp:2dmodel_energy}. 
We consider the case that only $v_F$ and $t_1$ are kept in the pentalayer Hamiltonian (Eq.~\ref{eqapp:chiralham}), a $SO(2)$-symmetric $v_3,v_4$ correction to the dispersion is included in perturbation theory, and the exponential form factor approximation is used (Eq.~\ref{eqapp:Mkq_exp}).
In this case, the orientation of the BZ does not affect the results. For all results, we shift the Wigner crystal solutions in real-space so that they satisfy the emergent intravalley $C_6$ and $M_1\mathcal{T}$ symmetries with origin at $\bm{r}=0$ for convenience. 

We outline the information presented in each of Figs.~\ref{fig:HFwavefunction_N36_t0.77_Hrad1.301_forcerad-0.601_U0.024_pstrt1_v0_only+1+1+1+exp+none+0_epsr5.00_C1} to \ref{fig:HFwavefunction_N36_t0.77_Hrad1.701_forcerad-0.601_U0.024_pstrt1_v0_only+1+1+1+exp+none+0_epsr5.00_C0}. The (logarithm of the) expectation value of the occupation $n(\bm{k})$ in the HF ground state is the same as that shown in App.~\ref{secapp:HFbandstructure}. Here, we also plot $n(\bm{k})$ as a function of $|\bm{k}|$. $O(\bm{k},\bm{k}+\bm{b}_1)=\langle \gamma^\dagger_{\bm{k}}\gamma_{\bm{k}+\bm{b}_1} \rangle$ gives the translation symmetry-breaking order parameter at the primitive RLV $\bm{b}_1$. Its magnitude indicates the regions in momentum space that contribute most to the translation symmetry-breaking, while its phase reveals information about the electronic topology of the Wigner crystal. Note that $O(\bm{k},\bm{k}+\bm{b}_1)$ is only defined when both $\bm{k}$ and $\bm{k}+\bm{b}_1$ lie within the cutoff. All other order parameters with a different momentum transfer can be found using $C_6$ symmetry, which is satisfied by both the $C=0$ and $C=1$ solutions. 

Figs.~\ref{fig:HFwavefunction_N36_t0.77_Hrad1.301_forcerad-0.601_U0.024_pstrt1_v0_only+1+1+1+exp+none+0_epsr5.00_C1} and \ref{fig:HFwavefunction_N36_t0.77_Hrad1.701_forcerad-0.601_U0.024_pstrt1_v0_only+1+1+1+exp+none+0_epsr5.00_C1} show results for the $C=1$ phase with relatively small 1.3$q_1$ and relatively large 1.7$q_1$ cutoff respectively. The two calculations exhibit similar behaviors in the ground state wavefunctions. For example, the occupation $n(k)$ is quickly suppressed beyond $|k|\sim q_2$. Furthermore, $|O(\bm{k},\bm{k}+\bm{b}_1)|$ is concentrated at the edge of the BZ, rapidly decays away from the BZ boundary, and takes values $1/2$ and $1/3$ at the $M_M$ and $K_M$ points respectively. In addition, the phase of $O(\bm{k},\bm{k}+\bm{b}_1)$ rotates smoothly when traversing between two BZ corners via one of the $M_M$ points. These phases are fixed according to the $C_6$ symmetry and the Chern number, as described in App.~\ref{secapp:2dmodel_HF}. 

Figs.~\ref{fig:HFwavefunction_N36_t0.77_Hrad1.301_forcerad-0.601_U0.024_pstrt1_v0_only+1+1+1+exp+none+0_epsr5.00_C0} and \ref{fig:HFwavefunction_N36_t0.77_Hrad1.701_forcerad-0.601_U0.024_pstrt1_v0_only+1+1+1+exp+none+0_epsr5.00_C0} show corresponding results for the $C=0$ phase. The key difference with the $C=1$ results is that the phase of $O(\bm{k},\bm{k}+\bm{b}_1)$ is nearly constant. The phases of $O(\bm{k},\bm{k}+\bm{b}_1)$ at high symmetry points are fixed according to the $C_6$ symmetry and the Chern number, as described in App.~\ref{secapp:2dmodel_HF}. 

In Fig.~\ref{fig:g1_N36_t0.77_Hrad1.301_forcerad-0.601_U0.024_pstrt1_v0_only+1+1+1+exp+none+0_epsr5.00_C1} (for the $C=1$ solution) and \ref{fig:g1_N36_t0.77_Hrad1.301_forcerad-0.601_U0.024_pstrt1_v0_only+1+1+1+exp+none+0_epsr5.00_C0} (for the $C=0$ solution), we show line-cuts in momentum space of the magnitude of the hybridization mean-field $g_1(\bm{k})$ using the $1.3q_1$ cutoff. We find that unlike the order parameter $O(\bm{k},\bm{k}+\bm{b}_1)$ which depends strongly on the momentum perpendicular to the BZ boundary, $g_1(\bm{k})$ varies much more slowly with $k_x$. This justifies the thin sliver approximation introduced in later sections when performing an analytical treatment of the Berry Trashcan model. We also find that the magnitude $|g_1(\bm{k})|$ only varies moderately around the BZ boundary (i.e.~consider $k_x=-b_1/2$ and $-q_2/2<k_y<q_2/2$).

\begin{figure}
    \centering
    \includegraphics[width = 1.0\linewidth]{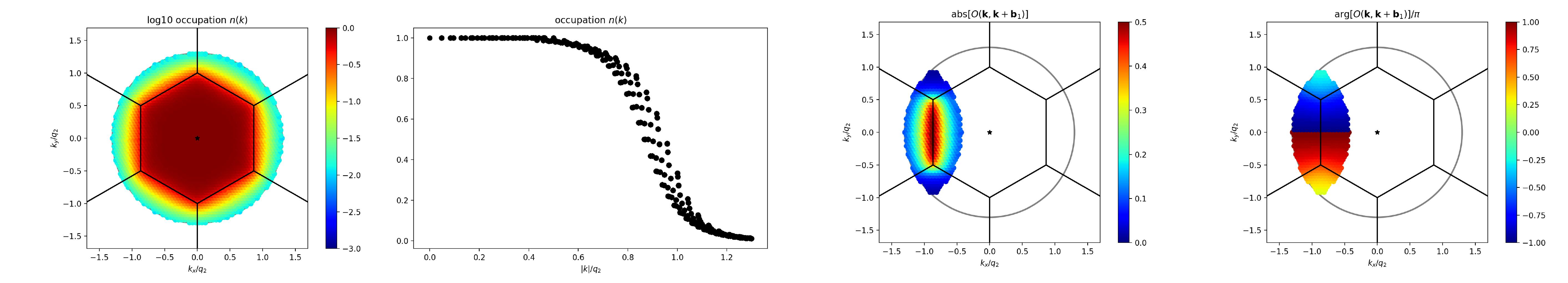}
    \caption{Details of the HF wavefunction of the $C=1$ insulator. In the R$5$G Hamiltonian (Eq.~\ref{eq:H_K}), only the $v_F,t_1$ terms are kept. A $SO(2)$-symmetric dispersion correction from $v_3,v_4$ (Eq.~\ref{eqapp:v3v4_dispersion}) is included. The exponential form factor approximation (Eq.~\ref{eqapp:Mkq_exp}) is used. Only valley $K$ and spin $\uparrow$ is included. The Hilbert space is made of conduction band states truncated based on a circular momentum cutoff. \\\textbf{System parameters}: $N_1=N_2=36$; Hilbert space cutoff radius $1.3q_1$; CN interaction scheme; $\epsilon=5$; $\xi=20\,\text{nm}$; $\theta=0.77^\circ$; $V=24\,\text{meV}$}
    \label{fig:HFwavefunction_N36_t0.77_Hrad1.301_forcerad-0.601_U0.024_pstrt1_v0_only+1+1+1+exp+none+0_epsr5.00_C1}
\end{figure}

\begin{figure}
    \centering
    \includegraphics[width = 1.0\linewidth]{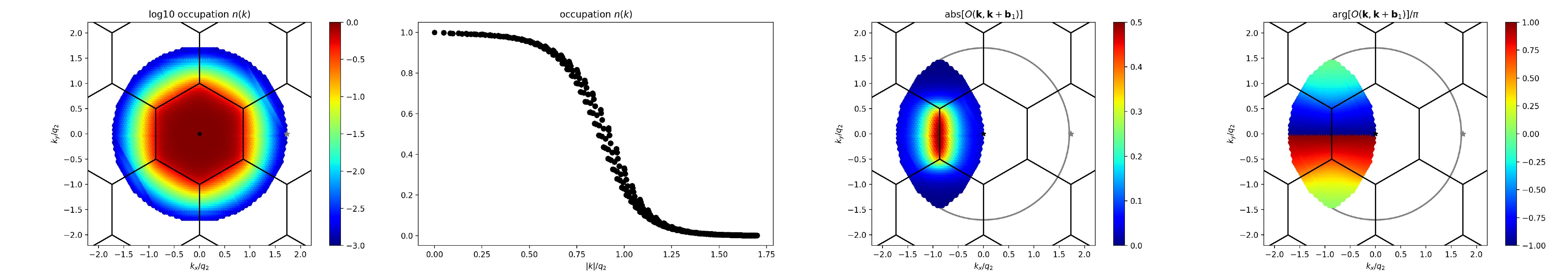}
    \caption{Details of the HF wavefunction of the $C=1$ insulator. In the R$5$G Hamiltonian (Eq.~\ref{eq:H_K}), only the $v_F,t_1$ terms are kept. A $SO(2)$-symmetric dispersion correction from $v_3,v_4$ (Eq.~\ref{eqapp:v3v4_dispersion}) is included. The exponential form factor approximation (Eq.~\ref{eqapp:Mkq_exp}) is used. Only valley $K$ and spin $\uparrow$ is included. The Hilbert space is made of conduction band states truncated based on a circular momentum cutoff. \\\textbf{System parameters}: $N_1=N_2=36$; Hilbert space cutoff radius $1.7q_1$; CN interaction scheme; $\epsilon=5$; $\xi=20\,\text{nm}$; $\theta=0.77^\circ$; $V=24\,\text{meV}$}
    \label{fig:HFwavefunction_N36_t0.77_Hrad1.701_forcerad-0.601_U0.024_pstrt1_v0_only+1+1+1+exp+none+0_epsr5.00_C1}
\end{figure}

\begin{figure}
    \centering
    \includegraphics[width = 1.0\linewidth]{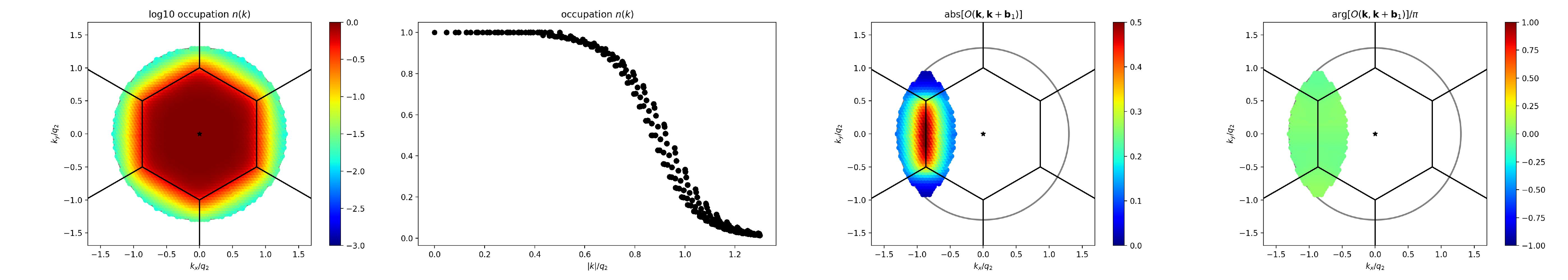}
    \caption{Details of the HF wavefunction of the $C=0$ insulator. In the R$5$G Hamiltonian (Eq.~\ref{eq:H_K}), only the $v_F,t_1$ terms are kept. A $SO(2)$-symmetric dispersion correction from $v_3,v_4$ (Eq.~\ref{eqapp:v3v4_dispersion}) is included. The exponential form factor approximation (Eq.~\ref{eqapp:Mkq_exp}) is used. Only valley $K$ and spin $\uparrow$ is included. The Hilbert space is made of conduction band states truncated based on a circular momentum cutoff. \\\textbf{System parameters}: $N_1=N_2=36$; Hilbert space cutoff radius $1.3q_1$; CN interaction scheme; $\epsilon=5$; $\xi=20\,\text{nm}$; $\theta=0.77^\circ$; $V=24\,\text{meV}$}
    \label{fig:HFwavefunction_N36_t0.77_Hrad1.301_forcerad-0.601_U0.024_pstrt1_v0_only+1+1+1+exp+none+0_epsr5.00_C0}
\end{figure}

\begin{figure}
    \centering
    \includegraphics[width = 1.0\linewidth]{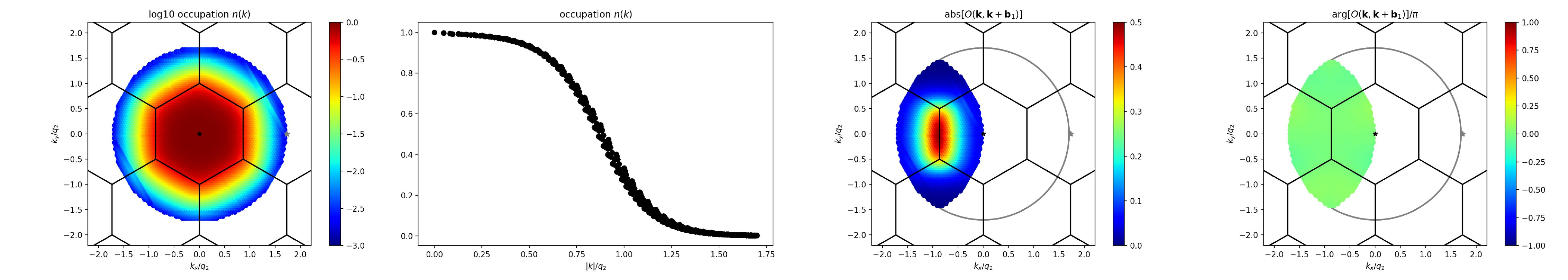}
    \caption{Details of the HF wavefunction of the $C=0$ insulator. In the R$5$G Hamiltonian (Eq.~\ref{eq:H_K}), only the $v_F,t_1$ terms are kept. A $SO(2)$-symmetric dispersion correction from $v_3,v_4$ (Eq.~\ref{eqapp:v3v4_dispersion}) is included. The exponential form factor approximation (Eq.~\ref{eqapp:Mkq_exp}) is used. Only valley $K$ and spin $\uparrow$ is included. The Hilbert space is made of conduction band states truncated based on a circular momentum cutoff. \\\textbf{System parameters}: $N_1=N_2=36$; Hilbert space cutoff radius $1.7q_1$; CN interaction scheme; $\epsilon=5$; $\xi=20\,\text{nm}$; $\theta=0.77^\circ$; $V=24\,\text{meV}$}
    \label{fig:HFwavefunction_N36_t0.77_Hrad1.701_forcerad-0.601_U0.024_pstrt1_v0_only+1+1+1+exp+none+0_epsr5.00_C0}
\end{figure}

\begin{figure}
    \centering
    \includegraphics[width = 0.7\linewidth]{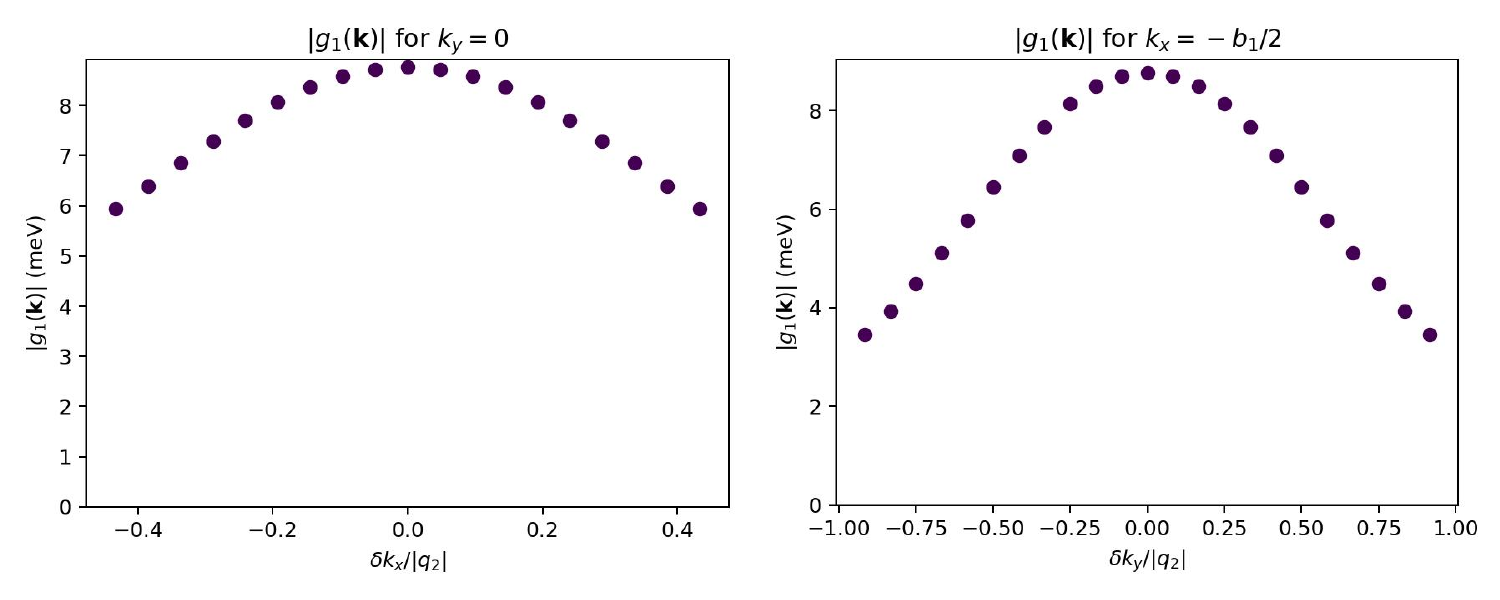}
    \caption{Hybridization mean-field $g_1(\bm{k})$ of the $C=1$ insulator. In the first plot, $\delta k_x$ is measured relative to the $M_M$ point. In the R$5$G Hamiltonian (Eq.~\ref{eq:H_K}), only the $v_F,t_1$ terms are kept. A $SO(2)$-symmetric dispersion correction from $v_3,v_4$ (Eq.~\ref{eqapp:v3v4_dispersion}) is included. The exponential form factor approximation (Eq.~\ref{eqapp:Mkq_exp}) is used. Only valley $K$ and spin $\uparrow$ is included. The Hilbert space is made of conduction band states truncated based on a circular momentum cutoff. \\\textbf{System parameters}: $N_1=N_2=36$; Hilbert space cutoff radius $1.3q_1$; CN interaction scheme; $\epsilon=5$; $\xi=20\,\text{nm}$; $\theta=0.77^\circ$; $V=24\,\text{meV}$}
    \label{fig:g1_N36_t0.77_Hrad1.301_forcerad-0.601_U0.024_pstrt1_v0_only+1+1+1+exp+none+0_epsr5.00_C1}
\end{figure}

\begin{figure}
    \centering
    \includegraphics[width = 0.7\linewidth]{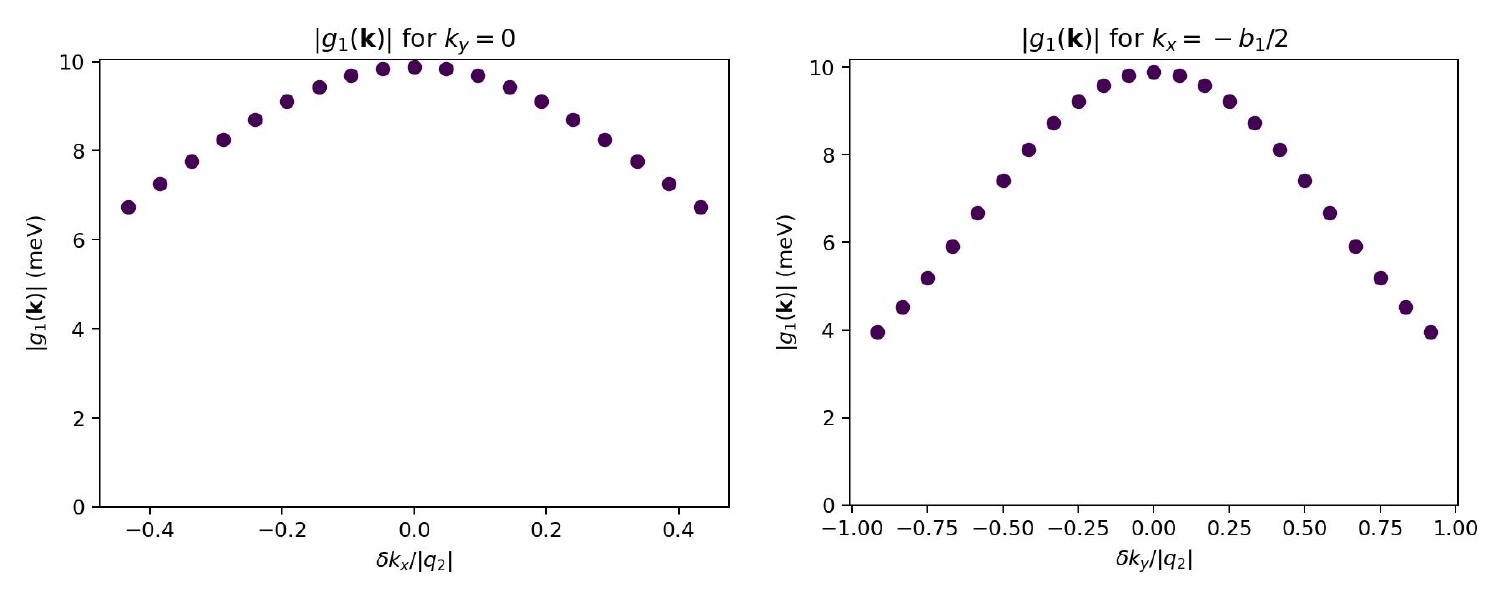}
    \caption{Hybridization mean-field $g_1(\bm{k})$ of the $C=0$ insulator. In the first plot, $\delta k_x$ is measured relative to the $M_M$ point. In the R$5$G Hamiltonian (Eq.~\ref{eq:H_K}), only the $v_F,t_1$ terms are kept. A $SO(2)$-symmetric dispersion correction from $v_3,v_4$ (Eq.~\ref{eqapp:v3v4_dispersion}) is included. The exponential form factor approximation (Eq.~\ref{eqapp:Mkq_exp}) is used. Only valley $K$ and spin $\uparrow$ is included. The Hilbert space is made of conduction band states truncated based on a circular momentum cutoff. \\\textbf{System parameters}: $N_1=N_2=36$; Hilbert space cutoff radius $1.3q_1$; CN interaction scheme; $\epsilon=5$; $\xi=20\,\text{nm}$; $\theta=0.77^\circ$; $V=24\,\text{meV}$}
    \label{fig:g1_N36_t0.77_Hrad1.301_forcerad-0.601_U0.024_pstrt1_v0_only+1+1+1+exp+none+0_epsr5.00_C0}
\end{figure}

\clearpage

\section{1D Trashcan Model: Setup}\label{secapp:1dtrashcansetup}

In preparation for the involved analytical treatment of the 2D Berry Trashcan model (see App.~\ref{secapp:2dmodel_setup}), we first consider a simpler 1D problem that captures some of the salient qualitative features. In this appendix section, we describe the Hamiltonian.

\subsection{Setup}\label{subsecapp:1dtrashcanmodel}

We consider the setup shown in Fig.~\ref{fig:1dtrashcanzones}. The choice of reduced BZ (this coincides with lowest shell of momenta, referred to as BZ 0) is parameterized by the reciprocal lattice vector (RLV) $G$ that will characterize the putative translation symmetry breaking. We consider an electronic density that allows for the following description. Within BZ 0, the non-interacting dispersion (blue) is very flat, but steeply rises for momenta outside BZ 0. For the trashcan model, the dispersion exactly vanishes within BZ 0, but outside it quickly increases linearly starting from the BZ edge. This motivates introducing a momentum cutoff which extends beyond the boundaries of the BZ by a small amount. In particular, in BZ 1 (BZ 2) that lies to the right (left) of BZ 0, we only keep a thin window of states adjacent to the BZ boundary. Only in narrow slivers around the BZ boundaries are momenta able to hybridize at wavevector $\pm G$.

\begin{figure}
    \centering
    \includegraphics[width = 0.4\linewidth]{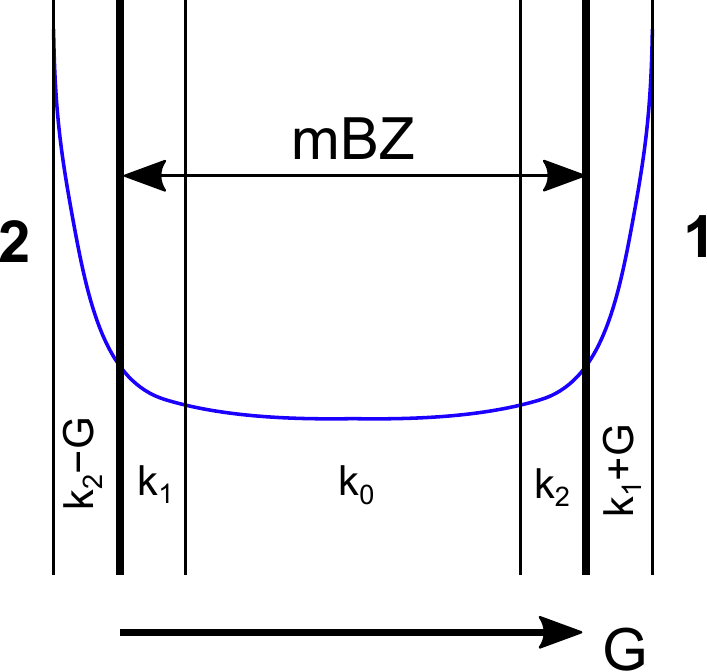}
    \caption{1D trashcan model with three BZ's 0, 1, and 2. The reduced BZ coincides with BZ 0, which is bordered by the thick lines. The band dispersion (blue) is nearly flat within BZ 0 ($|k|\leq \frac{G}{2}$), and quickly disperses outside. Owing to the sharp dispersion, only small regions of momentum space outside BZ 0 are kept in the model. These consist of regions labelled by $k_1+G$ and $k_2-G$, which fold onto regions labelled by $k_1$ and $k_2$ respectively within the BZ. $k_0$ denotes the rest of the BZ.}
    \label{fig:1dtrashcanzones}
\end{figure}

The general form of the Hamiltonian (without the momentum cutoff) takes the same form as Eq.~\ref{eqapp:Hintproject}
\begin{equation}\label{eqapp:1dtrashcanham}
    H=H_\text{kin}+H_\text{int}=\sum_k E(\mbf{k}) \gamma_{\mbf{k}}^\dagger \gamma_{\mbf{k}} + \frac{1}{2\Omega_{tot}}  \sum_{\bm{q}, \bm{k}, \bm{k'}} V_{\mbf{q}} M_{\mbf{k},\mbf{q}} M_{\mbf{k}',-\mbf{q}} \gamma_{\mbf{k}+ \mbf{q}}^\dagger \gamma_{\mbf{k}'- \mbf{q}}^\dagger \gamma_{\mbf{k}'} \gamma_{\mbf{k}}
\end{equation}
where $\gamma^\dagger_{\mbf{k}}$ is the band creation operator, $E(\mbf{k})$ is the band dispersion, $V_{\mbf{q}}$ is the interaction potential, and $M_{\mbf{k},\mbf{q}}=\langle \mbf{k}+\mbf{q}|\mbf{k} \rangle$ is the form factor. We introduce the inversion operator $I$ that acts on momenta as $I\mbf{k} = -\mbf{k}$. In the limit of the trashcan dispersion, $E(\bm{k})$ vanishes within BZ 0, and sharply rises linearly outside.

To account for the momentum cutoff and the momentum regions indicated in Fig.~\ref{fig:1dtrashcanzones}, we introduce a short-hand notation. The reduced BZ (which coincides with BZ 0), which consists of momenta where $|\mbf{k}|\leq \frac{G}{2}$, is divided into three disjoint regions. Momenta $\mbf{k}_0$ within the BZ do not have any partners at $\mbf{k}_0\pm\mbf{G}$ that lie within the cutoff. On the other hand, momenta $\mbf{k}_1$ ($\mbf{k}_2$) within the BZ have one partner $\mbf{k}_1+\mbf{G}$ ($\mbf{k}_2-\mbf{G}$) that also lies within the cutoff, but lives in BZ 1 (2). We introduce a notation for electron operators in BZ 0
\begin{equation}
    \gamma_{0,\mbf{k}_0}=\gamma_{\mbf{k}_0},\quad \gamma_{0,\mbf{k}_1}=\gamma_{\mbf{k}_1},\quad \gamma_{0,\mbf{k}_2}=\gamma_{\mbf{k}_2},
\end{equation}
where the subscript $0$ refers to BZ 0. We can then introduce a notation for electron operators that lie within the cutoff in BZ 1 and 2
\begin{equation}
    \gamma_{1,\mbf{k}_1}=\gamma_{\mbf{k}_1+\mbf{G}},\quad \gamma_{2,\mbf{k}_2}=\gamma_{\mbf{k}_2-\mbf{G}}.
\end{equation}
Using the inversion operator $I$, this can be summarized more succinctly as
\begin{equation}
        \gamma_{i,\mbf{k_i}} = \gamma_{\mbf{k_i }+ I^{i-1}\mbf{G}}= \gamma_{\mbf{k_i }- I^{i}\mbf{G}}
\end{equation}
where $i=1,2$. 

In the HF analysis, the order parameter is given by the one-body density matrix for the Slater determinant state
\begin{equation}\label{eqapp:1dtrashcanOP}
    O_{\mbf{k},\mbf{k}'}=\langle \gamma^\dagger_{\mbf{k}}\gamma_{\mbf{k}'}\rangle.
\end{equation}
The continuous translation invariance of the Hamiltonian (Eq.~\ref{eqapp:1dtrashcanham}) is preserved if $O_{\mbf{k},\mbf{k}'}\sim \delta_{\mbf{k},\mbf{k}'}$. However, we are primarily interested in investigating HF states which can break this symmetry down to a discrete translation group characterized by the primitive RLV $\mbf{G}$. In this case, the order parameter is allowed to be non-vanishing as long as $\mbf{k}$ and $\mbf{k}'$ are identical modulo a RLV. We can hence write down the decomposition
\begin{align}\label{eqapp:1dOPdecomposition}
    O_{\mbf{k},\mbf{k}'}=&
    \sum_{i=0}^2O_{00}(\mbf{k}_i)\delta_{\mbf{k},\mbf{k}'}\delta_{\mbf{k},\mbf{k}_i}+\sum_{i=1}^2O_{ii}(\mbf{k}_i)\delta_{\mbf{k},\mbf{k}'}\delta_{\mbf{k},\mbf{k}_i+I^{i-1}\mbf{G}}+\sum_{i=1}^2\left( O_{0i}(\mbf{k}_i)\delta_{\mbf{k},\mbf{k}_i}\delta_{\mbf{k}',\mbf{k}_i+I^{i-1}\mbf{G}}+\text{h.c.}\right),
\end{align}
where we have introduced the notation $O_{ij}(\mbf{k}_l)$ which denotes the element of the density matrix with reduced momentum $\mbf{k}_l$ between BZ $i$ and BZ $j$. Care should be taken to prevent duplicate entries at the BZ edges. Consider the BZ momentum $-\bm{G}/2$ which is equivalent to $\bm{G}/2$ when folding by a RLV. To maintain the inversion symmetry of the momentum regions, it will be convenient to include $-\bm{G}/2$ in momentum region $\bm{k}_1$ and $\bm{G}/2$ in momentum region $\bm{k}_2$. This means that the order parameter entry $O_{-\bm{G}/2,\bm{G}/2}$ could be represented by either $O_{01}(-\bm{G}/2)$ or $O_{20}(\bm{G}/2)$. This subtlety will be addressed in more detail below Eq.~\ref{eqapp:1dtrashcanHHFparameterized}.

For future purposes, it will also be useful to write
\begin{equation}
    O_{\mbf{k}+\mbf{q},\mbf{k}}=
    \sum_{i=0}^2O_{00}(\mbf{k}_i)\delta_{\mbf{q},0}\delta_{\mbf{k},\mbf{k}_i}+\sum_{i=1}^2O_{ii}(\mbf{k}_i)\delta_{\mbf{q},0}\delta_{\mbf{k},\mbf{k}_i+I^{i-1}\mbf{G}}+\sum_{i=1}^2\left( O_{0i}(\mbf{k}_i)\delta_{\mbf{q},-I^{i-1}\mbf{G}}\delta_{\mbf{k},\mbf{k}_i+I^{i-1}\mbf{G}}+\text{h.c.}\right).
\end{equation}

\subsubsection{Inversion symmetry}\label{secapp:1dtrashcaniinversionsym}
In some cases, we will impose inversion symmetry on the Hamiltonian and the mean-field solutions. On the plane wave creation operators $c^\dagger_{\mbf{k},\alpha}$, where $\alpha$ is some orbital index, inversion acts as
\begin{equation}
    I c_{\mbf{k}\alpha}^\dagger I^{-1} = \sum_{\beta}c_{-\mbf{k} \beta}^\dagger D_{\beta \alpha}.
\end{equation}
Inversion symmetry of the single-particle Bloch Hamiltonian $h(\mbf{k})$, from which the lowest band $\gamma_{\mbf{k}}^\dagger =  c_{\mbf{k}}^\dagger u_{\mbf{k}}$ is extracted with $u_{\mbf{k}}$ the corresponding Bloch function, leads to
\begin{gather}
    h(\mbf{k}) = D^\dagger h(- \mbf{k}) D, \quad h(\mbf{k}) u_{\mbf{k}} = E_{\mbf{k}}u_{\mbf{k}}\\
   \implies Du_{\mbf{k}} = e^{i \phi_{\mbf{k}}} u_{-\mbf{k}}\\
   \implies I \gamma_{\mbf{k}}^\dagger I^{-1} = I  c_{\mbf{k}}^\dagger I^{-1} u_{\mbf{k}} = c_{-\mbf{k}}^\dagger  D u_{\mbf{k}}= e^{i \phi_{\mbf{k}}}  c_{-\mbf{k}}^\dagger  u_{-\mbf{k}}= e^{i \phi_{\mbf{k}}}\gamma_{-\mbf{k}}^\dagger.
\end{gather}
Since $I^2=1$, we have $\phi_{\mbf{k}}+ \phi_{-\mbf{k}}=2 \pi n_{\mbf{k}}$ for some integer $n_{\mbf{k}}$. This implies $I  e^{-i\frac{1}{2}  \phi_{\mbf{k} } }  \gamma_{\mbf{k}}^\dagger I^{-1} = e^{i \pi n_{\mbf{k}}} e^{-i\frac{1}{2}  \phi_{-\mbf{k} } } \gamma^\dagger_{- \mbf{k}}$.
If we redefine the gauge on the band operators $e^{-i\frac{1}{2}  \phi_{\mbf{k} } }  \gamma_{\mbf{k}}^\dagger \rightarrow \gamma_{\mbf{k}}^\dagger$, then we have $I  \gamma_{\mbf{k}}^\dagger I^{-1} = (-1)^{n_{\mbf{k}}} \gamma^\dagger_{- \mbf{k}}$. The remaining sign $(-1)^{n_{\mbf{k}}}$ cannot fluctuate in a smooth gauge so it is independent of $\mbf{k}$. It can then be absorbed by a gauge transformation in the transformation matrix $D\rightarrow (-1)^{n_{\mbf{k}}} D$, leading to 
\begin{equation}
    I  \gamma_{\mbf{k}}^\dagger I^{-1} =  \gamma^\dagger_{- \mbf{k}}.
\end{equation}
This is always possible since in our continuum model, the only relevant inversion symmetric momentum is $\mbf{k}=0$. With this gauge choice, we have the following properties
\begin{gather}
    Du_{\mbf{k}} =  u_{-\mbf{k}},\quad \langle- \mbf{k}|-\mbf{k'}\rangle= \langle \mbf{k}|\mbf{k'}\rangle,\quad M_{\mbf{k},\mbf{q}}=M_{-\mbf{k},-\mbf{q}}\\
    \quad I  \gamma_{\mbf{k}}^\dagger I^{-1} =  \gamma^\dagger_{- \mbf{k}}\\
    I  \gamma_{0,\mbf{k}_i}^\dagger I^{-1} =  \gamma^\dagger_{0,- \mbf{k}_i}\,\,\text{ for }i=0,1,2\\
    I  \gamma_{i,\mbf{k}_i}^\dagger I^{-1} =  \gamma^\dagger_{\bar{i},-\mbf{k}_i}\,\,\text{ for }i=1,2\\
    O_{00, \mbf{k}_i}= O_{00, -\mbf{k}_i}\,\,\text{ for }i=0,1,2 \\
    O_{ii, \mbf{k_i}}= O_{\bar{i}\bar{i} ,-\mbf{k_i}}, \quad O_{0i,  \mbf{k_i}}= O_{0\bar{i}, -\mbf{k_i}}, \quad O_{i0, \mbf{k_i}}=  O_{\bar{i}0, -\mbf{k_i}}\,\,\text{ for }i=1,2,
\end{gather}
where $\bar{i}=2,1$ for $i=1,2$ respectively.

\clearpage

\section{1D Trashcan Model: Hartree-Fock Analysis}\label{secapp:1dtrashcan_HFanalysis}
In this appendix section, we perform a HF analysis of the 1D trashcan Hamiltonian in Eq.~\ref{eqapp:1dtrashcanham} with the order parameter in Eq.~\ref{eqapp:1dtrashcanOP}. Note that we do not claim that mean-field theory is valid for this 1D model, but the calculations here will be useful warm-up for the calculation in the 2D model in later appendix sections.

We first consider the mean-field decoupling of the interaction term. Wick's theorem yields 
\begin{align}
\gamma_{\mbf{k}+ \mbf{q}}^\dagger \gamma_{\mbf{k}'- \mbf{q}}^\dagger \gamma_{\mbf{k}'}\gamma_{\mbf{k}}
=&
 O_{\mbf{k} + \mbf{q},\mbf{k}}   \gamma_{\mbf{k}'- \mbf{q}}^\dagger \gamma_{\mbf{k'}}+ 
 O_{\mbf{k'} - \mbf{q},\mbf{k'}}   \gamma_{\mbf{k}
+ \mbf{q}}^\dagger \gamma_{\mbf{k}} -  O_{\mbf{k} + \mbf{q},\mbf{k'}}   \gamma_{\mbf{k}'- \mbf{q}}^\dagger \gamma_{\mbf{k}} -  O_{\mbf{k'} - \mbf{q},\mbf{k}}   \gamma_{\mbf{k}+ \mbf{q}}^\dagger \gamma_{\mbf{k'}}\\ & - O_{\mbf{k} + \mbf{q},\mbf{k}}  O_{\mbf{k'} - \mbf{q},\mbf{k'}} + O_{\mbf{k} + \mbf{q},\mbf{k'}} O_{\mbf{k'} - \mbf{q},\mbf{k}}. 
\end{align}
The decoupling of the interaction term is then
\begin{align}
    H_{\text{int}}^{\text{HF}}-E^{\text{HF,int}}_\text{tot}=& \frac{1}{\Omega_{tot}}\sum_{\mbf{k,k',q}}\left( V(q) M_{\mbf{k}, \mbf{q}} M_{\mbf{k'}, -\mbf{q}}  O_{\mbf{k'} - \mbf{q},\mbf{k'}}   \gamma_{\mbf{k}
+ \mbf{q}}^\dagger \gamma_{\mbf{k}} - V(q) M_{\mbf{k}, \mbf{q}} M_{\mbf{k'}, -\mbf{q}}  O_{\mbf{k'} - \mbf{q},\mbf{k}}   \gamma_{\mbf{k}
+ \mbf{q}}^\dagger \gamma_{\mbf{k'}}\right)\\ & +   \frac{1}{2\Omega_{tot}}\sum_{\mbf{k,k',q}} \left(V(|\mbf{k'}-\mbf{k}- \mbf{q}| ) M_{\mbf{k}, \mbf{k'}- \mbf{k}- \mbf{q}} M_{\mbf{k'},\mbf{k}- \mbf{k'} + \mbf{q}}  -V(q) M_{\mbf{k}, \mbf{q}} M_{\mbf{k'}, -\mbf{q}}\right)O_{\mbf{k} + \mbf{q},\mbf{k}}  O_{\mbf{k'} - \mbf{q},\mbf{k'}} ,
\end{align}
where $E^{\text{HF,int}}_\text{tot}$ is the interacting contribution to the total HF energy.
We now focus on the Hartree and Fock terms of the one-body mean-field Hamiltonian $H_{\text{int}}^{\text{HF}}$, in the case where continuous translation is broken to a discrete subgroup parameterized by the RLV $G$. For the Hartree term, we have three contributions corresponding to momentum transfers $\mbf{q}=0,\mbf{G},-\mbf{G}$. For the Fock term, we have three contributions corresponding to momentum transfers $\mbf{q}=\mbf{k}'-\mbf{k},\mbf{k}'-\mbf{k}+\mbf{G},\mbf{k}'-\mbf{k}-\mbf{G}$. The interaction part of the HF Hamiltonian is then
\begin{align}\label{eqapp:1dHHFintfirst}
    H_{\text{int}}^{\text{HF}}=&\frac{1}{\Omega_{tot}}\sum_{\mbf{k,k'}} V(0)   O_{\mbf{k'} ,\mbf{k'}}   \gamma_{\mbf{k}}^\dagger \gamma_{\mbf{k}}\\
    +&\frac{1}{\Omega_{tot}}\sum_{\mbf{k,k'}} V(G) \langle \mbf{k+G} | \mbf{k} \rangle \langle \mbf{k'-G} | \mbf{k'} \rangle     O_{\mbf{k'} - \mbf{G},\mbf{k'}}   \gamma_{\mbf{k}
+ \mbf{G}}^\dagger \gamma_{\mbf{k}}\\
+ &\frac{1}{\Omega_{tot}}\sum_{\mbf{k,k'}} V(G) \langle \mbf{k-G} | \mbf{k} \rangle \langle \mbf{k'+G} | \mbf{k'} \rangle     O_{\mbf{k'} + \mbf{G},\mbf{k'}}   \gamma_{\mbf{k}
- \mbf{G}}^\dagger \gamma_{\mbf{k}} \\
-&\frac{1}{\Omega_{tot}}\sum_{\mbf{k,k'}}V(|\mbf{k-k'}|) |\langle \mbf{k'}|\mbf{k} \rangle |^2 O_{\mbf{k'},\mbf{k'}}   \gamma_{\mbf{k}}^\dagger \gamma_{\mbf{k}}\\
-&\frac{1}{\Omega_{tot}}\sum_{\mbf{k,k'}} V(|\mbf{k-k'+G}|) \langle \mbf{k+G}|\mbf{k'} \rangle \langle \mbf{k'-G}|\mbf{k}\rangle 
  O_{\mbf{k'} - \mbf{G},\mbf{k'}}   \gamma_{\mbf{k}
+ \mbf{G}}^\dagger \gamma_{\mbf{k}}\\
-&\frac{1}{\Omega_{tot}}\sum_{\mbf{k,k'}} V(|\mbf{k-k'-G}|) \langle \mbf{k-G}|\mbf{k'} \rangle \langle \mbf{k'+G}|\mbf{k}\rangle 
  O_{\mbf{k}' + \mbf{G},\mbf{k}'}   \gamma_{\mbf{k}
- \mbf{G}}^\dagger \gamma_{\mbf{k}} ,\label{eqapp:1dHHFintlast}
\end{align}
where we have explicitly written out the form factors in terms of the Bloch states. Note that the second line is the complex conjugate of the third line, and the fifth line is the complex conjugate of the sixth line.  

The HF Hamiltonian (including the kinetic part) of the 1D trashcan model can be generally parameterized as
\begin{gather}\label{eqapp:1dtrashcanHHFparameterized}
    H^{\text{HF}}=\sum_{i=0}^2\sideset{}{'}\sum_{\mbf{k}_i}(E_{0,\mbf{k}_i}+\mu_{0,\mbf{k}_i})\gamma_{0,\mbf{k}_i}^\dagger\gamma_{0,\mbf{k}_i}+\sum_{i=1}^2\sideset{}{'}\sum_{\mbf{k}_i}(E_{i,\mbf{k}_i}+\mu_{i,\mbf{k}_i})\gamma_{i,\mbf{k}_i}^\dagger\gamma_{i,\mbf{k}_i}+\sum_{i=1}^2\sideset{}{'}\sum_{\mbf{k}_i}\left(t_{i,\mbf{k}_i}\gamma_{i,\mbf{k}_i}^\dagger\gamma_{0,\mbf{k}_i}+\text{h.c.}\right),
\end{gather}
where $E_{i,\mbf{k}}$ is the kinetic energy at reduced momentum $\mbf{k}$ in BZ $i$, and we have introduced the momentum-dependent band renormalization mean fields $\mu_i$ and hybridization mean fields $t_i$. The summations over $\bm{k}_i$ are primed, which indicates that `multiplicities' at the BZ boundary need to be accounted for in order to prevent overcounting terms. This arises because the regions $\bm{k}_1$ and $\bm{k}_2$ are mapped into each other by inversion $I$, which necessarily means that both regions contain momenta that fold onto $\bm{G}/2$ (which is equivalent to $-\bm{G}/2$) in the BZ. As an example of potential overcounting, note that $\gamma^\dagger_{0,\bm{k}_1=-\bm{G}/2}$ and $\gamma^\dagger_{2,\bm{k}_2=\bm{G}/2}$ are creation operators for the same  Bloch basis state. To account for this, the summation $\sideset{}{'}\sum_{\bm{k}_i}$ comes with a factor of $1/2$ if the summation momentum is $\bm{k}_1=-\bm{G}/2$ or $\bm{k}_2=\bm{G}/2$. Note that the errors incurred from neglecting this subtlety vanish as the momentum spacing goes to zero in the thermodynamic limit.

From Eqs.~\ref{eqapp:1dHHFintfirst}-\ref{eqapp:1dHHFintlast}, we obtain
{
\begin{align}\label{eqapp:1dtrashcanparamsfirst}
    \mu_{0, \mbf{k}_i}=& \frac{1}{\Omega_{tot}}  \left[\sum_{j=0}^{2}\sideset{}{'}\sum_{\mbf{k_j}'}\left( V_0 -V_{|\mbf{k}_i-\mbf{k_j}'|} |\langle \mbf{k}_i|\mbf{k_j}'\rangle|^2 \right)O_{00,\mbf{k_j}'} + \sum_{j=1}^2 \sideset{}{'}\sum_{\mbf{k_j}'}  \left(V_0-V_{|\mbf{k}_i- \mbf{k_j}'+ I^{j}\mbf{G} |} |\langle \mbf{k}_i| \mbf{k_j}'- I^{j}\mbf{G}\rangle|^2 \right)  O_{jj,\mbf{k_j}'}  \right] \\
    \mu_{i, \mbf{k}_i}= &\frac{1}{\Omega_{tot}}  \bigg[\sum_{j=0}^{2}\sideset{}{'}\sum_{\mbf{k_j}'}\left( V_0 -V_{|\mbf{k}_i-\mbf{k_j}'-I^i\mbf{G}|} |\langle \mbf{k}_i-I^i\mbf{G}|\mbf{k_j}'\rangle|^2 \right)O_{00,\mbf{k_j}'} \\
    &+ \sum_{j=1}^2 \sideset{}{'}\sum_{\mbf{k_j}'}  \left(V_0-V_{|\mbf{k}_i- \mbf{k_j}'-I^i\mbf{G}+ I^{j}\mbf{G} |} |\langle \mbf{k}_i-I^i\mbf{G}| \mbf{k_j}'- I^{j}\mbf{G}\rangle|^2 \right)  O_{jj,\mbf{k_j}'}  \bigg]\\
    t_{1,\mbf{k}_1}=& \frac{1 }{\Omega_{tot}}   \bigg[     \sideset{}{'}\sum_{\mbf{k'_1}}  \left( V_G\langle \mbf{k_1+ G}| \mbf{k_1} \rangle  \langle \mbf{k'_1}| \mbf{k'_1+G} \rangle - V_{\mbf{k_1-k_1'}} \langle\mbf{k_1'}|\mbf{k_1}\rangle \langle{\mbf{k_1+ G}} |\mbf{k_1'+ G}\rangle \right)     O_{01\mbf{k'_1}}  \\ &+  \sideset{}{'}\sum_{\mbf{k'_2}} \left( V_G\langle \mbf{k_1+ G}| \mbf{k_1} \rangle \langle \mbf{k'_2- G}| \mbf{k'_2} \rangle  - V_{|\mbf{k_1-k'_2+ G}|} \langle\mbf{k_1+G}|\mbf{k'_2}\rangle \langle{\mbf{k'_2- G}} |\mbf{k_1}\rangle \right)  O_{20 \mbf{k'_{2}}}     \bigg]\\
    t_{2,\mbf{k}_2}=& \frac{1 }{\Omega_{tot}}   \bigg[     \sideset{}{'}\sum_{\mbf{k'_2}}  \left( V_G\langle \mbf{k_2- G}| \mbf{k_2} \rangle  \langle \mbf{k'_2}| \mbf{k'_2-G} \rangle - V_{\mbf{k_2-k_2'}} \langle\mbf{k_2'}|\mbf{k_2}\rangle \langle{\mbf{k_2- G}} |\mbf{k_2'- G}\rangle \right)     O_{02\mbf{k'_2}}  \\ &+  \sideset{}{'}\sum_{\mbf{k'_1}} \left( V_G\langle \mbf{k_2- G}| \mbf{k_2} \rangle \langle \mbf{k'_1+ G}| \mbf{k'_1} \rangle  - V_{|\mbf{k_2-k'_1- G}|} \langle\mbf{k_2-G}|\mbf{k'_1}\rangle \langle{\mbf{k'_1+ G}} |\mbf{k_2}\rangle \right)  O_{10 \mbf{k'_{1}}}     \bigg],\label{eqapp:1dtrashcanparamslast}
\end{align}
}
where we recall that the primed momentum summations prevent overcounting contributions from the BZ edges.

Note that in the presence of inversion symmetry (App.~\ref{secapp:1dtrashcaniinversionsym}), we have
\begin{equation}\label{eqapp:1dtrashcaninvsym}
    E_{\mbf{k}}= E_{-\mbf{k}},\;\;\;     \mu_{0, \mbf{k}}= \mu_{0,-\mbf{k}},  \;\;\;    \mu_{1, \mbf{k_1}}= \mu_{2,-\mbf{k_1}},  \;\;\;    t_{1, \mbf{k_1}}= t_{2,-\mbf{k_1}}, \;\;\; \langle \mbf{k} |\mbf{k'} \rangle =\langle- \mbf{k} |- \mbf{k'} \rangle. 
\end{equation}

\subsection{Trivial form factors}
In this appendix subsection, we first address the case of trivial form factors where $\langle \mbf{k}|\mbf{k}'\rangle=1$ for all $\mbf{k},\mbf{k}'$. Eqs.~\ref{eqapp:1dtrashcanparamsfirst}-\ref{eqapp:1dtrashcanparamslast} in Eq.~\ref{eqapp:1dtrashcanHHFparameterized} become
{
\begin{align}
    \mu_{0, \mbf{k}_i}=& \frac{1}{\Omega_{tot}}  \left[\sum_{j=0}^{2}\sideset{}{'}\sum_{\mbf{k_j}'}\left( V_0 -V_{|\mbf{k}_i-\mbf{k_j}'|}\right)O_{00,\mbf{k_j}'} + \sum_{j=1}^2 \sideset{}{'}\sum_{\mbf{k_j}'}  \left(V_0-V_{|\mbf{k}_i- \mbf{k_j}'+ I^{j}\mbf{G} |} \right)  O_{jj,\mbf{k_j}'}  \right] \\
    \mu_{i, \mbf{k}_i}= &\frac{1}{\Omega_{tot}}  \bigg[\sum_{j=0}^{2}\sideset{}{'}\sum_{\mbf{k_j}'}\left( V_0 -V_{|\mbf{k}_i-\mbf{k_j}'-I^i\mbf{G}|}  \right)O_{00,\mbf{k_j}'} + \sum_{j=1}^2 \sideset{}{'}\sum_{\mbf{k_j}'}  \left(V_0-V_{|\mbf{k}_i- \mbf{k_j}'-I^i\mbf{G}+ I^{j}\mbf{G} |} |\right)  O_{jj,\mbf{k_j}'}  \bigg]\\
    t_{1,\mbf{k}_1}=& \frac{1 }{\Omega_{tot}}   \bigg[     \sideset{}{'}\sum_{\mbf{k'_1}}  \left( V_G - V_{\mbf{k_1-k_1'}}  \right)     O_{01\mbf{k'_1}} +  \sideset{}{'}\sum_{\mbf{k'_2}} \left( V_G  - V_{|\mbf{k_1-k'_2+ G}|} \right)  O_{20 \mbf{k'_{2}}}     \bigg]\\
    t_{2,\mbf{k}_2}=& \frac{1 }{\Omega_{tot}}   \bigg[     \sideset{}{'}\sum_{\mbf{k'_2}}  \left( V_G - V_{\mbf{k_2-k_2'}} \right)     O_{02\mbf{k'_2}} +  \sideset{}{'}\sum_{\mbf{k'_1}} \left( V_G - V_{|\mbf{k_2-k'_1- G}|} \right)  O_{10 \mbf{k'_{1}}}     \bigg].
\end{align}
}

If the dispersion in BZ 1 and 2 is sufficiently steep, then we expect that the momentum cutoff can be made very tight around the BZ boundary (the `thin-sliver' approximation). This means that $|\mbf{k}_1-\mbf{k}_1'|,|\mbf{k}_1-\mbf{k}_2'+G|\ll G$ etc., and we can approximate the hybridization fields $t_{i,\mbf{k}_i}$ as
\begin{align}
    t_{1,\mbf{k}_1}=& \frac{\left( V_G - V_{0}  \right) }{\Omega_{tot}}   \bigg[     \sideset{}{'}\sum_{\mbf{k'_1}}       O_{01\mbf{k'_1}} +  \sideset{}{'}\sum_{\mbf{k'_2}}  O_{20 \mbf{k'_{2}}}     \bigg]\\
    t_{2,\mbf{k}_2}=& \frac{\left( V_G - V_{0}  \right) }{\Omega_{tot}}   \bigg[     \sideset{}{'}\sum_{\mbf{k'_2}}      O_{02\mbf{k'_2}} +  \sideset{}{'}\sum_{\mbf{k'_1}} O_{10 \mbf{k'_{1}}}     \bigg].
\end{align}
Note that in taking e.g.~$V_{\bm{k}_1-\bm{k}_1'}\simeq V_0$, we assume that the interaction potential near momentum transfer $\bm{q}\simeq 0$ is smooth and regularized to be finite, for example due to screening by an external metallic gate.
In the presence of inversion symmetry, these hybridization fields become real and momentum-independent
\begin{equation}
    t_{1,\mbf{k}_1}=t_{2,\mbf{k}_2}=\frac{2\left( V_G - V_{0}  \right) }{\Omega_{tot}}\sideset{}{'}\sum_{\mbf{k'_1}}      \Re O_{01\mbf{k'_1}}\eqcolon  t.
\end{equation}
We note that in this case, the mean-field Hamiltonian is purely real and possesses an emergent time-reversal symmetry that is local in $k$, regardless of the order parameter. This time-reversal symmetry arises from both the inversion symmetry and the thin-sliver approximation. To see this, we relax the thin-sliver approximation to obtain
\begin{align}
 t_{1,\mbf{k_1}}&= \frac{1 }{\Omega_{tot}}  \left[      \sideset{}{'}\sum_{\mbf{k'_1}}  ( V_G - V_{\mbf{k_1-k_1'}}  )     O_{01\mbf{k'_1}}     + \sideset{}{'}\sum_{\mbf{k'_1}} ( V_G - V_{|\mbf{k_1+k'_1+ G}|} )  O_{10 \mbf{k'_{1}}}     \right] \\ &= \frac{1 }{\Omega_{tot}}   \left[      \sideset{}{'}\sum_{\mbf{k'_1}}  ( 2 V_G - V_{\mbf{k_1-k_1'}}-  V_{|\mbf{k_1+k'_1+ G}|}  )     \Re O_{01\mbf{k'_1}}     + i \sideset{}{'}\sum_{\mbf{k'_1}} ( V_{|\mbf{k_1+k'_1+ G}|} -V_{\mbf{k_1-k_1'}} )  \Im O_{01 \mbf{k'_{1}}}     \right] \\
 &= \frac{1 }{\Omega_{tot}}  \left[     \sideset{}{'}\sum_{\mbf{k'_1}}  ( 2 V_G - V_{|\mbf{\delta{k_1}-\delta{k_1'}}|}-  V_{|\mbf{\delta{k_1}+\delta{k'_1}}|}  )     \Re O_{01\mbf{k'_1}}     + i \sideset{}{'}\sum_{\mbf{k'_1}} ( V_{|\mbf{\delta{k_1}+\delta{k'_1}}|} -V_{|\mbf{\delta{k_1}-\delta{k_1'}}|} )  \Im O_{01 \mbf{k'_1}}   \right],
\end{align}
where $\mbf{k}_1= - \frac{\mbf{G}}{2} + \mbf{\delta{{k_1}}} $. This shows that if we do not approximate $\mbf{\delta{{k_1}}}=0$, then there is generally an imaginary part to the mean-field Hamiltonian.

We return to the band renormalization fields $\mu_i$. The thin-sliver approximation also motivates taking $\mbf{k}_i\simeq \frac{1}{2}I^i\mbf{G}$ for $i=1,2$, leading to
\begin{align}
    \mu_{0, \mbf{k}_0}=& \frac{1}{\Omega_{tot}}  \left[\sum_{\mbf{k_0}'}\left( V_0 -V_{|\mbf{k}_0-\mbf{k_0}'|}\right)O_{00,\mbf{k_0}'}+\sum_{i=1}^2\left( V_0 -V_{\mbf{k}_0 -I^i \frac{\mbf{G}}{2} } \right) \left( \sideset{}{'}\sum_{\mbf{k_i'} } O_{00\mbf{k_i'}}+\sideset{}{'}\sum_{\mbf{k'_{\bar{i}}}}   O_{\bar{i}\bar{i}\mbf{k'_{\bar{i}}}}\right)\right]  \\
    \mu_{0, \mbf{k}_i}=&\frac{1}{\Omega_{tot}}  \left[\sum_{\mbf{k'}_0}\left( V_0 -V_{|\frac{I^i\mbf{G}}{2}- \mbf{k'}_0|} \right) O_{00,\mbf{k'}_0} +    \left( V_0 -V_{\mbf{G}} \right)\left( \sideset{}{'}\sum_{\mbf{k_{\bar{i}}'} } O_{00\mbf{k_{\bar{i}}'}} +  \sideset{}{'}\sum_{\mbf{k_i'}}   O_{ii\mbf{k_i'}} \right) \right]\eqcolon \mu_{0,i}\,\,\text{ for }i=1,2 \\
    \mu_{i,\mbf{k_i}} =& \frac{1}{\Omega_{tot}}   \left[\sum_{\mbf{k}'_0 } \left(V_0- V_{|- \mbf{k}'_0 -\frac{ \mbf{I^iG}}{2}|}\right) O_{00, \mbf{k}_0'}  + (V_0- V_{\mbf{G}}) \left( \sideset{}{'}\sum_{\mbf{k'_{\bar{i}}}}    O_{\bar{i}\bar{i}, \mbf{k'_{\bar{i}}}}  + \sideset{}{'}\sum_{\mbf{k'_i}}    O_{00 ,\mbf{k'_i}}\right) \right] \eqcolon\mu_{i,i}.
\end{align}
In the presence of inversion symmetry, we have
\begin{gather}
    \mu_{0,\bm{k}_0}=\frac{1}{\Omega_{tot}}  \left[\sum_{\mbf{k_0}'}\left( V_0 -V_{|\mbf{k}_0-\mbf{k_0}'|}\right)O_{00,\mbf{k_0}'}+\left( 2V_0 -V_{\mbf{k}_0 -\frac{\mbf{G}}{2} }-V_{\mbf{k}_0 +\frac{\mbf{G}}{2} } \right)  \sideset{}{'}\sum_{\mbf{k_1'} } \left(O_{00\mbf{k_1'}}+  O_{11,\mbf{k'_{1}}}\right)\right]\\
    \mu_{0,1}=\mu_{0,2}=\mu_{1,1}=\mu_{2,2}=\frac{1}{\Omega_{tot}}  \left[\sum_{\mbf{k'}_0}\left( V_0 -V_{|-\frac{\mbf{G}}{2}- \mbf{k'}_0|} \right) O_{00,\mbf{k'}_0} +    \left( V_0 -V_{\mbf{G}} \right)\sideset{}{'}\sum_{\mbf{k_{1}'} } \left( O_{00\mbf{k_{1}'}} +    O_{11\mbf{k_1'}} \right) \right].
\end{gather}
So in the presence of inversion symmetry, the only independent mean-field parameters in the mean-field Hamiltonian are $t,\mu_{0,\bm{k}_0},\mu_{0,1}$ (which are all real)
\begin{align}\label{eqapp:1dtrashcanHHFrealparams}
    H^{\text{HF}}=&\sum_{\mbf{k}_0}(E_{0,\mbf{k}_0}+\mu_{0,\mbf{k}_0})\gamma_{0,\mbf{k}_0}^\dagger\gamma_{0,\mbf{k}_0}+\sideset{}{'}\sum_{\mbf{k}_1}(E_{0,\mbf{k}_1}+\mu_{0,1})\gamma_{0,\mbf{k}_1}^\dagger\gamma_{0,\mbf{k}_1}+\sideset{}{'}\sum_{\mbf{k}_2}(E_{0,\mbf{k}_2}+\mu_{0,1})\gamma_{0,\mbf{k}_2}^\dagger\gamma_{0,\mbf{k}_2}\\
    &+\sideset{}{'}\sum_{\mbf{k}_1}(E_{1,\mbf{k}_1}+\mu_{0,1})\gamma_{1,\mbf{k}_1}^\dagger\gamma_{1,\mbf{k}_1}+\sideset{}{'}\sum_{\mbf{k}_2}(E_{2,\mbf{k}_2}+\mu_{0,1})\gamma_{2,\mbf{k}_2}^\dagger\gamma_{2,\mbf{k}_2}\\
    &+t\sideset{}{'}\sum_{\mbf{k}_1}\left(\gamma_{1,\mbf{k}_1}^\dagger\gamma_{0,\mbf{k}_1}+\gamma_{0,\mbf{k}_1}^\dagger\gamma_{1,\mbf{k}_1}\right)+t\sideset{}{'}\sum_{\mbf{k}_2}\left(\gamma_{2,\mbf{k}_2}^\dagger\gamma_{0,\mbf{k}_2}+\gamma_{0,\mbf{k}_2}^\dagger\gamma_{2,\mbf{k}_2}\right).
\end{align}

To make further analytical progress, we assume that the interaction potential decays weakly, and assume the following expansion (quadratic potential approximation)
\begin{equation}\label{eqapp:1dtrashcanquadraticpotential}
    V(q)=V_0(1-\alpha q^2).
\end{equation}
Note that since $V(q)$ in 1D has dimensions of [energy]$\times$[length], then $V_0$ also has dimensions of [energy]$\times$[length].

Using the identity $\sum_{\mbf{k}_0}(\mbf{k}\cdot\mbf{k}_0) O_{00,\mbf{k}_0}=0$ valid in the presence of inversion, we obtain
\begin{gather}
        \mu_{0,\bm{k}_0}=\frac{\alpha V_0G^2}{4\Omega_{tot}} \left[ \sum_{\mbf{k_0}'}\left( \frac{4k_0^2}{G^2}+\frac{4k_0'^2}{G^2}\right)O_{00,\mbf{k_0}'}+2\left( 1+\frac{4k_0^2}{G^2}  \right)  \sideset{}{'}\sum_{\mbf{k_1'} } \left(O_{00\mbf{k_1'}}+  O_{11,\mbf{k'_{1}}}\right)\right]\\
        \mu_{0,1}=\frac{\alpha V_0G^2}{4\Omega_{tot}} \left[ \sum_{\mbf{k_0}'}\left( 1+\frac{4k_0'^2}{G^2}\right)O_{00,\mbf{k_0}'}+4\sideset{}{'}\sum_{\mbf{k_1'} } \left(O_{00\mbf{k_1'}}+  O_{11,\mbf{k'_{1}}}\right)\right]\\
         t = -\frac{2\alpha V_0G^2}{\Omega_{tot}}  \sideset{}{'}\sum_{\mbf{k'_1}}   \Re O_{10\mbf{k'_{1}}}.  \label{eqapp:1dtrashcantquadraticV}
\end{gather}
For convenience, we shift the chemical potential to measure energies from $\mu_{0,1}$. To this end, we define a shifted band renormalization field for $\mbf{k}_0$ in the BZ
\begin{equation}\label{eqapp:1dtrashcanmu'0l0}
    \mu'_{0 ,\mbf{k}_0}\coloneq \mu_{0 ,\mbf{k}_0}  - \mu_{0,1}=\frac{\alpha V_0 G^2}{4 \Omega_{tot}}  \left(  \frac{4 k_0^2}{G^2} -1 \right) \left( \sum_{\mbf{k_0}'}O_{00,\mbf{k_0'}}+2 \sideset{}{'}\sum_{\mbf{k_1'} } \left(O_{00,\mbf{k_1'}}+  O_{11,\mbf{k'_{1}}}\right) \right).
\end{equation}

Due to inversion symmetry, we can focus on the mean-field Hamiltonian for reduced momenta $\mbf{k}_0$ and $\mbf{k}_1$ (the Hamiltonian for $\mbf{k}_2$ is related to $\mbf{k}_1$ by symmetry)
\begin{gather}\label{eqapp:1dtrashcanHHFk0inversion}
    H^{\text{HF}}_{\mbf{k}_0}=\sum_{\mbf{k}_0}(E_{0,\mbf{k}_0}+\mu'_{0,\mbf{k}_0})\gamma_{0,\mbf{k}_0}^\dagger\gamma_{0,\mbf{k}_0}\\
    H^{\text{HF}}_{\mbf{k}_1}=\sideset{}{'}\sum_{\mbf{k}_1}\begin{pmatrix}\gamma^\dagger_{0, \mbf{k_1}}   &\gamma^\dagger_{1, \mbf{k_1}} \end{pmatrix}
    \left[
    \frac{1}{2}E_{1,\mbf{k}_1}\mathbbm{1}_2-\frac{1}{2}E_{1,\mbf{k}_1}\sigma_z + t\sigma_x
    \right]
    \begin{pmatrix}\gamma_{0, \mbf{k_1}}   \\\gamma_{1, \mbf{k_1}} \end{pmatrix},\label{eqapp:1dtrashcanHHFk1inversion}
\end{gather}
where $\mathbbm{1}_2$ is the $2\times 2$ identity matrix, $\sigma_x,\sigma_y,\sigma_z$ are the Pauli matrices, and we are in the flat-bottom (trashcan) limit $E_{0,\mbf{k}}=0$. We have used the primed momentum summation in $H^{\text{HF}}_{\mbf{k}_1}$ so that the mean-field term involving $-\bm{G}/2$ and $\bm{G}/2$ is not double-counted when the symmetry-related $H^{\text{HF}}_{\mbf{k}_2}$ is included in the total mean-field Hamiltonian. Note that at filling factor $\nu=1$, we necessarily have $O_{00,\mbf{k}_0}=1$ in the ground state if $\alpha>0$ (meaning that $V_0>V_G$ from Eq.~\ref{eqapp:1dtrashcanquadraticpotential}), i.e.~the region corresponding to $\mbf{k}_0$ in BZ 0 is fully occupied. This is because $\mu'_{0,\mbf{k}_0}<0$ for $\alpha>0$, such that the only way $\mbf{k}_0$ can be depopulated is if both eigenvalues for some $\mbf{k}_1$ are less than $0$. However, the latter is clearly not possible by inspecting $H^{\text{HF}}_{\mbf{k}_1}$. Explicitly, the eigenvalues in $\mbf{k}_1$ are
\begin{equation}
    E_{\pm,\mbf{k_1}} = \frac{1}{2} E_{1,\mbf{k_1} }\pm \sqrt{\left(\frac{1}{2} E_{\mbf{1,k_1}}\right)^2 + t^2 }= \frac{1}{2} E_{1,\mbf{k_1} }\pm d_{\mbf{k_1}},\;\;\;d_{\mbf{k_1}}= \sqrt{\left(\frac{1}{2} E_{\mbf{1,k_1}}\right)^2 + t^2 }.
\end{equation}
Since the kinetic energy $E_{1,\mbf{k_1} }$ vanishes at the BZ boundary $\mbf{k}_1=-\frac{\mbf{G}}{2}$ in the trashcan limit, we find the opening of a gap of at least $2t$, with minimum value $2t$ at $\mbf{k}_1=-\frac{\mbf{G}}{2}$. Letting 
\begin{equation}
    d_{x,\mbf{k}_1}=t,\quad d_{z,\mbf{k}_1}=-\frac{1}{2}E_{1,\mbf{k}_1},\quad \hat{d}_{\alpha,\mbf{k}_1}=\frac{d_{\alpha,\mbf{k}_1}}{d_{\mbf{k}_1}},\quad d_{\mbf{k}_1}=\sqrt{\sum_{\alpha}d_{\alpha,\mbf{k}_1}^2}
\end{equation}
where $\alpha=x,y,z$, we find the corresponding eigenvectors
\begin{equation}
    \psi_{\pm,\mbf{k}_1}=  \frac{1}{\sqrt{2}}\left(\begin{matrix}
          \sqrt{1\pm \hat{ d}_{z,\mbf{k}_1}} \\
    \pm  \frac{\hat{d}_{x,\mbf{k}_1} }{\sqrt{1\pm  \hat{ d}_{z,\mbf{k}_1}}}
     \end{matrix} \right),
\end{equation}
associated with annihilation operators
 \begin{gather}
     a_{\pm,\mbf{k_1}}= [\psi_{\pm,\mbf{k}_1}^*]_0 \gamma_{0 ,\mbf{k_1}} + [\psi_{\pm,\mbf{k}_1}^*]_1 \gamma_{1 ,\mbf{k_1}}\\  \gamma_{0, \mbf{k_1}} = [\psi_{+,\mbf{k_1}}]_0  a_{+, \mbf{k_1}}+[\psi_{-,\mbf{k_1}}]_0  a_{-, \mbf{k_1}}\\  \gamma_{1, \mbf{k_1}} = [\psi_{+,\mbf{k_1}}]_1  a_{+, \mbf{k_1}}+[\psi_{-,\mbf{k_1}}]_1  a_{-, \mbf{k_1}} .
 \end{gather}
For $\mbf{k}_1$, we occupy the negative energy states $E_{-,\mbf{k}_1}$ associated with $a_{-,\mbf{k}_1}^\dagger$, and evaluate the resulting order parameters 
\begin{equation}
O_{10,\mbf{k_1}}= - \frac{1}{2} \hat{d}_{x \mbf{k}_1},\quad 
O_{00,\mbf{k_1}}=\frac{1}{2} (1- \hat{d}_{z\mbf{k_1}}),\quad  O_{11,\mbf{k_1}}=\frac{1}{2} \frac{\hat{d}_{x\mbf{k_1}} ^2}{ 1- \hat{d}_{z\mbf{k_1}}} ,
\end{equation}
which remain purely real, and satisfy $O_{11,\mbf{k_1'}} +O_{00,\mbf{k_1'}}=1$ consistent with occupying one state per momentum. Inserting $O_{10,\mbf{k_1}}$ into the expression for $t$ (Eq.~\ref{eqapp:1dtrashcantquadraticV}) yields the self-consistency equation
\begin{equation}\label{eqapp:1dtrashcantselfconsistency}
    t = \frac{\alpha V_0 G^2 }{\Omega_{tot}}   \sideset{}{'}\sum_{\mbf{k_1}} \frac{t}{ \sqrt{(\frac{1}{2} E_{1,\mbf{k_1}})^2 + t^2 }}.
\end{equation}

\subsubsection{Case 1: $\alpha>0$}
For a non-trivial solution of Eq.~\ref{eqapp:1dtrashcantselfconsistency}, we require $\alpha>0$ (note that $V_0>0$) meaning that within the quadratic potential approximation, $V_0>V_G$ (see Eq.~\ref{eqapp:1dtrashcanquadraticpotential}). For a non-zero $t$, the self-consistency condition reads
\begin{equation}
    \frac{1}{\alpha V_0 G^2} = \frac{1 }{\Omega_{tot}}   \sideset{}{'}\sum_{\mbf{k_1}} \frac{1}{ \sqrt{(\frac{1}{2} E_{1,\mbf{k_1}})^2 + t^2 }} =\frac{1}{2\pi} \int_{-\frac{G}{2}}^{-\frac{G}{2} + \Lambda} dk_1 \frac{1}{ \sqrt{(\frac{1}{2} E_{1,\mbf{k_1}})^2 + t^2 }} 
\end{equation}
where $\Lambda$ is the momentum cutoff in BZ 1 away from the BZ edge at $-G/2$, and we have taken the thermodynamic limit in the second equality (where we do not need to worry about overcounting the contribution from $\bm{k}_1=-\bm{G}/2$). Assuming a linear dispersion $E_{1,\mbf{k_1}}= v\left( \frac{G}{2} + k_1\right)  $ away from the BZ with velocity $v$ (in the approximations taken so far, we do not include the interaction-induced renormalization of the velocity, which is negligible for sufficiently large bare kinetic velocity $v$), we have 
\begin{gather}
    \frac{1}{\alpha V_0 G^2} = \frac{1}{\pi v} \int_{0}^{ \frac{v\Lambda}{2}} dx \frac{1}{ \sqrt{x^2 + t^2 }} = \frac{1}{\pi v}  \text{arctanh}  \left[\frac{1}{ \sqrt{1 + \left(\frac{2t}{v \Lambda }\right) ^2 }} \right] \\
    \rightarrow t= \pm\frac{\Lambda v}{2} \sqrt{\frac{1}{\left(\tanh{ \frac{\pi v}{\alpha V_0 G^2}}\right)^2 }-1  }.
\end{gather}
Note that for large $v$, the mean-field hybridization $t$ decreases. This is because a steeper dispersion reduces the density of states. We cannot take the limit $v\rightarrow 0$, since this would be inconsistent with the thin-sliver approximation.

The two signs of $t$ lead to states that are degenerate (the HF eigenvalues only depend on the square of $t$) but with different topology, as we now demonstrate. The $\Gamma_M$ point has inversion eigenvalue $+1$. Consider the transformation of the lower (filled) band at $\mbf{k}_1=-\frac{\mbf{G}}{2}$
\begin{gather}
    a_{-,-\frac{\mbf{G}}{2}}^\dagger= \frac{1}{\sqrt{2}} \left(\sqrt{1- \hat{d}_{z, -\frac{G}{2} }} \gamma_{0,-\frac{G}{2}}^\dagger   -  \frac{\hat{d}_{x,\frac{-G}{2}} }{\sqrt{1{\color{blue}-}  \hat{ d}_{z,-\frac{G}{2}} }}  \gamma_{1,-\frac{G}{2}}^\dagger \right)= \frac{1}{\sqrt{2}} \left( \gamma_{-\frac{G}{2}}^\dagger   - \text{sgn} ({t })  \gamma_{\frac{G}{2}}^\dagger \right)\\
    \rightarrow 
    Ia_{-, -{\frac{G}{2}}}^\dagger I^{-1}=  \frac{1}{\sqrt{2}} \left( \gamma_{\frac{G}{2}}^\dagger   - \text{sgn} ({t })  \gamma_{-\frac{G}{2}}^\dagger \right) = - \text{sgn} ({t }) a_{-, {-\frac{G}{2}}}^\dagger, 
\end{gather}
which means that the product of the inversion eigenvalues in the BZ is $-\text{sgn}(t)$. Therefore for $t<0$, the HF Wigner crystal is associated with $s$ orbitals located at the $1a$ Wyckoff position of the reconstructed unit cell, while for $t>0$ it is associated with $s$ orbitals located at the $1b$ Wyckoff position. With inversion symmetry, $t>0$ and $t<0$ are separated by a phase transition. Of course, since the original Hamiltonian has continuous translation invariance, these two solutions are simply related by a real-space shift. In fact, the continuous translation symmetry generates a continuous manifold of degenerate HF solutions.

\subsubsection{Case 2: $\alpha<0$}
For $\alpha<0$, we need to revisit the result mentioned below Eq.~\ref{eqapp:1dtrashcanHHFk1inversion} that the region $\mbf{k}_0$ is fully occupied. We recall the definition of $\mu'_{0,\mbf{k}_0}$
\begin{gather}
    \mu'_{0 ,\mbf{k}_0}=-\frac{\alpha V_0 G^2}{4 \Omega_{tot}}  \left(1-  \frac{4 k_0^2}{G^2} \right) \left( \sum_{\mbf{k_0}'}O_{00,\mbf{k_0'}}+2 \sideset{}{'}\sum_{\mbf{k_1'} } \left(O_{00,\mbf{k_1'}}+  O_{11,\mbf{k'_{1}}}\right) \right)=A\left(1-  \frac{4 k_0^2}{G^2} \right)\\
    A=-\frac{\alpha V_0 G^2}{4 \Omega_{tot}} \left( \sum_{\mbf{k_0}'}O_{00,\mbf{k_0'}}+2 \sideset{}{'}\sum_{\mbf{k_1'} } \left(O_{00,\mbf{k_1'}}+  O_{11,\mbf{k'_{1}}}\right) \right),
\end{gather}
where $A>0$ for the case $\alpha<0$ of interest here. We find that $\mu'_{0,\mbf{k}_0}$ (Eq.~\ref{eqapp:1dtrashcanmu'0l0}) is largest at $\mbf{k}_0=0$ and decreases to zero as $\mbf{k}_0$ approaches the BZ boundary. For a given $t$, the eigenenergies at $\mbf{k}_1=- \mbf{G}/2$ are $\pm t$, and if $|t|< \mu_{0,\mbf{k}_0=0}'$, then some of the $\mbf{k}_0$ states will be emptied in favor of  some of the positive energy states in the $\mbf{k}_1$ sliver. For concreteness, consider a global chemical potential $\mu$. 
\begin{itemize}
\item
If $\mu>|t|$, the $+$ band for $\mbf{k}_1$ is partially or fully occupied (the $-$ band is fully occupied). The Fermi point ${\mbf{k}_{1F}}$ (which is negative and measured from the $\Gamma_M$ point) is determined by $\mu= E_{+,\mbf{k_{1F}}} = \frac{1}{2} E_{1,\mbf{k_{1F}} }+ \sqrt{(\frac{1}{2} E_{1,\mbf{k_{1F}}})^2 + t^2 }$, leading to
\begin{equation}
    E_{1,\mbf{k_{1F}}} = \frac{\mu^2- t^2}{\mu}.
\end{equation}
\item 
If $A>\mu$, then the $\mbf{k}_0$ region is not fully occupied. There is a Fermi point $\mbf{k}_{0F}$ according to the condition $\mu = A ( 1 -\frac{4 k_{0F}^2}{G^2}  )$, i.e.
\begin{equation}
    k_{0F} =\frac{G}{2} \sqrt{1-\frac{\mu}{A}}.
\end{equation}
\item 
For $A>\mu>|t|$, to ensure filling $\nu=1$, we have the constraint
\begin{equation}
    k_{0F} = \frac{G}{2}+ k_{1F}.
\end{equation}
\end{itemize}
We now need to derive the self-consistent equations, which require various expectation values. For the region of the BZ with reduced momentum $\mbf{k}_0$ and $\mbf{k}_1$, the ground state is $\prod_{|\mbf{k}_0|>k_{0F}}a_{\mbf{k_0}}^\dagger \prod_{\mbf{k}_1} a_{-,\mbf{k_1}}^\dagger  \prod_{\mbf{k}_1' < \mbf{k}_{1F}} a_{+,\mbf{k_1'}}^\dagger\ket{0} $ (note that we do not duplicate the action of $a^\dagger_{\pm,\bm{k}_1=-\bm{G}/2}$ and $a^\dagger_{\pm,\bm{k}_2=\bm{G}/2}$ when considering the full many-body wavefunction including all momentum regions). We obtain
\begin{gather}
    O_{10\mbf{k_1}}=- \frac{1}{2} \hat{d}_{x \mbf{k}_1}( 1-\Theta_{k_{1F}- k_1})\\
    O_{00\mbf{k_1}}=\frac{1}{2} (1- \hat{d}_{z\mbf{k_1}}) +  \frac{1}{2} (1+ \hat{d}_{z\mbf{k_1}}) \Theta_{k_{1F}- k_1}\\
    O_{11\mbf{k_1}}=\frac{1}{2} \frac{\hat{d}_{x\mbf{k_1}} ^2}{ 1- \hat{d}_{z\mbf{k_1}}}  + \frac{1}{2} \frac{\hat{d}_{x\mbf{k_1}} ^2}{ 1{+} \hat{d}_{z\mbf{k_1}}}   \Theta_{k_{1F}- k_1} \\
    O_{11\mbf{k_1'}} +O_{00\mbf{k_1'}}= 1 + \Theta_{k_{1F}- k_1}\\
    O_{00\mbf{k}_0} = 1- \Theta_{k_{0F} - |k_0| } \\
    \sum_{\mbf{k_0}} O_{00\mbf{k_0}} +    2  \sideset{}{'}\sum_{\mbf{k_1} }   \left(O_{00\mbf{k_1}}+O_{11\mbf{k_1}}\right) =  \frac{G \Omega_{tot}}{2\pi}
\end{gather}
where $\Theta_x$ is the Heaviside function that is equal to 1 when $x>0$, and 0 otherwise. The last equation simply reflects the filling factor $\nu=1$ of the state. The self-consistency equation for the hybridization $t$ for non-zero $t$ is then (Eq.~\ref{eqapp:1dtrashcantquadraticV})
\begin{equation}
    1=-  \frac{\alpha V_0 G^2 }{\Omega_{tot}}     \sideset{}{'}\sum_{\mbf{k_1}} \frac{1}{ \sqrt{(\frac{1}{2} E_{\mbf{k_1+G}})^2 + t^2 }} ( \Theta_{k_{1F}- k_1}-1) .
\end{equation}
Note that the LHS is positive, while the RHS is negative (recall that $\alpha<0$). Hence, a non-trivial solution with $t\neq0$ is not possible for trivial form factors.

\subsection{Non-trivial form factors}
We now consider the case with non-trivial form factors $\langle\bm{k}|\bm{k}'\rangle$ with inversion symmetry, i.e.~Eqs.~\ref{eqapp:1dtrashcanparamsfirst}-\ref{eqapp:1dtrashcaninvsym}. We continue to take the thin-sliver approximation. This means that $|\mbf{k}_1-\mbf{k}_1'|,|\mbf{k}_1-\mbf{k}_2'+G|\ll G$ etc., as well as $\mbf{k}_i\simeq \frac{1}{2}I^i\mbf{G}$ for $i=1,2$. The manipulations are similar to the previous subsection with trivial form factors, and lead to
{\small
\begin{gather}
        \mu_{0,\bm{k}_0}=\frac{1}{\Omega_{tot}}  \left[\sum_{\mbf{k_0}'}\left( V_0 -V_{|\mbf{k}_0-\mbf{k_0}'|}|\langle\mbf{k}_0|\mbf{k}'_0\rangle|^2\right)O_{00,\mbf{k_0}'}+\left( 2V_0 -V_{\mbf{k}_0 -\frac{\mbf{G}}{2} }|\langle\mbf{k}_0|\frac{\mbf{G}}{2}\rangle|^2-V_{\mbf{k}_0 +\frac{\mbf{G}}{2} } |\langle\mbf{k}_0|-\frac{\mbf{G}}{2}\rangle|^2\right)  \sideset{}{'}\sum_{\mbf{k_1'} } \left(O_{00\mbf{k_1'}}+  O_{11,\mbf{k'_{1}}}\right)\right]\\
    \mu_{0,1}=\mu_{0,2}=\mu_{1,1}=\mu_{2,2}=\frac{1}{\Omega_{tot}}  \left[\sum_{\mbf{k'}_0}\left( V_0 -V_{|-\frac{\mbf{G}}{2}- \mbf{k'}_0|}|\langle\mbf{k}'_0|\frac{\mbf{G}}{2}\rangle|^2 \right) O_{00,\mbf{k'}_0} +    \left( V_0 -V_{\mbf{G}} |\langle-\frac{\mbf{G}}{2}|\frac{\mbf{G}}{2}\rangle|^2\right)\sideset{}{'}\sum_{\mbf{k_{1}'} } \left( O_{00\mbf{k_{1}'}} +    O_{11\mbf{k_1'}} \right) \right]\\
    t_{1,\mbf{k}_1}=t_{2,\mbf{k}_2}=\frac{2\left( V_G |\langle-\frac{\mbf{G}}{2}|\frac{\mbf{G}}{2}\rangle|^2- V_{0}  \right) }{\Omega_{tot}}\sideset{}{'}\sum_{\mbf{k'_1}}      \Re O_{01\mbf{k'_1}}\eqcolon  t,\label{eqapp:1dtrashcantnontrivialform}
\end{gather}}
according to the parameterization of the mean-field Hamiltonian in Eq.~\ref{eqapp:1dtrashcanHHFrealparams}.
Reshifting the band renormalization fields relative to $\mu_{0,1}$ yields
\begin{align}
    \mu'_{0,\mbf{k}_0}=&\frac{1 }{\Omega_{tot}}\Bigg[
    \sum_{\mbf{k'_0}}\left(V_{|\frac{\mbf{G}}{2} - \mbf{k'_0}|} |\langle \frac{\mbf{G}}{2} |\mbf{k'_0}\rangle|^2 - V_{|\mbf{k_0}- \mbf{k'_0}|} |\langle \mbf{k_0}|\mbf{k'_0}\rangle|^2 \right)O_{00\mbf{k'_0}}\\
    &+\left(V_0+ V_{ G } |\langle -\frac{\mbf{G}}{2} | \frac{\mbf{G}}{2} \rangle|^2-V_{|\mbf{k_0}-\frac{\mbf{G}}{2}  |} |\langle \mbf{k_0}| \frac{\mbf{G}}{2} \rangle|^2  - V_{|\mbf{k_0}+\frac{ \mbf{G}} {2}  |} |\langle \mbf{k_0}| -\frac{\mbf{G}}{2} \rangle|^2\right) \sideset{}{'}\sum_{\mbf{k_1}}( O_{11\mbf{k_1}}+    O_{00\mbf{k_1}} )\Bigg].
\end{align}
Within the momentum regions $\mbf{k}_0$ and $\mbf{k}_1$, the mean-field Hamiltonian is parameterized according to Eq.~\ref{eqapp:1dtrashcanHHFk0inversion} and \ref{eqapp:1dtrashcanHHFk1inversion} respectively, while the Hamiltonian in the $\mbf{k}_2$ region can be obtained using inversion. From Eq.~\ref{eqapp:1dtrashcantnontrivialform}, we conclude that a non-trivial solution with $t\neq 0$ requires as a necessary condition
\begin{equation}
    V_G |\langle-\frac{\mbf{G}}{2}|\frac{\mbf{G}}{2}\rangle|^2<V_{0},
\end{equation}
because $O_{01,\bm{k}_1}$ has the opposite sign to $t$. Note that this is not a sufficient condition for obtaining a gapped state, since for example there could be Fermi surfaces depending on the details of $\mu_{0,\bm{k}_0}$, etc. Furthermore, unlike in the 2D model discussed in later Appendix sections, the phases of the form factors do not enter the analysis here.

\clearpage

\section{2D Berry Trashcan Model: Setup}\label{secapp:2dmodel_setup}

We consider a 2D model Hamiltonian that captures the main features of pristine R$n$G at low energies. In this appendix section, we describe the Hamiltonian, which generalizes the 1D setup of App.~\ref{secapp:1dtrashcansetup} to a 2D system with a hexagonal BZ.

\subsection{Setup}
In the presence of an appropriately-tuned interlayer potential $V$, the dispersion of the lowest conduction band of R$n$G (in the absence of the moir\'e potential) is gapped from the valence band and becomes very flat within a circle in momentum space centered at the Dirac momentum.  Beyond this circle, the band disperses quickly to higher energies. The band dispersion hence resembles a trashcan with a flat bottom and steep sides. The extraction of the kinetic parameters of the Berry Trashcan model based on R$n$G is presented in App.~\ref{subsecapp:trashcan_parameterization}

 \begin{figure}
 \centering
\includegraphics[width=0.6\columnwidth]{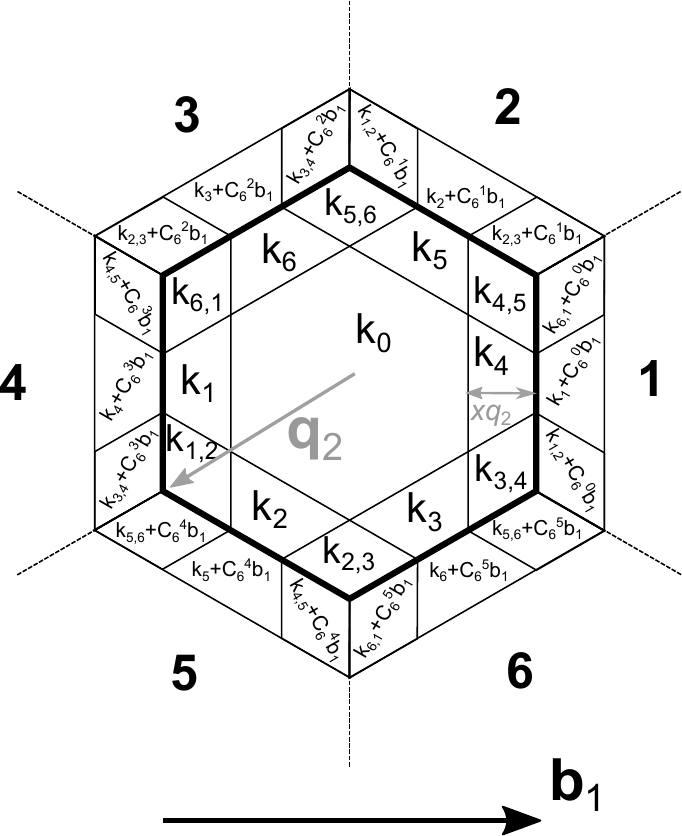} 
\caption{2D Berry Trashcan model with states belonging to seven BZ's 0, 1, 2, 3, 4, 5 and 6. (The choice of BZ sets the periodicity of the putative Wigner crystal state.) The momenta of the reduced BZ coincide with BZ 0, which is indicated by the thick hexagon. $\mbf{b}_1$ is a primitive RLV. We also indicate $\bm{q}_2$ which connects the $\Gamma_M$ point with one of the BZ corners (which corresponds to $K_M$). The band dispersion is nearly flat within BZ 0, and quickly disperses outside. Owing to the sharp dispersion, only small regions of momentum space outside BZ 0 need to be kept in the model. In particular, only states within the outermost hexagonal cutoff are kept in the single-particle Hilbert space. This cutoff forms a `sleeve' of width $\Lambda=x q_2$ outside BZ 0. The BZ is divided into non-overlapping regions depending on how many BZ's contain states within the cutoff that fold into that region. For instance, the region $\mbf{k}_0$ is associated only with states from BZ 0, the region $\mbf{k}_1$ is associated with states from BZ 0 and 1, and the region $\mbf{k}_{1,2}$ is associated with states from BZ 0, 1 and 2. Analogous statements hold for the other regions of the BZ. For BZs other than BZ 0, we also indicate the (unfolded) momentum in terms of the folded BZ momentum. $C_6$ corresponds to counterclockwise rotation by $\pi/3$. For instance, states in BZ 1 within the cutoff can be divided into those whose momenta are of the form $\bm{k}_1+\bm{b}_1$, $\bm{k}_{6,1}+\bm{b}_1$ or $\bm{k}_{1,2}+\bm{b}_1$.}
\label{figapp:2dtrashcanBZ}
\end{figure} 

For appropriately tuned electronic densities, the radius of the flat bottom approximately coincides the Fermi wavevector. This occurs for example for R5G for an effective filling factor of $\nu=1$ corresponding to twist angle $\theta\simeq 0.77^\circ$. For such situations, we consider the setup shown in Fig.~\ref{figapp:2dtrashcanBZ}. Since we are interested in phases which may break the continuous translation symmetry spontaneously, or may form in the presence of a extrinsic moir\'e potential, we specify a hexagonal BZ characterized by a primitive RLV $\mbf{b}_1$. This partitions momentum space into BZs, where BZ 0 is centered at the graphene Dirac momentum. The dispersion is relatively flat near the center of BZ 0, but increases rapidly for larger momenta. Hence, we impose a hexagonal momentum cutoff on allowed single-particle states (Fig.~\ref{figapp:2dtrashcanBZ}). (Note that we can consider more general cutoff geometries, such as a circular cutoff used in the numerical HF calculations in App.~\ref{secapp:HFphasediagrams}. The cutoff should preserve the symmetries of the Hamiltonian.) Due to the sharp trashcan-like dispersion, the cutoff can be chosen such that only narrow regions of states in BZs 1, 2, 3, 4, 5 and 6 are kept (no states in any higher BZs are included in the theory, which imposes a maximum size on the width $\Lambda=xq_2$ in momentum space within which states outside BZ 0 are kept). As illustrated in Fig.~\ref{figapp:2dtrashcanBZ}, the BZ momenta are divided into non-overlapping regions depending on how many unfolded momenta within the cutoff fold onto them.

The general form of the Hamiltonian (without the momentum cutoff) takes the same form as Eq.~\ref{eqapp:Hintproject}
\begin{equation}\label{eqapp:2dtrashcanham}
    H=\sum_k E(\mbf{k}) \gamma_{\mbf{k}}^\dagger \gamma_{\mbf{k}} + \frac{1}{2\Omega_{tot}}  \sum_{\bm{q}, \bm{k}, \bm{k'}} V_{\mbf{q}} M_{\mbf{k},\mbf{q}} M_{\mbf{k}',-\mbf{q}} \gamma_{\mbf{k}+ \mbf{q}}^\dagger \gamma_{\mbf{k}'- \mbf{q}}^\dagger \gamma_{\mbf{k}'} \gamma_{\mbf{k}}
\end{equation}
where $\gamma^\dagger_{\mbf{k}}$ is the band creation operator, $E(\mbf{k})$ is the band dispersion, $V_{\mbf{q}}$ is the interaction potential, and $M_{\mbf{k},\mbf{q}}=\langle \mbf{k}+\mbf{q}|\mbf{k} \rangle$ is the form factor. 

With the momentum cutoff shown in Fig.~\ref{figapp:2dtrashcanBZ}, we introduce a convenient notation (which we call BZ notation or `folded' notation) for electron operators. The electron operators in BZ 0 are spanned by
\begin{equation}
    \gamma_{0,\bm{k}_0},\gamma_{0,\bm{k}_1},\ldots,\gamma_{0,\bm{k}_6},\gamma_{0,\bm{k}_{1,2}},\ldots,\gamma_{0,\bm{k}_{5,6}},\gamma_{0,\bm{k}_{6,1}},
\end{equation}
where the subscript 0 indicates BZ 0. The corresponding regions of the BZ are indicated in Fig.~\ref{figapp:2dtrashcanBZ}. As a reminder, $\bm{k}_0$ indicates BZ momenta which only contain states from BZ 0, $\bm{k}_j$ for $j\neq 0$ indicates BZ momenta which contain states from BZ 0 and BZ $j$, and $\bm{k}_{i,j}$ indicates BZ momenta which contain states BZ 0, BZ $i$ and BZ $j$. Since the BZ coincides with BZ 0, then the physical momentum for BZ 0 coincides with the BZ momentum.

The electron operators in BZ $j$ for $j=1,\ldots,6$ are spanned by
\begin{equation}
    \gamma_{j,\bm{k}_j},\gamma_{j,\bm{k}_{j,j+1}},\gamma_{j,\bm{k}_{j-1,j}}
\end{equation}
where $j-1$ and $j+1$ in this notation will always looped back mod 6 into the range $1,\ldots 6$. More generally in any subscript, $j-1$ and $j+1$ will always be looped back mod 6 into the range $1,\ldots,6$ mod 6 unless otherwise stated. The corresponding physical momenta are $\bm{k}_j+C_6^{j-1}\bm{b}_1,\bm{k}_{j,j+1}+C_6^{j-1}\bm{b}_1,\bm{k}_{j-1,j}+C_6^{j-1}\bm{b}_1$, as indicated in Fig.~\ref{figapp:2dtrashcanBZ}. Here, $C_6$ refers to a counterclockwise rotation by $\pi/3$.

In the HF analysis, the order parameter is given by the one-body density matrix for the Slater determinant state
\begin{equation}\label{appeq:Okk'}
    O_{\mbf{k},\mbf{k}'}=\langle \gamma^\dagger_{\mbf{k}}\gamma_{\mbf{k}'}\rangle.
\end{equation}
The continuous translation invariance of the Hamiltonian (Eq.~\ref{eqapp:1dtrashcanham}) is preserved if $O_{\mbf{k},\mbf{k}'}\sim \delta_{\mbf{k},\mbf{k}'}$. However, we are primarily interested in investigating HF states which can break this symmetry down to a discrete translation group characterized by the primitive RLV $\bm{b}_1$ and its symmetry-related partners. In this case, the order parameter is allowed to be non-vanishing as long as $\mbf{k}$ and $\mbf{k}'$ are identical modulo a RLV. We can hence write down the decomposition
\begin{gather}\label{eqapp:2dtrashcanOPkknosym}
     O_{\mbf{k}\mbf{k'}}= O_{00}(\mbf{k_0}) \delta_{\mbf{k}, \mbf{k'}} \delta_{\mbf{k, k_0}} + \sum_{i=1}^6 \bigg[ O_{00}(\mbf{k_i}) \delta_{\mbf{k}, \mbf{k'}} \delta_{\mbf{k, k_i}}+ O_{ii}(\mbf{k_i})  \delta_{\mbf{k}, \mbf{k'}}  \delta_{\mbf{k}, \mbf{k_i}+ C_6^{i-1} \mbf{b_{1}}} \\  + O_{00}(\mbf{k_{i,i+1}}) \delta_{\mbf{k}, \mbf{k'}} \delta_{{\mbf{k, k_{i,i+1}}}}+  O_{ii}(\mbf{k_{i,i+1}})  \delta_{\mbf{k}, \mbf{k'}}  \delta_{\mbf{k}, \mbf{k_{i,i+1}}+ C_6^{i-1} \mbf{b_{1}}} +  O_{ii}(\mbf{k_{i-1,i}})  \delta_{\mbf{k}, \mbf{k'}}  \delta_{\mbf{k}, \mbf{k_{i-1,i}}+ C_6^{i-1} \mbf{b_{1}}}  \\ +  O_{0i}(\mbf{k_i})  \delta_{\mbf{k}, \mbf{k_i}}  \delta_{\mbf{k}', \mbf{k_i}+ C_6^{i-1} \mbf{b_{1}}} + O_{0i}(\mbf{k_{i,i+1}})  \delta_{\mbf{k}, \mbf{k_{i,i+1}}}  \delta_{\mbf{k}', \mbf{k_{i,i+1} }+ C_6^{i-1} \mbf{b_{1}}} +  O_{0i}(\mbf{k_{i-1,i}})  \delta_{\mbf{k}, \mbf{k_{i-1,i}}}  \delta_{\mbf{k}', \mbf{k_{i-1,i} }+ C_6^{i-1} \mbf{b_{1}}}\\ + O_{i0}(\mbf{k_i})  \delta_{\mbf{k}', \mbf{k_i}}  \delta_{\mbf{k}, \mbf{k_i}+ C_6^{i-1} \mbf{b_{1}}} +O_{i0}(\mbf{k_{i,i+1}})  \delta_{\mbf{k}', \mbf{k_{i,i+1}}}  \delta_{\mbf{k}, \mbf{k_{i,i+1}}+ C_6^{i-1} \mbf{b_{1}}} + O_{i0}(\mbf{k_{i-1,i}})  \delta_{\mbf{k}', \mbf{k_{i-1,i}}}  \delta_{\mbf{k}, \mbf{k_{i-1,i}}+ C_6^{i-1} \mbf{b_{1}}} \\  + O_{ii+1}(\mbf{k_{i,i+1}})  \delta_{\mbf{k}, \mbf{k_{i,i+1}}+ C_6^{i-1} \mbf{b_{1}}}  \delta_{\mbf{k}', \mbf{k_{i,i+1}}+ C_6^{i} \mbf{b_{1}}} + O_{i+1 i }(\mbf{k_{ i,i+1}})  \delta_{\mbf{k}, \mbf{k_{i,i+1 }}+ C_6^{i} \mbf{b_{1}}}  \delta_{\mbf{k}', \mbf{k_{ i,i+1 }}+ C_6^{i-1} \mbf{b_{1}}}\bigg],
\end{gather} 
where we have introduced the complex order parameter components $O_{ij}(\bm{k})\equiv \langle \gamma^\dagger_{\bm{k}+C^{i-1}_6\bm{b}_1}\gamma_{\bm{k}+C^{j-1}_6\bm{b}_1}\rangle$. For BZ momenta $\bm{k}_0$, we only have $O_{00}$. For BZ momenta $\bm{k}_j$ with $j\neq 0$, we have $O_{00},O_{0j},O_{j0},O_{jj}$. For BZ momenta $\bm{k}_{j,j+1}$, we have $O_{00}, O_{j0},  O_{0j},O_{jj}, O_{j,j+1},O_{j+1,j}{,O_{j+1,0},O_{0,j+1},O_{j+1,j+1}}$. Analogous to the discussion regarding the order parameter decomposition in the 1D case (see Eq.~\ref{eqapp:1dOPdecomposition}), care needs to be taken with equivalent entries of the order parameter at the boundary of the BZ that are not due to symmetry. For example, $O_{00}(\bm{k}_{1,2}=\bm{q}_2),O_{44}(\bm{k}_{3,4}=C_6^2\bm{q}_2)$, and $O_{55}(\bm{k}_{5,6}=C^{-2}_6\bm{q}_2)$ are equivalent and involve momenta at the BZ corner. Similarly, $O_{00}(\bm{k}_{4}=\bm{b}_1/2)$ and $O_{11}(\bm{k}_{1}=-\bm{b}_1/2)$ are equivalent and involve momenta at the BZ edge. This multiplicity subtlety will be discussed in more detail below Eq.~\ref{eqapp:2dtrashcanq0meanfield}.

\subsubsection{Symmetries}
We introduce the unitary (intravalley) $\frac{2\pi}{6}$-rotation operator $C_6$ and the antiunitary operator $M_1\mathcal{T}$ which is a combination of a mirror that flips $x$ and time-reversal (see App.~\ref{subsecapp:symmetryRnG/hBN}). For models like the holomorphic limit described in App.~\ref{subsecapp:holomorphicwfns} of primary interest in this work, these are indeed symmetries of the Hamiltonian. In this case, we choose the action on the band creation operators at physical momentum $\bm{k}$ as
\begin{gather}\label{eqapp:2dtrashcanC6op}
    C_6 \gamma_{\mbf{k}}^\dagger C_6^{-1} = e^{i\frac{\pi n}{6} } \gamma_{C_6 \mbf{k}}^\dagger\\
    M_1\mathcal{T}\gamma_{\mbf{k}}^\dagger (M_1 \mathcal{T})^{-1} = \gamma_{M_1\mathcal{T}\mbf{k}}^\dagger
\end{gather}
where $n$ is the number of layers. We also define $\frac{2\pi}{3}$- and $\pi$-rotation operators $C_3=C_6^2$ and $C_2=C_6^3$ respectively. Note that the microscopic lattice $\frac{2\pi}{6}$- and $\pi$-rotation operators, which are different from the ones considered here, would flip the valley index in graphene. These are also not symmetries for pristine rhombohedral graphene. Hence, the $C_6$ symmetry introduced in Eq.~\ref{eqapp:2dtrashcanC6op} is an emergent intravalley symmetry of the low-lying continuum Hamiltonian. Note that the full pristine rhombohedral graphene continuum model only has $C_3$ and $M_1\mathcal{T}$. At the level of form factor overlaps, $C_6$- and $M_1\mathcal{T}$-symmetry lead to
\begin{equation}
    \langle\mbf{k}|\mbf{k'}\rangle= \langle C_{6} \mbf{k}|C_{6} \mbf{k'}\rangle = \langle M_1\mathcal{T}\mbf{k'}|M_1\mathcal{T}\mbf{k}\rangle.
\end{equation}

For the creation operators in folded (BZ) notation, we have
\begin{gather}
C_6 \gamma^\dagger_{0, \mbf{k_0}} C_6^{-1}={e^{i\frac{\pi n}{6} }}\gamma^\dagger_{0, C_6 \mbf{k_{ 0}} }\\
    C_6 \gamma^\dagger_{j, \mbf{k_j}} C_6^{-1}={e^{i\frac{\pi n}{6} }}\gamma^\dagger_{j+1, C_6 \mbf{k_{ j}} }\,\text{   for  }j\neq 0\\
    C_6 \gamma^\dagger_{j, \mbf{k_{j,j+1}}} C_6^{-1}={e^{i\frac{\pi n}{6} }}\gamma^\dagger_{j+1, C_6 \mbf{k_{j,j+1}} }\,\text{   for  }j\neq 0\\
    C_6 \gamma^\dagger_{j, \mbf{k_{j-1,j}}} C_6^{-1}={e^{i\frac{\pi n}{6} }}\gamma^\dagger_{j+1, C_6 \mbf{k_{j-1,j}} }\,\text{   for  }j\neq 0\\
    M_1\mathcal{T} \gamma^\dagger_{0, \mbf{k_0}} (M_1\mathcal{T} )^{-1}=\gamma^\dagger_{0, M_1\mathcal{T} \mbf{k_{ 0}} }\\
    M_1\mathcal{T} \gamma^\dagger_{j, \mbf{k_j}} (M_1\mathcal{T} )^{-1}=\gamma^\dagger_{2-j, M_1\mathcal{T} \mbf{k_{ j}} }\,\text{   for  }j\neq 0\\
     M_1\mathcal{T} \gamma^\dagger_{j, \mbf{k_{j,j+1}}} (M_1\mathcal{T} )^{-1}=\gamma^\dagger_{2-j, M_1\mathcal{T} \mbf{k_{j,j+1}} }\,\text{   for  }j\neq 0\\
     M_1\mathcal{T} \gamma^\dagger_{j, \mbf{k_{j-1,j}}} (M_1\mathcal{T} )^{-1}=\gamma^\dagger_{2-j, M_1\mathcal{T} \mbf{k_{j-1,j}} }\,\text{   for  }j\neq 0
\end{gather}
where $2-j$ is looped back mod 6 into the range $1,\ldots,6$. 

Imposing $C_6$ and $M_1\mathcal{T}$ symmetries on the order parameter leads to 
\begin{equation}
    O_{\bm{k},\bm{k}'}=O_{C_6\bm{k},C_6\bm{k}'},\quad O_{\bm{k},\bm{k}'}=O^*_{M_1\mathcal{T}\bm{k},M_1\mathcal{T}\bm{k}'}
\end{equation}
or in folded notation $(j\neq 0)$
\begin{gather}
    O_{0,0,\mbf{k}_0}=O_{0,0,C_6\mbf{k}_0},\quad O_{0,0,\mbf{k}_{j}}=O_{0,0,C_6\mbf{k}_j},\quad O_{0,0,\mbf{k}_{j,j+1}}=O_{0,0,C_6\mbf{k}_{j,j+1}}, \quad O_{0,0,\mbf{k}_{j-1,j}}=O_{0,0,C_6\mbf{k}_{j-1,j}}\\
    O_{j,j,\mbf{k}_j}=O_{j+1,j+1,C_6\mbf{k}_j},\quad O_{j,j,\mbf{k}_{j,j+1}}=O_{j+1,j+1,C_6\mbf{k}_{j,j+1}},\quad O_{j,j,\mbf{k}_{j-1,j}}=O_{j+1,j+1,C_6\mbf{k}_{j-1,j}}\\
    O_{0,j,\mbf{k}_j}=O_{0,j+1,C_6\mbf{k}_j},\quad O_{0,j,\mbf{k}_{j,j+1}}=O_{0,j+1,C_6\mbf{k}_{j,j+1}},\quad O_{0,j,\mbf{k}_{j-1,j}}=O_{0,j+1,C_6\mbf{k}_{j-1,j}}\\
    O_{j,0,\mbf{k}_j}=O_{j+1,0,C_6\mbf{k}_j},\quad O_{j,0,\mbf{k}_{j,j+1}}=O_{j+1,0,C_6\mbf{k}_{j,j+1}},\quad O_{j,0,\mbf{k}_{j-1,j}}=O_{j+1,0,C_6\mbf{k}_{j-1,j}}\\
    O_{0,0,\mbf{k}_0}=O^*_{0,0,M_1\mathcal{T}\mbf{k}_0},\quad O_{0,0,\mbf{k}_{j}}=O^*_{0,0,M_1\mathcal{T}\mbf{k}_j},\quad O_{0,0,\mbf{k}_{j,j+1}}=O^*_{0,0,M_1\mathcal{T}\mbf{k}_{j,j+1}}, \quad O_{0,0,\mbf{k}_{j-1,j}}=O^*_{0,0,M_1\mathcal{T}\mbf{k}_{j-1,j}}\\
    O_{j,j,\mbf{k}_j}=O^*_{2-j,2-j,M_1\mathcal{T}\mbf{k}_j},\quad O_{j,j,\mbf{k}_{j,j+1}}=O^*_{2-j,2-j,M_1\mathcal{T}\mbf{k}_{j,j+1}},\quad O_{j,j,\mbf{k}_{j-1,j}}=O^*_{2-j,2-j,M_1\mathcal{T}\mbf{k}_{j-1,j}}\\
    O_{0,j,\mbf{k}_j}=O^*_{0,2-j,M_1\mathcal{T}\mbf{k}_j},\quad O_{0,j,\mbf{k}_{j,j+1}}=O^*_{0,2-j,M_1\mathcal{T}\mbf{k}_{j,j+1}},\quad O_{0,j,\mbf{k}_{j-1,j}}=O^*_{0,2-j,M_1\mathcal{T}\mbf{k}_{j-1,j}}\\
    O_{j,0,\mbf{k}_j}=O^*_{2-j,0,M_1\mathcal{T}\mbf{k}_j},\quad O_{j,0,\mbf{k}_{j,j+1}}=O^*_{2-j,0,M_1\mathcal{T}\mbf{k}_{j,j+1}},\quad O_{j,0,\mbf{k}_{j-1,j}}=O^*_{2-j,0,M_1\mathcal{T}\mbf{k}_{j-1,j}}.
\end{gather}
Using these relations and $\mbf{k}_i=C_6^{i-1} \mbf{k}_1,\,\mbf{k}_{i,i+1} = C_6^{i-1} \mbf{k}_{1,2}$, the decomposition of the full order parameter in Eq.~\ref{eqapp:2dtrashcanOPkknosym} can be reduced to quantities defined over just $\bm{k}_0,\bm{k}_1,\bm{k}_{1,2}$ in the BZ
\begin{gather}
    O_{\mbf{k}\mbf{k'}}= O_{00}(\mbf{k_0}) \delta_{\mbf{k}, \mbf{k'}} \delta_{\mbf{k, k_0}} + \sum_{i=1}^6 \bigg[ O_{00}(\mbf{k_1}) \delta_{\mbf{k}, \mbf{k'}} \delta_{\mbf{k, C_6^{i-1}k_1}}+ O_{11}(\mbf{k_1})  \delta_{\mbf{k}, \mbf{k'}}  \delta_{\mbf{k}, C_6^{i-1}(\mbf{k_1}+  \mbf{b_{1}})}\\  + O_{00}(\mbf{k_{1,2}}) \delta_{\mbf{k}, \mbf{k'}} \delta_{C_6^{i-1} \mbf{k_{1,2}}}+  O_{11}(\mbf{k_{1,2}})  \delta_{\mbf{k}, \mbf{k'}}  \delta_{\mbf{k}, C_6^{i-1}(\mbf{k_{1,2}}+  \mbf{b_{1}})} +  O_{22}(\mbf{k_{1,2}})  \delta_{\mbf{k}, \mbf{k'}}  \delta_{\mbf{k},  C_6^{i-1}( C_6^{-1} \mbf{k_{1,2}}+  \mbf{b_{1}})} \\ +  O_{01}(\mbf{k_1})  \delta_{\mbf{k}, C_6^{i-1}\mbf{k_1}}  \delta_{\mbf{k}', C_6^{i-1}(\mbf{k_1}+  \mbf{b_{1}})} + O_{01}(\mbf{k_{1,2}})  \delta_{\mbf{k}, C_6^{i-1}\mbf{k_{1,2}}}  \delta_{\mbf{k}',C_6^{i-1}( \mbf{k_{1,2} }+  \mbf{b_{1}})} +  O_{02}(\mbf{k_{1,2}})  \delta_{\mbf{k}, C_6^{i-2}\mbf{k_{1,2}}}  \delta_{\mbf{k}',C_6^{i-1} (C_6^{-1}\mbf{k_{1,2} }+ \mbf{b_{1}})}\\ + O_{10}(\mbf{k_1})  \delta_{\mbf{k}', C_6^{i-1} \mbf{k_1}}  \delta_{\mbf{k}, C_6^{i-1}( \mbf{k_1}+  \mbf{b_{1}})} +O_{10}(\mbf{k_{1,2}})  \delta_{\mbf{k}', C_6^{i-1}\mbf{k_{1,2}}}  \delta_{\mbf{k}, C_6^{i-1}( \mbf{k_{1,2}}+  \mbf{b_{1}})} + O_{20}(\mbf{k_{1,2}})  \delta_{\mbf{k}', C_6^{i-2}\mbf{k_{1,2}}}  \delta_{\mbf{k}, C_6^{i-1}( C_6^{-1}\mbf{k_{1,2}}+  \mbf{b_{1}})} \\  + O_{12}(\mbf{k_{1,2}})  \delta_{\mbf{k}, C_6^{i-1}(\mbf{k_{1,2}}+  \mbf{b_{1}})}  \delta_{\mbf{k}', C_6^{i-1}(\mbf{k_{1,2}}+ C_6 \mbf{b_{1}})} + O_{2 1 }(\mbf{k_{ 1,2}})  \delta_{\mbf{k}, C_6^{i-1} (\mbf{k_{1,2 }}+ C_6 \mbf{b_{1}})}  \delta_{\mbf{k}', C_6^{i-1}( \mbf{k_{ 1,2 }}+  \mbf{b_{1}})}\bigg].
\end{gather}

\clearpage

\section{2D Berry Trashcan Model: Hartree-Fock Analysis}\label{secapp:2dmodel_HF}

We first consider the mean-field decoupling of the interaction term. Wick's theorem yields 
\begin{align}
\gamma_{\mbf{k}+ \mbf{q}}^\dagger \gamma_{\mbf{k}'- \mbf{q}}^\dagger \gamma_{\mbf{k}'}\gamma_{\mbf{k}}
=&O_{\mbf{k} + \mbf{q},\mbf{k}}   \gamma_{\mbf{k}'- \mbf{q}}^\dagger \gamma_{\mbf{k'}}+ 
 O_{\mbf{k'} - \mbf{q},\mbf{k'}}   \gamma_{\mbf{k}
+ \mbf{q}}^\dagger \gamma_{\mbf{k}} -  O_{\mbf{k} + \mbf{q},\mbf{k'}}   \gamma_{\mbf{k}'- \mbf{q}}^\dagger \gamma_{\mbf{k}} -  O_{\mbf{k'} - \mbf{q},\mbf{k}}   \gamma_{\mbf{k}+ \mbf{q}}^\dagger \gamma_{\mbf{k'}}\\ & - O_{\mbf{k} + \mbf{q},\mbf{k}}  O_{\mbf{k'} - \mbf{q},\mbf{k'}} + O_{\mbf{k} + \mbf{q},\mbf{k'}} O_{\mbf{k'} - \mbf{q},\mbf{k}}. 
\end{align}
The decoupling of the interaction term is then
\begin{align}
    H_{\text{int}}^{\text{HF}}-E^{\text{HF,int}}_\text{tot}=& \frac{1}{\Omega_{tot}}\sum_{\mbf{k,k',q}}\left( V(q) M_{\mbf{k}, \mbf{q}} M_{\mbf{k'}, -\mbf{q}}  O_{\mbf{k'} - \mbf{q},\mbf{k'}}   \gamma_{\mbf{k}
+ \mbf{q}}^\dagger \gamma_{\mbf{k}} - V(q) M_{\mbf{k}, \mbf{q}} M_{\mbf{k'}, -\mbf{q}}  O_{\mbf{k'} - \mbf{q},\mbf{k}}   \gamma_{\mbf{k}
+ \mbf{q}}^\dagger \gamma_{\mbf{k'}}\right)\\ & +   \frac{1}{2\Omega_{tot}}\sum_{\mbf{k,k',q}} \left(V(|\mbf{k'}-\mbf{k}- \mbf{q}| ) M_{\mbf{k}, \mbf{k'}- \mbf{k}- \mbf{q}} M_{\mbf{k'},\mbf{k}- \mbf{k'} + \mbf{q}}  -V(q) M_{\mbf{k}, \mbf{q}} M_{\mbf{k'}, -\mbf{q}}\right)O_{\mbf{k} + \mbf{q},\mbf{k}}  O_{\mbf{k'} - \mbf{q},\mbf{k'}} ,
\end{align}
where $E^{\text{HF,int}}_\text{tot}$ is the interacting contribution to the total HF energy.
We now focus on the Hartree and Fock terms of the one-body mean-field Hamiltonian $H_{\text{int}}^{\text{HF}}$, assuming continuous translation symmetry is broken to a discrete subgroup characterized by the primitive RLV $\bm{b}_1$ and its symmetry related partners. We are interested in situations where the key physics can be captured with a cutoff that only extends slightly outside the BZ (i.e.~$\Lambda$ is small relative to $q_2$). This constrains the possible momentum transfers $\bm{q}$ of the order parameter $O_{\bm{k},\bm{k}'}$ to $\mbf{0}$ and $C_6^{j-1}\mbf{b}_1$ for $j=1,\ldots,6$. Hence for the Hartree term above, the summation over $\bm{q}$ is restricted to $\bm{q}=\mbf{0}$ and $\bm{q}=C_6^{j-1}\mbf{b}_1$ for $j=1,\ldots,6$, and for the Fock term, the summation over $\bm{q}$ is restricted to $\mbf{q}= \mbf{k'-k}$ and $ \mbf{q}= \mbf{k'-k} + C_6^{j-1} \mbf{b_{1}}$ for $j=1,\ldots, 6$. This leads to
\begin{align}\label{eqapp:2dtrashcanHHF_Hartree0}
    H_{\text{int}}^{\text{HF}}= \frac{1}{\Omega_{tot}}\sum_{\mbf{k,k'}}\Bigg[&V(0)   O_{\mbf{k'} ,\mbf{k'}}   \gamma_{\mbf{k}}^\dagger \gamma_{\mbf{k}}\\\label{eqapp:2dtrashcanHHF_HartreeG}
    +\sum_{j=1}^6 &V(b_{1}) \langle \mbf{k+C_6^{j-1}b_{1}} | \mbf{k} \rangle \langle \mbf{k'-C_6^{j-1}b_{1}} | \mbf{k'} \rangle O_{\mbf{k'} - C_6^{j-1}\mbf{b_{1}},\mbf{k'}} \gamma_{\mbf{k}
+ C_6^{j-1}\mbf{b_{1}}}^\dagger \gamma_{\mbf{k}}\\\label{eqapp:2dtrashcanHHF_Fock0}
    -&V(|\mbf{k-k'}|) |\langle \mbf{k'}|\mbf{k} \rangle |^2 O_{\mbf{k'},\mbf{k'}}   \gamma_{\mbf{k}}^\dagger \gamma_{\mbf{k}}\\\label{eqapp:2dtrashcanHHF_FockG}
    -\sum_{j=1}^6 &V(|\mbf{k-k'+C_6^{j-1} b_{1}}|) \langle \mbf{k+C_6^{j-1} b_{1}}|\mbf{k'} \rangle \langle \mbf{k'-C_6^{j-1} b_{1}}|\mbf{k} \rangle O_{\mbf{k'} - \mbf{C_6^{j-1} b_{1}},\mbf{k'}}   \gamma_{\mbf{k}
+ \mbf{C_6^{j-1} b_{1}}}^\dagger \gamma_{\mbf{k}} \Bigg].
\end{align}

\subsection{General parameterization}\label{secapp:2dtrashcan_genparam}
We now massage the above equations into a more manageable form. In particular, we will separate the interacting part of the mean-field Hamiltonian into a $\bm{q}=0$ part (Eq.~\ref{eqapp:2dtrashcanHHFfparam}) and a $\bm{q}\neq 0$ part (Eq.~\ref{eqapp:2dtrashcanHHFgparam}). The kinetic term is given in Eq.~\ref{eqapp:2dtrashcanHHFkinparam}.

\subsubsection{$\mathbf{q}=0$ mean-field terms}
We first consider Eqs.~\ref{eqapp:2dtrashcanHHF_Hartree0} and \ref{eqapp:2dtrashcanHHF_Fock0}, which correspond to terms $\sim \gamma^\dagger_{\mbf{k}}\gamma_{\mbf{k}}$ that are diagonal in the BZ index. We find
\begin{align}\label{eqapp:2dtrashcanq0meanfield}
    \frac{1}{\Omega_{tot}}&\sum_{\mbf{k,k'}}   \left(V_0 O_{\mbf{k'} ,\mbf{k'}}   -V(|\mbf{k-k'}|) |\langle \mbf{k'}|\mbf{k} \rangle |^2 O_{\mbf{k'},\mbf{k'}}\right)   \gamma_{\mbf{k}}^\dagger \gamma_{\mbf{k}}\\  &=  \frac{1}{\Omega_{tot}} \sum_{\mbf{k}}\Bigg[\sum_{\mbf{k'_0}} \bigg( V_0 - V_{|\mbf{k-k'_0}|} |\langle \mbf{k'_0}|\mbf{k} \rangle |^2  \bigg)O_{00,\mbf{k'_0}}\\ 
    &+ \sum_{i=1}^6  \sideset{}{'}\sum_{\mbf{k'_i}} \bigg( (V_0- V_{|\mbf{k-k'_i}|} |\langle \mbf{k'_i}|\mbf{k} \rangle |^2 )O_{00,\mbf{k'_i}} +( V_0 - V_{|\mbf{k}-\mbf{k'_i}- C_6^{i-1} \mbf{b_{1}}|} |\langle \mbf{k'_i}+ C_6^{i-1} \mbf{b_{1}}|\mbf{k} \rangle |^2 )O_{ii,\mbf{k'_i}}  \bigg) \\  
    &+\sum_{i=1}^6  \sideset{}{'}\sum_{\mbf{k'_{ii+1}}} \bigg( (V_0-V_{|\mbf{k-k'_{i,i+1}}|} |\langle \mbf{k'_{i,i+1}}|\mbf{k} \rangle |^2) O_{00,\mbf{k'_{i,i+1}}}   
    \\
    & \quad + (V_0- V_{|\mbf{k}-\mbf{k'_{i,i+1}}- C_6^{i-1} \mbf{b_{1}}|}  |\langle \mbf{k'_{i,i+1}}+ C_6^{i-1} \mbf{b_{1}}|\mbf{k} \rangle |^2 ) O_{ii,\mbf{k'_{i,i+1}}}  \\ 
    & \quad +  (V_0-V_{|\mbf{k}-\mbf{k'_{i,i+1}}- C_6^{i} \mbf{b_{1}}|} |\langle \mbf{k'_{i,i+1}}+ C_6^{i} \mbf{b_{1}}|\mbf{k} \rangle |^2) O_{i+1i+1,\mbf{k'_{i,i+1}}} \bigg)   \Bigg] \gamma_{\mbf{k}}^\dagger \gamma_{\mbf{k}}.
\end{align}
The summations over $\bm{k}_i$ and $\bm{k}_{i,i+1}$ are primed, which indicates that `multiplicities' at the BZ boundary need to be accounted for in order to prevent overcounting terms. As an example of potential overcounting, note that the two order parameter entries $O_{00,\bm{k}'_1=-\bm{G}/2}$ and $O_{44,\bm{k}'_4=\bm{G}/2}$ correspond to the same correlator $\langle \gamma^\dagger_{-\bm{G}/2}\gamma_{-\bm{G}/2} \rangle$, despite both order parameters formally belonging to different momentum regions in Fig.~\ref{figapp:2dtrashcanBZ}. This double equivalence applies to any correlator that involves momenta which both lie at the straight edge segments of the BZ. Similarly, correlators where both momenta lie at the BZ corners can be represented by three different order parameter entries from different momentum regions.
To account for this, the summations $\sideset{}{'}\sum_{\bm{k}_i}$ and $\sideset{}{'}\sum_{\bm{k}_{i,i+1}}$ come with a factor of $1/2$ if the summation momentum lies on a BZ edge. Furthermore, $\sideset{}{'}\sum_{\bm{k}_{i,i+1}}$ comes with a factor of $1/3$ if the summation momentum is at a BZ corner. Note that the errors incurred from neglecting this vanish as the momentum spacing goes to zero in the thermodynamic limit.

Using the decomposition
\begin{eqnarray}
    &\gamma_{\mbf{k}
}^\dagger \gamma_{\mbf{k}} =   \gamma_{0\mbf{k_0}}^\dagger \gamma_{0\mbf{k_0}}\delta_{\mbf{k,k_0}} +\sum_{i=1}^6\big(\gamma_{0\mbf{k_i}}^\dagger \gamma_{0\mbf{k_i}} \delta_{\mbf{k, k_i}} + \gamma_{i\mbf{k_i}}^\dagger \gamma_{i\mbf{k_i}} \delta_{\mbf{k, k_i+ C_6^{i-1}b_{1}}} \\ &+ \gamma_{0\mbf{k_{i,i+1}}}^\dagger \gamma_{0\mbf{k_{i,i+1}}} \delta_{\mbf{k, k_{i,i+1}}}  + \gamma_{i\mbf{k_{i,i+1}}}^\dagger \gamma_{i\mbf{k_{i,i+1}}}\delta_{\mbf{k, k_{i,i+1} + C_6^{i-1} b_{1}}}  + \gamma_{i+1\mbf{k_{i,i+1}}}^\dagger \gamma_{i+ 1\mbf{k_{i,i+1}}}\delta_{\mbf{k, k_{i,i+1} + C_6^{i} b_{1}}} \big) ,
\end{eqnarray}
the $\bm{q}=0$ terms can be parameterized as follows
\begin{gather}\label{eqapp:2dtrashcanHHFfparam}
    \sum_{\bm{k}}f_{\bm{k}}\gamma^\dagger_{\bm{k}}\gamma_{\bm{k}}=\sum_{\mbf{k_0}} f_{\mbf{k_0}}\gamma_{0\mbf{k_0}}^\dagger \gamma_{0\mbf{k_0}}  +\sum_{i=1}^6 \sideset{}{'}\sum_{\mbf{k_i}} \left(f_{\mbf{k_i}}\gamma_{0\mbf{k_i}}^\dagger \gamma_{0\mbf{k_i}} + f_{\mbf{k_i+ C_6^{i-1}b_{1}}} \gamma_{i\mbf{k_i}}^\dagger \gamma_{i\mbf{k_i}}\right)\\ + \sum_{i=1}^6\sideset{}{'}\sum_{\mbf{k_{i,i+1}}} \left( f_{\mbf{k_{i,i+1}}} \gamma_{0\mbf{k_{i,i+1}}}^\dagger \gamma_{0\mbf{k_{i,i+1}}}+ f_{\mbf{k_{i,i+1} + C_6^{i-1} b_{1}}}\gamma_{i\mbf{k_{i,i+1}}}^\dagger \gamma_{i\mbf{k_{i,i+1}}} + f_{\mbf{k_{i,i+1} + C_6^{i} b_{1}}}\gamma_{i+1\mbf{k_{i,i+1}}}^\dagger \gamma_{i+ 1\mbf{k_{i,i+1}}} \right)
\end{gather}
where we have introduced the momentum-dependent interaction-induced band renormalization field
\begin{align}\label{eqapp:fk_start}
    f_{\mbf{k}} =&\frac{1}{\Omega_{tot}}\sum_{\bm{k}'}\left(V_0-V_{|\bm{k}-\bm{k}'|}|\langle \mbf{k'}|\mbf{k} \rangle |^2 \right)O_{\bm{k}',\bm{k}'}\\
    =&\frac{1}{\Omega_{tot}} \Bigg[\sum_{\mbf{k'_0}} ( V_0 - V_{|\mbf{k-k'_0}|} |\langle \mbf{k'_0}|\mbf{k} \rangle |^2  )O_{00,\mbf{k'_0}} \\ &+ \sum_{i=1}^6  \sideset{}{'}\sum_{\mbf{k'_i}} \left( (V_0- V_{|\mbf{k-k'_i}|} |\langle \mbf{k'_i}|\mbf{k} \rangle |^2 )O_{00,\mbf{k'_i}} +( V_0 - V_{|\mbf{k}-\mbf{k'_i}- C_6^{i-1} \mbf{b_{1}}|} |\langle \mbf{k'_i}+ C_6^{i-1} \mbf{b_{1}}|\mbf{k} \rangle |^2 )O_{ii,\mbf{k'_i}}  \right) \\ & +\sum_{i=1}^6  \sideset{}{'}\sum_{\mbf{k'_{ii+1}}} \bigg( (V_0-V_{|\mbf{k-k'_{i,i+1}}|} |\langle \mbf{k'_{i,i+1}}|\mbf{k} \rangle |^2) O_{00,\mbf{k'_{i,i+1}}}  \\
    &+(V_0- V_{|\mbf{k}-\mbf{k'_{i,i+1}}- C_6^{i-1} \mbf{b_{1}}|}  |\langle \mbf{k'_{i,i+1}}+ C_6^{i-1} \mbf{b_{1}}|\mbf{k} \rangle |^2 ) O_{ii,\mbf{k'_{i,i+1}}}  \\ &+  (V_0-V_{|\mbf{k}-\mbf{k'_{i,i+1}}- C_6^{i} \mbf{b_{1}}|} |\langle \mbf{k'_{i,i+1}}+ C_6^{i} \mbf{b_{1}}|\mbf{k} \rangle |^2) O_{i+1i+1,\mbf{k'_{i,i+1}}} 
    \bigg)   \Bigg].
\end{align}
This satisfies $f_{\bm{k}}=f^*_{\bm{k}}$, which is consistent with the Hermiticity of the mean-field Hamiltonian.

\subsubsection{$|\mathbf{q}|=|\mathbf{b}_1|$ mean-field terms}
We new consider Eqs.~\ref{eqapp:2dtrashcanHHF_HartreeG} and \ref{eqapp:2dtrashcanHHF_FockG}, which correspond to terms that are off-diagonal in the BZ index. We begin with the Hartree term (Eq.~\ref{eqapp:2dtrashcanHHF_HartreeG})
\begin{eqnarray}\label{eqapp:2dtrashcanHHF_Hartreerepeat}
    \frac{1}{\Omega_{tot}}\sum_{\mbf{k,k'}} \sum_{j=1}^6 V_{b_1} \langle \mbf{k}+C_6^{j-1}\mbf{b}_{1} | \mbf{k} \rangle \langle \mbf{k}'-C_6^{j-1}\mbf{b}_{1} | \mbf{k'} \rangle       O_{\mbf{k'} - C_6^{j-1}\mbf{b_{1}},\mbf{k'}} \gamma_{\mbf{k}
+ C_6^{j-1}\mbf{b_{1}}}^\dagger \gamma_{\mbf{k}}.
\end{eqnarray}
We decompose the order parameter $O_{\mbf{k'} - C_6^{j-1}\mbf{b_{1}},\mbf{k'}}$ above into its components in the different BZs
\begin{align}
    O_{\mbf{k'} - C_6^{j-1}\mbf{b_{1}},\mbf{k'}}&=     O_{0{j},\mbf{k_j'}}     \delta_{\mbf{k}', \mbf{k_j'}+ C_6^{j-1} \mbf{b_{1}}} +  O_{0j,\mbf{k_{j,j+1}'}}   \delta_{\mbf{k}', \mbf{k_{j,j+1}' }+ C_6^{j-1} \mbf{b_{1}}} +  O_{0j,\mbf{k_{j-1,j}'}}   \delta_{\mbf{k}', \mbf{k_{j-1,j}' }+ C_6^{j-1} \mbf{b_{1}}}\\ &+ O_{j-3,0,\mbf{k_{j-3}'}}   \delta_{\mbf{k}', \mbf{k_{j-3}'}}   +O_{j-3,0,\mbf{k_{j-3,j-2}'}}  \delta_{\mbf{k}', \mbf{k_{j-3,j-2}'}}   + O_{j-3,0,\mbf{k_{j-4,j-3}'}}    \delta_{\mbf{k}', \mbf{k_{j-4,j-3}'}}  \\ & + O_{j-2,j-1,\mbf{k_{j-2,j-1}'}}   \delta_{\mbf{k}', \mbf{k_{j-2,j-1}'}+ C_6^{j-2} \mbf{b_{1}}} + O_{j-4, j-5 ,\mbf{k_{ j-5,j-4}'}}    \delta_{\mbf{k}', \mbf{k_{ j-5,j-4 }'}+ C_6^{j} \mbf{b_{1}}}. 
\end{align}
These eight terms can be deduced from inspecting Fig.~\ref{figapp:2dtrashcanBZ}. Substitution into Eq.~\ref{eqapp:2dtrashcanHHF_Hartreerepeat} yields
\begin{align}
    &\frac{1}{\Omega_{tot}}\sum_{\mbf{k,k'}} \sum_{j=1}^6 V_{b_1} \langle \mbf{k+C_6^{j-1}b_{1}} | \mbf{k} \rangle \langle \mbf{k'-C_6^{j-1}b_{1}} | \mbf{k'} \rangle       O_{\mbf{k'} - C_6^{j-1}\mbf{b_{1}},\mbf{k'}} \gamma_{\mbf{k}
+ C_6^{j-1}\mbf{b_{1}}}^\dagger \gamma_{\mbf{k}}\\
&=\frac{1}    {\Omega_{tot}}\sum_{\mbf{k}} \sum_{j=1}^6 V_{b_1} \langle \mbf{k+C_6^{j-1}b_{1}} | \mbf{k} \rangle \bigg[ \sideset{}{'}\sum_{\mbf{k'_j}}   \langle \mbf{{k'_j}} | \mbf{k'_j}+ C_6^{j-1} \mbf{b_{1}} \rangle  O_{0{j},\mbf{k'_j}}   +  \nonumber \\ 
& + \sideset{}{'}\sum_{\mbf{k'_{j,j+1}}}  \langle \mbf{{k'_{j,j+1} }} | \mbf{k'_{j,j+1} }+ C_6^{j-1} \mbf{b_{1}} \rangle  O_{0j,\mbf{k'_{j,j+1}}}   +  \nonumber \\ & + \sideset{}{'}\sum_{\mbf{k'_{j-1,j}}}  \langle \mbf{k'_{j-1,j} } |  \mbf{k'_{j-1,j} }+ C_6^{j-1} \mbf{b_{1}} \rangle  O_{0j,\mbf{k'_{j-1,j}}} + \nonumber\\ 
&+ \sideset{}{'}\sum_{\mbf{k'_{j-3}}}  \langle \mbf{{k'_{j-3}}-C_6^{j-1}b_{1}} |\mbf{k'_{j-3}} \rangle  O_{j-3,0,\mbf{k'_{j-3}}}     + \nonumber \\ 
& + \sideset{}{'}\sum_{\mbf{k'_{j-3,j-2}}}  \langle \mbf{{k'_{j-3,j-2}}-C_6^{j-1}b_{1}} |\mbf{k'_{j-3,j-2}} \rangle O_{j-3,0, \mbf{k'_{j-3,j-2}}}   + \nonumber \\ 
& +\sideset{}{'}\sum_{\mbf{k'_{j-4,j-3}}}   \langle \mbf{{k'_{j-4,j-3}}-C_6^{j-1}b_{1}} | \mbf{k'_{j-4,j-3}} \rangle O_{j-3,0, \mbf{k'_{j-4,j-3}}}      +\nonumber \\ 
& + \sideset{}{'}\sum_{\mbf{k'_{j-2,j-1}}}  \langle \mbf{k'_{j-2,j-1}}+ C_6^{j+3} \mbf{b_{1}} | \mbf{k'_{j-2,j-1}}+ C_6^{j-2} \mbf{b_{1}} \rangle  O_{j-2,j-1,\mbf{k'_{j-2,j-1}}}    + \nonumber \\ 
& + \sideset{}{'}\sum_{\mbf{k'_{ j-5,j-4 }}}   \langle \mbf{k'_{ j-5,j-4 }}+ C_6^{j+1} \mbf{b_{1}} | \mbf{k'_{ j-5,j-4 }}+ C_6^{j} \mbf{b_{1}} \rangle O_{j-4, j-5 ,\mbf{k'_{ j-5,j-4}}}    \bigg]  \gamma_{\mbf{k}
+ C_6^{j-1}\mbf{b_{1}}}^\dagger \gamma_{\mbf{k}}.
\end{align}

We continue with the finite-momentum Fock term (Eq.~\ref{eqapp:2dtrashcanHHF_FockG})
\begin{equation}
    -\frac{1}{\Omega_{tot}}\sum_{\bm{k},\bm{k}'}\sum_{j=1}^6 V_{|\mbf{k}-\mbf{k}'+C_6^{j-1} \mbf{b}_{1}|} \langle \mbf{k}+C_6^{j-1} \mbf{b}_{1}|\mbf{k}' \rangle \langle \mbf{k}'-C_6^{j-1} \mbf{b}_{1}|\mbf{k} \rangle O_{\mbf{k'} - C_6^{j-1} \mbf{b}_{1},\mbf{k'}}   \gamma_{\mbf{k}
+ {C_6^{j-1} \mbf{b}_{1}}}^\dagger \gamma_{\mbf{k}}.
\end{equation}
Using the decomposition of $O_{\mbf{k'} - C_6^{j-1}\mbf{b_{1}},\mbf{k'}}$, we obtain
\begin{align}
    &-\frac{1}{\Omega_{tot}}\sum_{\bm{k},\bm{k}'}\sum_{j=1}^6 V_{|\mbf{k-k'+C_6^{j-1} b_{1}}|} \langle \mbf{k+C_6^{j-1} b_{1}}|\mbf{k}' \rangle \langle \mbf{k'-C_6^{j-1} b_{1}}|\mbf{k} \rangle O_{\mbf{k'} - \mbf{C_6^{j-1} b_{1}},\mbf{k'}}   \gamma_{\mbf{k}+ \mbf{C_6^{j-1} b_{1}}}^\dagger \gamma_{\mbf{k}}\\
&=-\frac{1}{\Omega_{tot}}\sum_{\mbf{k}}\sum_{j=1}^6 \bigg[\sideset{}{'}\sum_{\mbf{k_j'}} V_{|\mbf{k}- \mbf{k'_j}|} \langle \mbf{k}+C_6^{j-1} \mbf{b_{1}}| \mbf{k'_j}+ C_6^{j-1} \mbf{b_{1}} \rangle \langle  \mbf{k'_j}|\mbf{k}\rangle O_{0j,\mbf{k'_j}} \\
 & + \sideset{}{'}\sum_{\mbf{k'_{j,j+1}}} V_{|\mbf{k}-\mbf{k'_{j,j+1} }|}\langle \mbf{k}+C_6^{j-1} \mbf{b_{1}}|\mbf{k'_{j,j+1} }+ C_6^{j-1} \mbf{b_{1}} \rangle \langle \mbf{k'_{j,j+1} }|\mbf{k}\rangle    O_{0j,\mbf{k'_{j,j+1}}}  \\ 
 & + \sideset{}{'}\sum_{\mbf{k'_{j-1,j}}} V_{|\mbf{k}- \mbf{k'_{j-1,j} }|}\langle \mbf{k}+C_6^{j-1} \mbf{b_{1}}|\mbf{{k'_{j-1,j} }}+ C_6^{j-1} \mbf{b_{1}} \rangle \langle \mbf{k'_{j-1,j} }|\mbf{k}\rangle   O_{0j,\mbf{k'_{j-1,j}}}   \\ 
 &+  \sideset{}{'}\sum_{\mbf{k'_{j-3}}} V_{|\mbf{k}- \mbf{k'_{j-3}}+C_6^{j-1} \mbf{b_{1}}|} \langle \mbf{k}+C_6^{j-1} \mbf{b_{1}}|\mbf{k'_{j-3}} \rangle \langle \mbf{k'_{j-3}}-C_6^{j-1} \mbf{b_{1}}|\mbf{k}\rangle  O_{j-3,0,\mbf{k'_{j-3}}}     \\ 
 & + \sideset{}{'}\sum_{\mbf{k'_{j-3,j-2}}} V_{|\mbf{k} - \mbf{k'_{j-3,j-2}}+C_6^{j-1} \mbf{b_{1}}|} \langle \mbf{k}+C_6^{j-1} \mbf{b_{1}}|\mbf{k'_{j-3,j-2}} \rangle \langle \mbf{k'_{j-3,j-2}}-C_6^{j-1} \mbf{b_{1}}|\mbf{k}\rangle  O_{j-3,0,\mbf{k'_{j-3,j-2}}}  \\ 
 & + \sideset{}{'}\sum_{\mbf{k'_{j-4,j-3}}} V_{|\mbf{k}- \mbf{k'_{j-4,j-3}}+C_6^{j-1} \mbf{b_{1}}|} \langle \mbf{k}+C_6^{j-1} \mbf{b_{1}}|\mbf{k'_{j-4,j-3}} \rangle \langle \mbf{k'_{j-4,j-3}}-C_6^{j-1} \mbf{b_{1}}|\mbf{k}\rangle  O_{j-3,0,\mbf{k'_{j-4,j-3}}}   \\
 & + \sideset{}{'}\sum_{\mbf{k'_{j-2,j-1}}} V_{|\mbf{k}- \mbf{k'_{j-2,j-1}}+C_6^{j} \mbf{b_{1}}|} \langle \mbf{k}+C_6^{j-1} \mbf{b_{1}}|\mbf{k'_{j-2,j-1}}+ C_6^{j-2} \mbf{b_{1}} \rangle \langle \mbf{k'_{j-2,j-1}}-C_6^{j} \mbf{b_{1}}|\mbf{k}\rangle   O_{j-2,j-1,\mbf{k'_{j-2,j-1}}}   \\
 & + \sideset{}{'}\sum_{\mbf{k'_{j-5,j-4}}} V_{|\mbf{k}- \mbf{k'_{ j-5,j-4 }}+C_6^{j+4} \mbf{b_{1}}|} \langle \mbf{k}+C_6^{j-1} \mbf{b_{1}}| \mbf{k'_{ j-5,j-4 }}+ C_6^{j} \mbf{b_{1}} \rangle \langle  \mbf{k'_{ j-5,j-4 }}-C_6^{j+4} \mbf{b_{1}}|\mbf{k}\rangle   O_{j-4, j-5 ,\mbf{k'_{ j-5,j-4}}}   \\
 &\quad\quad \bigg] \gamma_{\mbf{k}+ C_6^{j-1} \mbf{b_{1}}}^\dagger \gamma_{\mbf{k}}.
\end{align}

In both finite-$\bm{q}$ terms, the operator that appears is $\gamma_{\mbf{k}+ C_6^{j-1} \mbf{b_{1}}}^\dagger \gamma_{\mbf{k}}$, which admits a decomposition analogous to that of $O_{\mbf{k'} - C_6^{j-1}\mbf{b_{1}},\mbf{k'}}$:
\begin{eqnarray}
    &\gamma_{\mbf{k}
+ C_6^{j-1} \mbf{b_{1}}}^\dagger \gamma_{\mbf{k}}  = \gamma_{j,\mbf{k_j}
}^\dagger \gamma_{0\mbf{k_j}} \delta_{\mbf{k, k_j}}+ \gamma_{0,\mbf{k_{j+3}}
}^\dagger \gamma_{j+3,\mbf{k_{j+3}}} \delta_{\mbf{k}, \mbf{k_{j+3}} +C_6^{j+2} \mbf{b_{1}}} \\ & +\gamma_{j, \mbf{k_{j,j+1}}
}^\dagger \gamma_{0,\mbf{k_{j,j+1}}} \delta_{\mbf{k, k_{j,j+1}}}+\gamma_{0, \mbf{k_{j+2,j+3}}
}^\dagger \gamma_{j+3,\mbf{k_{j+2,j+3}}} \delta_{\mbf{k, k_{j+2,j+3}} + C_6^{j+2} \mbf{b_{1}}} \\ &+\gamma_{j,\mbf{k_{j-1,j}}
}^\dagger \gamma_{0,\mbf{k_{j-1,j}}} \delta_{\mbf{k, k_{j-1,j}}}+ \gamma_{0,\mbf{k_{j+3,j+4}}
}^\dagger \gamma_{j+3,\mbf{k_{j+3,j+4}}} \delta_{\mbf{k, k_{j+3,j+4}} +C_6^{j+2} \mbf{b_{1}} } \\
&{+\gamma^\dagger_{j+1,\mbf{k_{j+1,j+2}}}\gamma_{j+2,\mbf{k_{j+1,j+2}}}
\delta_{\mbf{k},\mbf{k_{j+1,j+2}}+C_6^{j+1}\mbf{b_{1}}}+
\gamma^\dagger_{j-1,\mbf{k_{j-2,j-1}}}\gamma_{j-2,\mbf{k_{j-2,j-1}}}
\delta_{\mbf{k},\mbf{k_{j-2,j-1}}+C_6^{j+3}\mbf{b_{1}}}
} .
 \end{eqnarray}
The finite-$\bm{q}$ terms can then be parameterized as
\begin{gather}\label{eqapp:2dtrashcanHHFgparam}
    \sum_{j=1}^6\sum_{\bm{k}}g_{j,\bm{k}}\gamma_{\mbf{k}+ C_6^{j-1} \mbf{b_{1}}}^\dagger \gamma_{\mbf{k}}=\sum_{j=1}^6 \bigg(
\sideset{}{'}\sum_{\mbf{k_j}} g_{j,\mbf{k_j}} \gamma_{j,\mbf{k_j}
}^\dagger \gamma_{0\mbf{k_j}} +\sideset{}{'}\sum_{\mbf{k_{j+3}}} g_{j,\mbf{k_{j+3}} +C_6^{j+2} \mbf{b_{1}}}  \gamma_{0,\mbf{k_{j+3}}
}^\dagger \gamma_{j+3,\mbf{k_{j+3}}}  \\  + \sideset{}{'}\sum_{\mbf{k_{j,j+1}}} g_{j,\mbf{k_{j,j+1}}} \gamma_{j, \mbf{k_{j,j+1}}
}^\dagger \gamma_{0,\mbf{k_{j,j+1}}} + \sideset{}{'}\sum_{\mbf{k_{j+2,j+3}}} g_{j,\mbf{k_{j+2,j+3}} + C_6^{j+2} \mbf{b_{1}}}  \gamma_{0, \mbf{k_{j+2,j+3}}
}^\dagger \gamma_{j+3,\mbf{k_{j+2,j+3}}} \\ +\sideset{}{'}\sum_{\mbf{ k_{j-1,j}}} g_{j,\mbf{ k_{j-1,j}}}  \gamma_{j,\mbf{k_{j-1,j}}
}^\dagger \gamma_{0,\mbf{k_{j-1,j}}} + \sideset{}{'}\sum_{\mbf{k_{j+3,j+4}}} g_{j,\mbf{k_{j+3,j+4}} +C_6^{j+2} \mbf{b_{1}}}   
\gamma_{0,\mbf{k, k_{j+3,j+4}}
}^\dagger \gamma_{j+3,\mbf{k_{j+3,j+4}}}  \\
+\sideset{}{'}\sum_{\mbf{k_{j+1,j+2}}} g_{j,\mbf{k_{j+1,j+2}}+C_6^{j+1}\mbf{b_{1}}}
\gamma^\dagger_{j+1,\mbf{k_{j+1,j+2}}}\gamma_{j+2,\mbf{k_{j+1,j+2}}}+
\sideset{}{'}\sum_{\mbf{k_{j-2,j-1}}} g_{j,\mbf{k_{j-2,j-1}}+C_6^{j+3}\mbf{b_{1}}} 
\gamma^\dagger_{j-1,\mbf{k_{j-2,j-1}}}\gamma_{j-2,\mbf{k_{j-2,j-1}}}\bigg) 
\end{gather}
where we have introduced the interaction-induced hybridization potential
\begin{eqnarray}\label{eqapp:gk_first}
    &g_{j,\mbf{k}}=\frac{1}{\Omega_{tot}}\sum_{\bm{k}'}\bigg[V_{b_{1}} \langle \mbf{k+C_6^{j-1}b_{1}} | \mbf{k} \rangle \langle \mbf{k'-C_6^{j-1}b_{1}} | \mbf{k'} \rangle \\
    &-V_{|\mbf{k-k'+C_6^{j-1} b_{1}}|} \langle \mbf{k+C_6^{j-1} b_{1}}|\mbf{k'} \rangle \langle \mbf{k'-C_6^{j-1} b_{1}}|\mbf{k} \rangle  \bigg]O_{\mbf{k'} - \mbf{C_6^{j-1} b_{1}},\mbf{k'}}\\
    &= \frac{1}    {\Omega_{tot}}  \bigg[ \sideset{}{'}\sum_{\mbf{k'_j}}\big(V_{b_{1}}    \langle \mbf{k+C_6^{j-1}b_{1}} | \mbf{k} \rangle \langle \mbf{{k'_j}} | \mbf{k'_j}+ C_6^{j-1} \mbf{b_{1}} \rangle\\ &- V_{|\mbf{k}- \mbf{k'_j}|} \langle \mbf{k}+C_6^{j-1} \mbf{b_{1}}| \mbf{k'_j}+ C_6^{j-1} \mbf{b_{1}} \rangle \langle  \mbf{k'_j}|\mbf{k}\rangle \big)
 O_{0j,\mbf{k'_j}}        \\ & + \sideset{}{'}\sum_{\mbf{k'_{j,j+1}}} \big(V_{b_{1}} \langle \mbf{k}+C_6^{j-1}\mbf{b_{1}} | \mbf{k} \rangle \langle \mbf{{k'_{j,j+1} }} | \mbf{k'_{j,j+1} }+ C_6^{j-1} \mbf{b_{1}} \rangle \\ &-V_{|\mbf{k}-\mbf{k'_{j,j+1} }|} \langle \mbf{k}+C_6^{j-1} \mbf{b_{1}}|\mbf{k'_{j,j+1} }+ C_6^{j-1} \mbf{b_{1}} \rangle \langle \mbf{k'_{j,j+1} }|\mbf{k}\rangle   \big)   O_{0j,\mbf{k'_{j,j+1}}}   \\ & + \sideset{}{'}\sum_{\mbf{k'_{j-1,j}}}  \big(V_{b_{1}} \langle \mbf{k}+C_6^{j-1}\mbf{b_{1}} | \mbf{k} \rangle \langle \mbf{k'_{j-1,j} } |  \mbf{k'_{j-1,j} }+ C_6^{j-1} \mbf{b_{1}} \rangle\\ &-V_{|\mbf{k}- \mbf{k'_{j-1,j} }|} \langle \mbf{k}+C_6^{j-1} \mbf{b_{1}}|\mbf{k'_{j-1,j} }+ C_6^{j-1} \mbf{b_{1}} \rangle \langle \mbf{k'_{j-1,j} }|\mbf{k}\rangle\big)  O_{0j,\mbf{k'_{j-1,j}}}\\ &+ \sideset{}{'}\sum_{\mbf{k'_{j-3}}} \big( V_{b_{1}}\langle \mbf{k+C_6^{j-1}b_{1}} | \mbf{k} \rangle \langle \mbf{{k'_{j-3}}-C_6^{j-1}b_{1}} |\mbf{k'_{j-3}} \rangle  \\ &- V_{|\mbf{k}- \mbf{k'_{j-3}}+C_6^{j-1} \mbf{b_{1}}|} \langle \mbf{k}+C_6^{j-1} \mbf{b_{1}}|\mbf{k'_{j-3}} \rangle \langle \mbf{k'_{j-3}}-C_6^{j-1} \mbf{b_{1}}|\mbf{k}\rangle \big) O_{j-3,0,\mbf{k'_{j-3}}}     \\ & + \sideset{}{'}\sum_{\mbf{k'_{j-3,j-2}}}  \big(V_{b_{1}}\langle \mbf{k+C_6^{j-1}b_{1}} | \mbf{k} \rangle \langle \mbf{{k'_{j-3,j-2}}-C_6^{j-1}b_{1}} |\mbf{k'_{j-3,j-2}}\rangle \\ &-V_{|\mbf{k} - \mbf{k'_{j-3,j-2}}+C_6^{j-1} \mbf{b_{1}}|} \langle \mbf{k}+C_6^{j-1} \mbf{b_{1}}|\mbf{k'_{j-3,j-2}} \rangle \langle \mbf{k'_{j-3,j-2}}-C_6^{j-1} \mbf{b_{1}}|\mbf{k}\rangle  \big)  O_{j-3,0, \mbf{k'_{j-3,j-2}}}  \\ & +\sideset{}{'}\sum_{\mbf{k'_{j-4,j-3}}}  \big(V_{b_{1}} \langle \mbf{k+C_6^{j-1}b_{1}} | \mbf{k} \rangle \langle \mbf{{k'_{j-4,j-3}}-C_6^{j-1}b_{1}} | \mbf{k'_{j-4,j-3}} \rangle \\ &-V_{|\mbf{k}- \mbf{k'_{j-4,j-3}}+C_6^{j-1} \mbf{b_{1}}|} \langle \mbf{k}+C_6^{j-1} \mbf{b_{1}}|\mbf{k'_{j-4,j-3}} \rangle \langle \mbf{k'_{j-4,j-3}}-C_6^{j-1} \mbf{b_{1}}|\mbf{k}\rangle \big) O_{j-3,0, \mbf{k'_{j-4,j-3}}}     \\ & + \sideset{}{'}\sum_{\mbf{k'_{j-2,j-1}}}  \big(V_{b_{1}} \langle \mbf{k+C_6^{j-1}b_{1}} | \mbf{k} \rangle \langle \mbf{k'_{j-2,j-1}}+ C_6^{j+3} \mbf{b_{1}} | \mbf{k'_{j-2,j-1}}+ C_6^{j-2} \mbf{b_{1}} \rangle \\ &- V_{|\mbf{k}- \mbf{k'_{j-2,j-1}}+C_6^{j} \mbf{b_{1}}|} \langle \mbf{k}+C_6^{j-1} \mbf{b_{1}}|\mbf{k'_{j-2,j-1}}+ C_6^{j-2} \mbf{b_{1}} \rangle \langle \mbf{k'_{j-2,j-1}}-C_6^{j} \mbf{b_{1}}|\mbf{k}\rangle  \big)   O_{j-2,j-1,\mbf{k'_{j-2,j-1}}} \\ & + \sideset{}{'}\sum_{\mbf{k'_{ j-5,j-4 }}}  \big(V_{b_{1}}\langle \mbf{k+C_6^{j-1}b_{1}} | \mbf{k} \rangle \langle \mbf{k'_{ j-5,j-4 }}+ C_6^{j+1} \mbf{b_{1}} | \mbf{k'_{ j-5,j-4 }}+ C_6^{j} \mbf{b_{1}}\rangle   \\ &- V_{|\mbf{k}- \mbf{k'_{ j-5,j-4 }}+C_6^{j+4} \mbf{b_{1}}|} \langle \mbf{k}+C_6^{j-1} \mbf{b_{1}}| \mbf{k'_{ j-5,j-4 }}+ C_6^{j} \mbf{b_{1}} \rangle \langle  \mbf{k'_{ j-5,j-4 }}-C_6^{j+4} \mbf{b_{1}}|\mbf{k}\rangle \big)  O_{j-4, j-5 ,\mbf{k'_{ j-5,j-4}}}   \bigg],
\end{eqnarray}
where $j=1,\dots,6$. Through straightforward algebra, it can be shown that $g_{j-3,\mbf{k_{j}} +C_6^{j-1} \mbf{b_{1}}}  = g_{j,\mbf{k_j}}^\star$,  $ g_{j-3,\mbf{k_{j,j+1}} +C_6^{j-1} \mbf{b_{1}}} =g_{j,\mbf{k_{j,j+1}}}^\star$, $g_{j-2,\mbf{k_{j,j+1}} + C_6^{j} \mbf{b_{1}}}  =g_{j+1,\mbf{ k_{j,j+1}}} ^\star$ and $g_{j-1,\mbf{k_{j,j+1}}+C_6^{j}\mbf{b_{1}}}=  g_{j+2,\mbf{k_{j,j+1}}+C_6^{j-1}\mbf{b_{1}}}^\star$. In other words
\begin{equation}\label{eqapp:2dtrashcangherm}
    g^*_{j+3,\bm{k}+C_6^{j-1}\bm{b}_1}=g_{j,\bm{k}}.
\end{equation}
These ensure that the mean-field Hamiltonian is Hermitian.

\subsubsection{Kinetic term}
For completeness, we also decompose the one-body non-interacting kinetic term $\sum_{\bm{k}}E_{\bm{k}}\gamma^\dagger_{\bm{k}}\gamma_{\bm{k}}$
\begin{gather}\label{eqapp:2dtrashcanHHFkinparam}
    \sum_{\bm{k}}E_{\bm{k}}\gamma^\dagger_{\bm{k}}\gamma_{\bm{k}}=\sum_{\mbf{k_0}} E_{\mbf{k_0}}\gamma_{0\mbf{k_0}}^\dagger \gamma_{0\mbf{k_0}}  +\sum_{i=1}^6 \sideset{}{'}\sum_{\mbf{k_i}} \left(E_{\mbf{k_i}}\gamma_{0\mbf{k_i}}^\dagger \gamma_{0\mbf{k_i}} + E_{\mbf{k_i+ C_6^{i-1}b_{1}}} \gamma_{i\mbf{k_i}}^\dagger \gamma_{i\mbf{k_i}}\right)\\ + \sum_{i=1}^6\sideset{}{'}\sum_{\mbf{k_{i,i+1}}} \left( E_{\mbf{k_{i,i+1}}} \gamma_{0\mbf{k_{i,i+1}}}^\dagger \gamma_{0\mbf{k_{i,i+1}}}+ E_{\mbf{k_{i,i+1} + C_6^{i-1} b_{1}}}\gamma_{i\mbf{k_{i,i+1}}}^\dagger \gamma_{i\mbf{k_{i,i+1}}} + E_{\mbf{k_{i,i+1} + C_6^{i} b_{1}}}\gamma_{i+1\mbf{k_{i,i+1}}}^\dagger \gamma_{i+ 1\mbf{k_{i,i+1}}} \right).
\end{gather}

\subsection{Symmetries of the General Model}\label{subsecapp:symmetries_general_model}
We now outline the properties of the general parameterization in Sec.~\ref{secapp:2dtrashcan_genparam} in the presence of $C_6$ and $M_1\mathcal{T}$ symmetries. 

We first consider $C_6$ symmetry. This leads to
\begin{equation}
    E_{\bm{k}}=E_{C_6\bm{k}},\quad f_{\bm{k}}=f_{C_6\bm{k}},\quad g_{j,\bm{k}}=g_{j+1,C_6\bm{k}}.
\end{equation}

We now consider $M_1\mathcal{T}$ symmetry, which leads to
\begin{equation}
    E_{\bm{k}}=E_{M_1\mathcal{T}\bm{k}},\quad f_{\bm{k}}=f_{M_1\mathcal{T}\bm{k}},\quad g^*_{j,\bm{k}}=g_{M_1\mathcal{T}j,M_1\mathcal{T}\bm{k}}.
\end{equation}

We combine the symmetries and Hermiticity to show the following identities. For example
\begin{gather}
    g_{j,-C^{j-1}_6\frac{\bm{b}_1}{2}+\delta\bm{k}_j}=g^*_{j+3,C^{j-1}_6\frac{\bm{b}_1}{2}+\delta\bm{k}_j}=g^*_{j,-C^{j-1}_6\frac{\bm{b}_1}{2}-\delta\bm{k}_j},
\end{gather}
where we used Hermiticity in the first equality, and assumed $C_2$ symmetry in the second equality. The above suggests that in the momentum region $\bm{k}_j$, it is useful to work in terms of the relative momentum $\delta \bm{k}_j=\bm{k}_j+C^{j-1}\bm{b}_1$ from the basepoint $-C^{j-1}\bm{b}_1$, since there is an emergent `time-reversal symmetry' that takes $\delta \bm{k}_j\leftrightarrow -\delta \bm{k}_j$. Note though that for a generic $-C^{j-1}\bm{b}_1+\delta{\bm{k}_j}$ that lies within BZ 0 (and hence coincides with the conventional labelling of momenta within the reduced BZ), $-C^{j-1}\bm{b}_1-\delta{\bm{k}_j}$ does not also lie within BZ 0 unless it sits along the BZ edge.

We also have
\begin{align}
    g_{j,C^{j-1}_6\bm{q}_2+\delta\bm{k}_{j,j+1}}&=g^*_{j+3,C^{j-1}_6(\bm{q}_2+\bm{b}_1)+\delta\bm{k}_{j,j+1}}=g^*_{j+3,C^{j+1}_6\bm{q}_2+\delta\bm{k}_{j,j+1}}=g^*_{j+1,C^{j-1}_6\bm{q}_2-C_6\delta\bm{k}_{j,j+1}}\\
    &=g^*_{j,C^{j-2}_6\bm{q}_2-\delta\bm{k}_{j,j+1}}=g_{M_1\mathcal{T}j,-M_1C^{j-2}_6\bm{q}_2+M_1\delta\bm{k}_{j,j+1}},
\end{align}
where we used Hermiticity in the first equality, used $\bm{q}_2+\bm{b}_1=C^2_6\bm{q}_2$ in the second equality, assumed $C_3$ symmetry in the third and fourth equalities, and used $M_1\mathcal{T}$ symmetry in the fifth equality. The above suggests that in the momentum region $\bm{k}_{j,j+1}$, it is useful to work in terms of the relative momentum $\delta\bm{k}_{j,j+1}=\bm{k}_{j,j+1}-C^{j-1}_6\bm{b}_1$. Similar comments as for the previous identity regarding the position of momentum arguments within or outside BZ 0 apply here as well.

We summarize the symmetry- and Hermiticity-induced constraints for a fixed $j$, focusing in the vicinity of the momentum regions $\bm{k}_1$ and $\bm{k}_{1,2}$ for $j=1$
\begin{gather}\label{eqapp:2dtrashcang1k1cond}
    g_{1,-\frac{\bm{b}_1}{2}+\delta\bm{k}_1}=g^*_{1,-\frac{\bm{b}_1}{2}-\delta\bm{k}_1}=g_{1,-\frac{\bm{b}_1}{2}+M_1\delta\bm{k}_1}\\
    g_{1,\bm{q}_2+\delta\bm{k}_{1,2}}=g_{1,\bm{q}_2+M_1\delta\bm{k}_{1,2}}.
\end{gather}
We also have the following identity in the vicinity of region $\bm{k}_{1,2}$ that relates $j=1$ and $2$
\begin{equation}
    g_{1,\bm{q}_2+\delta\bm{k}_{1,2}}=g^*_{2,\bm{q}_2-C_6\delta\bm{k}_{1,2}}.
\end{equation}
The above can be transformed to conditions on other $j$'s by applying $C_6$.

\subsection{Symmetry-induced Constraints on Chern Numbers}\label{subsecapp:symChern}

In this subsection, we assume that we have obtained a gapped mean-field insulator that preserves $C_6$ and $M_1\mathcal{T}$ symmetries, and study how the Chern number $C$ is constrained by the order parameters at the high symmetry  points of the BZ~\cite{fang2012bulk}. Note that because the Berry Trashcan model has continuous rotation and translation invariance, the symmetry constraints on the Chern number derived below also hold for degenerate HF solutions related by translation and/or rotation (alternatively, one can consider a new appropriate origin/orientation for the symmetry operators and derive the analogous constraints). Later in App.~\ref{subsecapp:full_Chern}, we will discuss how the full Chern number (without modding by an integer) can be obtained using knowledge of the mean-field Hamiltonian around the entire BZ boundary.

\subsubsection{$M_M$ points}
We first consider the mean-field problem at the $M_M$ points. With $C_6$ symmetry, we can just study the physics at one $M_M$ point. For concreteness, consider the situation in the region $\bm{k}_1$ with momentum deviation $\delta\bm{k}_1=0$ (a more general solution of the mean-field problem in region $\bm{k}_1$ will be presented in App.~\ref{subsecapp:gensolnk1}). Up to an overall constant $E_{-\frac{\bm{b}_1}{2}}+f_{-\frac{\bm{b}_1}{2}}$ which we ignore here, the $2\times 2$ mean-field Hamiltonian at $\bm{k}=-\frac{\bm{b}_1}{2}$ takes the form
\begin{equation}
    H^\text{HF}_{-\frac{\bm{b}_1}{2}}=g_{1,-\frac{\bm{b}_1}{2}}\gamma^\dagger_{1,-\frac{\bm{b}_1}{2}}\gamma_{0,-\frac{\bm{b}_1}{2}}+\text{h.c.}=\begin{pmatrix}
        \gamma^\dagger_{0,-\frac{\bm{b}_1}{2}}&\gamma^\dagger_{1,-\frac{\bm{b}_1}{2}}
    \end{pmatrix}
    \begin{pmatrix}
        0 & g_{1,-\frac{\bm{b}_1}{2}}^*\\
        g_{1,-\frac{\bm{b}_1}{2}} & 0
    \end{pmatrix}
    \begin{pmatrix}
        \gamma_{0,-\frac{\bm{b}_1}{2}}\\\gamma_{1,-\frac{\bm{b}_1}{2}}
    \end{pmatrix}.
\end{equation}

The lower eigenstate is associated with the creation operator
\begin{equation}
    a^\dagger_{-,-\frac{\mbf{b_{1}}}{2}}=   \frac{1}{\sqrt{2}} \left(  \gamma^\dagger_{0, -\frac{\mbf{b_{1}}}{2}}  -\frac{g_{1, \frac{-\mbf{b_{1}}}{2}}}{ |g_{1, -\frac{\mbf{b_{1}}}{2}}|}     \gamma^\dagger_{ 1,-\frac{\mbf{b_{1}}}{2}}\right)=\frac{1}{\sqrt{2}} \left(  \gamma^\dagger_{0, -\frac{\mbf{b_{1}}}{2}}  -\text{sgn}\left(g_{1, -\frac{\mbf{b_{1}}}{2}}\right)     \gamma^\dagger_{ 1,-\frac{\mbf{b_{1}}}{2}}\right)
\end{equation}
where in the last equality, we have used Eq.~\ref{eqapp:2dtrashcang1k1cond}. Note that this implies the order parameter
\begin{equation}
    O_{10,-\frac{\bm{b}_1}{2}}=-\frac{1}{2}\text{sgn}\left(g_{1, -\frac{\mbf{b_{1}}}{2}}\right).
\end{equation}
The $C_2$ eigenvalue of the lower mean-field Bloch state is then
\begin{equation}
    C_2 a^\dagger_{-,-\frac{\mbf{b_{1}}}{2}} C_2^{-1}=-\text{sgn}\left(g_{1, -\frac{\mbf{b_{1}}}{2}}\right)e^{i\frac{n\pi}{2}}a^\dagger_{-,-\frac{\mbf{b_{1}}}{2}} =\text{sgn}\left(O_{10,-\frac{\bm{b}_1}{2}}\right)e^{i\frac{n\pi}{2}}a^\dagger_{-,-\frac{\mbf{b_{1}}}{2}}. 
\end{equation}
Note that the $C_2$ eigenvalue of the $\Gamma_M$ point, which must be occupied for an insulating state, is just $e^{i\frac{n\pi}{2}}$. Considering the $C_2$ eigenvalues of the $\Gamma_M$ point and the three $M_M$ points in a $C_6$ symmetric situation yields the parity of the Chern number~\cite{fang2012bulk} 
\begin{equation}\label{eqapp:MChern}
    (-1)^C=-\text{sgn}\left(g_{1, -\frac{\mbf{b_{1}}}{2}}\right)=\text{sgn}\left(O_{10,-\frac{\bm{b}_1}{2}}\right).
\end{equation}

\subsubsection{$K_M$ and $K'_M$ points}\label{subsubsecapp:KK'sym}

We now consider the mean-field problem at the $K_M$ and $K_M'$ points. With $C_6$ symmetry, we can just study one of these points. For concreteness, consider the situation in region $\bm{k}_{1,2}$ with momentum deviation $\delta\bm{k}_{1,2}=0$ (a more general solution of the mean-field problem in the region $\bm{k}_{1,2}$ will be presented in App.~\ref{subsecapp:gensolnk12}). Up to an overall constant $E_{\bm{q}_2}+f_{\bm{q}_2}$ which we ignore here, the $3\times 3$ mean-field Hamiltonian at $\bm{k}=\bm{q}_2$ takes the form
\begin{align}
    H^\text{HF}_{\bm{q}_2}&=\begin{pmatrix}
        \gamma^\dagger_{0,\bm{q}_2}&\gamma^\dagger_{1,\bm{q}_2}&\gamma^\dagger_{2,\bm{q}_2}
    \end{pmatrix}
    \begin{pmatrix}
        0 & g^*_{1,\bm{q}_2} & g^*_{2,\bm{q}_2} \\
        g_{1,\bm{q}_2} & 0 & g_{6,\bm{q}_2+C_6\bm{b}_1} \\
        g_{2,\bm{q}_2} & g^*_{6,\bm{q}_2+C_6\bm{b}_1} & 0
    \end{pmatrix}
    \begin{pmatrix}
        \gamma_{0,\bm{q}_2}\\\gamma_{1,\bm{q}_2}\\\gamma_{2,\bm{q}_2}
    \end{pmatrix}\\
    &=\begin{pmatrix}
        \gamma^\dagger_{0,\bm{q}_2}&\gamma^\dagger_{1,\bm{q}_2}&\gamma^\dagger_{2,\bm{q}_2}
    \end{pmatrix}
    \begin{pmatrix}
        0 & g_{12}^* & g_{12} \\
        g_{12} & 0 & g_{12}^* \\
        g_{12}^* & g_{12} & 0
    \end{pmatrix}
    \begin{pmatrix}
        \gamma_{0,\bm{q}_2}\\\gamma_{1,\bm{q}_2}\\\gamma_{2,\bm{q}_2}
    \end{pmatrix}
\end{align}
where we have used $g_{12}\equiv g_{1,\bm{q}_2}=g^*_{2,\bm{q}_2}$ and $g_{6,\bm{q}_2+C_6\bm{b}_1}=g_{6,C_6^{-2}\bm{q}_2}=g_{1,C_6^{-1}\bm{q}_2}=g^*_{1,\bm{q}_2}$ in the second equality. The three eigenstates and their eigenenergies and wavefunction ratios $\lambda_{\bm{q}_2}=[\psi]_{2}/[\psi]_0$, where $[\psi]_j$ refers to the $j$'th component of $\psi$, are
\begin{gather}
    \psi_{A}=\frac{1}{\sqrt{3}}(1,1,1),\quad E_A=2\Re g_{12}, \quad \lambda_{\bm{q}_2,A}=1\\
    \psi_B=\frac{1}{\sqrt{3}}(1,e^{i\frac{2\pi}{3}},e^{i\frac{4\pi}{3}}),\quad E_B= -\Re g_{12} + \sqrt{3}\Im g_{12}, \quad \lambda_{\bm{q}_2,B}=e^{i\frac{4\pi}{3}}\\
    \psi_C=\frac{1}{\sqrt{3}}(1,e^{i\frac{4\pi}{3}},e^{i\frac{2\pi}{3}}), \quad E_C=-\Re g_{12} - \sqrt{3}\Im g_{12}, \quad \lambda_{\bm{q}_2,C}=e^{i\frac{2\pi}{3}}.
\end{gather}
The corresponding $C_3$ eigenvalue is $\theta_K=e^{i\frac{n\pi}{3}}\lambda_K$.
The lowest energy solution corresponds to the value of $m$ satisfying $\text{arg}(g_{12}e^{-i\frac{2(m+1)\pi}{3}})\in[0,\frac{2\pi}{3}]$, which constrains $C=m\mod 3$ and leads to $O_{01,\bm{q}_2}+O_{20,\bm{q}_2}+O_{12,\bm{q}_2}=e^{i\frac{2\pi C}{3}}$.

We can now constrain the Chern number $C$ mod $6$ using
\begin{equation}
    e^{i \frac{\pi}{3} C} =(-1)^F \eta_{\Gamma} \theta_{K} \xi_{M} 
\end{equation}
where $\eta_\Gamma$ is the $C_6$ eigenvalue at $\Gamma_M$, $\theta_{K}$ is the $C_3$ eigenvalue at $K_M$, $\xi_{M}$ is the $C_2$ eigenvalue at $M_M$, and $F$ defined via $C_6^6= (-1)^F$ is equal to $n$~\cite{fang2012bulk}. Using the results above, we have
\begin{equation}\label{eqapp:KMChern}
    e^{i \frac{\pi}{3} C} =-\text{sgn}\left(g_{1, \frac{-\mbf{b_{1}}}{2}}\right) \lambda_{\bm{q}_2} =\text{sgn}\left(O_{10,-\frac{\bm{b}_1}{2}}\right)\lambda_{\bm{q}_2}
\end{equation}
with $\lambda_{\bm{q}_2} $ the wavefunction ratio corresponding to the lowest energy state at $\bm{q}_2$. 

For a $C=0$ solution, we hence have $\text{sgn}\left(O_{10,-\frac{\bm{b}_1}{2}}\right)>0$ and the lowest eigenstate at the $K_M$ point is $\psi_A$.

For a $C=1$ solution, we hence have $\text{sgn}\left(O_{10,-\frac{\bm{b}_1}{2}}\right)<0$ and the lowest eigenstate at the $K_M$ point is $\psi_B$.

It will be useful later to parameterize the order parameter at $\delta \bm{k}_{1,2}=0$ in terms of the Chern number. In particular, consider the combination
\begin{equation}\label{eqapp:O012_k12_definition}
    O_{012,\bm{k}_{1,2}}\equiv O_{01,\bm{k}_{1,2}}+O_{20,\bm{k}_{1,2}}+O_{12,\bm{k}_{1,2}}.
\end{equation}
Note that without knowledge of the $C_2$ eigenvalue at the $M_M$ points, or if $C_2$ symmetry is not present, we can only constrain the Chern number mod $3$ as $e^{i\frac{2\pi}{3}C}=\lambda_{\bm{q}_2}^2$, i.e.~solution $A,B,C$ corresponds to $C=0,1,-1$ mod $3$ respectively. Using the results above, we obtain
\begin{equation}
    O_{012,\delta\bm{k}_{12}=0}=e^{i\frac{2\pi}{3}C}.
\end{equation}

\subsection{$f_{\mathbf{k}}$ in more detail}\label{subsecapp:fkmoredetail}

In this subsection, we consider the momentum-dependent band renormalization field $f_{\bm{k}}$ (Eq.~\ref{eqapp:fk_start}) in more detail, and consider various approximations. For convenience, we repeat Eq.~\ref{eqapp:fk_start} here
\begin{align}
    f_{\mbf{k}} =&\frac{1}{\Omega_{tot}}\sum_{\bm{k}'}\left(V_0-V_{|\bm{k}-\bm{k}'|}|\langle \mbf{k'}|\mbf{k} \rangle |^2 \right)O_{\bm{k}',\bm{k}'}\\
    =&\frac{1}{\Omega_{tot}} \Bigg[\sum_{\mbf{k'_0}} ( V_0 - V_{|\mbf{k-k'_0}|} |\langle \mbf{k'_0}|\mbf{k} \rangle |^2  )O_{00,\mbf{k'_0}} \\ &+ \sum_{i=1}^6  \sideset{}{'}\sum_{\mbf{k'_i}} \left( (V_0- V_{|\mbf{k-k'_i}|} |\langle \mbf{k'_i}|\mbf{k} \rangle |^2 )O_{00,\mbf{k'_i}} +( V_0 - V_{|\mbf{k}-\mbf{k'_i}- C_6^{i-1} \mbf{b_{1}}|} |\langle \mbf{k'_i}+ C_6^{i-1} \mbf{b_{1}}|\mbf{k} \rangle |^2 )O_{ii,\mbf{k'_i}}  \right) \\ & +\sum_{i=1}^6  \sideset{}{'}\sum_{\mbf{k'_{ii+1}}} \bigg( (V_0-V_{|\mbf{k-k'_{i,i+1}}|} |\langle \mbf{k'_{i,i+1}}|\mbf{k} \rangle |^2) O_{00,\mbf{k'_{i,i+1}}}  \\
    &+(V_0- V_{|\mbf{k}-\mbf{k'_{i,i+1}}- C_6^{i-1} \mbf{b_{1}}|}  |\langle \mbf{k'_{i,i+1}}+ C_6^{i-1} \mbf{b_{1}}|\mbf{k} \rangle |^2 ) O_{ii,\mbf{k'_{i,i+1}}}  \\ &+  (V_0-V_{|\mbf{k}-\mbf{k'_{i,i+1}}- C_6^{i} \mbf{b_{1}}|} |\langle \mbf{k'_{i,i+1}}+ C_6^{i} \mbf{b_{1}}|\mbf{k} \rangle |^2) O_{i+1i+1,\mbf{k'_{i,i+1}}} 
    \bigg)   \Bigg].
\end{align}
If we assume $C_6$ symmetry in the Hamiltonian and the order parameter, we obtain
\begin{align}
    f_{\mbf{k}} =&\frac{1}{\Omega_{tot}}\sum_{\bm{k}'}\left(V_0-V_{|\bm{k}-\bm{k}'|}|\langle \mbf{k'}|\mbf{k} \rangle |^2 \right)O_{\bm{k}',\bm{k}'}\\
    =&\frac{1}{\Omega_{tot}} \Bigg[\sum_{\mbf{k'_0}} ( V_0 - V_{|\mbf{k-k'_0}|} |\langle \mbf{k'_0}|\mbf{k} \rangle |^2  )O_{00,\mbf{k'_0}} \\ &+ \sum_{i=1}^6  \sideset{}{'}\sum_{\mbf{k'_1}} \left( (V_0- V_{|\mbf{k}-C_6^{i-1}\mbf{k'_1}|} |\langle C_6^{i-1}\mbf{k'_1}|\mbf{k} \rangle |^2 )O_{00,\mbf{k'_1}} +( V_0 - V_{|\mbf{k}-C_6^{i-1}\mbf{k'_1}- C_6^{i-1} \mbf{b_{1}}|} |\langle C_6^{i-1}\mbf{k'_1}+ C_6^{i-1} \mbf{b_{1}}|\mbf{k} \rangle |^2 )O_{11,\mbf{k'_1}}  \right) \\ & +\sum_{i=1}^6  \sideset{}{'}\sum_{\mbf{k'_{12}}} \bigg( (V_0-V_{|\mbf{k}-C_6^{i-1}\mbf{k'_{1,2}}|} |\langle C_6^{i-1}\mbf{k'_{1,2}}|\mbf{k} \rangle |^2) O_{00,\mbf{k'_{1,2}}}  \\
    &+(V_0- V_{|\mbf{k}-C_6^{i-1}\mbf{k'_{1,2}}- C_6^{i-1} \mbf{b_{1}}|}  |\langle C_6^{i-1}\mbf{k'_{1,2}}+ C_6^{i-1} \mbf{b_{1}}|\mbf{k} \rangle |^2 ) O_{11,\mbf{k'_{1,2}}}  \\ &+  (V_0-V_{|\mbf{k}-C_6^{i-1}\mbf{k'_{1,2}}- C_6^{i} \mbf{b_{1}}|} |\langle C_6^{i-1}\mbf{k'_{1,2}}+ C_6^{i} \mbf{b_{1}}|\mbf{k} \rangle |^2) O_{22,\mbf{k'_{1,2}}} 
    \bigg)   \Bigg].
\end{align}

\subsubsection{GMP limit and exponential interaction}\label{subsubsecapp:f_GMP_exp}

We consider the GMP limit of the form factor (Eq.~\ref{eqapp:Mkq_exp})
\begin{equation}
    \langle \mbf{k}|\mbf{k'} \rangle= e^{-\frac{v_F^2}{2 t^2} |\mbf{k-k'}|^2  -\frac{v_F^2}{2 t^2} 2i \mbf{k}\times \mbf{k'}}  = e^{- \frac{\beta}{2} |\mbf{k-k'}|^2  - \beta i \mbf{k}\times k'},\;\;\;\; \beta= \frac{v_F^2}{t^2},
\end{equation}
where $2\beta$ is the (uniform) Berry curvature, as well as an exponential interaction parameterized by exponent $\alpha$
\begin{equation}\label{eqapp:exponential_interaction}
    V_{\bm{k}}=V_0e^{-\alpha|\bm{k}|^2}.
\end{equation}
\begin{figure}
    \centering
    \includegraphics[width = 0.3\linewidth]{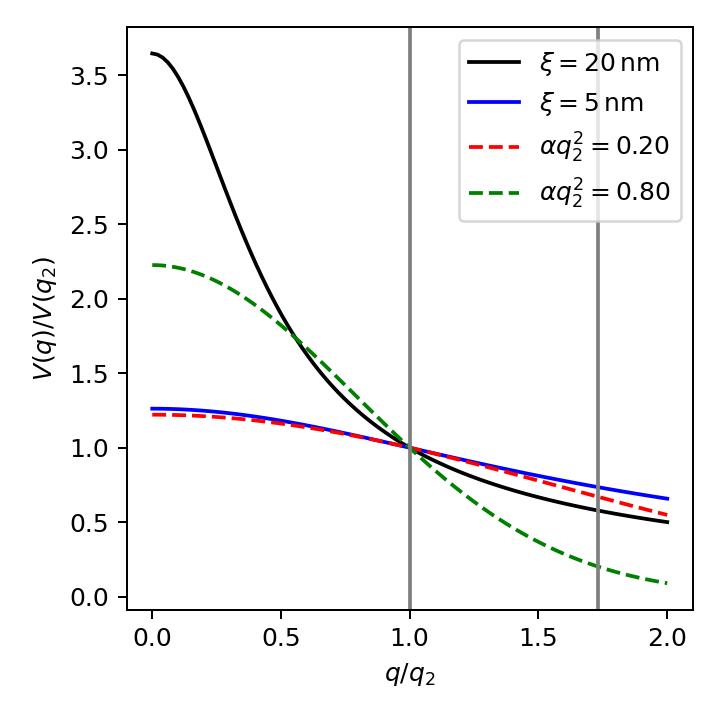}
    \caption{Comparison of the exponential interaction $V_q\propto e^{-\alpha q^2}$ and the gate-screened interaction $V_q\propto \frac{\tanh\frac{q\xi}{2}}{q}$ for representative values of $\alpha $ and $\xi$. $q_2\simeq 0.364\,\text{nm}^{-1}$ is set by the BZ corresponding to R$n$G/hBN at twist angle $\theta\simeq 0.77^\circ$. Vertical lines correspond to $q=q_2$ and $q=b_1$.}
    \label{figapp:interaction_func}
\end{figure}
In Fig.~\ref{figapp:interaction_func}, we compare the exponential interaction and a physical dual gate-screened interaction parameterized by gate-to-gate distance $\xi$. Note that for the mean-field analysis of the trashcan model, the maximum relevant momentum transfer that enters the finite-momentum terms (i.e.~the $g_{j,\mathbf{k}}$ terms) is $q=b_1$, which corresponds to the Hartree contribution. For the zero-momentum terms (i.e.~the $f_{\mathbf{k}}$ terms), the maximum momentum transfer is $q\simeq 2q_2$. For short gate distances $\xi\lesssim 5\,$nm, the gate-screened interaction can be modelled well by the exponential interaction for all relevant $q$. For larger gate distances such as $\xi\simeq 20\,$nm, the exponential interaction decays rapidly compared to the $1/q$ tail of the gate-screened Coulomb, so quantitative agreement cannot be obtained over the whole range of relevant momentum transfers.

The square modulus of the form factor and interaction potential will often appear together, so with the exponential interaction we have 
\begin{equation}
    V_{|\mbf{k-k'}|}| \langle \mbf{k}|\mbf{k'} \rangle|^2= V_0e^{-(\alpha + \beta)  |\mbf{k-k'}|^2}=V_0e^{-\phi  |\mbf{k-k'}|^2},\quad \phi=\alpha+\beta,
\end{equation}
which defines $\phi$.
Using the above leads to
\begin{align}
    f_{\mbf{k}} =&\frac{V_0}{\Omega_{tot}}\sum_{\bm{k}'}\left(1-e^{-\phi|\bm{k}-\bm{k}'|^2}\right)O_{\bm{k}',\bm{k}'}\\
    =&\frac{V_0}{\Omega_{tot}} \Bigg[\sum_{\mbf{k'_0}} (1 - e^{-\phi|\bm{k}-\bm{k_0}'|^2})O_{00,\mbf{k'_0}} \\ &+ \sum_{i=1}^6  \sideset{}{'}\sum_{\mbf{k'_1}} \left( (1- e^{-\phi|C_6^{1-i}\bm{k}-\bm{k_1}'|^2})O_{00,\mbf{k'_1}} +( 1 - e^{-\phi|C_6^{1-i}\bm{k}-\bm{k_1}'-\bm{b}_{1}|^2} )O_{11,\mbf{k'_1}}  \right) \\ & +\sum_{i=1}^6  \sideset{}{'}\sum_{\mbf{k'_{1,2}}} \bigg( (1-e^{-\phi|C_6^{1-i}\bm{k}-\bm{k_{1,2}}'|^2}) O_{00,\mbf{k'_{1,2}}}  \\
    &+(1- e^{-\phi|C_6^{1-i}\bm{k}-\bm{k_{1,2}}'-\bm{b}_{1}|^2}) O_{11,\mbf{k'_{1,2}}}  \\ &+  (1-e^{-\phi|C_6^{1-i}\bm{k}-\bm{k_{1,2}}'-C_6\bm{b}_{1}|^2}) O_{22,\mbf{k'_{1,2}}} 
    \bigg)   \Bigg].
\end{align}
We now introduce the parameterization
\begin{equation}\label{eqapp:dk_param}
    \mbf{k_1'}= -\frac{1}{2} \mbf{b_{1}} + \mbf{\delta k_1' },\;\;\; \mbf{k_{12}'} = \mbf{q_2}+ \mbf{\delta k_{12}'}
\end{equation}
so that $\mbf{\delta k_1' }$ denotes the deviation from the $M_M$-point and $\mbf{\delta k_{12}' }$ denotes the deviation from the $K_M$-point, leading to
\begin{align}\label{eqapp:fkexp}
    f_{\mbf{k}} =&\frac{V_0}{\Omega_{tot}}\sum_{\bm{k}'}\left(1-e^{-\phi|\bm{k}-\bm{k}'|^2}\right)O_{\bm{k}',\bm{k}'}\\
    =&\frac{V_0}{\Omega_{tot}} \Bigg[\sum_{\mbf{k'_0}} (1 - e^{-\phi|\bm{k}-\bm{k_0}'|^2})O_{00,\mbf{k'_0}} \\ &+ \sum_{j=1}^6  \sideset{}{'}\sum_{\delta\mbf{k'_1}} \left( (1- e^{-\phi|C_6^{1-j}\bm{k}-\delta \mbf{k'_1}+ \frac{1}{2}\mbf{b_{1}}|^2})O_{00,\mbf{k'_1}} +( 1 - e^{-\phi|C_6^{1-j}\bm{k}-\delta \mbf{k'_1}- \frac{1}{2}\mbf{b_{1}}|^2} )O_{11,\mbf{k'_1}}  \right) \\ & +\sum_{j=1}^6  \sideset{}{'}\sum_{\delta\mbf{k'_{1,2}}} \bigg( (1-e^{-\phi|C_6^{1-j}\bm{k}- \mbf{\delta k_{12}'}-\mbf{q_2}|^2}) O_{00,\mbf{k'_{1,2}}}  \\
    &+(1- e^{-\phi|C_6^{1-j}\bm{k}- \mbf{\delta k_{12}'}-C_6^2\mbf{q_2}|^2}) O_{11,\mbf{k'_{1,2}}}  \\ &+  (1-e^{-\phi|C_6^{1-j}\bm{k}- \mbf{\delta k_{12}'}-C_6^{-2}\mbf{q_2}|^2}) O_{22,\mbf{k'_{1,2}}} 
    \bigg)   \Bigg].
\end{align}

For later purposes, it will be useful to specify $f_{\bm{k}}$ explicitly in the different momentum regions
\begin{itemize}
    \item $f_{\bm{k}_0}$
    \item $f_{\mbf{k_i } }= f_{C_6^{i-1}\mbf{k_1 }} =   f_{C_6^{i-1}(\mbf{\delta k_1 -  \frac{1}{2} b_{1}})}$
    \item $f_{\mbf{k_i + C_6^{i-1} b_{1}}} = f_{C_6^{i-1}(\mbf{k_1 +  b_{1}})} =   f_{C_6^{i-1}(\mbf{\delta k_1 +  \frac{1}{2} b_{1}})}$
    \item $f_{\mbf{k_{i,i+1}}}= f_{C_6^{i-1}\mbf{k_{1,2}}}= f_{C_6^{i-1}(\mbf{q_2+ \delta k_{1,2}})}$
    \item $f_{\mbf{k_{i,i+1}} + C_6^{i-1} \mbf{b_{1}}}= f_{C_6^{i-1}(\mbf{k_{1,2}} + \mbf{b_{1}}) }= f_{C_6^{i-1}(\mbf{ \delta k_{1,2} + C_6^2 q_2})}$
    \item $f_{\mbf{k_{i,i+1}} + C_6^{i} \mbf{b_{1}}}= f_{C_6^{i-1}(\mbf{k_{1,2}} + C_6 \mbf{b_{1}}) }= f_{C_6^{i-1}(\mbf{ \delta k_{1,2} + C_6^{-2} q_2})}$.
\end{itemize}
We now make a further approximation that consists of expanding the exponential in Eq.~\ref{eqapp:fkexp} to first order in $\phi$
{
\begin{align}
        f_{\mbf{k_0}}=&\frac{V_0\phi }{\Omega_{tot}}\bigg[\sum_{\mbf{k'_0}} ( k^2_0+ k_0'^2  )O_{00,\mbf{k'_0}} + 6  ( k^2_0+ ( \frac{1}{2} b_{1})^2 ) \sideset{}{'}\sum_{\mbf{\delta k'_1}}  (O_{00,\mbf{k'_1}}+ O_{11,\mbf{k'_1}}  ) \\
        &+6 ( k_0^2+ q_2^2)\sideset{}{'}\sum_{\mbf{\delta k'_{1,2}}}     (O_{00,\mbf{k'_{1,2}}} +O_{11,\mbf{k'_{1,2}}}  + O_{22,\mbf{k'_{1,2}}} )+ 6 \sideset{}{'}\sum_{\mbf{\delta k'_1}}   \mbf{\delta k'_1}\cdot  \mbf{b_{1}}(-O_{00,\mbf{k'_1}} +O_{11,\mbf{k'_1}}  ) \\
        &+12 \sideset{}{'}\sum_{\mbf{\delta k'_{1,2}}}   \mbf{q_2}\cdot( \mbf{\delta k'_{1,2}}  O_{00,\mbf{k'_{1,2}}} + C_6^{-2} \mbf{\delta k'_{1,2}} O_{11,\mbf{k'_{1,2}}}  +   C_6^2 \mbf{\delta k_{1,2}'}O_{22,\mbf{k'_{1,2}}})\\
        &+ 6 \sideset{}{'}\sum_{\mbf{\delta k'_1}}   \delta k_1'^2( O_{00,\mbf{k'_1}} +O_{11,\mbf{k'_1}}  )  +6 \sideset{}{'}\sum_{\mbf{\delta k'_{1,2}}}   \delta k_{1,2}'^2  (O_{00,\mbf{k'_{1,2}}} +  O_{11,\mbf{k'_{1,2}}}  +    O_{22,\mbf{k'_{1,2}}}     )\bigg]
\end{align}
\begin{align}
    f_{\mbf{k_i }}=&\frac{V_0\phi }{\Omega_{tot}}\bigg[\sum_{\mbf{k'_0}}  ((  \frac{1}{2} b_{1})^2+ k_0'^2  )O_{00,\mbf{k'_0}} + \ 3 b_{1}^2\sideset{}{'}\sum_{\mbf{\delta k'_1}} ( O_{00,\mbf{k'_1}} +O_{11,\mbf{k'_1}}  )\\
    &+6  (( \frac{1}{2} b_{1})^2+q_2^2) \sideset{}{'}\sum_{\mbf{\delta k'_{1,2}}}   ( O_{00,\mbf{k'_{1,2}}} +   O_{11,\mbf{k'_{1,2}}}  + O_{22,\mbf{k'_{1,2}}} )-\mbf{\delta k_1} \cdot  \mbf{ b_{1}}\sum_{\mbf{k'_0}}   O_{00,\mbf{k'_0}}\\
    &+  6 \sideset{}{'}\sum_{\mbf{\delta k'_1}}   ( - (\mbf{\delta k_1}\cdot \mbf{ b_{1}} +\mbf{\delta k'_1}\cdot \mbf{b_{1}} )O_{00,\mbf{k'_1}} +( -\mbf{\delta k_1}\cdot  \mbf{ b_{1}}+\mbf{\delta k'_1}\cdot \mbf{b_{1}}  )O_{11,\mbf{k'_1}}  )\\
    &+6 \sideset{}{'}\sum_{\mbf{\delta k'_{1,2}}} \big(   ( -\mbf{\delta k_1} \cdot \mbf{ b_{1}}+2\mbf{\delta k'_{1,2}} \cdot{ q_2}) O_{00,\mbf{k'_{1,2}}} +  ( -\mbf{\delta k_1} \cdot  \mbf{b_{1}}+2 \mbf{\delta k'_{1,2}}\cdot  C_6^2 \mbf{q_2}  ) O_{11,\mbf{k'_{1,2}}}  \\
    &+   (-\mbf{\delta k_1}\cdot \mbf{b_{1}}+ 2 \mbf{\delta k'_{1,2}}\cdot C^{-2}_6 \mbf{q_2} ) O_{22,\mbf{k'_{1,2}}} \big)+\sum_{\mbf{k'_0}}  \delta k_1^2 O_{00,\mbf{k'_0}} +  6 \sideset{}{'}\sum_{\mbf{\delta k'_1}}   (\delta k_1^2+ \delta k_1'^2) (O_{00,\mbf{k'_1}}+ O_{11,\mbf{k'_1}}  )\\ 
    &+6 \sideset{}{'}\sum_{\mbf{\delta k'_{1,2}}}(\delta k_1^2+ \delta k_{12}'^2) (   O_{00,\mbf{k'_{1,2}}} +O_{11,\mbf{k'_{1,2}}}  + O_{22,\mbf{k'_{1,2}}} ) \bigg]
\end{align}
\begin{align}
    f_{\mbf{k_i + C_6^{i-1} b_{1}}}=&\frac{V_0\phi }{\Omega_{tot}}\bigg[\sum_{\mbf{k'_0}} (  ( \frac{1}{2} b_{1})^2+ k_0'^2  )O_{00,\mbf{k'_0}} +  3 b_{1}^2 \sideset{}{'}\sum_{\mbf{\delta k'_1}}  ( O_{00,\mbf{k'_1}} +O_{11,\mbf{k'_1}}  )\\
    &+6( (   \frac{1}{2} b_{1})^2+q_2^2) \sideset{}{'}\sum_{\mbf{\delta k'_{1,2}}}   ( O_{00,\mbf{k'_{1,2}}} +  O_{11,\mbf{k'_{1,2}}}  + O_{22,\mbf{k'_{1,2}}} )  + \mbf{\delta k_1 \cdot  b_{1}} \sum_{\mbf{k'_0}}     O_{00,\mbf{k'_0}} \\
    &+   6 \sideset{}{'}\sum_{\mbf{\delta k'_1}}  \big( ( \mbf{\delta k_1-\delta k_1'}) \cdot   \mbf{b_{1}}O_{00,\mbf{k'_1}}+ ( \mbf{\delta k_1+\delta k_1'}) \cdot   \mbf{b_{1}}O_{11,\mbf{k'_1}} \big)\\
    &+6 \sideset{}{'}\sum_{\mbf{\delta k'_{1,2}}}   \big(  (\mbf{\delta k_1} \cdot   \mbf{b_{1}}+2 \mbf{\delta k'_{1,2}}\cdot  \mbf{q_2}) O_{00,\mbf{k'_{1,2}}} +  (\mbf{\delta k_1 }\cdot  \mbf{b_{1}}+2 \mbf{\delta k'_{1,2}}\cdot  C_6^2 \mbf{q_2}  ) O_{11,\mbf{k'_{1,2}}}\\
    &+  ( \mbf{\delta k_1 }\cdot  \mbf{b_{1}}+2 \mbf{\delta k'_{1,2}}\cdot C^{-2}_6 \mbf{q_2} ) O_{22,\mbf{k'_{1,2}}} \big)+\sum_{\mbf{k'_0}}   \delta k_1^2 O_{00,\mbf{k'_0}}+   6 \sideset{}{'}\sum_{\mbf{\delta k'_1}}   (\delta k_1^2 +\delta k_1'^2 )(O_{00,\mbf{k'_1}} + O_{11,\mbf{k'_1}}  )\\
    &+6 \sideset{}{'}\sum_{\mbf{\delta k'_{1,2}}}    (\delta k_1^2+\delta k_{1,2}'^2) (O_{00,\mbf{k'_{1,2}}} +   O_{11,\mbf{k'_{1,2}}}  +O_{22,\mbf{k'_{1,2}}} )\bigg]
\end{align}
\begin{align}
    f_{\mbf{k_{i,i+1}}}=&\frac{V_0\phi }{\Omega_{tot}}\bigg[\sum_{\mbf{k'_0}}   (q_2^2+k_0'^2  )O_{00,\mbf{k'_0}} + 6 (   ( \frac{1}{2} b_{1})^2 +q_2^2)\sideset{}{'}\sum_{\mbf{\delta k'_1}}  (O_{00,\mbf{k'_1}} +O_{11,\mbf{k'_1}}  )\\
    &+   12q_2^2 \sideset{}{'}\sum_{\mbf{\delta k'_{1,2}}}    ( O_{00,\mbf{k'_{1,2}}} + O_{11,\mbf{k'_{1,2}}}  + O_{22,\mbf{k'_{1,2}}} )+\sum_{\mbf{k'_0}}   2 \mbf{q_2}\cdot  \mbf{\delta k_{1,2} } O_{00,\mbf{k'_0}}\\
    &+ 6 \sideset{}{'}\sum_{\mbf{\delta k'_1}}  \big( (2\mbf{q_2}\cdot\mbf{ \delta k_{1,2}}-\mbf{\delta k'_1}\cdot \mbf{b_{1}} )O_{00,\mbf{k'_1}}+(  2 \mbf{q_2}\cdot \mbf{ \delta k_{1,2}}+ \mbf{\delta k'_1}\cdot \mbf{b_{1}})O_{11,\mbf{k'_1}}  \big)\\
    &+ 12 \sideset{}{'}\sum_{\mbf{\delta k'_{1,2}}}    \big( (\mbf{q_2}\cdot \mbf{\delta k_{1,2}}+\mbf{q_2}\cdot \mbf{\delta k'_{1,2}} )  O_{00,\mbf{k'_{1,2}}} +  (\mbf{q_2}\cdot \mbf{\delta k_{1,2}}+C_6^2 \mbf{q_2}\cdot \mbf{\delta k'_{1,2}} ) O_{11,\mbf{k'_{1,2}}} \\
    &+  (\mbf{q_2}\cdot \mbf{\delta k_{1,2}}+C_6^{-2}\mbf{q_2}\cdot \mbf{\delta k'_{1,2}} ) O_{22,\mbf{k'_{1,2}}} \big)+\sum_{\mbf{k'_0}}   \delta k_{1,2}^2 O_{00,\mbf{k'_0}} +  6 \sideset{}{'}\sum_{\mbf{\delta k'_1}}    (\delta k_{1,2}^2+\delta k_1'^2 )(O_{00,\mbf{k'_1}} + O_{11,\mbf{k'_1}}  )\\
    &+ 6\sideset{}{'}\sum_{\mbf{\delta k'_{1,2}}}     \big(\delta k_{1,2}^2+\delta k_{1,2}'^2) (O_{00,\mbf{k'_{1,2}}} + O_{11,\mbf{k'_{1,2}}}  +  O_{22,\mbf{k'_{1,2}}} )  \big)\bigg]
\end{align}
\begin{align}
    f_{\mbf{k_{i,i+1}} + C_6^{i-1} \mbf{b_{1}}}=&\frac{V_0\phi}{\Omega_{tot}}\bigg[\sum_{\mbf{k'_0}} (   q_2^2+k_0'^2 )O_{00,\mbf{k'_0}} +  6 ( ( \frac{1}{2} b_{1})^2 +  q_2^2) \sideset{}{'}\sum_{\mbf{\delta k'_1}}   ( O_{00,\mbf{k'_1}} +O_{11,\mbf{k'_1}}  )\\
    &+12 q_2^2 \sideset{}{'}\sum_{\mbf{\delta k'_{1,2}}}   (  O_{00,\mbf{k'_{1,2}}} +  O_{11,\mbf{k'_{1,2}}}  +  O_{22,\mbf{k'_{1,2}}} )+\sum_{\mbf{k'_0}} ( 2 C_6^2 \mbf{q_2}\cdot \mbf{ \delta k_{1,2} } )O_{00,\mbf{k'_0}}\\
    &+ 6 \sideset{}{'}\sum_{\mbf{\delta k'_1}}   \big(  (2 \mbf{ \delta k_{1,2}} \cdot C_6^2 \mbf{q_2} -\mbf{\delta k'_1}\cdot \mbf{b_{1}})O_{00,\mbf{k'_1}} + (2 \mbf{ \delta k_{1,2} }\cdot C_6^2 \mbf{q_2}+  \mbf{\delta k'_1}\cdot \mbf{b_{1}})  O_{11,\mbf{k'_1}}  \big)\\
    &+12 \sideset{}{'}\sum_{\mbf{\delta k'_{1,2}}}   \big( (\mbf{ \delta k_{1,2}} \cdot C_6^2\mbf{ q_2}+\mbf{\delta k'_{1,2}} \cdot \mbf{q_2}) O_{00,\mbf{k'_{1,2}}}+(\mbf{ \delta k_{1,2} }\cdot  C_6^2 \mbf{q_2}+\mbf{\delta k'_{1,2}}\cdot  C_6^2 \mbf{q_2}   ) O_{11,\mbf{k'_{1,2}}} \\
    &+  (\mbf{ \delta k_{1,2}} \cdot C_6^2 \mbf{q_2}+\mbf{\delta k'_{1,2}}\cdot  C^{-2}_6 \mbf{q_2} ) O_{22,\mbf{k'_{1,2}}}\big)+ \sum_{\mbf{k'_0}}  \delta k_{1,2}^2 O_{00,\mbf{k'_0}}\\
    &+ 6 \sideset{}{'}\sum_{\mbf{\delta k'_1}}    ( \delta k_{1,2}^2+\delta k_1'^2 )(O_{00,\mbf{k'_1}}+ O_{11,\mbf{k'_1}}  )+6 \sideset{}{'}\sum_{\mbf{\delta k'_{1,2}}}   (  \delta k_{1,2}^2+ \delta k_{1,2}'^2) (O_{00,\mbf{k'_{1,2}}} + O_{11,\mbf{k'_{1,2}}}  + O_{22,\mbf{k'_{1,2}}} ) \bigg]
\end{align}

\begin{align}
    f_{\mbf{k_{i,i+1}} + C_6^{i} \mbf{b_{1}}}=&\frac{V_0\phi}{\Omega_{tot}}\bigg[\sum_{\mbf{k'_0}} ( ( q_2^2+k_0'^2  )O_{00,\mbf{k'_0}} + 6 (\frac{1}{2} b_{1})^2 + q_2^2) \sideset{}{'}\sum_{\mbf{\delta k'_1}}   ( O_{00,\mbf{k'_1}} + O_{11,\mbf{k'_1}}  )\\
    &+ 12q^2_2\sideset{}{'}\sum_{\mbf{\delta k'_{1,2}}}   ( O_{00,\mbf{k'_{1,2}}} + O_{11,\mbf{k'_{1,2}}}  +O_{22,\mbf{k'_{1,2}}} )+2 C_6^{-2}\mbf{ q_2}\cdot \mbf{ \delta k_{1,2}} O_{00,\mbf{k'_0}}\\
    &+ 6 \sideset{}{'}\sum_{\mbf{\delta k'_1}}   \big( (2\mbf{ \delta k_{1,2}} \cdot C_6^{-2} \mbf{q_2} -\mbf{\delta k'_1}\cdot  \mbf{b_{1}} )O_{00,\mbf{k'_1}} +(2 \mbf{ \delta k_{1,2} }\cdot C_6^{-2} \mbf{q_2} + \mbf{\delta k'_1}\cdot \mbf{b_{1}}  )O_{11,\mbf{k'_1}}  \big)\\
    &+12\sideset{}{'}\sum_{\mbf{\delta k'_{1,2}}}   \big( (\mbf{ \delta k_{1,2} }\cdot C_6^{-2}\mbf { q_2}+ \mbf{\delta k'_{1,2}} \cdot \mbf{q_2}) O_{00,\mbf{k'_{1,2}}} +  (\mbf{ \delta k_{1,2}} \cdot C_6^{-2} \mbf{q_2}+\mbf{\delta k'_{1,2}}\cdot C_6^2 \mbf{q_2}   ) O_{11,\mbf{k'_{1,2}}}\\
    &+   (\mbf{ \delta k_{1,2} }\cdot C_6^{-2} \mbf{q_2}+ \mbf{\delta k'_{1,2}}\cdot C^{-2}_6 \mbf{q_2} ) O_{22,\mbf{k'_{1,2}}} \big)+\sum_{\mbf{k'_0}}   \delta k_{1,2}^2O_{00,\mbf{k'_0}}\\
    &+ 6 \sideset{}{'}\sum_{\mbf{\delta k'_1}}    ( \delta k_{1,2}^2 + \delta k_1'^2 )(O_{00,\mbf{k'_1}} +O_{11,\mbf{k'_1}}  )+6\sideset{}{'}\sum_{\mbf{\delta k'_{1,2}}}   (  \delta k_{1,2}^2+ \delta k_{1,2}'^2)( O_{00,\mbf{k'_{1,2}}} +O_{11,\mbf{k'_{1,2}}}  + O_{22,\mbf{k'_{1,2}}} ) \bigg]
\end{align}
}

We define the following quantities
\begin{align}
    f_{\mbf{k_0}}^0 =& \frac{V_0 \phi}{\Omega_{tot}} \bigg[\sum_{\mbf{k'_0}}   (k_0^2+ k_0'^2)O_{00,\mbf{k'_0}} + 6 (k_0^2+ (\frac{b_{1}}{2})^2) \sideset{}{'}\sum_{\mbf{k'_1}}   (O_{00,\mbf{k'_1}}
 + O_{11,\mbf{k'_1}} ) \\& +6 ( k_0^2+q_2^2 ) \sideset{}{'}\sum_{\mbf{k'_{1,2}}} (O_{00,\mbf{k'_{1,2}}} +O_{11,\mbf{k'_{1,2}}}  +  O_{2,2,\mbf{k'_{1,2}}}  )\bigg]\\
 f_{1} =& \frac{V_0\phi}{\Omega_{tot}} \bigg[\sum_{\mbf{k'_0}}   ((\frac{b_{1}}{2})^2+ k_0'^2)O_{00,\mbf{k'_0}} + 3 b_{1}^2 \sideset{}{'}\sum_{\mbf{k'_1}}   (O_{00,\mbf{k'_1}}
 + O_{11,\mbf{k'_1}} ) \\& +6  ((\frac{b_{1}}{2})^2+q_2^2 ) \sideset{}{'}\sum_{\mbf{k'_{1,2}}}  (O_{00,\mbf{k'_{1,2}}} + O_{11,\mbf{k'_{1,2}}}  + O_{2,2,\mbf{k'_{1,2}}} ) \bigg]\\
 f_{{1,2}} =& \frac{V_0\phi}{\Omega_{tot}} \bigg[\sum_{\mbf{k'_0}}   (q_2^2+ k_0'^2)O_{00,\mbf{k'_0}} + 6 (q_2^2+(\frac{b_{1}}{2})^2) \sideset{}{'}\sum_{\mbf{k'_1}}   (O_{00,\mbf{k'_1}}
 + O_{11,\mbf{k'_1}} )  \\& +12 q_2^2 \sideset{}{'}\sum_{\mbf{k'_{1,2}}} ( O_{00,\mbf{k'_{1,2}}} + O_{11,\mbf{k'_{1,2}}}  + O_{2,2,\mbf{k'_{1,2}}} ) \bigg]\\
 \delta f =& 6\frac{V_0\phi}{\Omega_{tot}} \bigg[\mbf{b_{1}}\cdot \sideset{}{'}\sum_{\mbf{\delta k'_1}}    \mbf{\delta k'_1}   (-O_{00,\mbf{k'_1}} +O_{11,\mbf{k'_1}}  )\\
 &+2\mbf{q_2} \cdot \sideset{}{'}\sum_{\mbf{\delta k'_{1,2}}} (    \mbf{\delta k'_{1,2}}  O_{00,\mbf{k'_{1,2}}} + C_6^{-2}  \mbf{\delta k'_{1,2}}    O_{11,\mbf{k'_{1,2}}}  + C_6^2  \mbf{\delta k'_{1,2}}  O_{22,\mbf{k'_{1,2}}} )\bigg]\\
 \delta^2f=&  6\frac{V_0\phi }{\Omega_{tot}} \bigg[ \sideset{}{'}\sum_{\mbf{\delta k'_1}}      \delta k_1'^2 (O_{00,\mbf{k'_1}} +O_{11,\mbf{k'_1}}  ) +\sideset{}{'}\sum_{\mbf{\delta k'_{1,2}}}    \delta k_{1,2}'^2( O_{00,\mbf{k'_{1,2}}} +O_{11,\mbf{k'_{1,2}}}  + O_{22,\mbf{k'_{1,2}}} )   \bigg]\\
 \mathcal{O}=&  \frac{V_0\phi }{\Omega_{tot}} \bigg[\sum_{\mbf{k'_0}} O_{00,\mbf{k'_0}}+ 6 \sideset{}{'}\sum_{\mbf{\delta k'_1}}    (O_{00,\mbf{k'_1}} +O_{11,\mbf{k'_1}}  ) + 6\sideset{}{'}\sum_{\mbf{\delta k'_{1,2}}}   ( O_{00,\mbf{k'_{1,2}}} +O_{11,\mbf{k'_{1,2}}}  + O_{22,\mbf{k'_{1,2}}} )   \bigg]. 
\end{align}
Note that $\mathcal{O}$ is proportional to $\phi$ and the total particle number. Arranging in powers of the momentum deviation $\delta$, we obtain
\begin{align}
    &f_{\mbf{k_0}}= f_{\mbf{k_0}}^0   +   \delta f+ \delta^2 f  \\ & f_{\mbf{k_i}} = f_1 +\delta f-  \mbf{\delta k_1 \cdot  b_{1}} \mathcal{O}+ \delta^2 f+ \delta k_1^2 \cdot\mathcal{O}  ,  \\ & f_{\mbf{k_i + C_6^{i-1} b_{1}}}= f_1 +\delta f+  \mbf{\delta k_1 \cdot  b_{1}} \mathcal{O}+ \delta^2 f+ \delta k_1^2 \cdot \mathcal{O}     \\ &f_{\mbf{k_{i,i+1}}}=f_{1,2} + \delta f+2\mbf{\delta k_{1,2}}\cdot \mbf{q_2} \mathcal{O}+ \delta^2f + \delta k_{1,2}^2 \cdot \mathcal{O}     \\ &   f_{\mbf{k_{i,i+1}} + C_6^{i-1} \mbf{b_{1}}}  =f_{1,2}+ \delta f+2\mbf{\delta k_{1,2}}\cdot C_6^2 \mbf{q_2} \mathcal{O}+  \delta^2f + \delta k_{1,2}^2 \cdot \mathcal{O}     \\ &   f_{\mbf{k_{i,i+1}} + C_6^{i} \mbf{b_{1}}}= f_{1,2} + \delta f+2\mbf{\delta k_{1,2}}\cdot C_6^{-2} \mbf{q_2} \mathcal{O} + \delta^2f + \delta k_{1,2}^2 \cdot \mathcal{O}.
\end{align}
We define corresponding primed quantities as the above quantities measured with respect to $f_{1,2} + \delta f+ \delta f^2$, which corresponds to the constant part in the $\bm{k}_{i,i+1}$ region
\begin{align}\label{eqapp:fprime}
    &f'_{\mbf{k_0}}= f_{\mbf{k_0}}^0  -f_{1,2}=( k_0^2-q_2^2 ) \mathcal{O}  \\ 
    & f'_{\mbf{k_i}} = f_1 - f_{1,2}+ (-  \mbf{\delta k_1 \cdot  b_{1}} + \delta k_1^2 )\mathcal{O} =\left(\left({b_{1}}/{2}\right)^2- q_2^2   - \mbf{\delta k_1 \cdot  b_{1}} +  \delta k_1^2\right) \mathcal{O}  \\ 
    & f'_{\mbf{k_i + C_6^{i-1} b_{1}}}= f_1 - f_{1,2} +  (\mbf{\delta k_1 \cdot  b_{1}} +  \delta k_1^2 ) \mathcal{O} =\left(\left({b_{1}}/{2}\right)^2- q_2^2   + \mbf{\delta k_1 \cdot  b_{1}} +  \delta k_1^2\right) \mathcal{O}     \\ 
    &f'_{\mbf{k_{i,i+1}}}=(2\mbf{\delta k_{1,2}}\cdot \mbf{q_2}+ \delta k_{1,2}^2)  \mathcal{O}     \\ 
    &   f'_{\mbf{k_{i,i+1}} + C_6^{i-1} \mbf{b_{1}}}  = (2\mbf{\delta k_{1,2}}\cdot C_6^2 \mbf{q_2} + \delta k_{1,2}^2) \mathcal{O}     \\ 
    &   f'_{\mbf{k_{i,i+1}} + C_6^{i} \mbf{b_{1}}}= (2\mbf{\delta k_{1,2}}\cdot C_6^{-2} \mbf{q_2}  + \delta k_{1,2}^2 ) \mathcal{O}.
\end{align}

\subsection{$g_{j,\mathbf{k}}$ in more detail}

In this subsection, we consider the interaction-induced hybridization (Eq.~\ref{eqapp:gk_first}) in more detail. For convenience, we repeat Eq.~\ref{eqapp:gk_first} here
\begin{eqnarray}
    &g_{j,\mbf{k}}=\frac{1}{\Omega_{tot}}\sum_{\bm{k}'}\bigg[V_{b_{1}} \langle \mbf{k+C_6^{j-1}b_{1}} | \mbf{k} \rangle \langle \mbf{k'-C_6^{j-1}b_{1}} | \mbf{k'} \rangle \\
    &-V_{|\mbf{k-k'+C_6^{j-1} b_{1}}|} \langle \mbf{k+C_6^{j-1} b_{1}}|\mbf{k'} \rangle \langle \mbf{k'-C_6^{j-1} b_{1}}|\mbf{k} \rangle  \bigg]O_{\mbf{k'} - \mbf{C_6^{j-1} b_{1}},\mbf{k'}}\\
    &= \frac{1}    {\Omega_{tot}}  \bigg[ \sideset{}{'}\sum_{\mbf{k'_j}}\big(V_{b_{1}}    \langle \mbf{k+C_6^{j-1}b_{1}} | \mbf{k} \rangle \langle \mbf{{k'_j}} | \mbf{k'_j}+ C_6^{j-1} \mbf{b_{1}} \rangle\\ &- V_{|\mbf{k}- \mbf{k'_j}|} \langle \mbf{k}+C_6^{j-1} \mbf{b_{1}}| \mbf{k'_j}+ C_6^{j-1} \mbf{b_{1}} \rangle \langle  \mbf{k'_j}|\mbf{k}\rangle \big)
 O_{0j,\mbf{k'_j}}        \\ & + \sideset{}{'}\sum_{\mbf{k'_{j,j+1}}} \big(V_{b_{1}} \langle \mbf{k}+C_6^{j-1}\mbf{b_{1}} | \mbf{k} \rangle \langle \mbf{{k'_{j,j+1} }} | \mbf{k'_{j,j+1} }+ C_6^{j-1} \mbf{b_{1}} \rangle \\ &-V_{|\mbf{k}-\mbf{k'_{j,j+1} }|} \langle \mbf{k}+C_6^{j-1} \mbf{b_{1}}|\mbf{k'_{j,j+1} }+ C_6^{j-1} \mbf{b_{1}} \rangle \langle \mbf{k'_{j,j+1} }|\mbf{k}\rangle   \big)   O_{0j,\mbf{k'_{j,j+1}}}   \\ & + \sideset{}{'}\sum_{\mbf{k'_{j-1,j}}}  \big(V_{b_{1}} \langle \mbf{k}+C_6^{j-1}\mbf{b_{1}} | \mbf{k} \rangle \langle \mbf{k'_{j-1,j} } |  \mbf{k'_{j-1,j} }+ C_6^{j-1} \mbf{b_{1}} \rangle\\ &-V_{|\mbf{k}- \mbf{k'_{j-1,j} }|} \langle \mbf{k}+C_6^{j-1} \mbf{b_{1}}|\mbf{k'_{j-1,j} }+ C_6^{j-1} \mbf{b_{1}} \rangle \langle \mbf{k'_{j-1,j} }|\mbf{k}\rangle\big)  O_{0j,\mbf{k'_{j-1,j}}}\\ &+ \sideset{}{'}\sum_{\mbf{k'_{j-3}}} \big( V_{b_{1}}\langle \mbf{k+C_6^{j-1}b_{1}} | \mbf{k} \rangle \langle \mbf{{k'_{j-3}}-C_6^{j-1}b_{1}} |\mbf{k'_{j-3}} \rangle  \\ &- V_{|\mbf{k}- \mbf{k'_{j-3}}+C_6^{j-1} \mbf{b_{1}}|} \langle \mbf{k}+C_6^{j-1} \mbf{b_{1}}|\mbf{k'_{j-3}} \rangle \langle \mbf{k'_{j-3}}-C_6^{j-1} \mbf{b_{1}}|\mbf{k}\rangle \big) O_{j-3,0,\mbf{k'_{j-3}}}     \\ & + \sideset{}{'}\sum_{\mbf{k'_{j-3,j-2}}}  \big(V_{b_{1}}\langle \mbf{k+C_6^{j-1}b_{1}} | \mbf{k} \rangle \langle \mbf{{k'_{j-3,j-2}}-C_6^{j-1}b_{1}} |\mbf{k'_{j-3,j-2}}\rangle \\ &-V_{|\mbf{k} - \mbf{k'_{j-3,j-2}}+C_6^{j-1} \mbf{b_{1}}|} \langle \mbf{k}+C_6^{j-1} \mbf{b_{1}}|\mbf{k'_{j-3,j-2}} \rangle \langle \mbf{k'_{j-3,j-2}}-C_6^{j-1} \mbf{b_{1}}|\mbf{k}\rangle  \big)  O_{j-3,0, \mbf{k'_{j-3,j-2}}}  \\ & +\sideset{}{'}\sum_{\mbf{k'_{j-4,j-3}}}  \big(V_{b_{1}} \langle \mbf{k+C_6^{j-1}b_{1}} | \mbf{k} \rangle \langle \mbf{{k'_{j-4,j-3}}-C_6^{j-1}b_{1}} | \mbf{k'_{j-4,j-3}} \rangle \\ &-V_{|\mbf{k}- \mbf{k'_{j-4,j-3}}+C_6^{j-1} \mbf{b_{1}}|} \langle \mbf{k}+C_6^{j-1} \mbf{b_{1}}|\mbf{k'_{j-4,j-3}} \rangle \langle \mbf{k'_{j-4,j-3}}-C_6^{j-1} \mbf{b_{1}}|\mbf{k}\rangle \big) O_{j-3,0, \mbf{k'_{j-4,j-3}}}     \\ & + \sideset{}{'}\sum_{\mbf{k'_{j-2,j-1}}}  \big(V_{b_{1}} \langle \mbf{k+C_6^{j-1}b_{1}} | \mbf{k} \rangle \langle \mbf{k'_{j-2,j-1}}+ C_6^{j+3} \mbf{b_{1}} | \mbf{k'_{j-2,j-1}}+ C_6^{j-2} \mbf{b_{1}} \rangle \\ &- V_{|\mbf{k}- \mbf{k'_{j-2,j-1}}+C_6^{j} \mbf{b_{1}}|} \langle \mbf{k}+C_6^{j-1} \mbf{b_{1}}|\mbf{k'_{j-2,j-1}}+ C_6^{j-2} \mbf{b_{1}} \rangle \langle \mbf{k'_{j-2,j-1}}-C_6^{j} \mbf{b_{1}}|\mbf{k}\rangle  \big)   O_{j-2,j-1,\mbf{k'_{j-2,j-1}}} \\ & + \sideset{}{'}\sum_{\mbf{k'_{ j-5,j-4 }}}  \big(V_{b_{1}}\langle \mbf{k+C_6^{j-1}b_{1}} | \mbf{k} \rangle \langle \mbf{k'_{ j-5,j-4 }}+ C_6^{j+1} \mbf{b_{1}} | \mbf{k'_{ j-5,j-4 }}+ C_6^{j} \mbf{b_{1}}\rangle   \\ &- V_{|\mbf{k}- \mbf{k'_{ j-5,j-4 }}+C_6^{j+4} \mbf{b_{1}}|} \langle \mbf{k}+C_6^{j-1} \mbf{b_{1}}| \mbf{k'_{ j-5,j-4 }}+ C_6^{j} \mbf{b_{1}} \rangle \langle  \mbf{k'_{ j-5,j-4 }}-C_6^{j+4} \mbf{b_{1}}|\mbf{k}\rangle \big)  O_{j-4, j-5 ,\mbf{k'_{ j-5,j-4}}}   \bigg].
\end{eqnarray}
We first implement $C_6$ symmetry so that the order parameters and summation momenta are evaluated in regions $\bm{k}_1$ and $\bm{k}_{1,2}$
{\tiny
\begin{eqnarray}
    &g_{j,\bm{k}}=\frac{1}{\Omega_\text{tot}}\bigg[\sideset{}{'}\sum_{\mbf{k'_1}}\Big(\big(V_{b_{1}}    \langle \mbf{k+C_6^{j-1}b_{1}} | \mbf{k} \rangle \langle \mbf{k'_1} | \mbf{k'_1}+  \mbf{b_{1}} \rangle- V_{|C_6^{1-j} \mbf{k}- \mbf{k'_1}|}\langle C_6^{1-j} \mbf{k}+ \mbf{b_{1}}| \mbf{k'_1}+  \mbf{b_{1}} \rangle \langle  \mbf{k'_1}|C_6^{1-j}\mbf{k}\rangle \big)
 O_{01,\mbf{k'_1}} \\
    &+  \big( V_{b_{1}}\langle \mbf{k+C_6^{j-1}b_{1}} | \mbf{k} \rangle \langle {\mbf{k'_{1}}+b_{1}} |\mbf{k'_{1}} \rangle  - V_{|C_6^{1-j} \mbf{k}+ \mbf{k'_{1}}+\mbf{b_{1}}|} \langle C_6^{1-j}\mbf{k}+ \mbf{b_{1}}|-\mbf{k'_{1}} \rangle \langle -\mbf{k'_{1}}- \mbf{b_{1}}|C_6^{1-j} \mbf{k}\rangle\big) O_{1,0,\mbf{k'_{1}}}\Big)\\
    &+\sideset{}{'}\sum_{\mbf{k'_{1,2}}} \Big(\big(V_{b_{1}} \langle \mbf{k}+C_6^{j-1}\mbf{b_{1}} | \mbf{k} \rangle \langle {\mbf{k'_{1,2} }} | \mbf{k'_{1,2} }+  \mbf{b_{1}} \rangle-V_{|C_6^{1-j}\mbf{k}-\mbf{k'_{1,2} }|} \langle C_6^{1-j} \mbf{k}+ \mbf{b_{1}}|\mbf{k'_{1,2} }+  \mbf{b_{1}} \rangle \langle \mbf{k'_{1,2} }|C_6^{1-j}\mbf{k}\rangle   \big)   O_{01,\mbf{k'_{1,2}}}\\
    &+(V_{b_{1}}\langle \mbf{k+C_6^{j-1}b_{1}} | \mbf{k} \rangle \langle {\mbf{k'_{1,2}}+b_{1}} |\mbf{k'_{1,2}}\rangle -V_{|C_6^{1-j} \mbf{k} +\mbf{k'_{1,2}}+\mbf{b_{1}}|} \langle C_6^{1-j} \mbf{k}+ \mbf{b_{1}}|-\mbf{k'_{1,2}} \rangle \langle -\mbf{k'_{1,2}}- \mbf{b_{1}}|C_6^{1-j}\mbf{k}\rangle  \big) \rangle O_{1,0, \mbf{k'_{1,2}}}\\
    &+  \big(V_{b_{1}} \langle \mbf{k}+C_6^{j-1}\mbf{b_{1}} | \mbf{k} \rangle \langle  \mbf{k'_{1,2} } |   \mbf{k'_{1,2} }+  C_6\mbf{b_{1}} \rangle - V_{|C_6^{2-j}\mbf{k}-   \mbf{k'_{1,2} }|} \langle C_6^{2-j}\mbf{k}+C_6\mbf{b_{1}}|{  \mbf{k'_{1,2} }+  C_6\mbf{b_{1}}} \rangle \langle  \mbf{k'_{1,2} }|C_6^{2-j} \mbf{k}\rangle\big)  O_{02,\mbf{k'_{1,2}}} \\
    &+  \big(V_{b_{1}} \langle \mbf{k+C_6^{j-1}b_{1}} | \mbf{k} \rangle \langle {\mbf{k'_{1,2}}+C_6b_{1}} | \mbf{k'_{1,2}} \rangle -V_{|C_6^{-1-j}\mbf{k}- \mbf{k'_{1,2}}- C_6\mbf{b_{1}}|} \langle C_6^{-1-j}\mbf{k}- C_6\mbf{b_{1}}|\mbf{k'_{1,2}} \rangle \langle  \mbf{k'_{1,2}}+ C_6\mbf{b_{1}}|C_6^{-1-j} \mbf{k}\rangle \big) O_{2,0, \mbf{k'_{1,2}}} \\
    & +  \big(V_{b_{1}} \langle \mbf{k+C_6^{j-1}b_{1}} | \mbf{k} \rangle \langle \mbf{k'_{1,2}}+  \mbf{b_{1}} |   \mbf{k'_{1,2}}+  C_6\mbf{b_{1}} \rangle- V_{|C_6^{-j}\mbf{k}+  \mbf{k'_{1,2}}+ \mbf{b_{1}}|}\langle C_6^{-j} \mbf{k}- C_6^2\mbf{b_{1}}|-  \mbf{k'_{1,2}}-  C_6\mbf{b_{1}} \rangle \langle -  \mbf{k'_{1,2}}- \mbf{b_{1}}|C_6^{-j}\mbf{k}\rangle  \big)   O_{1,2,\mbf{k'_{1,2}}}\\
    &+\big(V_{b_{1}}\langle \mbf{k+C_6^{j-1}b_{1}} | \mbf{k} \rangle \langle  \mbf{k'_{ 1,2 }}+ C_6\mbf{b_{1}} | \mbf{k'_{ 1,2 }}+  \mbf{b_{1}}\rangle - V_{|C_6^{-j} \mbf{k}- \mbf{k'_{ 1,2 }}- C_6\mbf{b_{1}}|} \langle C_6^{-j} \mbf{k}- C_6^2\mbf{b_{1}}| \mbf{k'_{ 1,2 }}+ \mbf{b_{1}} \rangle \langle  \mbf{k'_{ 1,2 }}+ C_6\mbf{b_{1}}|C_6^{-j}\mbf{k}\rangle \big)  O_{2, 1 ,\mbf{k'_{ 1,2}}}   \bigg)\bigg].
\end{eqnarray}
}

We now use the parameterization of the momentum in Eq.~\ref{eqapp:dk_param} for the momentum summations
\begin{eqnarray}
    & g_{j, \mbf{k}}=  \frac{1}    {\Omega_{tot}}  \bigg[  \sideset{}{'}\sum_{\mbf{k'_1}}(V_{b_{1}}    \langle \mbf{k+C_6^{j-1}b_{1}} | \mbf{k} \rangle \langle {-\frac{\mbf{b_{1}}}{2}+ \mbf{\delta k'_1}} |\frac{\mbf{b_{1}}}{2}+  \mbf{\delta k'_1} \rangle\\ & -  V_{|C_6^{1-j} \mbf{k} + \frac{\mbf{b_{1}}}{2} - \mbf{\delta k'_1}|} \langle C_6^{1-j} \mbf{k}+ \mbf{b_{1}}|\frac{\mbf{b_{1}}}{2} + \mbf{\delta k'_1} \rangle \langle - \frac{\mbf{b_{1}}}{2}+  \mbf{\delta k'_1}|C_6^{1-j}\mbf{k}\rangle )
 O_{01,\mbf{k'_1}}   + \nonumber\\ &+  ( V_{b_{1}}\langle \mbf{k+C_6^{j-1}b_{1}} | \mbf{k} \rangle \langle  \frac{\mbf{b_{1}}}{2} + \mbf{\delta k'_{1}} |- \frac{\mbf{b_{1}}}{2} + \mbf{\delta k'_{1}} \rangle   \\ & - V_{|C_6^{1-j} \mbf{k}+ \frac{\mbf{b_{1}}}{2}+ \mbf{\delta k'_{1}}|} \langle C_6^{1-j}\mbf{k}+ \mbf{b_{1}}|\frac{\mbf{b_{1}}}{2}-\mbf{\delta k'_{1}} \rangle \langle - \frac{\mbf{b_{1}}}{2} -\mbf{\delta k'_{1}}|C_6^{1-j} \mbf{k}\rangle ) O_{1,0,\mbf{k'_{1}}}    \\ & + \sideset{}{'}\sum_{\mbf{k'_{1,2}}} (V_{b_{1}} \langle \mbf{k}+C_6^{j-1}\mbf{b_{1}} | \mbf{k} \rangle \langle \mbf{q_2+ {\delta k'_{1,2} }} | C_6^2 \mbf{q_2} + \mbf{\delta k'_{1,2} } \rangle \\ & - V_{|C_6^{1-j}\mbf{k}-\mbf{q_2}- \mbf{\delta k'_{1,2} }|} \langle C_6^{1-j} \mbf{k}+ \mbf{b_{1}}|C_6^2 \mbf{q_2}+ \mbf{\delta k'_{1,2} } \rangle \langle \mbf{q_2}+ \mbf{\delta k'_{1,2} }|C_6^{1-j}\mbf{k}\rangle   )   O_{01,\mbf{k'_{1,2}}}   \\ & +(V_{b_{1}}\langle \mbf{k+C_6^{j-1}b_{1}} | \mbf{k} \rangle \langle C_6^2 \mbf{q_2}+  \mbf{\delta k'_{1,2}} |\mbf{q_2+ \delta k'_{1,2}}\rangle \\ &- V_{|C_6^{1-j} \mbf{k} +C_6^2 \mbf{q_2}+ \mbf{\delta k'_{1,2}}|} \langle C_6^{1-j} \mbf{k}+ \mbf{b_{1}}|-\mbf{q_2}- \mbf{\delta k'_{1,2}} \rangle \langle -C_6^2 \mbf{q_2} - \mbf{\delta k'_{1,2}}|C_6^{1-j}\mbf{k}\rangle  ) \rangle O_{1,0, \mbf{k'_{1,2}}} \\ & +  (V_{b_{1}} \langle \mbf{k}+C_6^{j-1}\mbf{b_{1}} | \mbf{k} \rangle \langle  \mbf{q_2+ \delta k'_{1,2} } |  C_6^{-2}\mbf{q_2}+  \mbf{\delta k'_{1,2} }\rangle \\ & - V_{|C_6^{2-j}\mbf{k}- \mbf{q_2} - \mbf{\delta k'_{1,2} }|} \langle C_6^{2-j}\mbf{k}+C_6\mbf{b_{1}}|{ C_6^{-2}\mbf{q_2} +  \mbf{\delta k'_{1,2} }} \rangle \langle  \mbf{q_2+ \delta k'_{1,2} }|C_6^{2-j} \mbf{k}\rangle)  O_{02,\mbf{k'_{1,2}}}      \\ & +  (V_{b_{1}} \langle \mbf{k+C_6^{j-1}b_{1}} | \mbf{k} \rangle \langle C_6^{-2} \mbf{q_2}+  \mbf{\delta k'_{1,2}} | \mbf{q_2+ \delta k'_{1,2}} \rangle  \\ & -V_{|C_6^{-1-j}\mbf{k}-C_6^{-2}\mbf{q_2}- \mbf{\delta k'_{1,2}}|} \langle C_6^{-1-j}\mbf{k}- C_6\mbf{b_{1}}|\mbf{q_2+ \delta k'_{1,2}} \rangle \langle C_6^{-2}\mbf{q_2}+ \mbf{\delta k'_{1,2}}|C_6^{-1-j} \mbf{k}\rangle ) O_{2,0, \mbf{k'_{1,2}}}     \\ & +  (V_{b_{1}} \langle \mbf{k+C_6^{j-1}b_{1}} | \mbf{k} \rangle \langle C_6^2 \mbf{q_2}+ \mbf{\delta k'_{1,2}} |  C_6^{-2} \mbf{q_2} +  \mbf{\delta k'_{1,2}} \rangle-\nonumber \\ & - V_{|C_6^{-j}\mbf{k}+ C_6^2\mbf{q_2}+  \mbf{\delta k'_{1,2}}|}\langle C_6^{-j} \mbf{k}- C_6^2\mbf{b_{1}}|-C_6^{-2} \mbf{q_2} - \mbf{\delta k'_{1,2}} \rangle \langle - C_6^2 \mbf{q_2} -  \mbf{\delta k'_{1,2}}|C_6^{-j}\mbf{k}\rangle  )   O_{1,2,\mbf{k'_{1,2}}}   \\ & +   V_{b_{1}}\langle \mbf{k+C_6^{j-1}b_{1}} | \mbf{k} \rangle \langle  C_6^{-2}\mbf{q_2}+ \mbf{\delta k'_{ 1,2 }} | C_6^2 \mbf{q_2}+ \mbf{\delta k'_{ 1,2 }} \rangle \\ & - V_{|C_6^{-j} \mbf{k}- C_6^{-2} \mbf{q_2}- \mbf{\delta k'_{ 1,2 }}|} \langle C_6^{-j} \mbf{k}- C_6^2\mbf{b_{1}}|C_6^2 \mbf{q_2}+  \mbf{\delta k'_{ 1,2 }} \rangle \langle  C_6^{-2} \mbf{q_2} + \mbf{\delta k'_{ 1,2 }}|C_6^{-j}\mbf{k}\rangle )  O_{2, 1 ,\mbf{k'_{ 1,2}}}   \bigg].
\end{eqnarray}

For later convenience, it will be useful to specify the hybridization function explicitly in the momentum regions. In particular, we are interested in $g_{1,\bm{k}_1},g_{1,\bm{k}_{1,2}},g_{2,\bm{k}_{1,2}}$, since the others can be obtained using $C_6$ symmetry. We first focus on $g_{1,\bm{k}_1}$
\begin{eqnarray}
    & g_{1, \mbf{k_1}}=  \frac{1}    {\Omega_{tot}}  \bigg[  \sideset{}{'}\sum_{\mbf{k'_1}}(V_{b_{1}}    \langle \frac{\mbf{b_{1}}}{2}+ \mbf{\delta k_1} |- \frac{\mbf{b_{1}}}{2}+  \mbf{\delta k_1} \rangle \langle {-\frac{\mbf{b_{1}}}{2}+ \mbf{\delta k'_1}} |\frac{\mbf{b_{1}}}{2}+  \mbf{\delta k'_1} \rangle -\nonumber \\\ & -  V_{| \mbf{\delta k_1}  - \mbf{\delta k'_1}|} \langle \frac{\mbf{b_{1}}}{2}+ \mbf{\delta k_1}|\frac{\mbf{b_{1}}}{2} + \mbf{\delta k'_1} \rangle \langle - \frac{\mbf{b_{1}}}{2}+  \mbf{\delta k'_1}|- \frac{\mbf{b_{1}}}{2}+ \mbf{\delta k_1}\rangle )
 O_{01,\mbf{k'_1}}   \\ &+  ( V_{b_{1}}\langle \frac{\mbf{b_{1}}}{2}+  \mbf{\delta k_1} |- \frac{\mbf{b_{1}}}{2}+ \mbf{\delta k_1} \rangle \langle  \frac{\mbf{b_{1}}}{2} + \mbf{\delta k'_{1}} |- \frac{\mbf{b_{1}}}{2} + \mbf{\delta k'_{1}} \rangle \\ & - V_{|\mbf{\delta k_1}+ \mbf{\delta k'_{1}}|} \langle \frac{\mbf{b_{1}}}{2}+ \mbf{\delta k_1}|\frac{\mbf{b_{1}}}{2}-\mbf{\delta k'_{1}} \rangle \langle \frac{\mbf{b_{1}}}{2} +\mbf{\delta k'_{1}}|\frac{\mbf{b_{1}}}{2}-  \mbf{\delta k_1}\rangle ) O_{1,0,\mbf{k'_{1}}}    \\ & + \sideset{}{'}\sum_{\mbf{k'_{1,2}}} (V_{b_{1}} \langle \frac{\mbf{b_{1}}}{2}+ \mbf{\delta k_1} | -\frac{\mbf{b_{1}}}{2} + \mbf{\delta k_1} \rangle \langle \mbf{q_2+ {\delta k'_{1,2} }} | C_6^2 \mbf{q_2} + \mbf{\delta k'_{1,2} } \rangle\\ & - V_{|\mbf{\delta k_1}- \frac{1}{2} C_{6}\mbf{q_2}- \mbf{\delta k'_{1,2} }|} \langle \frac{\mbf{b_{1}}}{2}+ \mbf{\delta k_1}|C_6^2 \mbf{q_2}+ \mbf{\delta k'_{1,2} } \rangle \langle \mbf{q_2}+ \mbf{\delta k'_{1,2} }|- \frac{\mbf{b_{1}}}{2} + \mbf{\delta k_1}\rangle   )   O_{01,\mbf{k'_{1,2}}}   \\ & +(V_{b_{1}}\langle \frac{\mbf{b_{1}}}{2}+  \mbf{\delta k_1} | - \frac{\mbf{b_{1}}}{2}+ \mbf{\delta k_1} \rangle \langle C_6^2 \mbf{q_2}+  \mbf{\delta k'_{1,2}} |\mbf{q_2+ \delta k'_{1,2}}\rangle\\ &- V_{| \mbf{\delta k_1}+\frac{1}{2} C_6 \mbf{q_2}+ \mbf{\delta k'_{1,2}}|} \langle \frac{\mbf{b_{1}}}{2}+  \mbf{\delta k_1}|-\mbf{q_2}- \mbf{\delta k'_{1,2}} \rangle \langle C_6^2 \mbf{q_2} + \mbf{\delta k'_{1,2}}| \frac{\mbf{b_{1}}}{2}- \mbf{\delta k_1}\rangle  ) \rangle O_{1,0, \mbf{k'_{1,2}}}   \\ & +  (V_{b_{1}} \langle \frac{\mbf{b_{1}}}{2}+ \mbf{\delta k_1} | - \frac{\mbf{b_{1}}}{2} + \mbf{\delta k_1} \rangle \langle  \mbf{q_2+ \delta k'_{1,2} } |  -C_6\mbf{q_2}+  \mbf{\delta k'_{1,2} }\rangle \\ & - V_{|C_6\mbf{\delta k_1 } - \frac{1}{2} C_{6}^{-1}\mbf{q_2} - \mbf{\delta k'_{1,2} }|} \langle C_6\frac{\mbf{b_{1}}}{2}+ C_6\mbf{\delta k_1}|-C_6\mbf{q_2} +  \mbf{\delta k'_{1,2} } \rangle \langle  \mbf{q_2+ \delta k'_{1,2} }|- C_6\frac{\mbf{b_{1}}}{2} +C_6 \mbf{\delta k_1}\rangle)  O_{02,\mbf{k'_{1,2}}}     \\ & +  (V_{b_{1}} \langle \frac{\mbf{b_{1}}}{2}+ \mbf{\delta k_1} | - \frac{\mbf{b_{1}}}{2}+ \mbf{\delta k_1} \rangle \langle - C_6 \mbf{q_2}+  \mbf{\delta k'_{1,2}} | \mbf{q_2+ \delta k'_{1,2}} \rangle  \\ & -V_{|-C_6\mbf{\delta k_1} -\frac{1}{2} C_6^{-1}\mbf{q_2}- \mbf{\delta k'_{1,2}}|} \langle -C_6 \mbf{\delta k_1 }- C_6\frac{\mbf{b_{1}}}{2} |\mbf{q_2+ \delta k'_{1,2}} \rangle \langle -C_6\mbf{q_2}+ \mbf{\delta k'_{1,2}}| C_6\frac{\mbf{b_{1}}}{2}-C_6  \mbf{\delta k_1}\rangle ) O_{2,0, \mbf{k'_{1,2}}}      +\nonumber \\ & +  (V_{b_{1}} \langle  \frac{\mbf{b_{1}}}{2} + \mbf{\delta k_1} | -\frac{\mbf{b_{1}}}{2}+  \mbf{\delta  k_1} \rangle \langle C_6^2 \mbf{q_2}+ \mbf{\delta k'_{1,2}} |  C_6^{-2} \mbf{q_2} +  \mbf{\delta k'_{1,2}} \rangle-\nonumber \\ & - V_{|C_6^{-1}\mbf{\delta k_1} - \frac{1}{2} \mbf{q_2}+  \mbf{\delta k'_{1,2}}|}\langle C_6^{-1} \mbf{\delta k_1}-C_6^2 \frac{\mbf{b_{1}}}{2} |-C_6^{-2} q_2- \mbf{\delta k'_{1,2}} \rangle \langle - C_6^2 \mbf{q_2} -  \mbf{\delta k'_{1,2}}|C_6^{-1}\mbf{\delta k_1 } +C_6^2\frac{\mbf{b_{1}}}{2}  \rangle  )   O_{1,2,\mbf{k'_{1,2}}}   \\ & +   V_{b_{1}}\langle \frac{\mbf{b_{1}}}{2}+ \mbf{\delta k_1} |- \frac{\mbf{b_{1}}}{2}+  \mbf{\delta k_1} \rangle \langle  C_6^{-2}\mbf{q_2}+ \mbf{\delta k'_{ 1,2 }} | C_6^2 \mbf{q_2}+ \mbf{\delta k'_{ 1,2 }} \rangle  \\ & - V_{|C_6^{-1} \mbf{\delta k_1}+ \frac{1}{2} \mbf{q_2} - \mbf{\delta k'_{ 1,2 }}|} \langle C_6^{-1} \mbf{\delta k_1 }  -C_6^2\frac{ \mbf{b_{1}}}{2}|C_6^2 \mbf{q_2}+  \mbf{\delta k'_{ 1,2 }} \rangle \langle  C_6^{-2} \mbf{q_2} + \mbf{\delta k'_{ 1,2 }}|C_6^2\frac{ \mbf{b_{1}}}{2}+ C_6^{-1}\mbf{\delta k_1}\rangle )  O_{2, 1 ,\mbf{k'_{ 1,2}}}   \bigg].
\end{eqnarray}
The Hartree and Fock terms can be collected together to obtain
\begin{eqnarray}\label{eqapp:g1_k1}
    & g_{1, \mbf{k_1}}=  \frac{1}    {\Omega_{tot}}  \bigg[ V_{b_{1}}    \langle \frac{\mbf{b_{1}}}{2}+ \mbf{\delta k_1} |- \frac{\mbf{b_{1}}}{2}+  \mbf{\delta k_1} \rangle  \Big(\sideset{}{'}\sum_{\mbf{k'_1}}( \langle {-\frac{\mbf{b_{1}}}{2}+ \mbf{\delta k'_1}} |\frac{\mbf{b_{1}}}{2}+  \mbf{\delta k'_1} \rangle  O_{01,\mbf{k'_1}} + h.c) \\ &  + \sideset{}{'}\sum_{\mbf{k'_{1,2}}} \big( \langle \mbf{q_2+ {\delta k'_{1,2} }} | C_6^2 \mbf{q_2} + \mbf{\delta k'_{1,2} } \rangle O_{01,\mbf{k'_{1,2}}} +  \langle  \mbf{q_2+ \delta k'_{1,2} } |  -C_6\mbf{q_2}+  \mbf{\delta k'_{1,2} }\rangle O_{02,\mbf{k'_{1,2}}}\\
    &+ \langle C_6^2 \mbf{q_2}+ \mbf{\delta k'_{1,2}} |  C_6^{-2} \mbf{q_2} +  \mbf{\delta k'_{1,2}} \rangle O_{1,2,\mbf{k'_{1,2}}} +h.c. \big)\Big)\\ & -  \sideset{}{'}\sum_{\mbf{k'_1}} \big(V_{| \mbf{\delta k_1}  - \mbf{\delta k'_1}|} \langle \frac{\mbf{b_{1}}}{2}+ \mbf{\delta k_1}|\frac{\mbf{b_{1}}}{2} + \mbf{\delta k'_1} \rangle \langle - \frac{\mbf{b_{1}}}{2}+  \mbf{\delta k'_1}|- \frac{\mbf{b_{1}}}{2}+ \mbf{\delta k_1}\rangle 
 O_{01,\mbf{k'_1}}  \\ & + V_{|\mbf{\delta k_1}+ \mbf{\delta k'_{1}}|} \langle \frac{\mbf{b_{1}}}{2}+ \mbf{\delta k_1}|\frac{\mbf{b_{1}}}{2}-\mbf{\delta k'_{1}} \rangle \langle -\frac{\mbf{b_{1}}}{2} -\mbf{\delta k'_{1}}|-\frac{\mbf{b_{1}}}{2}+  \mbf{\delta k_1}\rangle ) O_{1,0,\mbf{k'_{1}}} \big)   \\ & - \sideset{}{'}\sum_{\mbf{k'_{1,2}}} \big(V_{|\mbf{\delta k_1}- \frac{1}{2} C_{6}\mbf{q_2}- \mbf{\delta k'_{1,2} }|} \langle \frac{\mbf{b_{1}}}{2}+ \mbf{\delta k_1}|C_6^2 \mbf{q_2}+ \mbf{\delta k'_{1,2} } \rangle \langle \mbf{q_2}+ \mbf{\delta k'_{1,2} }|- \frac{\mbf{b_{1}}}{2} + \mbf{\delta k_1}\rangle   )   O_{01,\mbf{k'_{1,2}}}    \\ & + V_{| \mbf{\delta k_1}+\frac{1}{2} C_6 \mbf{q_2}+ \mbf{\delta k'_{1,2}}|} \langle \frac{\mbf{b_{1}}}{2}+  \mbf{\delta k_1}|-\mbf{q_2}- \mbf{\delta k'_{1,2}} \rangle \langle  C_6^{-1} \mbf{q_2} - \mbf{\delta k'_{1,2}}| -\frac{\mbf{b_{1}}}{2}+ \mbf{\delta k_1}\rangle  ) \rangle O_{1,0, \mbf{k'_{1,2}}}  \\ & + V_{|C_6\mbf{\delta k_1 } - \frac{1}{2} C_{6}^{-1}\mbf{q_2} - \mbf{\delta k'_{1,2} }|} \langle \frac{\mbf{b_{1}}}{2}+ \mbf{\delta k_1}|-\mbf{q_2} + C_6^{-1} \mbf{\delta k'_{1,2} } \rangle \langle  C_6^{-1}\mbf{q_2}+ C_6^{-1}\mbf{\delta k'_{1,2} }|- \frac{\mbf{b_{1}}}{2} + \mbf{\delta k_1}\rangle)  O_{02,\mbf{k'_{1,2}}} \\ & +V_{|-C_6\mbf{\delta k_1} -\frac{1}{2} C_6^{-1}\mbf{q_2}- \mbf{\delta k'_{1,2}}|} \langle \frac{\mbf{b_{1}}}{2}+ \mbf{\delta k_1 } |+C_6^{2} \mbf{q_2}+ C_6^{2} \mbf{\delta k'_{1,2}} \rangle \langle \mbf{q_2}+C_6^{2} \mbf{\delta k'_{1,2}}| -\frac{\mbf{b_{1}}}{2}+  \mbf{\delta k_1}\rangle ) O_{2,0, \mbf{k'_{1,2}}}  \\ & + V_{|C_6^{-1}\mbf{\delta k_1} - \frac{1}{2} \mbf{q_2}+  \mbf{\delta k'_{1,2}}|}\langle  \frac{\mbf{b_{1}}}{2} +\mbf{\delta k_1}|C_6^{2} q_2- C_6 \mbf{\delta k'_{1,2}} \rangle \langle  \mbf{q_2} -  C_6 \mbf{\delta k'_{1,2}}|- \frac{\mbf{b_{1}}}{2}+\mbf{\delta k_1 }   \rangle  )   O_{1,2,\mbf{k'_{1,2}}}    \\ & + V_{|C_6^{-1} \mbf{\delta k_1}+ \frac{1}{2} \mbf{q_2} - \mbf{\delta k'_{ 1,2 }}|} \langle \frac{ \mbf{b_{1}}}{2}+ \mbf{\delta k_1 }  |- \mbf{q_2}+ C_6 \mbf{\delta k'_{ 1,2 }} \rangle \langle  C_6^{-1} \mbf{q_2} + C_6 \mbf{\delta k'_{ 1,2 }}|-\frac{ \mbf{b_{1}}}{2}+ \mbf{\delta k_1}\rangle )  O_{2, 1 ,\mbf{k'_{ 1,2}}} \big)  \bigg].
\end{eqnarray}
A similar expression can be obtained for $g_{1,\bm{k}_{12}}$
\begin{eqnarray}\label{eqapp:g1_k12_HF}
    & g_{1, \mbf{k_{12}}}=  \frac{1}    {\Omega_{tot}}  \bigg[ V_{b_{1}}    \langle C_6^2 \mbf{q_2 }+ \mbf{\delta k_{12}} | \mbf{q_2+\delta k_{12}} \rangle  \Big( \sideset{}{'}\sum_{\mbf{k'_1}}( \langle \mbf{-\frac{{b_{1}}}{2}+ {\delta k'_1}} |\frac{\mbf{b_{1}}}{2}+  \mbf{\delta k'_1} \rangle  O_{01,\mbf{k'_1}} + h.c. )\\ & +  \sideset{}{'}\sum_{\mbf{k'_{1,2}}} \Big(  \langle \mbf{q_2+ {\delta k'_{1,2} }} | C_6^2 \mbf{q_2} + \mbf{\delta k'_{1,2} } \rangle  O_{01,\mbf{k'_{1,2}}} +  \langle  \mbf{q_2+ \delta k'_{1,2} } |  C_6^{-2}\mbf{q_2}+  \mbf{\delta k'_{1,2} }\rangle O_{02,\mbf{k'_{1,2}}} \\ &+ \langle C_6^2 \mbf{q_2}+ \mbf{\delta k'_{1,2}} |  C_6^{-2} \mbf{q_2} +  \mbf{\delta k'_{1,2}} \rangle  O_{1,2,\mbf{k'_{1,2}}}  + h.c.)\Big) \\ & - \sideset{}{'}\sum_{\mbf{k'_1}} \big(V_{|\frac{1}{2} C_6 \mbf{  q_2}+\mbf{\delta k_{12}} - \mbf{\delta k'_1}|} \langle C_6^2\mbf{  q_2}+\mbf{\delta k_{12}} |\frac{\mbf{b_{1}}}{2} + \mbf{\delta k'_1} \rangle \langle - \frac{\mbf{b_{1}}}{2}+  \mbf{\delta k'_1}|\mbf{  q_2}+\mbf{\delta k_{12}} \rangle 
 O_{01,\mbf{k'_1}}  \\ & + V_{|\frac{1}{2} C_6 \mbf{  q_2}+\mbf{\delta k_{12}}+ \mbf{\delta k'_{1}}|} \langle C_6^2 \mbf{  q_2}+\mbf{\delta k_{12}}|\frac{\mbf{b_{1}}}{2}-\mbf{\delta k'_{1}} \rangle \langle - \frac{\mbf{b_{1}}}{2} -\mbf{\delta k'_{1}}|\mbf{  q_2}+\mbf{\delta k_{12}}\rangle  O_{1,0,\mbf{k'_{1}}} \big)  \\ & - \sideset{}{'}\sum_{\mbf{k'_{1,2}}} \big( V_{|\mbf{\delta k_{12}}- \mbf{\delta k'_{1,2} }|} \langle C_6^2 \mbf{  q_2}+\mbf{\delta k_{12}}|C_6^2 \mbf{q_2}+ \mbf{\delta k'_{1,2} } \rangle \langle \mbf{q_2}+ \mbf{\delta k'_{1,2} }|\mbf{  q_2}+\mbf{\delta k_{12}}\rangle      O_{01,\mbf{k'_{1,2}}}    \\ &+ V_{|\mbf{\delta k_{12}} +C_6 \mbf{q_2}+ \mbf{\delta k'_{1,2}}|} \langle C_6^2 \mbf{  q_2}+\mbf{\delta k_{12}}|-\mbf{q_2}- \mbf{\delta k'_{1,2}} \rangle \langle C_6^{-1} \mbf{q_2} - \mbf{\delta k'_{1,2}}|\mbf{  q_2}+\mbf{\delta k_{12}}\rangle   O_{1,0, \mbf{k'_{1,2}}}  \\ & + V_{|C_6^2\mbf{  q_2}+C_6 \mbf{\delta k_{12}} - \mbf{\delta k'_{1,2} }|} \langle C_6^2 \mbf{  q_2}+ \mbf{\delta k_{12}}|- \mbf{q_2} + C_6^{-1} \mbf{\delta k'_{1,2} } \rangle \langle  C_6^{-1} \mbf{q_2}+C_6^{-1} \mbf{ \delta k'_{1,2} }| \mbf{q_2}+ \mbf{\delta k_{12}}\rangle)  O_{02,\mbf{k'_{1,2}}}      \\ & +V_{|C_{6}^{-2} \mbf{\delta k_{12}}- \mbf{\delta k'_{1,2}}|} \langle C_6^2 \mbf{  q_2}+\mbf{\delta k_{12}}|C_6^2 \mbf{q_2}+ C_6^2 \mbf{ \delta k'_{1,2}} \rangle \langle \mbf{q_2}+ C_6^2 \mbf{\delta k'_{1,2}}|\mbf{  q_2}+  \mbf{\delta k_{12}}\rangle ) O_{2,0, \mbf{k'_{1,2}}}          \\ & + V_{|C_6^{-1}\mbf{\delta k_{12}}+  \mbf{\delta k'_{1,2}}|}\langle C^2_6 \mbf{  q_2}+\mbf{\delta k_{12}}|C_6^2 q_2 - C_6 \mbf{\delta k'_{1,2}} \rangle \langle  \mbf{q_2} -  C_6 \mbf{\delta k'_{1,2}}|\mbf{  q_2}+ \mbf{\delta k_{12}}\rangle  )   O_{1,2,\mbf{k'_{1,2}}}    \\ &+ V_{|\mbf{  q_2}+C_6^{-1} \mbf{\delta k_{12}}- \mbf{\delta k'_{ 1,2 }}|} \langle C_6^2 \mbf{  q_2}+ \mbf{\delta k_{12}}|- \mbf{q_2}+  C_6 \mbf{\delta k'_{ 1,2 }} \rangle \langle  C_6^{-1} \mbf{q_2} + C_6 \mbf{\delta k'_{ 1,2 }}|\mbf{  q_2}+\mbf{\delta k_{12}}\rangle )  O_{2, 1 ,\mbf{k'_{ 1,2}}} \big)  \bigg]
\end{eqnarray}

$g_{2,\bm{k}_{1,2}}$ can be obtained from above by the following relation valid with $C_6$ symmetry
\begin{equation}
    g_{1,\mbf{\bm{q}_2+\delta k_{12}}} = g_{2,\bm{q}_2- C_6\mbf{\delta k_{12}}} ^*.
\end{equation}

\subsection{General solution in $\mathbf{k}_0$}\label{subsecapp:gensolnk0}

In this subsection, we consider the solution of the HF equations in momenta $\bm{k}_0$
\begin{equation}
    H_{\bm{k}_0}=\sum_{\bm{k}_0}(E_{\bm{k}_0}+f_{\bm{k}_0}-\mu)\gamma^\dagger_{0,\bm{k}_0}\gamma_{0,\bm{k}_0}
\end{equation}
where we have added a uniform chemical potential $\mu$, which will also be applied to the other momentum regions. Since $\bm{k}_0$ only involves states in BZ 0, the Hamiltonian is already diagonal, and the remaining task is to analyze when there will be Fermi surfaces in this region. For the general form above, all states will be occupied if $\mu>\text{max}_{\bm{k}_0}[E_{\bm{k}_0}+f_{\bm{k}_0}]$.

To proceed further, we utilize the approximations for $f_{\bm{k}}$ introduced in Sec.~\ref{subsecapp:fkmoredetail}. In particular, we use $C_6$ symmetry, the Gaussian interaction, the GMP limit, and the limit of small $\phi$. We also perform an overall shift with respect to $f_{1,2}+\delta f+\delta^2 f$ (this is the constant part in region $\bm{k}_{1,2}$ --- see Eq.~\ref{eqapp:fprime}), leading to
\begin{gather}\label{eqapp:H'k0}
    H_{\bm{k}_0}'=\sum_{\mbf{k_0}}( E_{\mbf{k_0}}+ f_{\mbf{k_0}}' - \mu) \gamma_{0,\mbf{k_0}}^\dagger \gamma_{0,\mbf{k_0}}\\
    f'_{\bm{k}_0}=(k_0^2-q_2^2)\mathcal{O}.
\end{gather}
Note that $k_0<q_2$ always, and we expect $\phi>0$ due to the decay of the GMP form factor and interaction with momentum. Furthermore, $\mathcal{O}>0$ since the filling factor is greater than 1. Hence, $f'_{\bm{k}_0}$ is always negative. The $\bm{k}_0$ region is fully occupied if $\mu>\text{max}_{\bm{k}_0}[E_{\bm{k}_0}+f'_{\bm{k}_0}]$. For dispersion of the Berry trashcan, we have $E_{\bm{k}_0}=0$ since this corresponds to momenta within the flat bottom. This means that there are no Fermi pockets in the $\bm{k}_0$ region if $\mu>x(x-2)q_2^2\mathcal{O}$, where we have used the fact that $\bm{k}_0$ has maximum magnitude $(1-x)q_2$ (see Fig.~\ref{figapp:2dtrashcanBZ}).

\subsection{General solution in $\mathbf{k}_1$}\label{subsecapp:gensolnk1}

In this subsection, we consider the solution of the HF equations for momenta $\bm{k}_1$
\begin{equation}
    H_{\bm{k}_1} = \sideset{}{'}\sum_{\mbf{k_1}} \bigg[(E_{\mbf{k_1}}+f_{\mbf{k_1}}-\mu) \gamma_{0\mbf{k_1}}^\dagger \gamma_{0\mbf{k_1}}  + ( E_{\mbf{k_1}+ \mbf{b_{1}}} +f_{\mbf{k_1+ b_{1}}}-\mu ) \gamma_{1\mbf{k_1}}^\dagger \gamma_{1\mbf{k_1}} + (g_{1,\mbf{k_1}} \gamma_{1,\mbf{k_1}
}^\dagger \gamma_{0\mbf{k_1}} + h.c. )\bigg].
\end{equation}
This is a $2\times 2$ matrix that couples BZ 0 and BZ 1, and can be parameterized as 
\begin{gather}
    H_{\bm{k}_1}=\sideset{}{'}\sum_{\bm{k}_1}\begin{pmatrix}
        \gamma^\dagger_{0,\bm{k}_1} & \gamma^\dagger_{1,\bm{k}_1}
    \end{pmatrix}[d_{0,\bm{k}_1}+\bm{d}_{\bm{k}_1}\cdot\bm{\sigma}]\begin{pmatrix}
        \gamma_{0,\bm{k}_1} \\ \gamma_{1,\bm{k}_1}
    \end{pmatrix}\\
    d_{0,\bm{k}_1}=\frac{1}{2}\left(E_{\mbf{k_1}} +f_{\mbf{k_1}}+E_{\mbf{k_1}+ \mbf{b_{1}}} +f_{\mbf{k_1+ b_{1}}}\right)-\mu\\
    d_{z,\bm{k}_1}=\frac{1}{2}\left(E_{\mbf{k_1}} +f_{\mbf{k_1}}-E_{\mbf{k_1}+ \mbf{b_{1}}} -f_{\mbf{k_1+ b_{1}}}\right)\\
    d_{x,\bm{k}_1}=\text{Re}\,g_{1,\bm{k}_1},\quad d_{y,\bm{k}_1}=\text{Im}\,g_{1,\bm{k}_1}
\end{gather}
with solutions
\begin{gather}
    \psi_{\pm,\bm{k}_1}=   \frac{1}{\sqrt{2}}\left(\begin{matrix}
          \sqrt{1\pm \hat{d}_{z,\bm{k}_1} }\\
    \pm \frac{\hat{d}_{x,\bm{k}_1} + i \hat{d}_{y,\bm{k}_1}}{\sqrt{1\pm \hat{ d}_{z,\bm{k}_1}}}
     \end{matrix} \right) , \quad \hat{d}_{i,\bm{k}_1}= \frac{{d}_{i,\bm{k}_1}}{|\bm{d}_{\bm{k}_1}|}\\
     E_{\pm,\mbf{k_1}}= d_{0 ,\mbf{k_1}} \pm |\bm{d}_{\bm{k}_1}|=  d_{0,\mbf{k_1}} \pm \sqrt{ |g_{1,\mbf{k_1}}|^2+ d_{z, \mbf{k_1}}^2}.
\end{gather}
These states are associated with operators
\begin{gather}
     a_{\pm, \mbf{k_1}}= [\psi_{\pm,\bm{k}_1}]_0^\star \gamma_{0, \mbf{k_1}} + [\psi_{\pm,\bm{k}_1}]_1^\star \gamma_{1 ,\mbf{k_1}}\\  \gamma_{0, \mbf{k_1}} = [\psi_{+,\bm{k}_1}]_0  a_{+ ,\mbf{k_1}}+[\psi_{-,\bm{k}_1}]_0  a_{- ,\mbf{k_1}}\\  \gamma_{1 ,\mbf{k_1}} = [\psi_{+,\bm{k}_1}]_1  a_{+, \mbf{k_1}}+[\psi_{-,\bm{k}_1}]_1  a_{- ,\mbf{k_1}}.
 \end{gather} 
We are interested in insulating states where the lower band ($-$) is fully occupied and the upper band ($+$) is fully unoccupied so that there are no Fermi pockets within the $\bm{k}_1$ momenta. This is the case if $\text{max}_{\bm{k}_1}[E_{-,\bm{k}_1}]<\mu<\text{min}_{\bm{k}_1}[E_{+,\bm{k}_1}]$. Combined with the condition that the $\bm{k}_0$ momenta are fully filled, we also have the necessary constraint $\text{max}_{\bm{k}_0}[E_{\bm{k}_0}+f_{\bm{k}_0}]<\text{min}_{\bm{k}_1}[E_{+,\bm{k}_1}]$. If these are satisfied, then the state in the $\bm{k}_1$ region is $\prod_{\bm{k}_1}a^\dagger_{-,\bm{k}_1}\ket{\text{vac}}$, leading to the following order parameters
\begin{gather}\label{eqapp:O10_k1}
    O_{10,\mbf{k_1}}= [\psi_{-,\bm{k}_1}]_1^\star [\psi_{-,\bm{k}_1}]_0= - \frac{1}{2} (\hat{d}_{x \mbf{k}_1}- i \hat{d}_{y \mbf{k}_1})= - \frac{1}{2} \frac{g_{1\mbf{k_1}}^\star}{|\bm{d}_{k_1}|}  \\  
  O_{00,\mbf{k_1}}=  [\psi_{-,\bm{k}_1}]_0^\star [\psi_{-,\bm{k}_1}]_0= \frac{1}{2} (1- \hat{d}_{z\mbf{k_1}})  \\ 
  O_{11,\mbf{k_1}}=  [\psi_{-,\bm{k}_1}]_1^\star [\psi_{-,\bm{k}_1}]_1= \frac{1}{2} \frac{\hat{d}_{x\mbf{k_1}} ^2+ \hat{d}_{y\mbf{k_1}} ^2}{ 1- \hat{d}_{z\mbf{k_1}} }  \\  O_{11,\mbf{k_1}} +O_{00,\mbf{k_1}}= \frac{1}{2} \frac{1}{1- \hat{d}_{z\mbf{k_1}} } ((1- \hat{d}_{z\mbf{k_1}})^2+ \hat{d}_{x\mbf{k_1}} ^2+ \hat{d}_{y\mbf{k_1}} ^2)= 1.
\end{gather}
Note that all the manipulations above are exact so far.

To proceed further, we utilize the approximations for $f_{\bm{k}}$ introduced in Sec.~\ref{subsecapp:fkmoredetail}. In particular, we use $C_6$ symmetry, the exponential interaction, the GMP limit, and the limit of small $\phi$. We also perform an overall shift with respect to $f_{1,2}+\delta f+\delta^2 f$ (see Eq.~\ref{eqapp:fprime}). 

The approximations described above lead to
\begin{gather}
    d_{0,\mbf{k_1}}= \frac{1}{2}\left(E_{\mbf{k_1}}+ E_{\mbf{k_1}+ \mbf{b_{1}}} +2\left(\left({b_{1}}/{2}\right)^2- q_2^2   +  \delta k_1^2\right)\mathcal{O}\right)  \\
    d_{z,\mbf{k_1}}= \frac{1}{2}\left(E_{\mbf{k_1}}-E_{\mbf{k_1}+ \mbf{b_{1}}}- 2 \mbf{\delta k_1} \cdot \mbf{b_{1}}\mathcal{O}\right).
\end{gather}
Note that $\mbf{\delta k_1} \cdot \mbf{b_{1}}>0$, and for the trashcan dispersion we have $E_{\mbf{k_1}}-E_{\mbf{k_1}+ \mbf{b_{1}}}<0$, so that $d_{z,\bm{k}_1}<0$. If we set $g_{1,\bm{k}_1}=0$, the interaction-induced part proportional to $\mathcal{O}$ acts to steepen the velocity of the bare dispersion. More concretely, consider $E_{\bm{k}_1}=0$ and $E_{\bm{k}_1+\bm{b}_1}=v\delta{k_{1x}}$, i.e.~we only take the steep component of the dispersion perpendicular to the BZ boundary. Then we have
\begin{gather}\label{eqapp:steepd0dz}
    d_{0,\mbf{k_1}}= \frac{1}{2}v\delta {k_{1x}}+
    \left(\left({b_{1}}/{2}\right)^2- q_2^2   +  \delta k_1^2\right)\mathcal{O}\\
    d_{z,\mbf{k_1}}= -\frac{1}{2}\left(v+ 2 {b_{1}}\mathcal{O}\right)\delta k_{1x}\\
    E_{\pm,\bm{k}_1}=\frac{1}{2}v\delta {k_{1x}}+
    \left(\left({b_{1}}/{2}\right)^2- q_2^2   +  \delta k_1^2\right)\mathcal{O}\pm\sqrt{\left(\frac{1}{2}\left(v+ 2 {b_{1}}\mathcal{O}\right)\delta k_{1x}\right)^2+|g_{1,\bm{k}_1}|^2}.
\end{gather}
For intermediate $\delta k_{1x}$ where $\frac{1}{2}\left(v+ 2 {b_{1}}\mathcal{O}\right)\delta k_{1x}\gg |g_{1,\bm{k}_1}|$, we find
\begin{gather}
    E_{+,\bm{k}_1}\simeq (v+b_{1}\mathcal{O})\delta k_{1x}+
    \left(\left({b_{1}}/{2}\right)^2- q_2^2   +  \delta k_1^2\right)\mathcal{O}\\
    E_{-,\bm{k}_1}\simeq -b_{1}\mathcal{O}\delta k_{1x} +
    \left(\left({b_{1}}/{2}\right)^2- q_2^2   +  \delta k_1^2\right)\mathcal{O},
\end{gather}
showing that the velocity is enhanced.

{
Since $E_{+,\bm{k}_1}\geq d_{0,\bm{k}_1}-d_{z,\bm{k}_1}$ 
(this is saturated if $g_{\bm{1,\bm{k}_1}}=0$), we have the following sufficient condition to guarantee that $\text{max}_{\bm{k}_0}[E_{\bm{k}_0}+f'_{\bm{k}_0}]<\text{min}_{\bm{k}_1}[E_{+,\bm{k}_1}]$ (i.e.~the highest state in $\bm{k}_0$ is below the upper band in $\bm{k}_1$)
\begin{gather}
    \text{min}_{\bm{k}_1}[d_{0,\bm{k}_1}-d_{z,\bm{k}_1}]=\text{min}_{\bm{k}_1}[E_{\mbf{k_1}+ \mbf{b_{1}}}+\left(\left({b_{1}}/{2}\right)^2- q_2^2 +\mbf{\delta k_1} \cdot \mbf{b_{1}} +  \delta k_1^2\right)\mathcal{O}]>\text{max}_{\bm{k}_0}[E_{\bm{k}_0}+f'_{\bm{k}_0}].
\end{gather}
Note that in the LHS above, $E_{\mbf{k_1}+ \mbf{b_{1}}}$ is an increasing function of $\delta k_{1x}$ for the dispersion of the Berry Trashcan, and the part proportional to $\mathcal{O}$ is increasing in $|\delta \bm{k}_1|$. Hence, it suffices to evaluate the LHS at $\delta \bm{k}_1=0$ (i.e. $\bm{k}_1=-\bm{b}_1/2$). Assuming that the dispersion is flat within BZ 0 so that $E_{\bm{k}_1}=0$, we obtain
\begin{gather}
    \left({b_{1}}/{2}\right)^2- q_2^2 >x(x-2)q_2^2
\end{gather}
where we have substituted the result for $\text{max}_{\bm{k}_0}[E_{\bm{k}_0}+f'_{\bm{k}_0}]$ in the same approximations. Using $q_2 = b_{1}/\sqrt{3}$, we obtain the constraint $4- 2 \sqrt{3} <\frac{1}{x} < 4+ 2 \sqrt{3} $. Since we must have $x<1$, this leads to
\begin{equation}
    x>\frac{1}{4+ 2 \sqrt{3}}\simeq 0.13,
\end{equation}
which we emphasize is a sufficient but not necessary condition for $\text{max}_{\bm{k}_0}[E_{\bm{k}_0}+f'_{\bm{k}_0}]<\text{min}_{\bm{k}_1}[E_{+,\bm{k}_1}]$.

To guarantee the absence of Fermi pockets in $\bm{k}_1$, we must also satisfy $\text{max}_{\bm{k}_1}[E_{-,\bm{k}_1}]<\text{min}_{\bm{k}_1}[E_{+,\bm{k}_1}]$. We cannot use the same $g_{\bm{k}_1}=0$ limit as above, because in this case (and using the trashcan dispersion) we have $E_{-,\bm{k}_1}=\left(\left({b_{1}}/{2}\right)^2- q_2^2 -\mbf{\delta k_1} \cdot \mbf{b_{1}} +  \delta k_1^2\right)\mathcal{O}$
}. Hence, $E_{-,\bm{k}_1}$ can rise above $\text{min}_{\bm{k}_1}[E_{+,\bm{k}_1}]=\left({b_{1}}/{2}\right)^2- q_2^2$, and some finite threshold on the magnitude of $g_{\bm{k}_1}$ is needed to open a full indirect gap. 

In the following, we will assume that either such a threshold has been reached, or the effective dispersion has been suitably deformed so that a full indirect gap is opened in the $\bm{k}_1$ region. We will also assume that an analogous threshold has been reached to guarantee compatibility with an insulating gap in the $\bm{k}_{12}$ region. If such thresholds have not been reached, the mean-field Hamiltonian would still have direct gaps, and the topology of the state constructed by occupying the lowest HF band would be unchanged as the gap is increased towards the fully insulating regime, as long as no gap closings are encountered.

\subsection{General solution in $\mathbf{k}_{1,2}$}\label{subsecapp:gensolnk12}

In this subsection, we consider the solution of the HF equations for $\bm{k}_{1,2}$
\begin{gather}\label{eqapp:k12_3x3matrix}
    H_{\bm{k}_2}=\sideset{}{'}\sum_{\mbf{k_{1,2}}} \begin{pmatrix}\gamma_{0, \mbf{k_{1,2}}
}^\dagger & \gamma_{1, \mbf{k_{1,2}}
}^\dagger& \gamma_{2, \mbf{k_{1,2}}
}^\dagger\end{pmatrix} \begin{bmatrix}
E_{\mbf{k_{1,2}}}+ f_{\mbf{k_{1,2}}}  &  g_{1,\mbf{k_{1,2}}}^\star  &  g_{2,\mbf{k_{1,2}}}^\star \\
 g_{1,\mbf{k_{1,2}}}  &   E_{\mbf{k_{1,2}} +\mbf{b_{1}}}+ f_{\mbf{k_{1,2} +  b_{1}}}  & g_{6,\mbf{k_{1,2}}+C_6\mbf{b_{1}}}\\
 g_{2,\mbf{k_{1,2}}}  & g_{6,\mbf{k_{1,2}}+C_6\mbf{b_{1}}}^\star & E_{\mbf{k_{1,2}} + C_6 \mbf{b_{1}}}+ f_{\mbf{k_{1,2} + C_6 b_{1}}}
\end{bmatrix}\begin{pmatrix}
\gamma_{0,\mbf{k_{1,2}}}  \\
\gamma_{1,\mbf{k_{1,2}}}  \\
\gamma_{2,\mbf{k_{1,2}}}  
\end{pmatrix}.
\end{gather}
This $3\times 3$ matrix does not generally have an analytic solution for the eigenvalues and eigenvectors. We instead derive implicit expressions for these in terms of the eigenvalue $E_{i,\bm{k}_{1,2}}$, where $i=0,1,2$ indexes the three eigenvalues. We use the following shorthand notation for convenience
\begin{gather}
E_{i,\bm{k}_{1,2}}\rightarrow E\\
    E_{\mbf{k_{1,2}}}+ f_{\mbf{k_{1,2}}}\rightarrow E_0,\quad E_{\mbf{k_{1,2}}+\bm{b}_1}+ f_{\mbf{k_{1,2}}+\bm{b}_1}\rightarrow E_1,\quad E_{\mbf{k_{1,2}}+C_6\bm{b}_1}+ f_{\mbf{k_{1,2}}+C_6\bm{b}_1}\rightarrow E_2\\
    g_{1,\mbf{k_{1,2}}} \rightarrow g_1,\quad g_{2,\mbf{k_{1,2}}} \rightarrow g_2,\quad g_{6,\mbf{k_{1,2}} +C_6\mbf{b_{1}}}\rightarrow g_0.
\end{gather}
Similarly, we let $\psi_0,\psi_1,\psi_2$ be the components of the eigenvector of Eq.~\ref{eqapp:k12_3x3matrix}. The eigenvalue problem written out is
\begin{gather}
    g_1^*\psi_1+g_2^*\psi_2=(E-E_0)\psi_0\\
    g_1\psi_0+g_0\psi_2=(E-E_1)\psi_1\\
    g_2\psi_0+g_0^*\psi_1=(E-E_2)\psi_2.
\end{gather}
We write $\psi_1$ and $\psi_2$ in terms of the first component, and also use the normalization and the secular equation to express $|\psi_0|^2$ in terms of the eigenvalue $E$ (remembering that all quantities are still functions of $\bm{k}_{12}$)
\begin{gather}
    \psi_1 = \frac{(E-E_2) g_1 + g_0 g_2  }{ (E- E_1)(E-E_2) - |g_0|^2}\psi_0 ,\quad \psi_2= \frac{(E-E_1) g_2 + g_1g_0^\star}{ (E-E_1) (E- E_2) - |g_0|^2} \psi_0 \\
    (E-E_0)(E-E_1) (E-E_2) - (E-E_0)|g_0|^2 - (E-E_2)|g_1|^2 -(E-E_1)|g_2|^2 - g_0 g_1^\star g_2 - g_0^\star g_1 g_2^\star =0\\
    |\psi_0|^2 = \frac{ (E-E_1) (E-E_2) - |g_0|^2 }{ (E-E_0)(E- E_1) +(E-E_0)(E- E_2) + (E-E_1)(E- E_2) - (|g_0|^2+|g_1|^2+ |g_2|^2 )}.
\end{gather}
The annihilation operator $b_i$ for band $i$ can be written
\begin{gather}
b_{i} = \sum_{j=0,1,2} [\psi_{i}^*]_j \gamma_{j}\\      \gamma_{j}= \sum_{i=0,1,2}   [\psi_i]_jb_{i}
\end{gather}
where $[\psi_i]_j$ is the $j$'th component of the eigenvector for band $i$. 

For an insulating state corresponding to filling $\nu=1$ of the BZ, we leave the higher two bands empty and fully fill the lowest band (for which we drop the band label).  The order parameter in region $\bm{k}_{12}$ is (again leaving the momentum and band labels implicit)
\begin{equation}
    O_{jj'}=\langle \gamma^\dagger_j \gamma_{j'}\rangle=\psi^*_{j}\psi_{j'}.
\end{equation}
Using the expressions in terms of the eigenvalue $E$ and $O_{00}=|\psi_0|^2$, we obtain
\begin{eqnarray}
    & O_{01} = O_{10}^\star= \psi_0^\star \psi_1 =  \frac{(E-E_2) g_1 + g_0 g_2  }{ (E- E_1)(E-E_2) - |g_0|^2}|\psi_0|^2 \\ &  O_{02} = O_{20}^\star= \psi_0^\star \psi_2= \frac{(E-E_1) g_2 + g_1g_0^\star}{ (E-E_1) (E- E_2) - |g_0|^2} |\psi_0|^2 \\ & O_{12} = O_{21}^\star= \psi_1^\star \psi_2= \frac{((E-E_2) g_1^\star + g_0^\star g_2^\star  ) ((E-E_1) g_2 + g_1g_0^\star)}{ ((E-E_1) (E- E_2) - |g_0|^2)^2} |\psi_0|^2 =\frac{1}{|\psi_0|^2} O_{02} O_{10} \\ &{  O_{11}=\psi_1^\dagger \psi_1 = \frac{((E-E_2) g_1^\star + g_0^\star g_2^\star ) ( (E-E_2) g_1 + g_0 g_2 ) }{ ((E- E_1)(E-E_2) - |g_0|^2)^2} |\psi_0|^2  }.
\end{eqnarray}
After some algebra, these can be further reduced to expressions in terms of $E$ alone
\begin{eqnarray}
    & O_{01} =   \frac{(E-E_2) g_1 + g_0 g_2  }{ { (E-E_0)(E- E_1) +(E-E_0)(E- E_2) + (E-E_1)(E- E_2) - (|g_0|^2+|g_1|^2+ |g_2|^2 ) }} \\ &  O_{02} = \frac{(E-E_1) g_2 + g_1g_0^\star}{ { (E-E_0)(E- E_1) +(E-E_0)(E- E_2) + (E-E_1)(E- E_2) - (|g_0|^2+|g_1|^2+ |g_2|^2 ) }}\\ & O_{12}= \frac{(E-E_0) g_0^\star + g_1^\star g_2   }{ { (E-E_0)(E- E_1) +(E-E_0)(E- E_2) + (E-E_1)(E- E_2) - (|g_0|^2+|g_1|^2+ |g_2|^2 ) }}  \\ &  O_{11}= \frac{  (E-E_0) (E- E_2) - | g_2|^2  }{ (E-E_0)(E- E_1) +(E-E_0)(E- E_2) + (E-E_1)(E- E_2) - (|g_0|^2+|g_1|^2+ |g_2|^2 )}.
\end{eqnarray}

\clearpage

\section{2D Berry Trashcan Model: Total mean-field energy and self-consistency}\label{secapp:2dmodel_energy}

In this appendix section, we consider the expression for the total mean-field energy $E_\text{tot}$
\begin{align}
    E_\text{tot}=\langle H_{\text{kin}}\rangle+\langle H_{\text{int}}\rangle=\langle H_{\text{kin}}\rangle+\frac{1}{2}\langle H^\text{HF}_{\text{int}}\rangle
\end{align}
where the expectation value is taken in the many-body mean-field ground state, and the interacting part of the mean-field Hamiltonian $H^{\text{HF}}_{\text{int}}$ is evaluated in this same state. We also analyze the conditions for self-consistency in certain limits.

Using $C_6$ symmetry of the Hamiltonian and the order parameter, we obtain
\begin{align}
    E_\text{tot}&=  \sum_{\mbf{k_0}} (E_{\mbf{k_0}} +\frac{1}{2} f_{\mbf{k_0}}) O_{00 \mbf{k_0}}  \\ &+
    6 \sideset{}{'}\sum_{\mbf{k_1}} \bigg[ (E_{\mbf{k_1}} + \frac{1}{2} f_{\mbf{k_1}}) O_{00 \mbf{k_1}}+ (E_{\mbf{k_1+ b_{1}}} + \frac{1}{2} f_{\mbf{k_1+ b_{1}}}) O_{11 \mbf{k_1}} + \frac{1}{2} g_{1,\mbf{k_1}} O_{10 \mbf{k_1}}  + \frac{1}{2} g_{1,\mbf{k_1}}^\star  O_{01 \mbf{k_1}}\bigg]\\ & +
   6 \sideset{}{'}\sum_{\mbf{k_{12}}} \bigg[(E_{\mbf{k_{12}}} + \frac{1}{2} f_{\mbf{k_{12}}}) O_{00 \mbf{k_{12}}}   + (E_{\mbf{k_{12}+ b_{1}}} + \frac{1}{2} f_{\mbf{k_{12}+ b_{1}}}) O_{11 \mbf{k_{12}}}  + (E_{\mbf{k_{12}+ C_6 b_{1}}} + \frac{1}{2} f_{\mbf{k_{12}+ C_6 b_{1}}})   O_{22 \mbf{k_{12}}}  \\ &+  \frac{1}{2} g_{1,\mbf{k_{12}}} O_{10 \mbf{k_{12}}} + \frac{1}{2} g_{1,\mbf{k_{12}}}^\star O_{01 \mbf{k_{12}}}  +  \frac{1}{2} g_{2,\mbf{k_{12}}} O_{20 \mbf{k_{12}}}   + \frac{1}{2} g_{2,\mbf{k_{12}}}^\star   O_{02 \mbf{k_{12}}} 
  +  \frac{1}{2} g_{6,\mbf{k_{12} + C_6 b_{1}}} O_{12 \mbf{k_{12}}} + \frac{1}{2} g_{6,\mbf{k_{12}+ C_6 b_{1}}}^\star O_{21 \mbf{k_{12}}}\bigg]
\end{align}
where we do not include the contribution from the overall chemical potential $\mu$. The overall factors of 6 arise from $C_6$ symmetry which relates the $\bm{k}_1$ and $\bm{k}_{12}$ regions with the other $\bm{k}_j$ and $\bm{k}_{j,j+1}$ regions for $j=2,\ldots 6$. We will mostly be concerned with mean-field states which are insulating at $\nu=1$, corresponding to an occupation number of 1 per momentum in the BZ. This leads to the total particle number $\sum_{\bm{k}}O_{\bm{k},\bm{k}}$ (see Eq.~\ref{appeq:Okk'} for the definition of $O_{\bm{k},\bm{k}'}$)
being equal to the number of moir\'e unit cells $N_M$ (i.e.~the number of discrete momenta in the BZ), as well as $O_{00,\bm{k}_0}=O_{00,\bm{k}_1}+O_{11,\bm{k}_1}=O_{00,\bm{k}_{12}}+O_{11,\bm{k}_{12}}+O_{22,\bm{k}_{12}}=1$.

\subsection{$\mathbf{k}_{12}$-only limit}\label{subsecapp:Etot_k12_only_limit}
We consider first only keeping terms that involve the $\bm{k}_{12}$ regions (see Refs.~\cite{soejima2024AHC2,dong2024stability,crepel2024efficientpredictionsuperlatticeanomalous} for previous works that have considered similar approaches in the context of R$n$G). Hence, we discard contributions arising from the $\bm{k}_0$ and $\bm{k}_1$ regions (and any symmetry related regions). Because we retain just the high symmetry points at the BZ corners, we can only distinguish Chern numbers mod 3. Note that doing so neglects the $\bm{k}_1$ regions which contain the majority of the gapless regions around the vicinity of the entire BZ boundary. Hence, a proper treatment of Wigner crystallization requires accounting for these regions, as done later in App.~\ref{subsecapp:k1_k12_limit} and beyond.

Keeping just $\bm{k}_{12}$ leads to
\begin{align}
    E_\text{tot}&=6\sideset{}{'}\sum_{\bm{k}_{12}}\bigg[\frac{1}{2}f_{\bm{k}_{12}}+E_{\bm{k}_{12}}O_{00\bm{k}_{12}}+\frac{1}{2}g_{12}^*O_{012,\bm{k}_{12}}+\frac{1}{2}g_{12}O^*_{012,\bm{k}_{12}}\\
    &+(E_{\mbf{k_{12}+ b_{1}}} + \frac{1}{2} f_{\mbf{k_{12}+ b_{1}}}-\frac{1}{2}f_{\mbf{k_{12}}}) O_{11 \mbf{k_{12}}}  + (E_{\mbf{k_{12}+ C_6 b_{1}}} + \frac{1}{2} f_{\mbf{k_{12}+ C_6 b_{1}}}-\frac{1}{2}f_{\mbf{k_{12}}})   O_{22 \mbf{k_{12}}}\bigg],
\end{align}
where $O_{012,\bm{k}_{12}}$ is defined in Eq.~\ref{eqapp:O012_k12_definition}, and we have assumed an insulating state at $\nu=1$. Above, we have further taken the limit that the hybridization fields $g_{1,\mathbf{k}_{12}},g_{2,\mathbf{k}_{12}},g_{6,\mathbf{k}_{12}+C_6\bm{b}_1}$ in region $\bm{k}_{12}$ are independent of $\bm{k}_{12}$. In particular, we take the values $g_{1,\mathbf{k}_{12}}=g_{2,\mathbf{k}_{12}}^*=g_{6,\mathbf{k}_{12}+C_6\bm{b}_1}^*=g_{12}$ appropriate at $\delta \bm{k}_{12}=0$, i.e.~at the BZ corner. This approximation is expected to be valid when the momentum region $\bm{k}_{12}$ is small. Note however that this does not assume that the order parameter itself is unchanging in the region $\bm{k}_{12}$, since it could still vary appreciably due to the (interaction-induced) dispersion. 
Consider also approximating the mean-field potential for momenta that fold onto the region $\bm{k}_{12}$ to be equal the value at $\delta\bm{k}_{12}=0$, i.e.~$f_{\bm{k}_{12}}=f_{\bm{k}_{12}+\bm{b}_1}=f_{\bm{k}_{12}+C_6\bm{b}_1}=f_{12}$, so that
\begin{equation}\label{eqapp:Etot_k12only}
    E_\text{tot}=6\sideset{}{'}\sum_{\bm{k}_{12}}\bigg[\frac{1}{2}f_{12}+E_{\bm{k}_{12}}O_{00\bm{k}_{12}}+E_{\mbf{k_{12}+ b_{1}}}O_{11\bm{k}_{12}}+E_{\mbf{k_{12}+ C_6b_{1}}}O_{22\bm{k}_{12}}+\frac{1}{2}g_{12}^*O_{012,\bm{k}_{12}}+\frac{1}{2}g_{12}O^*_{012,\bm{k}_{12}}\bigg].
\end{equation}

Our objective is investigate how $E_\text{tot}$ depends on the nature of the insulating mean-field state. Since we have retained information on just the $\bm{k}_{12}$ region, we can only resolve the Chern number $C$ mod 3. To understand the energetic competition between the different Chern states, we analyze the mean fields $g_{12}$ and $f_{12}$. We begin with $g_{12}$ from Eq.~\ref{eqapp:g1_k12_HF}. We set all $\delta \bm{k}_{12},\delta\bm{k}'_{12}$ that appear in the form factors and interaction potential to zero in $g_{12}$. Again, this is justified in the limit of a small $\bm{k}_{12}$ region, where these quantities are not expected to vary significantly. We find
\begin{gather}
    g_{12}=\frac{\langle C_6^2\bm{q}_2|\bm{q}_2\rangle}{\Omega_{\text{tot}}} \bigg[V_{b_{1}}\langle \mbf{q_2} | C_6^2 \mbf{q_2} \rangle \sideset{}{'}\sum_{\mbf{k'_{1,2}}} O_{012,\bm{k}'_{12}}+h.c.\bigg]\\
    -\frac{1}{\Omega_\text{tot}}\bigg[
    V_0\sideset{}{'}\sum_{\mbf{k'_{1,2}}}O_{012,\bm{k}'_{12}}+V_{q_2}\langle C_6^2\bm{q}_2|-\bm{q}_2\rangle^2\sideset{}{'}\sum_{\mbf{k'_{1,2}}}O^*_{012,\bm{k}'_{12}}
    \bigg].
\end{gather}
Note that $V_0$ should be regularized if $V(\bm{q}\rightarrow0)$ diverges, which is not an issue in the case of a screened interaction.
Since the remaining dependence on the momentum $\bm{k}_{12}$ is in the order parameter, we define
\begin{equation}
    \mathcal{O}_{012}\equiv\sideset{}{'}\sum_{\mbf{k_{1,2}}}(O_{01,\mbf{k_{1,2}}}+O_{20,\mbf{k_{1,2}}}+O_{12,\mbf{k_{1,2}}})
\end{equation}
leading to
\begin{gather}\label{eqapp:g12_k12only}
    g_{12}=\frac{1}{\Omega_\text{tot}}\bigg[
    \left(V_{b_{1}}|\langle C_6^2 \mbf{q_2} | \mbf{q_2} \rangle|^2 -V_0\right)\mathcal{O}_{012}+\left(V_{b_{1}}\langle C_6^2 \mbf{q_2} | \mbf{q_2}\rangle^2 -V_{q_2}\langle C_6^2\bm{q}_2|-\bm{q}_2\rangle^2\right)\mathcal{O}_{012}^*
    \bigg].
\end{gather}
Above, the $V_{b_1}$ terms correspond to Hartree decouplings while the $V_0$ and $V_{q_2}$ terms correspond to Fock decouplings.

We now turn to the evaluation of $f_{12}$, again setting $\delta \bm{k}_{12},\delta\bm{k}'_{12}=0$ in all form factors and interaction potentials
\begin{gather}
    f_{12}=\frac{1}{\Omega_{tot}}\bigg[
    \sideset{}{'}\sum_{\mbf{k'_{12}}} \sum_{i=1}^6  \Big( (V_0-V_{|\mbf{q_2-C_6^{i-1} q_2}|} |\langle C_6^{i-1} \bm{q}_2|\mbf{q_2} \rangle |^2) O_{00,\mbf{k'_{1,2}}} \\ +  (V_0- V_{|\mbf{q_2-C_6^{i-1}( q_2+  {b_{1}}})|}  |\langle C_6^{i-1}(\mbf{q_2}+  \mbf{b_{1}})|\mbf{q_2} \rangle |^2 ) O_{11,\mbf{k'_{12}}}\\ +  (V_0-V_{|\mbf{q_2}-C_6^{i-1} (\bm{q}_2+ C_6 \mbf{b_{1}})|} |\langle C_6^{i-1} (\mbf{q_2}+ C_6 \mbf{b_{1}})|\mbf{q_2} \rangle |^2) O_{22,\mbf{k'_{12}}} \Big) \bigg]\\
    =\frac{1}{\Omega_{tot}}\bigg[
    \sideset{}{'}\sum_{\mbf{k'_{12}}} \sum_{i=1}^6 (V_0-V_{|\mbf{q_2-C_6^{i-1} q_2}|} |\langle C_6^{i-1} \mbf{q_2}|\mbf{q_2} \rangle |^2)\bigg],
\end{gather}
where we have used $O_{00\mbf{k_{12}}}+O_{11\mbf{k_{12}}}+ O_{22\mbf{k_{12}}}  =1$ for an insulating state. Hence, $f_{12}$ is independent of the order parameter with the approximations used here. 

In Eq.~\ref{eqapp:Etot_k12only}, the terms that depend on the kinetic energy $E_{\bm{k}}$ are still difficult to handle, since simple expressions for $O_{00,\bm{k}_{12}},O_{11,\bm{k}_{12}},O_{22,\bm{k}_{12}}$ away from $\bm{k}_{12}=\bm{q}_2$ are not forthcoming (see App.~\ref{subsecapp:gensolnk12}). To proceed, we assume that these contributions do not strongly differ between the various gapped mean-field solutions. The latter all satisfy $O_{00,\bm{q}_{2}}=O_{11,\bm{q}_{2}}=O_{22,\bm{q}_{2}}=\frac{1}{3}$. Effectively, we discard deviations at $\delta\bm{k}_{12}\neq 0$ owing to the kinetic energy. Therefore, the part of the total energy that differentiates between the mean-field solutions is
\begin{align}
    E_\text{tot}&=3(g_{12}^*\mathcal{O}_{012}+g_{12}\mathcal{O}_{012}^*)\\
    &=\frac{6}{\Omega_\text{tot}}\bigg[\left(V_{b_{1}}|\langle C_6^2 \mbf{q_2} | \mbf{q_2} \rangle|^2 -V_0\right)|\mathcal{O}_{012}|^2+\text{Re}\left[\left(V_{b_{1}}\langle \mbf{q_2} | C_6^2 \mbf{q_2}\rangle^2 -V_{q_2}\langle -\bm{q}_2|C_6^2\bm{q}_2\rangle^2\right)\mathcal{O}_{012}^2\right]\bigg].
\end{align}
We also neglect the dependence of the finite-momentum components of the order parameter $O_{ij,\bm{k}_{12}}$ with non-zero $\delta\bm{k}_{12}$. Then we have that $\mathcal{O}_{012}=N_{1,2}e^{i\frac{2\pi}{3}C}$, where $N_{1,2}$ is the number of momenta in the $\bm{k}_{1,2}$ region and $C$ is the Chern number (see App.~\ref{subsubsecapp:KK'sym}). This leads to
\begin{align}\label{eqapp:Etot_k12_g12O12}
    E_\text{tot}
    &=\frac{6N_{1,2}^2}{\Omega_\text{tot}}\bigg[\left(V_{b_{1}}|\langle C_6^2 \mbf{q_2} | \mbf{q_2} \rangle|^2 -V_0\right)+\text{Re}\left[\left(V_{b_{1}}\langle \mbf{q_2} | C_6^2 \mbf{q_2}\rangle^2 -V_{q_2}\langle -\bm{q}_2|C_6^2\bm{q}_2\rangle^2\right)e^{-i\frac{2\pi}{3}C}\right]\bigg].
\end{align}
Extracting only the part that depends on the Chern number $C$ (mod 3) of the insulating state gives
\begin{equation}\label{eqapp:Etot_k12_g12O12_Conly}
    E_\text{tot}=\frac{6N_{1,2}^2}{\Omega_\text{tot}}\text{Re}\left[\left(V_{b_{1}}\langle \mbf{q_2} | C_6^2 \mbf{q_2}\rangle^2 -V_{q_2}\langle -\bm{q}_2|C_6^2\bm{q}_2\rangle^2\right)e^{-i\frac{2\pi}{3}C}\right].
\end{equation}
The first term captures the Hartree penalty (this is positive when combined with the neglected $V_{b_{1}}|\langle C_6^2 \mbf{q_2} | \mbf{q_2} \rangle|^2$ term in Eq.~\ref{eqapp:Etot_k12_g12O12}) for inducing charge density order at wavevector $\bm{b}_1$ (and symmetry-related wavevectors), while the second term captures the Fock exchange energy. Note that for completely trivial form factors $\langle \bm{k}|\bm{k}'\rangle=1$ and realistic interactions such as the Coulomb potential where $V_{b_{1}}<V_{q_2}$, the energy is lowest for $C=0$ mod 3. Therefore for such interactions, non-trivial form factors are required to stabilize other Chern numbers. 

We also ask when a given solution in the $\bm{k}_{12}$ region can be self-consistent. This requires that $g_{12}$ (Eq.~\ref{eqapp:g12_k12only}) satisfies certain inequalities depending on $C$ mod 3. In particular (see App.~\ref{subsubsecapp:KK'sym})
\begin{itemize}
    \item 
    For $C=0$ mod 3, we require $\sqrt{3}\text{Re}\,g_{12}<-|\text{Im}\,g_{12}|$, i.e.~$\frac{2\pi}{3}<\text{arg}\,g_{12}$ or $\text{arg}\,g_{12}<-\frac{2\pi}{3}$\\
    \item 
    For $C=1$ mod 3, we require $\text{Im}\,g_{12}<0$ and $\text{Im}\,g_{12}<\sqrt{3}\text{Re}\,g_{12}$, i.e.~$-\frac{2\pi}{3}<\text{arg}\,g_{12}<0$\\
    \item 
    For $C=-1$ mod 3, we require $\text{Im}\,g_{12}>0$ and $\text{Im}\,g_{12}>-\sqrt{3}\text{Re}\,g_{12}$, i.e.~$0<\text{arg}\,g_{12}<\frac{2\pi}{3}$
\end{itemize}
where we take $\text{arg}\,g_{12}$ to lie in the interval $[-\pi,\pi)$. The Chern number $C=m\mod 3$ of the lowest energy solution corresponds to the value of $m$ satisfying $\text{arg}(g_{12}e^{-i\frac{2(m+1)\pi}{3}})\in[0,\frac{2\pi}{3}]$.

\subsubsection{Parallel transport approximation (PTA)}
We have identified that the hybridization $g_{12}$ (Eq.~\ref{eqapp:g12_k12only}) and the $C$-dependent part of energy $E_\text{tot}$ (Eq.~\ref{eqapp:Etot_k12_g12O12}) rely on not just the magnitude of the form factors $\langle \mbf{q_2} | C_6^2 \mbf{q_2}\rangle$ and $\langle -\bm{q}_2|C_6^2\bm{q}_2\rangle$, but also their phases. We now relate the latter to the Berry curvature $\Omega(\bm{k})$ under the parallel transport approximation (PTA). Using $\langle \psi_{\mbf{k}} \ket{\psi_{\mbf{k + dk}}}= 1 + \mbf{dk} \cdot \langle \psi_{\mbf{k}}  \ket{\mbf{\partial}_{\mbf{k}}\psi_{\mbf{k}}} \approx e^{\mbf{dk} \cdot \langle \psi_{\mbf{k}}  \ket{\mbf{\partial}_{\mbf{k}}\psi_{\mbf{k}}}}$, we approximate
\begin{equation}
    \langle \psi_{\mbf{k_1}} \ket{\psi_{\mbf{k_2}}} \approx |\langle \psi_{\mbf{k_1}} \ket{\psi_{\mbf{k_2}}}| e^{\int_{\mbf{k_1}}^{\mbf{k_2}} \mbf{dk} \cdot \langle \psi_{\mbf{k}}  \ket{\mbf{\partial}_{\mbf{k}}\psi_{\mbf{k}}} }
\end{equation}
where the momentum integral is taken along the shortest path connecting $\bm{k}_1$ and $\bm{k}_2$. 
To relate this to a momentum-space integral of the Berry curvature, consider the specific example
\begin{gather}
    \langle -\bm{q}_2|C_6^2\bm{q}_2\rangle=
    |\langle -\bm{q}_2|C_6^2\bm{q}_2\rangle|e^{\int_{-\mbf{q_2}}^{C_6^2\mbf{ q_2}} \mbf{dk} \cdot \langle \psi_{\mbf{k}}  \ket{\mbf{\partial}_{\mbf{k}}\psi_{\mbf{k}}} }\\
    =|\langle \mbf{-q_2}|C_6^2 \mbf{q_2 }  \rangle | e^{-\frac{1}{6}(\int^{-\mbf{q_2}}_{C_6^2\mbf{ q_2}} + \int_{-\mbf{q_2}}^{-C_6\mbf{ q_2}} + \int_{-C_6 \mbf{q_2}}^{C_6^{-1}\mbf{ q_2}} + \int_{C_6^{-1}\mbf{q_2}}^{\mbf{ q_2}} + \int_{\mbf{q_2}}^{C_6\mbf{ q_2}}  + \int_{C_6 \mbf{q_2}}^{C_6^2\mbf{ q_2}} )\mbf{dk} \cdot \langle \psi_{\mbf{k}}  \ket{\mbf{\partial}_{\mbf{k}}\psi_{\mbf{k}}} }=|\langle \mbf{-q_2}|C_6^2 \mbf{q_2 }  \rangle | e^{\frac{i}{6 }\varphi_{\text{BZ}}}
\end{gather}
where $\varphi_\text{BZ}=\int_{\text{BZ 0}} d\bm{k}\,\Omega(\bm{k})$ is the Berry curvature through BZ 0, and we have used $C_6$ to connect the line integrals into a closed (counter-clockwise) loop. This motivates the following generalization to arbitrary overlaps 
\begin{equation}
    \langle \bm{k} \ket{\bm{k}'}\approx |\langle \bm{k} \ket{\bm{k}'}|e^{- i \varphi_{\bm{k}\rightarrow \bm{k}'}}
\end{equation}
where $\varphi_{\bm{k}\rightarrow\bm{k}'}$ is the Berry connection integrated along the path $\bm{k}\rightarrow \bm{k}'\rightarrow 0 \rightarrow \bm{k}$, which is equal to the Berry curvature flux through the triangle with corners $\bm{k},\bm{k}',\bm{0}$ (with an additional sign if $\hat{z}\cdot \bm{k}\times \bm{k}'$ is negative). Note that the PTA is exact for the GMP limit where $\Omega(\bm{k})=2v_F^2/t_1^2$ is the uniform Berry curvature. For systems with relatively uniform Berry curvature, we can approximate
\begin{equation}\label{eqapp:PTA_uniform_Berry}
    \varphi_{\bm{k}\rightarrow \bm{k}'}=\frac{\Omega_{\bm{k}\rightarrow\bm{k}'}}{\Omega_{\text{BZ}}}\varphi_{\text{BZ}},
\end{equation}
where $\Omega_{\bm{k}\rightarrow\bm{k}'}$ is the signed area of the triangle with corners $\bm{k},\bm{k}',\bm{0}$. In other words, we consider the Berry curvature to take a uniform value corresponding to its averaged value over BZ 0. 

Taking the PTA and uniform Berry curvature approximation [these are not approximations if the GMP form factors are used where $\varphi_{\text{BZ}}=3\sqrt{3}(v_Fq_2/t_1)^2$], where $\langle \mbf{-q_2}|C_6^2 \mbf{q_2 }  \rangle =|\langle \mbf{-q_2}|C_6^2 \mbf{q_2 }  \rangle | e^{\frac{i}{6 }\varphi_{\text{BZ}}}$ and $\langle \mbf{q_2}|C_6^2 \mbf{q_2 }  \rangle =|\langle \mbf{q_2}|C_6^2 \mbf{q_2 }  \rangle | e^{-\frac{i}{6 }\varphi_{\text{BZ}}}$, we find
\begin{gather}
    g_{12}=\frac{N_{1,2}}{\Omega_\text{tot}}\bigg[
    \left(V_{b_{1}}|\langle C_6^2 \mbf{q_2} | \mbf{q_2} \rangle|^2 -V_0\right)e^{i\frac{2\pi}{3}C}+\left(V_{b_{1}}|\langle C_6^2 \mbf{q_2} | \mbf{q_2}\rangle|^2 e^{\frac{i}{3 }\varphi_{\text{BZ}}} -V_{q_2}|\langle C_6^2\bm{q}_2|-\bm{q}_2\rangle|^2e^{-\frac{i}{3 }\varphi_{\text{BZ}}}\right)e^{-i\frac{2\pi}{3}C}
    \bigg]
\end{gather}
\begin{align}
    E_\text{tot}&=\frac{6N_{1,2}^2}{\Omega_\text{tot}}\text{Re}\left[\left(V_{b_{1}}|\langle \mbf{q_2} | C_6^2 \mbf{q_2}\rangle|^2e^{-\frac{i}{3 }\varphi_{\text{BZ}}} -V_{q_2}|\langle -\bm{q}_2|C_6^2\bm{q}_2\rangle|^2e^{\frac{i}{3 }\varphi_{\text{BZ}}}\right)e^{-i\frac{2\pi}{3}C}\right]\\
    &=\frac{6N_{1,2}^2}{\Omega_\text{tot}}\bigg[\Big(V_{b_{1}}|\langle \mbf{q_2} | C_6^2 \mbf{q_2}\rangle|^2-V_{q_2}|\langle -\bm{q}_2|C_6^2\bm{q}_2\rangle|^2\Big)\cos\left(\frac{\varphi_\text{BZ}}{3}\right)\cos \left(\frac{2\pi C}{3}\right)\\
    &\quad -\Big(V_{b_{1}}|\langle \mbf{q_2} | C_6^2 \mbf{q_2}\rangle|^2+V_{q_2}|\langle -\bm{q}_2|C_6^2\bm{q}_2\rangle|^2\Big)\sin\left(\frac{\varphi_\text{BZ}}{3}\right)\sin \left(\frac{2\pi C}{3}\right)\bigg].
\end{align}
Note that the expressions above are invariant under changing $\varphi_{\text{BZ}}$ by multiples of $6\pi$ (keeping the form factor magnitudes unchanged).   For realistic interaction potentials and form factor magnitudes, we expect that $V_{b_{1}}|\langle \mbf{q_2} | C_6^2 \mbf{q_2}\rangle|^2<V_{q_2}|\langle -\bm{q}_2|C_6^2\bm{q}_2\rangle|^2$, so that the coefficients of both $\cos\left(\frac{\varphi_\text{BZ}}{3}\right)\cos \left(\frac{2\pi C}{3}\right)$ and $\sin\left(\frac{\varphi_\text{BZ}}{3}\right)\sin \left(\frac{2\pi C}{3}\right)$ are negative, with the former having equal or smaller magnitude. With this constraint, the $C=1$ mod 3 state has lower energy than $C=-1$ mod 3 for $0<\varphi_{\text{BZ}}<3\pi$, and vice versa for $3\pi<\varphi_{\text{BZ}}<6\pi$. Furthermore, $C=1$ mod 3 is guaranteed to have lower energy than $C=0$ mod 3 if $\pi<\varphi_{\text{BZ}}<3\pi$. At $\varphi_\text{BZ}=0$, $C=0$ is guaranteed to have the lowest energy. Hence, for fixed $V_{b_{1}}|\langle \mbf{q_2} | C_6^2 \mbf{q_2}\rangle|^2<V_{q_2}|\langle -\bm{q}_2|C_6^2\bm{q}_2\rangle|^2$, there will be a transition at $0<\varphi_\text{BZ}<\pi$ between $C=0$ mod 3 and $C=1$ mod 3 as $\varphi_{\text{BZ}}$ is increased from 0.

We can further specialize to the exponential interaction $V_{\bm{k}}=V_0e^{-\alpha|\bm{k}|^2}$ (Eq.~\ref{eqapp:exponential_interaction}) with $\phi=\alpha+\beta$, leading to
\begin{gather}\label{eqapp:k12only_g12_exponential}
    g_{12}=\frac{V_0N_{1,2}}{\Omega_\text{tot}}\bigg[
    \left(e^{-\phi b_1^2} -1\right)e^{i\frac{2\pi}{3}C}+\left(e^{-\phi b_1^2} e^{\frac{i}{3 }\varphi_{\text{BZ}}} -e^{-\phi q_2^2}e^{-\frac{i}{3 }\varphi_{\text{BZ}}}\right)e^{-i\frac{2\pi}{3}C}
    \bigg]
\end{gather}
\begin{align}\label{eqapp:k12only_Etot_exponential}
    E_\text{tot}
    &=\frac{6V_0N_{1,2}^2}{\Omega_\text{tot}}\bigg[\Big(e^{-\phi b_1^2}-e^{-\phi q_2^2}\Big)\cos\left(\frac{\varphi_\text{BZ}}{3}\right)\cos \left(\frac{2\pi C}{3}\right)-\Big(e^{-\phi b_1^2}+e^{-\phi q_2^2}\Big)\sin\left(\frac{\varphi_\text{BZ}}{3}\right)\sin \left(\frac{2\pi C}{3}\right)\bigg].
\end{align}
Note that $b_1$ and $q_2$ are tied to $\varphi_{\text{BZ}}$ via the relation $\varphi_{\text{BZ}}=3\sqrt{3}(v_Fq_2/t_1)^2$. $\phi=0$ corresponds to the `phases-only' model where $V_{b_{1}}|\langle \mbf{q_2} | C_6^2 \mbf{q_2}\rangle|^2=V_{q_2}|\langle -\bm{q}_2|C_6^2\bm{q}_2\rangle|^2$, leading to
\begin{gather}\label{eqapp:k12only_g12_phasesonly}
    g_{12}=\frac{2iV_0N_{1,2}}{\Omega_\text{tot}}\sin\left(\frac{\varphi_{\text{BZ}}}{3}\right)e^{-i\frac{2\pi}{3}C}
\end{gather}
\begin{align}\label{eqapp:k12only_Etot_phasesonly}
    E_\text{tot}
    &=-\frac{12V_0N_{1,2}^2}{\Omega_\text{tot}}\sin\left(\frac{\varphi_\text{BZ}}{3}\right)\sin \left(\frac{2\pi C}{3}\right).
\end{align}
$C=1$ mod 3 is the lowest energy state for $0<\varphi_{\text{BZ}}<3\pi$, while $C=-1$ mod 3 is the lowest energy state for $-3\pi<\varphi_{\text{BZ}}<0$ (with the results being invariant under shifting $\varphi_{\text{BZ}}$ by $6\pi$).

 \begin{figure}
 \centering
\includegraphics[width=1.0\columnwidth]{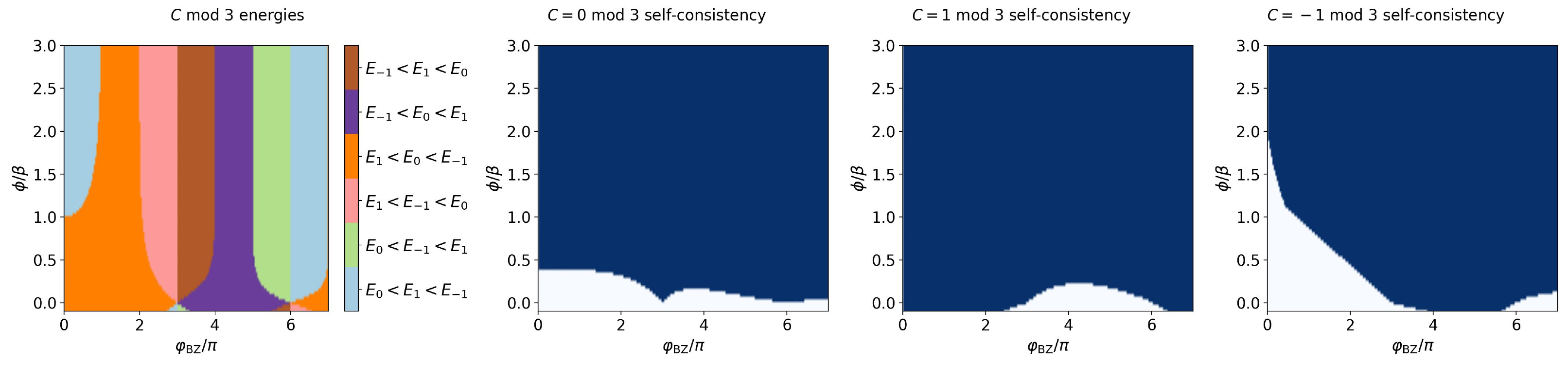} 
\caption{Chern energy ordering (see Eq.~\ref{eqapp:k12only_Etot_exponential}) for the $\bm{k}_{12}$-only limit, as a function of $\varphi_{\text{BZ}}$ and the parameter $\phi=\alpha+\beta$. We consider the GMP form factors (Eq.~\ref{eqapp:Mkq_exp}) where $\beta=\frac{v_F^2}{t_1^2}$, and $\varphi_{\text{BZ}}=3\sqrt{3}(v_Fq_2/t_1)^2$ also sets $q_2$ and $b_1$. Note that $\phi=0$ corresponds to the phases-only limit of the model. The exponential interaction is characterized by a parameter $\alpha$ (Eq.~\ref{eqapp:exponential_interaction}). $E_1$ should be understood as the energy of the $C=1$ mod 3 solution, etc. For the self-consistency plots (which is based on the value of $g_{12}$ in Eq.~\ref{eqapp:k12only_g12_exponential}), dark blue indicates that the Chern solution is self-consistent at the $K_\text{M}$ points.}
\label{figapp:k12_selfconsistency_energy_GMP_varphi}
\end{figure}

We now address the self-consistency conditions. Returning to the situation where only the PTA is invoked, the real and imaginary parts of $g_{12}$ are
\begin{align}
    \frac{\Omega_\text{tot}}{N_{1,2}}\text{Re}\,g_{12}&=\left[\left(V_{b_{1}}|\langle C_6^2 \mbf{q_2} | \mbf{q_2} \rangle|^2 -V_0\right)+\left(V_{b_{1}}|\langle C_6^2 \mbf{q_2} | \mbf{q_2}\rangle|^2  -V_{q_2}|\langle C_6^2\bm{q}_2|-\bm{q}_2\rangle|^2\right)\cos \left(\frac{\varphi_{\text{BZ}}}{3}\right)\right]\cos \left(\frac{2\pi C}{3}\right)\\
    &+\left[\left(V_{b_{1}}|\langle C_6^2 \mbf{q_2} | \mbf{q_2}\rangle|^2  +V_{q_2}|\langle C_6^2\bm{q}_2|-\bm{q}_2\rangle|^2\right)\sin \left(\frac{\varphi_{\text{BZ}}}{3}\right)\right]\sin \left(\frac{2\pi C}{3}\right)
    \\
    \frac{\Omega_\text{tot}}{N_{1,2}}\text{Im}\,g_{12}&=\left[\left(V_{b_{1}}|\langle C_6^2 \mbf{q_2} | \mbf{q_2} \rangle|^2 -V_0\right)+\left(-V_{b_{1}}|\langle C_6^2 \mbf{q_2} | \mbf{q_2}\rangle|^2  +V_{q_2}|\langle C_6^2\bm{q}_2|-\bm{q}_2\rangle|^2\right)\cos \left(\frac{\varphi_{\text{BZ}}}{3}\right)\right]\sin \left(\frac{2\pi C}{3}\right)\\
    &+\left[\left(V_{b_{1}}|\langle C_6^2 \mbf{q_2} | \mbf{q_2}\rangle|^2  +V_{q_2}|\langle C_6^2\bm{q}_2|-\bm{q}_2\rangle|^2\right)\sin \left(\frac{\varphi_{\text{BZ}}}{3}\right)\right]\cos \left(\frac{2\pi C}{3}\right).
\end{align}

\begin{itemize}
\item For the $C=1$ mod 3 state, we require $\text{Im}\,g_{12}<0$ and $\text{Im}\,g_{12}<\sqrt{3}\text{Re}\,g_{12}$:
\begin{gather}
    -\sqrt{3}V_0+V_{b_{1}}|\langle C_6^2 \mbf{q_2} | \mbf{q_2}\rangle|^2\left[\sqrt{3}\left(1-\cos \left(\frac{\varphi_{\text{BZ}}}{3}\right)\right)-\sin \left(\frac{\varphi_{\text{BZ}}}{3}\right)\right]+V_{q_2}|\langle C_6^2\bm{q}_2|-\bm{q}_2\rangle|^2\left[\sqrt{3}\cos \left(\frac{\varphi_{\text{BZ}}}{3}\right)-\sin \left(\frac{\varphi_{\text{BZ}}}{3}\right)\right]<0\\
    -\sqrt{3}V_0+V_{b_{1}}|\langle C_6^2 \mbf{q_2} | \mbf{q_2}\rangle|^2\left[\sqrt{3}-2\sin \left(\frac{\varphi_{\text{BZ}}}{3}\right)\right]-2V_{q_2}|\langle C_6^2\bm{q}_2|-\bm{q}_2\rangle|^2\sin \left(\frac{\varphi_{\text{BZ}}}{3}\right)<0.
\end{gather}
Consider the ordering $V_0\geq V_{q_2}|\langle C_6^2\bm{q}_2|-\bm{q}_2\rangle|^2\geq V_{b_{1}}|\langle C_6^2 \mbf{q_2} | \mbf{q_2}\rangle|^2$ (the inequalities are saturated in the phases-only model). For $\sin \left(\frac{\varphi_{\text{BZ}}}{3}\right)>0$, the second condition is always satisfied. 

We are interested in determining the values of $\varphi_{\text{BZ}}$ for which the first condition is guaranteed to be satisfied. Since $V_0$ on the LHS has a negative coefficient and helps to satisfy the inequality, we can set $V_0= V_{q_2}|\langle C_6^2\bm{q}_2|-\bm{q}_2\rangle|^2$ for our purposes. Then the inequality can be rewritten
\begin{equation}
    -\left(V_{q_2}|\langle C_6^2\bm{q}_2|-\bm{q}_2\rangle|^2+V_{b_{1}}|\langle C_6^2 \mbf{q_2} | \mbf{q_2}\rangle|^2\right)\sin \left(\frac{\varphi_{\text{BZ}}}{3}\right)-\left(V_{q_2}|\langle C_6^2\bm{q}_2|-\bm{q}_2\rangle|^2-V_{b_{1}}|\langle C_6^2 \mbf{q_2} | \mbf{q_2}\rangle|^2\right)\sqrt{3}\left(1-\cos \left(\frac{\varphi_{\text{BZ}}}{3}\right)\right)<0,
\end{equation}
which is guaranteed to be satisfied if $0<\varphi_{\text{BZ}}<3\pi$. Hence, the self-consistency condition for $C=1$ mod 3 is guaranteed to be satisfied if it is also the lowest energy solution.

\item For the $C=0$ mod 3 state, we require $\sqrt{3}\text{Re}\,g_{12}<-|\text{Im}\,g_{12}|$. We consider the physically reasonable ordering $V_0\geq V_{q_2}|\langle C_6^2\bm{q}_2|-\bm{q}_2\rangle|^2\geq V_{b_{1}}|\langle C_6^2 \mbf{q_2} | \mbf{q_2}\rangle|^2$. Again, increasing $V_0$ only helps the inequality so we can set $V_0=V_{q_2}|\langle C_6^2\bm{q}_2|-\bm{q}_2\rangle|^2$, leading to
\begin{gather}
   \left(V_{b_{1}}|\langle C_6^2 \mbf{q_2} | \mbf{q_2}\rangle|^2  -V_{q_2}|\langle C_6^2\bm{q}_2|-\bm{q}_2\rangle|^2\right)\sqrt{3}\left(1+\cos \left(\frac{\varphi_{\text{BZ}}}{3}\right)\right)+  \left(V_{b_{1}}|\langle C_6^2 \mbf{q_2} | \mbf{q_2}\rangle|^2  +V_{q_2}|\langle C_6^2\bm{q}_2|-\bm{q}_2\rangle|^2\right)\left|\sin \left(\frac{\varphi_{\text{BZ}}}{3}\right)\right|<0.
\end{gather}
The inequality is never guaranteed to be satisfied. In fact for the phases-only model, $\sqrt{3}\text{Re}\,g_{12}+|\text{Im}\,g_{12}|$ is non-negative for all $\varphi_{\text{BZ}}$, so that $C=0\mod 3$ is never self-consistent. However, if $C=0\,$mod 3 is the lowest energy solution, the the inequality is indeed satisfied.
\item The conditions for $C=-1\mod3$ can be obtained from $C=1\mod3$ by taking $\varphi_{\text{BZ}}\rightarrow -\varphi_{\text{BZ}}$.
\end{itemize}

In Fig.~\ref{figapp:k12_selfconsistency_energy_GMP_varphi}, we show an example of the energy ordering and self-consistency conditions of the three Chern states $C=0,1,-1$ mod 3 for the limit of GMP form factors (Eq.~\ref{eqapp:Mkq_exp}) and the exponential interaction (Eq.~\ref{eqapp:exponential_interaction}). In other words, we use the expressions in Eq.~\ref{eqapp:k12only_g12_exponential} and \ref{eqapp:k12only_Etot_exponential}. We note that for $\phi>0$, the minimum energy solution is also always self-consistent. For $\varphi_{\text{BZ}}<\pi$, we find that the lowest Chern solution switches from $C=0$ mod 3 to $C=1$ mod 3 as $\phi$ is decreased.   $\phi$ controls the relative strength of the Hartree $(V_{b_1})$ and Fock $(V_{q_2})$ contributions that distinguish the energies of the Chern states. A smaller $\phi$ increases the Hartree penalty and hence favors $C=1$ mod 3 for small $\varphi_{\text{BZ}}$.

\subsection{$\mathbf{k}_1$ and $\mathbf{k}_{12}$ limit}\label{subsecapp:k1_k12_limit}

We now consider including the $\bm{k}_1$ region as well in our mean-field energy and self-consistency analysis. Without neglecting contributions from the $\bm{k}_1$ region, the total mean-field energy reads
\begin{align}
    E_\text{tot}&=
    6 \sideset{}{'}\sum_{\mbf{k_1}} \bigg[ (E_{\mbf{k_1}} + \frac{1}{2} f_{\mbf{k_1}}) O_{00 \mbf{k_1}}+ (E_{\mbf{k_1+ b_{1}}} + \frac{1}{2} f_{\mbf{k_1+ b_{1}}}) O_{11 \mbf{k_1}} + \frac{1}{2} g_{1,\mbf{k_1}} O_{10 \mbf{k_1}}  + \frac{1}{2} g_{1,\mbf{k_1}}^\star  O_{01 \mbf{k_1}}\bigg]\\ & +
   6 \sideset{}{'}\sum_{\mbf{k_{12}}} \bigg[(E_{\mbf{k_{12}}} + \frac{1}{2} f_{\mbf{k_{12}}}) O_{00 \mbf{k_{12}}}   + (E_{\mbf{k_{12}+ b_{1}}} + \frac{1}{2} f_{\mbf{k_{12}+ b_{1}}}) O_{11 \mbf{k_{12}}}  + (E_{\mbf{k_{12}+ C_6 b_{1}}} + \frac{1}{2} f_{\mbf{k_{12}+ C_6 b_{1}}})   O_{22 \mbf{k_{12}}}\bigg]  \\ &+3(g_{12}^*\mathcal{O}_{012}+g_{12}\mathcal{O}^*_{012}),
\end{align}
where we have approximated the hybridization field $g_{1}$ as constant within the $\bm{k}_{12}$ region. If we make a similar approximation for the band renormalization field in the $\bm{k}_{12}$ region, and assume an insulating state, we obtain 
\begin{align}\label{eqapp:Etot_k1_k12_v2}
    E_\text{tot}&=
    6 \sideset{}{'}\sum_{\mbf{k_1}} \bigg[\frac{1}{2} f_{\mbf{k_1}}+ E_{\mbf{k_1}} O_{00 \mbf{k_1}}+ (E_{\mbf{k_1+ b_{1}}} + \frac{1}{2} f_{\mbf{k_1+ b_{1}}}-\frac{1}{2} f_{\mbf{k_1}}) O_{11 \mbf{k_1}} - \frac{1}{2} \frac{|g_{1,\mbf{k_1}}|^2}{|\bm{d}_{\bm{k}_1}|} \bigg]\\ & +
   6 \sideset{}{'}\sum_{\mbf{k_{12}}} \bigg[\frac{1}{2}f_{12}+E_{\mbf{k_{12}}} O_{00 \mbf{k_{12}}}   + E_{\mbf{k_{12}+ b_{1}}} O_{11 \mbf{k_{12}}}  + E_{\mbf{k_{12}+ C_6 b_{1}}}  O_{22 \mbf{k_{12}}}\bigg]  \\ &+3(g_{12}^*\mathcal{O}_{012}+g_{12}\mathcal{O}^*_{012}).
\end{align}

We re-evaluate $f_{12}$ to check its dependence on the $\bm{k}_1$ region
\begin{align}
    f_{12} =&\frac{1}{\Omega_{tot}} \Bigg[\sum_{i=1}^6  \sideset{}{'}\sum_{\mbf{k'_1}} \left( (V_0- V_{|\mbf{q}_2-C_6^{i-1}\mbf{k'_1}|} |\langle C_6^{i-1}\mbf{k'_1}|\mbf{q}_2 \rangle |^2 )O_{00,\mbf{k'_1}} +( V_0 - V_{|\mbf{q}_2-C_6^{i-1}\mbf{k'_1}- C_6^{i-1} \mbf{b_{1}}|} |\langle C_6^{i-1}\mbf{k'_1}+ C_6^{i-1} \mbf{b_{1}}|\mbf{q}_2 \rangle |^2 )O_{11,\mbf{k'_1}}  \right) \\ & +\sum_{i=1}^6  \sideset{}{'}\sum_{\mbf{k'_{12}}} \bigg( (V_0-V_{|\mbf{q}_2-C_6^{i-1}\mbf{q}_2|} |\langle C_6^{i-1}\mbf{q_{2}}|\mbf{q}_2 \rangle |^2) O_{00,\mbf{k'_{1,2}}}  \\
    &+(V_0- V_{|\mbf{q}_2-C_6^{i-1}\mbf{q}_2- C_6^{i-1} \mbf{b_{1}}|}  |\langle C_6^{i-1}\mbf{q}_2+ C_6^{i-1} \mbf{b_{1}}|\mbf{q}_2 \rangle |^2 ) O_{11,\mbf{k}'_{1,2}}  \\ &+  (V_0-V_{|\mbf{q}_2-C_6^{i-1}\mbf{q}_2- C_6^{i} \mbf{b_{1}}|} |\langle C_6^{i-1}\mbf{q}_2+ C_6^{i} \mbf{b_{1}}|\mbf{q}_2 \rangle |^2) O_{22,\mbf{k'_{1,2}}} 
    \bigg)   \Bigg]\\
    =&\frac{1}{\Omega_{tot}} \Bigg[\sum_{i=1}^6  \sideset{}{'}\sum_{\mbf{k'_1}}  (V_0- V_{|\mbf{q}_2-C_6^{i-1}\mbf{k'_1}|} |\langle C_6^{i-1}\mbf{k'_1}|\mbf{q}_2 \rangle |^2 ) +\sum_{i=1}^6  \sideset{}{'}\sum_{\mbf{k'_{12}}}  (V_0-V_{|\mbf{q}_2-C_6^{i-1}\mbf{q}_2|} |\langle C_6^{i-1}\mbf{q_{2}}|\mbf{q}_2 \rangle |^2) \\
    &+\sum_{i=1}^6  \sideset{}{'}\sum_{\mbf{k'_1}}(V_{|C_6^{1-i}\mbf{q}_2-\delta\mbf{k'_1}+\frac{1}{2}\bm{b}_1|}|\langle \delta\mbf{k'_1}-\frac{1}{2}\bm{b}_1|C^{1-i}\mbf{q}_2 \rangle |^2-V_{|C_6^{1-i}\mbf{q}_2-\delta\mbf{k'_1}-\frac{1}{2}\bm{b}_1|} |\langle \delta\bm{k'_1}+ \frac{1}{2}\mbf{b_{1}}|C_6^{1-i}\mbf{q}_2 \rangle |^2)O_{11,\mbf{k'_1}}\Bigg].
\end{align}
In the thin-sliver approximation, the form factors and interaction potential depends only weakly on the component $\delta k_{1x}$ perpendicular to the BZ. Neglecting this (but keeping the dependence on $\delta k_{1y}$) leads to
\begin{align}
    f_{12} =&\frac{1}{\Omega_{tot}} \Bigg[\sum_{i=1}^6  \sideset{}{'}\sum_{\mbf{k'_1}}  (V_0- V_{|\mbf{q}_2-C_6^{i-1}\mbf{k'_1}|} |\langle C_6^{i-1}\mbf{k'_1}|\mbf{q}_2 \rangle |^2 ) +\sum_{i=1}^6  \sideset{}{'}\sum_{\mbf{k'_{12}}}  (V_0-V_{|\mbf{q}_2-C_6^{i-1}\mbf{q}_2|} |\langle C_6^{i-1}\mbf{q_{2}}|\mbf{q}_2 \rangle |^2) \\
    &+\sum_{i=1}^6  \sideset{}{'}\sum_{\mbf{k'_1}}(V_{|C_6^{1-i}\mbf{q}_2-\delta {k'_{1y}}\hat{y}+\frac{1}{2}\bm{b}_1|}|\langle \delta {k'_{1y}}\hat{y}-\frac{1}{2}\bm{b}_1|C^{1-i}\mbf{q}_2 \rangle |^2-V_{|C_6^{1-i}\mbf{q}_2-\delta {k'_{1y}}\hat{y}-\frac{1}{2}\bm{b}_1|} |\langle \delta {k'_{1y}}\hat{y}+ \frac{1}{2}\mbf{b_{1}}|C_6^{1-i}\mbf{q}_2 \rangle |^2)O_{11,\mbf{k'_1}}\Bigg]\\
    =&\frac{1}{\Omega_{tot}} \Bigg[\sum_{i=1}^6  \sideset{}{'}\sum_{\mbf{k'_1}}  (V_0- V_{|\mbf{q}_2-C_6^{i-1}\mbf{k'_1}|} |\langle C_6^{i-1}\mbf{k'_1}|\mbf{q}_2 \rangle |^2 ) +\sum_{i=1}^6  \sideset{}{'}\sum_{\mbf{k'_{12}}}  (V_0-V_{|\mbf{q}_2-C_6^{i-1}\mbf{q}_2|} |\langle C_6^{i-1}\mbf{q_{2}}|\mbf{q}_2 \rangle |^2)\Bigg].
\end{align}
In the first term of the second line above, we relabelled $\bm{k}'_1\rightarrow M_1\mathcal{T}\bm{k}'_1$ (which changes the sign of $\delta k'_{1y}$) and $i\rightarrow i+3$, and used $M_1\mathcal{T}$ symmetry which constrains $O_{11,\bm{k}_1'}=O_{11,M_1\mathcal{T}\bm{k}_1'}$. This then cancels with the second term of the second line. Therefore, we find that $f_{12}$ is independent of the order parameter in the approximations considered here. 

The remaining dependence on the order parameter in $E_\text{tot}$ is then 
\begin{align}
    E_\text{tot}&=
    6 \sideset{}{'}\sum_{\mbf{k_1}} \bigg[\frac{1}{2} f_{\mbf{k_1}}+ E_{\mbf{k_1}} O_{00 \mbf{k_1}}+ (E_{\mbf{k_1+ b_{1}}} + \frac{1}{2} f_{\mbf{k_1+ b_{1}}}-\frac{1}{2} f_{\mbf{k_1}}) O_{11 \mbf{k_1}} - \frac{1}{2} \frac{|g_{1,{\delta k_{1y}}}|^2}{|\bm{d}_{\bm{k}_1}|} \bigg]\\ & +
   6 \sideset{}{'}\sum_{\mbf{k_{12}}} \bigg[E_{\mbf{k_{12}}} O_{00 \mbf{k_{12}}}   + E_{\mbf{k_{12}+ b_{1}}} O_{11 \mbf{k_{12}}}  + E_{\mbf{k_{12}+ C_6 b_{1}}}  O_{22 \mbf{k_{12}}}\bigg]  \\ &+3(g_{12}^*\mathcal{O}_{012}+g_{12}\mathcal{O}^*_{012}),
\end{align}
where we introduced $g_{1,{\delta k_{1y}}}\equiv g_{1,{-\frac{\bm{b}_1}{2}+\delta k_{1y}\hat{y}}}$, which reflects the fact that the hybridization field does not depend on $\delta k_{1x}$ in thin-sliver approximation, as shown below. We comment that the terms that may be expected to appreciably distinguish the insulating HF solutions with different Chern numbers are
\begin{align}\label{eqapp:Etot_distinguishC}
    E_\text{tot}&=
    6 \sideset{}{'}\sum_{\mbf{k_1}} \bigg[(E_{\mbf{k_1+ b_{1}}} + \frac{1}{2} f_{\mbf{k_1+ b_{1}}}-\frac{1}{2} f_{\mbf{k_1}}) O_{11 \mbf{k_1}}- \frac{1}{2} \frac{|g_{1,{\delta k_{1y}}}|^2}{|\bm{d}_{\bm{k}_1}|} \bigg]+
   3(g_{12}^*\mathcal{O}_{012}+g_{12}\mathcal{O}^*_{012}).
\end{align}
Above, we have specialized to the Berry Trashcan dispersion where the dispersion $E_{\bm{k}_1}=E_{\bm{k}_{12}}=0$ vanishes within the first BZ since it lies within the flat bottom, assumed that the order parameter in the $\bm{k}_{12}$ region is independent of $\bm{k}_{12}$ (so that $O_{ii,\bm{k}_{12}}=\frac{1}{3}$ regardless of $C$), and that $f_{\bm{k}_1}$ does not depend on $C$ (see Eq.~\ref{eqapp:fprime}).

We now study the expressions for $g_{1,{\delta k_{1y}}}$ and $g_{12}$. We first consider $g_{1,{\bm{k}_1}}$ (Eq.~\ref{eqapp:g1_k1}) and make the approximations corresponding to a small $\bm{k}_{12}$ region
\begin{eqnarray}
    & g_{1, \mbf{k_1}}=  \frac{1}    {\Omega_{tot}}  \bigg[ V_{b_{1}}    \langle \frac{\mbf{b_{1}}}{2}+ \mbf{\delta k_1} |- \frac{\mbf{b_{1}}}{2}+  \mbf{\delta k_1} \rangle  \Big(\sideset{}{'}\sum_{\mbf{k'_1}}( \langle {-\frac{\mbf{b_{1}}}{2}+ \mbf{\delta k'_1}} |\frac{\mbf{b_{1}}}{2}+  \mbf{\delta k'_1} \rangle  O_{01,\mbf{k'_1}} + h.c) \\ &  + \big( \langle \mbf{q_2} | C_6^2 \mbf{q_2} \rangle \mathcal{O}_{012}+h.c.\big)\Big)\\ & -  \sideset{}{'}\sum_{\mbf{k'_1}} \big(V_{| \mbf{\delta k_1}  - \mbf{\delta k'_1}|} \langle \frac{\mbf{b_{1}}}{2}+ \mbf{\delta k_1}|\frac{\mbf{b_{1}}}{2} + \mbf{\delta k'_1} \rangle \langle - \frac{\mbf{b_{1}}}{2}+  \mbf{\delta k'_1}|- \frac{\mbf{b_{1}}}{2}+ \mbf{\delta k_1}\rangle 
 O_{01,\mbf{k'_1}}  \\ & + V_{|\mbf{\delta k_1}+ \mbf{\delta k'_{1}}|} \langle \frac{\mbf{b_{1}}}{2}+ \mbf{\delta k_1}|\frac{\mbf{b_{1}}}{2}-\mbf{\delta k'_{1}} \rangle \langle -\frac{\mbf{b_{1}}}{2} -\mbf{\delta k'_{1}}|-\frac{\mbf{b_{1}}}{2}+  \mbf{\delta k_1}\rangle ) O_{1,0,\mbf{k'_{1}}} \big) \\&
 -V_{|\mbf{\delta k_1}- \frac{1}{2} C_{6}\mbf{q_2}|} \langle \frac{\mbf{b_{1}}}{2}+ \mbf{\delta k_1}|C_6^2 \mbf{q_2}\rangle \langle \mbf{q_2}|- \frac{\mbf{b_{1}}}{2} + \mbf{\delta k_1}\rangle   )\mathcal{O}_{012}\\&
 -V_{| \mbf{\delta k_1}+\frac{1}{2} C_6 \mbf{q_2}|} \langle \frac{\mbf{b_{1}}}{2}+  \mbf{\delta k_1}|-\mbf{q_2} \rangle \langle  C_6^{-1} \mbf{q_2}| -\frac{\mbf{b_{1}}}{2}+ \mbf{\delta k_1}\rangle  ) \rangle \mathcal{O}_{012}^*\bigg].
\end{eqnarray}
Indeed, the only dependence on $\delta k_{1x}$ in $g_{1,\bm{k}_1}$ is in the interaction potential and form factors, which vary only weakly with $k_{1x}$. We can then use $M_1\mathcal{T}$ symmetry to rewrite
\begin{eqnarray}
    & g_{1,{\delta k_{1y}}}=  \frac{1}    {\Omega_{tot}}  \bigg[ V_{b_{1}}    \langle \frac{\mbf{b_{1}}}{2}+ {\delta k_{1y}\hat{y}} |- \frac{\mbf{b_{1}}}{2}+  {\delta k_{1y}\hat{y}} \rangle  \Big(2\sideset{}{'}\sum_{\mbf{k'_1}}( \langle {-\frac{\mbf{b_{1}}}{2}+ {\delta k'_{1y}}\hat{y}} |\frac{\mbf{b_{1}}}{2}+  {\delta k'_{1y}\hat{y}} \rangle  O_{01,\mbf{k'_1}}) \\ &  + \big( \langle \mbf{q_2} | C_6^2 \mbf{q_2} \rangle \mathcal{O}_{012}+h.c.\big)\Big)\\ & -  \sideset{}{'}\sum_{\mbf{k'_1}} 2\big(V_{| {\delta k_{1y}\hat{y}}  - {\delta k'_{1y}\hat{y}}|} \langle \frac{\mbf{b_{1}}}{2}+ {\delta k_{1y}\hat{y}}|\frac{\mbf{b_{1}}}{2} + {\delta k'_{y1}\hat{y}} \rangle \langle - \frac{\mbf{b_{1}}}{2}+  {\delta k'_{1y}\hat{y}}|- \frac{\mbf{b_{1}}}{2}+ {\delta k_{1y}\hat{y}}\rangle 
 O_{01,\mbf{k'_1}}\big) \\&
 -V_{|{\delta k_{1y}\hat{y}}- \frac{1}{2} C_{6}\mbf{q_2}|} \langle \frac{\mbf{b_{1}}}{2}+ {\delta k_{1y}\hat{y}}|C_6^2 \mbf{q_2}\rangle \langle \mbf{q_2}|- \frac{\mbf{b_{1}}}{2} + {\delta k_{1y}\hat{y}}\rangle   )\mathcal{O}_{012}\\&
 -V_{| {\delta k_{1y}\hat{y}}+\frac{1}{2} C_6 \mbf{q_2}|} \langle \frac{\mbf{b_{1}}}{2}+  {\delta k_{1y}\hat{y}}|-\mbf{q_2} \rangle \langle  C_6^{-1} \mbf{q_2}| -\frac{\mbf{b_{1}}}{2}+ {\delta k_{1y}\hat{y}}\rangle  ) \rangle \mathcal{O}_{012}^*\bigg].
\end{eqnarray}
Introducing the short-hand notation
\begin{eqnarray}\label{eqapp:Mdeltak1y}
    & \langle \frac{\mbf{b_{1}}}{2}+ \delta k_{1y} \mbf{\hat{y}} |- \frac{\mbf{b_{1}}}{2}+  \delta k_{1y} \mbf{\hat{y}}\rangle = M_{ \delta k_{1y} }  =  M^*_{- \delta k_{1y} }  \\\label{eqapp:Ndeltak1y} & \langle \frac{\mbf{b_{1}}}{2}+ \delta k_{1y} \mbf{\hat{y}}|C_6^2 \mbf{q_2} \rangle \langle \mbf{q_2}|- \frac{\mbf{b_{1}}}{2} + \delta k_{1y} \mbf{\hat{y}}\rangle  = N_{ \delta k_{1y} } \\ &\label{eqapp:Pdeltak1y} \langle \frac{\mbf{b_{1}}}{2}+ \delta k_{1y} \mbf{\hat{y}}|\frac{\mbf{b_{1}}}{2} + \delta k'_{1y}\mbf{\hat{y}} \rangle\langle - \frac{\mbf{b_{1}}}{2}+  {\delta k'_{1y}\hat{y}}|- \frac{\mbf{b_{1}}}{2}+ {\delta k_{1y}\hat{y}}\rangle= P_{\delta k_{1y}, \delta k'_{1y}}
\end{eqnarray}
we obtain
\begin{eqnarray}
    &\Omega_{tot} g_{1,\delta k_{1y} } =   V_{b_{1}}     M_{ \delta k_{1y} }  \big(   2Re[\langle \mbf{q_2} | C_6^2 \mbf{q_2}  \rangle \mathcal{O}_{012} ] + 2 \sideset{}{'}\sum_{\mbf{k'_1}}  M^\star_{ \delta k_{1y}' }    O_{01,\mbf{k'_1}}  \big) \\ & -  V_{|\delta k_{1y} \mbf{\hat{y}}- \frac{1}{2} C_{6}\mbf{q_2}|} N_{\delta k_{1y}}   \mathcal{O}_{012}  - V_{| \delta k_{1y} \mbf{\hat{y}}+\frac{1}{2} C_6 \mbf{q_2}|}   N_{-\delta k_{1y}}^\star \mathcal{O}_{012}^\star  -  2 \sideset{}{'}\sum_{\mbf{k'_1}} V_{|\delta k_{1y} - \delta k'_{1y}|} P_{\delta k_{1y}, \delta k'_{1y}} O_{01,\mbf{k'_{1}}}.
\end{eqnarray} 
With the same approximations, we can obtain an analogous expression for $g_{12}$
\begin{eqnarray}  
    &\Omega_{tot} g_{12} =  (   V_{b_{1}}    |\langle C_6^2 \mbf{q_2 } | \mbf{q_2} \rangle |^2 -   V_{0})  \mathcal{O}_{012}+ ( V_{b_{1}}    \langle C_6^2 \mbf{q_2 } | \mbf{q_2} \rangle^2   - V_{ q_2} (\langle C_6^2 \mbf{  q_2}|-\mbf{q_2} \rangle)^2 )  \mathcal{O}_{012}^\star  \\ & + 2 V_{b_{1}}    \langle C_6^2 \mbf{q_2 } | \mbf{q_2} \rangle   \sideset{}{'}\sum_{\mbf{k'_1}} M_{\delta k_{1y}'}^\star   O_{01, \mbf{k'_1}}    - 2\sideset{}{'}\sum_{\mbf{k'_1}} V_{|\frac{1}{2} C_6 \mbf{  q_2} - \delta k'_{1y} \mbf{\hat{y}}|}  N_{\delta k_{1y}'}^\star    O_{01,\mbf{k'_{1}}} .
\end{eqnarray} 
It can be shown that $g_{12}$ is equal to $g_{1,\delta k_{1y}}$ evaluated at $\delta k_{1y}=-\frac{q_2}{2}$. Therefore, we will not need to derive separately the expression for $g_{12}$. The hybridization field $g_{1,\delta k_{1y} }$ is currently expressed in terms of the order parameters $\mathcal{O}_{012}$ and $O_{01, \mbf{k_1}}$. As in App.~\ref{subsecapp:Etot_k12_only_limit}, we will approximate $\mathcal{O}_{012}=\approx N_{1,2}e^{i\frac{2\pi}{3}C}$. We cannot make as simple of an approximation for $O_{01, \mbf{k_1}}$, since it may vary rapidly in the $\delta k_{1x}$ direction (despite the narrow cutoff), and will certainly vary appreciably across the height $-\frac{q_2}{2}\lesssim \delta k_{1y}\lesssim\frac{q_2}{2}$ of the $\bm{k}_1$ region. 

The presence of form factors in the expressions above that vary with $\delta k_{1y},\delta k'_{1y}$ in some generic manner, for example in the quantities $M_{\delta k_{1y}},N_{\delta k_{1y}},P_{\delta k_{1y},\delta k'_{1y}}$, prevents analytical progress. To proceed, we use the PTA with the approximation of uniform Berry curvature (Eq.~\ref{eqapp:PTA_uniform_Berry}), which will allow for explicit functional forms of the dependence of the phases on $\delta k_{1y},\delta k'_{1y}$. Eqs.~\ref{eqapp:Mdeltak1y}, \ref{eqapp:Ndeltak1y}, \ref{eqapp:Pdeltak1y} become
\begin{eqnarray}
    & M_{ \delta k_{1y} } =|\langle \frac{\mbf{b_{1}}}{2}+ \delta k_{1y} \mbf{\hat{y}} |- \frac{\mbf{b_{1}}}{2}+  \delta k_{1y} \mbf{\hat{y}}\rangle | e^{- i \frac{\varphi_\text{BZ}}{6} \frac{2 \delta k_{1y}}{q_2} }   \\ & N_{ \delta k_{1y} }= | \langle \frac{\mbf{b_{1}}}{2}+ \delta k_{1y} \mbf{\hat{y}}|C_6^2 \mbf{q_2} \rangle || \langle \mbf{q_2}|- \frac{\mbf{b_{1}}}{2} + \delta k_{1y} \mbf{\hat{y}}\rangle|  e^{ i \frac{\varphi_\text{BZ}}{6}(1+ \frac{2 \delta k_{1y}}{q_2} )}  \\ & P_{\delta k_{1y}, \delta k'_{1y}}= |\langle \frac{\mbf{b_{1}}}{2}+ \delta k_{1y} \mbf{\hat{y}}|\frac{\mbf{b_{1}}}{2} + \delta k'_{1y}\mbf{\hat{y}} \rangle |
    |\langle -\frac{\mbf{b_{1}}}{2}+ \delta k'_{1y} \mbf{\hat{y}}|-\frac{\mbf{b_{1}}}{2} + \delta k_{1y}\mbf{\hat{y}} \rangle |
    e^{-i \frac{\varphi_\text{BZ}}{6} (\frac{2\delta k_{1y}'}{q_2}- \frac{2\delta k_{1y}}{q_2})}.
\end{eqnarray}
To get explicit forms of the absolute values of the form factors, we further consider the limit of GMP form factors (in which case the PTA is exact)
\begin{eqnarray}
    & M_{ \delta k_{1y} } =e^{-\frac{\beta}{2}b_1^2} e^{- i \frac{\varphi_\text{BZ}}{6} \frac{2 \delta k_{1y}}{q_2} }   \\ & N_{ \delta k_{1y} }= e^{-\beta(\delta k_{1y}+\frac{q_2}{2})^2}  e^{ i \frac{\varphi_\text{BZ}}{6}(1+ \frac{2 \delta k_{1y}}{q_2} )}  \\ & P_{\delta k_{1y}, \delta k'_{1y}}= e^{-\beta(\delta k_{1y}-\delta k'_{1y})^2}
    e^{-i \frac{\varphi_\text{BZ}}{6} (\frac{2\delta k_{1y}'}{q_2}- \frac{2\delta k_{1y}}{q_2})}.
\end{eqnarray}
Combining this with the exponential interaction $V_{\bm{k}}=V_0e^{-\alpha|\bm{k}|^2}$ with $\phi=\alpha+\beta$, we obtain for $g_{1,\delta k_{1y} } $
\begin{eqnarray}
    &\frac{\Omega_{tot}}{V_0} g_{1,\delta k_{1y} } =   e^{-\phi b_1^2} e^{- i \frac{\varphi_\text{BZ}}{6} \frac{2 \delta k_{1y}}{q_2} }  \big(   2Re[e^{-i\frac{\varphi_{\text{BZ}}}{6}} \mathcal{O}_{012} ] + 2 \sideset{}{'}\sum_{\mbf{k'_1}}  e^{ i \frac{\varphi_\text{BZ}}{6} \frac{2 \delta k'_{1y}}{q_2} }     O_{01,\mbf{k'_1}}  \big) \\ & -  e^{-\phi(\delta k_{1y}+\frac{q_2}{2})^2}  e^{ i \frac{\varphi_\text{BZ}}{6}(1+ \frac{2 \delta k_{1y}}{q_2} )}   \mathcal{O}_{012}  - e^{-\phi(-\delta k_{1y}+\frac{q_2}{2})^2}  e^{ -i \frac{\varphi_\text{BZ}}{6}(1-\frac{2 \delta k_{1y}}{q_2} )} \mathcal{O}_{012}^\star 
    \\& -  2 \sideset{}{'}\sum_{\mbf{k'_1}} e^{-\phi(\delta k_{1y}-\delta k'_{1y})^2}
    e^{-i \frac{\varphi_\text{BZ}}{6} (\frac{2\delta k_{1y}'}{q_2}- \frac{2\delta k_{1y}}{q_2})} O_{01,\mbf{k'_{1}}}
    \\& =N_{1,2}\bigg[
    2 e^{-\phi b_1^2}e^{- i \frac{\varphi_\text{BZ}}{6} \frac{2 \delta k_{1y}}{q_2} } Re[e^{-i\frac{\varphi_{\text{BZ}}}{6}} e^{i\frac{2\pi}{3}C}]-  e^{-\phi(\delta k_{1y}+\frac{q_2}{2})^2}  e^{ i \frac{\varphi_\text{BZ}}{6}(1+ \frac{2 \delta k_{1y}}{q_2} )} e^{i\frac{2\pi}{3}C} - e^{-\phi(-\delta k_{1y}+\frac{q_2}{2})^2}  e^{ -i \frac{\varphi_\text{BZ}}{6}(1-\frac{2 \delta k_{1y}}{q_2} )} e^{-i\frac{2\pi}{3}C}
    \bigg]
    \\& + 2 \sideset{}{'}\sum_{\mbf{k'_1}} \left( e^{-\phi b_1^2}e^{i \frac{\varphi_\text{BZ}}{6} (\frac{2\delta k_{1y}'}{q_2}- \frac{2\delta k_{1y}}{q_2})} - e^{-\phi(\delta k_{1y}-\delta k'_{1y})^2}
    e^{-i \frac{\varphi_\text{BZ}}{6} (\frac{2\delta k_{1y}'}{q_2}- \frac{2\delta k_{1y}}{q_2})}  \right) O_{01,\mbf{k'_1}},
\end{eqnarray} 
where we have used $\mathcal{O}_{012}=N_{1,2}e^{i\frac{2\pi}{3}C}$.

We now address the order parameter $O_{01,\bm{k}}$. From Eq.~\ref{eqapp:O10_k1}, we have
\begin{equation}
    O_{01,\bm{k}_1}=- \frac{1}{2} \frac{g_{1,\mbf{k_1}}}{\sqrt{|g_{1,\bm{k}_1}|^2+d^2_{z,\bm{k}_1}}}=- \frac{1}{2} \frac{g_{1,\delta k_{1y}}}{\sqrt{|g_{1,\delta k_{1y}}|^2+d^2_{z,\bm{k}_1}}},
\end{equation}
where we have used the fact that $g_{1,\bm{k}_1}$ does not depend on $\delta k_{1x}$ in the thin sliver approximation used here. Since we expect that $d_{z,\bm{k}_1}$ will depend much more strongly on $\delta{k}_{1x}$ than $\delta{k}_{1y}$, we only retain the $\delta {k}_{1x}$ dependence and parameterize
\begin{equation}
    d_{z,\bm{k}_1}\approx -v'\delta{k}_{1x}
\end{equation}
in terms of an effective velocity $v'$. This means that flat bottom of the effective dispersion is treated as hexagonal, which is expected to only quantitatively influence the results compared to using a circular flat bottom for strong enough interactions. Note that $v'$ has a contribution $\frac{v}{2}$ from the bare kinetic energy, as well as additional interaction-induced contributions. From the approximate expression in Eq.~\ref{eqapp:steepd0dz}, where $v'=\frac{1}{2}(v+2b_1\mathcal{O})$, it can be seen that the value of $v'$ should not depend significantly on the Chern number of the mean-field insulator. The integrals over $\delta k_{1x}$ can now be performed analytically
\begin{align}
    O_{01,\delta k_{1y}}\equiv \sum_{\delta k_{1x}}O_{01,\bm{k}_1}&=-\frac{1}{2}g_{1,\delta k_{1y}}\frac{L_x}{2\pi}\int_0^\Lambda dk_x\frac{1}{\sqrt{v'^2 k_x^2+|g_{1,\delta k_{1y}}|^2}}
    =-\frac{1}{2}g_{1,\delta k_{1y}}\frac{L_x}{2\pi v'}\text{arctanh}\frac{1}{\sqrt{1+\left|\frac{g_{1,\delta k_{1y}}}{v'\Lambda}\right|^2}}\\
    &\approx -\frac{1}{2}g_{1,\delta k_{1y}}\frac{L_x}{2\pi v'}\log\left(\frac{2v'\Lambda}{|g_{1,\delta k_{1y}}|}\right)\label{eqapp:O01_log}
\end{align}
where the $\approx$ on the last line is accurate for small $\frac{|g_{1,\delta k_{1y}}|}{v'\Lambda}$. We further introduce a variable $b_{\delta k_{1y}}$ that is an even function of $\delta k_{1y}$ and depends on the hybridization field
\begin{gather}\label{eqapp:bk1y}
    b_{\delta k_{1y}}\equiv \frac{1}{(2\pi)^2 v'}\log\left(\frac{2v'\Lambda}{|g_{1,\delta k_{1y}}|}\right)\\
    O_{01,\delta k_{1y}}=-\frac{1}{2}b_{\delta k_{1y}}g_{1,\delta k_{1y}}2\pi L_x.
\end{gather}
For analytical progress later, we will further consider the approximation where we approximate $b_{\delta k_{1y}}\approx b_{\delta k_{1y}=0}\equiv b$ by its value at $\delta k_{1y}=0$
\begin{gather}
    b\equiv \frac{1}{(2\pi)^2 v'}\log\left(\frac{2v'\Lambda}{|g_{1,\delta k_{1y}=0}|}\right).
\end{gather}

We insert the above into the expressions for $g_{1,\delta k_{1y}}$ to obtain
\begin{eqnarray}\label{eq:g1_GMP_general}
    &\frac{\Omega_{tot}}{V_0} g_{1,\delta k_{1y} }=N_{1,2}\bigg[
    2 e^{-\phi b_1^2}e^{- i \frac{\varphi_\text{BZ}}{6} \frac{2 \delta k_{1y}}{q_2} } \cos\left(-\frac{\varphi_{\text{BZ}}}{6}+\frac{2\pi C}{3}\right) \\
    &-  e^{-\phi(\delta k_{1y}+\frac{q_2}{2})^2}  e^{ i \frac{\varphi_\text{BZ}}{6}(1+ \frac{2 \delta k_{1y}}{q_2} )} e^{i\frac{2\pi}{3}C} \\
    &- e^{-\phi(-\delta k_{1y}+\frac{q_2}{2})^2}  e^{ -i \frac{\varphi_\text{BZ}}{6}(1-\frac{2 \delta k_{1y}}{q_2} )} e^{-i\frac{2\pi}{3}C}
    \bigg]
    \\& -b\Omega_{tot}\int_{-\frac{q_2}{2}}^{\frac{q_2}{2}}d\delta k'_{1y} \left( e^{-\phi b_1^2}e^{i \frac{\varphi_\text{BZ}}{6} (\frac{2\delta k_{1y}'}{q_2}- \frac{2\delta k_{1y}}{q_2})} - e^{-\phi(\delta k_{1y}-\delta k'_{1y})^2}
    e^{-i \frac{\varphi_\text{BZ}}{6} (\frac{2\delta k_{1y}'}{q_2}- \frac{2\delta k_{1y}}{q_2})}  \right) g_{1,\delta k'_{1y}},
\end{eqnarray}
which is now a single complex integral equation. Effectively, the $\bm{k}_{12}$ region acts as a $C$-dependent known function in the integral equation for $g_{1,\delta k_{1y}}$.

We now revisit the $\bm{k}_1$ contribution $E_{\text{tot},1}$ to the total energy expression $E_{\text{tot}}$ in Eq.~\ref{eqapp:Etot_distinguishC}. Using $O_{11 \mbf{k_1}}= \frac{1}{2} (1+ \frac{d_{z\mbf{k_1}}}{\sqrt{ d_{z\mbf{k_1}}^2 +| g_{1,\mbf{k_1}}|^2}})$ from App.~\ref{subsecapp:gensolnk1}, we have
\begin{align}
    E_\text{tot,1}&=
    6 \sideset{}{'}\sum_{\mbf{k_1}} \bigg[(E_{\mbf{k_1+ b_{1}}} + \frac{1}{2} f_{\mbf{k_1+ b_{1}}}-\frac{1}{2} f_{\mbf{k_1}}) O_{11 \mbf{k_1}}- \frac{1}{2} \frac{|g_{1,{\delta k_{1y}}}|^2}{|\bm{d}_{\bm{k}_1}|} \bigg]\\
    &=\text{const.}+6 \sideset{}{'}\sum_{\mbf{k_1}} \bigg[-\frac{1}{2}v'(v'+\frac{v}{2})\delta {k^2_{1x}} \frac{1}{\sqrt{|g_{1,\mbf{k_1}}|^2+(v'\delta k_{1x})^2}}- \frac{1}{2} \frac{|g_{1,\mbf{k_1}}|^2}{\sqrt{|g_{1,\mbf{k_1}}|^2+(v'\delta k_{1x})^2}}\bigg]\\
    &=\text{const.}+6 \sideset{}{'}\sum_{\mbf{k_1}} \bigg[-\frac{1}{2} \sqrt{ (v' \delta k_{1x})^2 +| g_{1,\delta k_{1y}}|^2} -  \frac{1}{2}\frac{v}{2v'} \frac{ (v' \delta k_{1x})^2 }{\sqrt{ (v' \delta k_{1x})^2 +| g_{1,\delta k_{1y}}|^2}}\bigg]\\
    &=\text{const.}-    3 \frac{\Omega_{tot}}{ (2 \pi)^2} v' \int_{- \frac{q_2}{2}}^{ \frac{q_2}{2}} d\delta k_{1y}   \int_0^{\Lambda_x } d\delta k_{1x}  \bigg[ \sqrt{  \delta k_{1x}^2 +\frac{| g_{1,\delta k_{1y}}|^2}{v'^2} } +   \frac{v}{2v'}\frac{  \delta k_{1x}^2 }{\sqrt{  \delta k_{1x}^2 +\frac{ | g_{1,\delta k_{1y}}|^2}{v'^2}}}  \bigg]
\end{align}
where we have neglected terms that do not depend on the Chern number. As a reminder, $v$ is the bare kinetic velocity, while $2v'$ is the effective interaction-renormalized velocity. Using the integrals
\begin{eqnarray}\label{eqapp:Ek1_integrals}
   &\int_0^{\Lambda_x } d\delta k_{1x}   \sqrt{  \delta k_{1x}^2 +\frac{| g_{1,\delta k_{1y}}|^2}{v'^2} } =\frac{1}{2} \left(\Lambda_x \sqrt{\frac{ | g_{1,\delta k_{1y}}|^2}{v'^2}+\Lambda_x^2}+\frac{ | g_{1,\delta k_{1y}}|^2}{v'^2} \tanh ^{-1}\left(\frac{\Lambda_x}{\sqrt{\frac{ | g_{1,\delta k_{1y}}|^2}{v'^2}+\Lambda_x^2}}\right)\right) \\ & \int_0^{\Lambda_x } d\delta k_{1x}    \frac{  \delta k_{1x}^2 }{\sqrt{  \delta k_{1x}^2 +\frac{ | g_{1,\delta k_{1y}}|^2}{v'^2}}}  = \frac{1}{2} \left(\Lambda_x \sqrt{\frac{ | g_{1,\delta k_{1y}}|^2}{v'^2}+\Lambda_x^2}-\frac{ | g_{1,\delta k_{1y}}|^2}{v'^2} \tanh ^{-1}\left(\frac{\Lambda_x}{\sqrt{\frac{ | g_{1,\delta k_{1y}}|^2}{v'^2}+\Lambda_x^2}}\right)\right),  
\end{eqnarray}
we obtain
\begin{align}
    E_\text{tot,1}&=\text{const.}-    \frac{3}{2} \frac{\Omega_{tot}}{ (2 \pi)^2} v' \int_{- \frac{q_2}{2}}^{ \frac{q_2}{2}} d\delta k_{1y}    \left[ \left(1+\frac{v}{2v'}\right)\Lambda_x^2 \sqrt{\frac{ | g_{1,\delta k_{1y}}|^2}{v'^2\Lambda_x^2}+1}+\left(1-\frac{v}{2v'}\right)\frac{ | g_{1,\delta k_{1y}}|^2}{v'^2} \tanh ^{-1}\left(\frac{1}{\sqrt{\frac{ | g_{1,\delta k_{1y}}|^2}{v'^2\Lambda_x^2}+1}}\right) \right].
\end{align}
In the case where the kinetic velocity dominates over the interaction-induced contribution, we have $v'\simeq \frac{v}{2}$, leading to
\begin{align}\label{eqapp:Etot1_kineticv}
    E_\text{tot,1}&\simeq\text{const.}-        \frac{3}{2} \frac{\Omega_{tot}}{ (2 \pi)^2v'}  \int_{- \frac{q_2}{2}}^{ \frac{q_2}{2}} d\delta k_{1y}    | g_{1,\delta k_{1y}}|^2,
\end{align}
where we consider the regime  $|g_1|\ll \Lambda_x v'$.

For completeness, we also write down the $\bm{k}_{12}$ contribution $E_{\text{tot},12}$ to the total mean-field energy $E_{\text{tot}}=E_{\text{tot},1}+E_{\text{tot},12}$
\begin{align}\label{eqapp:Etot12_N12}
    E_{\text{tot},12}=6N_{1,2}
    \text{Re}\left[g_{12}e^{-i\frac{2\pi}{3}C}\right],
\end{align}
where we have used $\mathcal{O}_{012}=N_{1,2}e^{i\frac{2\pi}{3}C}$.

In App.~\ref{subsecapp:k1_k12_HF_phi0} to \ref{secapp:fockonly_linearphi}, we solve the self-consistency equation Eq.~\ref{eq:g1_GMP_general} and evaluate the energetic competition between the different Chern states in various limits. Note that in App.~\ref{subsecapp:full_Chern}, we discuss how the full Chern number $C$ (rather than just $C\mod6$) can be determined given the solution of the above mean-field problem.

\subsubsection{Hartree and Fock, $\phi=0$}
\label{subsecapp:k1_k12_HF_phi0}

We consider the limit $\phi=0$ of Eq.~\ref{eq:g1_GMP_general}
\begin{eqnarray}
    &\frac{\Omega_{tot}}{V_0} g_{1,\delta k_{1y} }=4N_{1,2}\bigg[
    \cos (\frac{\varphi_\text{BZ}}{6} \frac{2 \delta k_{1y}}{q_2}) \sin (\frac{\varphi_\text{BZ}}{6}) \sin (\frac{2\pi C}{3})-i\sin (\frac{\varphi_\text{BZ}}{6} \frac{2 \delta k_{1y}}{q_2}) \cos (\frac{\varphi_\text{BZ}}{6}) \cos (\frac{2\pi C}{3})
    \bigg]
    \\& -2ib\Omega_{tot}\int_{-\frac{q_2}{2}}^{\frac{q_2}{2}}d\delta k'_{1y}
    \sin  \left(\frac{\varphi_\text{BZ}}{6} (\frac{2\delta k_{1y}'}{q_2}- \frac{2\delta k_{1y}}{q_2})  \right) g_{1,\delta k'_{1y}}.
\end{eqnarray}
To simplify further, we split into real and imaginary parts, and use the fact that $\text{Re}\,g_{1,\delta k_{1y}}$ is even while $\text{Im}\,g_{1,\delta k_{1y}}$ is odd from $M_1 \mathcal{T}$ symmetry, leading to
\begin{eqnarray}
    &\frac{\Omega_{tot}}{V_0} \text{Re}\,g_{1,\delta k_{1y} }=4N_{1,2}
    \cos (\frac{\varphi_\text{BZ}}{6} \frac{2 \delta k_{1y}}{q_2}) \sin (\frac{\varphi_\text{BZ}}{6}) \sin (\frac{2\pi C}{3})
     +2b\Omega_{tot}\cos  (\frac{\varphi_\text{BZ}}{6} \frac{2\delta k_{1y}}{q_2} )\int_{-\frac{q_2}{2}}^{\frac{q_2}{2}}d\delta k'_{1y}
    \sin  (\frac{\varphi_\text{BZ}}{6} \frac{2\delta k_{1y}'}{q_2} ) \text{Im}\,g_{1,\delta k'_{1y}}\\
    &\frac{\Omega_{tot}}{V_0} \text{Im}\,g_{1,\delta k_{1y} }=-4N_{1,2}
    \sin (\frac{\varphi_\text{BZ}}{6} \frac{2 \delta k_{1y}}{q_2}) \cos(\frac{\varphi_\text{BZ}}{6}) \cos (\frac{2\pi C}{3})
     +2b\Omega_{tot}\sin  (\frac{\varphi_\text{BZ}}{6} \frac{2\delta k_{1y}}{q_2} )\int_{-\frac{q_2}{2}}^{\frac{q_2}{2}}d\delta k'_{1y}
    \cos  (\frac{\varphi_\text{BZ}}{6} \frac{2\delta k_{1y}'}{q_2} ) \text{Re}\,g_{1,\delta k'_{1y}}.
\end{eqnarray}
We notice that the real and imaginary parts of the hybridization field varies with $\delta k_{1y}$ according to $\cos (\frac{\varphi_\text{BZ}}{6} \frac{2 \delta k_{1y}}{q_2})$ and $\sin(\frac{\varphi_\text{BZ}}{6} \frac{2 \delta k_{1y}}{q_2})$ respectively. We therefore introduce a complex quantity $G$ which itself depends on the hybridization field
\begin{eqnarray}
    &\text{Re}\,g_{1,\delta k_{1y} }=2\text{Re}G \cos (\frac{\varphi_\text{BZ}}{6} \frac{2 \delta k_{1y}}{q_2}),\quad \text{Im}\,g_{1,\delta k_{1y} }=2\text{Im}G \sin (\frac{\varphi_\text{BZ}}{6} \frac{2 \delta k_{1y}}{q_2})\\
    &\frac{\Omega_{tot}}{V_0}\text{Re}G=2N_{1,2}
     \sin (\frac{\varphi_\text{BZ}}{6}) \sin (\frac{2\pi C}{3})
     +2b\Omega_{tot}\text{Im}G\int_{-\frac{q_2}{2}}^{\frac{q_2}{2}}d\delta k'_{1y}
    \sin^2  (\frac{\varphi_\text{BZ}}{6} \frac{2\delta k_{1y}'}{q_2} ) \\
    &\frac{\Omega_{tot}}{V_0}\text{Im}G=-2N_{1,2}
     \cos (\frac{\varphi_\text{BZ}}{6}) \cos (\frac{2\pi C}{3})
     +2b\Omega_{tot}\text{Re}G\int_{-\frac{q_2}{2}}^{\frac{q_2}{2}}d\delta k'_{1y}
    \cos^2  (\frac{\varphi_\text{BZ}}{6} \frac{2\delta k_{1y}'}{q_2} ) .
\end{eqnarray}
Evaluating the integrals, we find
\begin{eqnarray}
    &\begin{pmatrix}
\text{Re}G  \\
\text{Im}G 
\end{pmatrix}= \frac{1}{\text{Det}} \begin{pmatrix}
1 & V_0    b q_2         (  1-\frac{3  \sin \left(\frac{\varphi_{\text{BZ}}}{3}\right)}{ \varphi_{\text{BZ}}}) \\
 V_0 b q_2 (  1+\frac{3  \sin \left(\frac{\varphi_{\text{BZ}}}{3}\right)}{ \varphi_{\text{BZ}}}) & 1 
\end{pmatrix}\begin{pmatrix}
2\frac{N_{12} }{\Omega_{tot}}          V_0 \sin(\frac{2\pi}{3} C)\sin( \frac{\varphi_{\text{BZ}}}{6})   \\
  - 2\frac{N_{12}}{\Omega_{tot}} V_0        \cos(\frac{2\pi}{3} C)\cos( \frac{\varphi_{\text{BZ}}}{6}) 
\end{pmatrix} \\ & \text{Det}=1 - (V_0    b q_2 )^2         \left(  1- \left(\frac{3  \sin \left(\frac{\varphi_{\text{BZ}}}{3}\right)}{ \varphi_{\text{BZ}}}\right)^2\right).
\end{eqnarray}
Note that for $\varphi_\text{BZ}=0$, there is no non-trivial solution so that $g_{1,\delta k_{1y}}=0$, i.e.~a gapless state with no translation symmetry-breaking. For finite $\varphi_{\text{BZ}}$, the expression for $\text{Re}G$ is still implicit since $b$ depends on  $\text{Re}G$ according to
\begin{eqnarray}
&b = \frac{1}{(2\pi)^2 v'}\log\left(\frac{v'\Lambda}{|\text{Re}G|}\right).
\end{eqnarray}

\begin{figure}
    \centering
    \includegraphics[width = 1.0\linewidth]{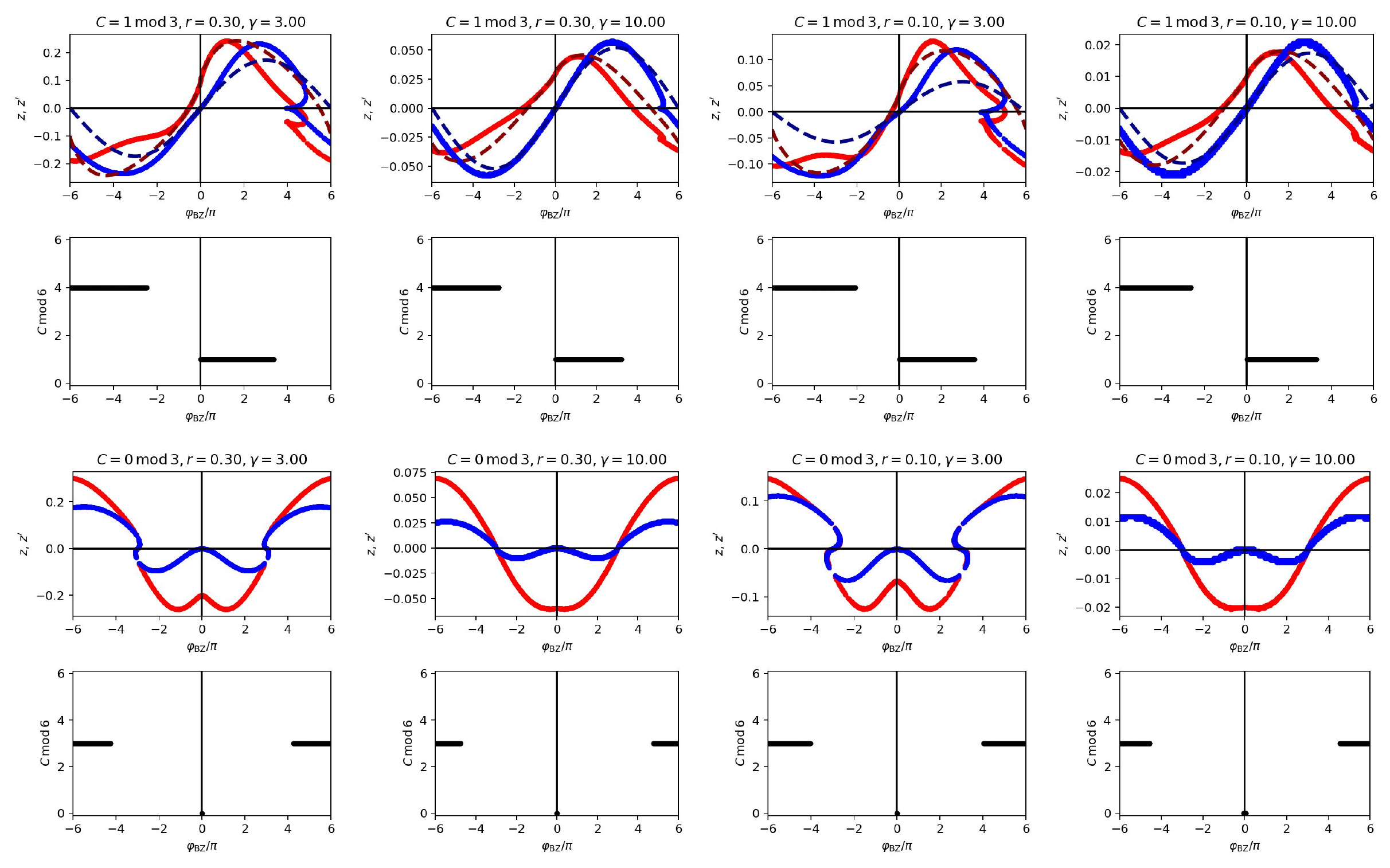}
    \caption{First and third rows show solutions for $z$ (blue, see Eq.~\ref{eqapp:k1_k12_phi0_z}) and $z'$ (red, see Eq.~\ref{eqapp:k1_k12_phi0_zprime}) as a function of $\varphi_{\text{BZ}}$ for different values of $C\mod 3$, $\gamma=\frac{(2\pi)^2v'}{q_2V_0}$ and $r=\frac{A_{k_{12}}}{A_{k_1}}$. For $C=1\mod 3$, the dashed lines indicate the corresponding approximations in Eq.~\ref{eqapp:HF_phi0_z_smallvarphi} and \ref{eqapp:HF_phi0_zprime_smallvarphi}. Second and fourth rows show corresponding values of $C\mod 6$, if self-consistent. Results for $C=-1\mod 3$ can be obtained by taking both $C\rightarrow -C$ and $\varphi_\text{BZ}\rightarrow -\varphi_\text{BZ}$. }
    \label{figapp:k1_k12_phi0_ReImg}
\end{figure}

We cast the equation for $\text{Re}G$ into dimensionless form. We define dimensionless quantities $z=\frac{\text{Re}G}{v'\Lambda}$, $\gamma=\frac{(2\pi)^2v'}{q_2V_0}$, and the momentum areas of the $\bm{k}_1$ and $\bm{k}_{12}$ regions as $A_{k_1}=q_2\Lambda$ and $A_{k_{12}}=N_{1,2}\frac{(2\pi)^2}{\Omega_{tot}}$, so that $V_0bq_2=-\frac{\log|z|}{\gamma}$. We let $r=\frac{A_{k_{12}}}{A_{k_1}}$ be the ratio of the momentum areas, which is expected to be small. We find after inserting the expression for $b$ the implicit equation
\begin{eqnarray}\label{eqapp:k1_k12_phi0_z}
    &z=\frac{\frac{2r}{\gamma}}{1-\left(\frac{\log |z|}{\gamma}\right)^2\left(1-\left(\frac{3  \sin \left(\frac{\varphi_{\text{BZ}}}{3}\right)}{ \varphi_{\text{BZ}}}\right)^2\right)}\left[\sin(\frac{2\pi}{3} C)\sin( \frac{\varphi_{\text{BZ}}}{6})+\cos(\frac{2\pi}{3} C)\cos( \frac{\varphi_{\text{BZ}}}{6})\frac{\log |z|}{\gamma}\left(1-\frac{3  \sin \left(\frac{\varphi_{\text{BZ}}}{3}\right)}{ \varphi_{\text{BZ}}}\right)  \right].
\end{eqnarray}
Recall the expression for $b$ used here is valid if $|z|\ll 1$, which motivates considering large $\gamma$. The solution is invariant under taking both $C\rightarrow -C$ and $\varphi_{\text{BZ}}\rightarrow -\varphi_{\text{BZ}}$. We also write the dimensionless expression for $z'=\frac{\text{Im}G}{v'\Lambda}$
\begin{eqnarray}\label{eqapp:k1_k12_phi0_zprime}
    &z'=-\frac{\frac{2r}{\gamma }}{1-\left(\frac{\log |z|}{\gamma}\right)^2\left(1-\left(\frac{3  \sin \left(\frac{\varphi_{\text{BZ}}}{3}\right)}{ \varphi_{\text{BZ}}}\right)^2\right)}\left[\sin(\frac{2\pi}{3} C)\sin( \frac{\varphi_{\text{BZ}}}{6})\frac{\log |z|}{\gamma}\left(1+\frac{3  \sin \left(\frac{\varphi_{\text{BZ}}}{3}\right)}{ \varphi_{\text{BZ}}}\right)+\cos(\frac{2\pi}{3} C)\cos( \frac{\varphi_{\text{BZ}}}{6})  \right],
\end{eqnarray}
which is also invariant under taking both $C\rightarrow -C$ and $\varphi_{\text{BZ}}\rightarrow -\varphi_{\text{BZ}}$. It can be shown that we recover the $\bm{k}_{12}$-only limit of Eq.~\ref{eqapp:k12only_g12_phasesonly} if we neglect all the $\log|z|$ terms arising from the $\bm{k}_1$ region.  Note that if $|z|=|z'|$, then the replacement of $b_{\delta k_{1y}}$ Eq.~\ref{eqapp:bk1y} with a $\delta k_{1y}$-independent $b$ does not incur any errors.
In the first and third rows of Fig.~\ref{figapp:k1_k12_phi0_ReImg}, we plot $z$ and $z'$ as a function of $\varphi_{\text{BZ}}$ for different values of $\gamma$, $r$, and $C\mod 3$. 

We also consider the self-consistency conditions at the high-symmetry points. At the $M_M$ points, we recall that even (odd) $C$ requires negative (positive) $g_{1,\delta k_{1y}=0}$. Hence, an even (odd) $C$ requires negative (positive) $z$. For the $K_M$ points, recall that the lowest energy solution of the mean-field Hamiltonian at $\bm{k}=\bm{q}_2$ corresponds to $C=m\mod 3$ with $m$ satisfying $\text{arg}(g_{12}e^{-i\frac{2(m+1)\pi}{3}})\in[0,\frac{2\pi}{3}]$. Here, we have $g_{12}=g_{1,\delta k_{1y}=-\frac{q_2}{2}}\propto z\cos( \frac{\varphi_{\text{BZ}}}{6})-iz'\sin( \frac{\varphi_{\text{BZ}}}{6})$. From these considerations, we can determine which values of $C\mod 6$ are self-consistent. In the second and fourth rows of Fig.~\ref{figapp:k1_k12_phi0_ReImg}, we plot the self-consistent value of $C\mod6$, assuming either $C=0$ or $1\mod3$ (the corresponding results for $C=-1\mod 3$ can be obtained by taking both $C\rightarrow -C$ and $\varphi_\text{BZ}\rightarrow -\varphi_\text{BZ}$). We note that $C=0\mod 3$ does not yield a self-consistent solution for small values of $\varphi_\text{BZ}$. For small/moderate values of $\varphi_{\text{BZ}}>0$, we find that $C=1\mod 6$ is the unique self-consistent solution. This means that for such values of $\varphi_{\text{BZ}}$, the gapped mean-field solution has $C=1\mod6$. 

In fact, by considering the analysis of App.~\ref{subsecapp:full_Chern}, we find that this $C=1\mod 6$ solution has Chern number $C=1$. To see this, consider the relative phase $\beta_{\delta k_{1y}}$ between the coefficients of the Bloch state at $\bm{k}=(-b_1/2,\delta k_{1y})$ and $(b_1/2,\delta k_{1y})$ in the HF wavefunction. Since $z,z'>0$,  $\beta_{\delta k_{1y}}$ starts at $\pi$ at $\delta k_{1y}=0$, and decreases to $\pi/2<\beta_{\delta k_{1y}}<\pi$ for $\delta k_{1y}\rightarrow -q_2/2$ in the $\bm{k}_1$ region. Hence, according to the discussion of App.~\ref{subsecapp:full_Chern}, $\beta_{\delta k_{1y}}$ winds from $2\pi/3$ to $4\pi/3$ as we traverse the entire left edge of the BZ vertically upwards, leading to $C=1$ (see Eq.~\ref{eqapp:full_C_beta}).

To obtain analytical approximations, we consider the case of small $\varphi_{\text{BZ}}$, where $\frac{3  \sin \left(\frac{\varphi_{\text{BZ}}}{3}\right)}{\varphi_{\text{BZ}}}\simeq 1$, and we can massage the expressions for $\text{Re}G$ and $\text{Im}G$ into a form that is consistent to linear order in $\varphi_{\text{BZ}}$
\begin{eqnarray}\label{eqapp:HF_phi0_z_smallvarphi}
&z\approx \frac{2r}{\gamma } \sin(\frac{2\pi}{3} C)\sin( \frac{\varphi_{\text{BZ}}}{6}) \\
&z'\approx -\frac{2r}{\gamma }\left( 2\frac{\log |z|}{\gamma}\sin(\frac{2\pi}{3} C)\sin( \frac{\varphi_{\text{BZ}}}{6})+\cos(\frac{2\pi}{3} C)\cos( \frac{\varphi_{\text{BZ}}}{6})  \right).\label{eqapp:HF_phi0_zprime_smallvarphi}
\end{eqnarray}
Note that in this limit, the $C=0\mod 3$ solution is gapless since $z$ vanishes. Therefore for $C=0\mod 3$, $z$ only develops a non-zero value  beyond first order in $\varphi_{\text{BZ}}$, as reflected in Fig.~\ref{figapp:k1_k12_phi0_ReImg}. For $C=1\mod 3$, we plot the values of $z$ and $z'$ with dashed lines, and find good agreement with the general numerical solution of Eq.~\ref{eqapp:k1_k12_phi0_z} and \ref{eqapp:k1_k12_phi0_zprime} for a range of $\varphi_{\text{BZ}}$. 

We now evaluate the part of the total mean-field energy that depends on the Chern number (see Eq.~\ref{eqapp:Etot_distinguishC}). We begin with the $\bm{k}_{12}$ contribution from Eq.~\ref{eqapp:Etot12_N12}
\begin{eqnarray}
    &E_{\text{tot},12}=12N_{1,2}\left(
    \text{Re}G \cos(\frac{2\pi}{3} C)\cos( \frac{\varphi_{\text{BZ}}}{6})  - \text{Im}G\sin(\frac{2\pi}{3} C)\sin( \frac{\varphi_{\text{BZ}}}{6})  
    \right).
\end{eqnarray}
Note that for $\phi=0$, we have that $v'=\frac{v}{2}$ is entirely set by the kinetic velocity (see App.~\ref{subsubsecapp:f_GMP_exp}), so  Eq.~\ref{eqapp:Etot1_kineticv} holds.  Considering the regime $|z|\ll 1$, we obtain for the $\bm{k}_1$ contribution to the mean-field energy 
\begin{eqnarray}
    &E_{\text{tot},1}=\text{const.}-  3 \frac{q_2\Omega_{tot}}{ (2 \pi)^2v'}  \left[(\text{Re}G)^2+(\text{Im}G)^2+\frac{3\sin(\frac{\varphi_{\text{BZ}}}{3})}{\varphi_{\text{BZ}}}\left((\text{Re}G)^2-(\text{Im}G)^2\right)\right].
\end{eqnarray}

\subsubsection{Hartree and Fock, linear order in $\phi$}\label{subsecapp:k1_k12_linearphi_HF}

In this section, we retain terms up to linear order in $\phi$ in Eq.~\ref{eq:g1_GMP_general}
\begin{eqnarray}
    &\frac{\Omega_{tot}}{V_0} g_{1,k}=N_{1,2}\bigg[
    2 (1-\phi b_1^2)e^{- i \frac{\varphi_\text{BZ}}{6} \frac{2 k}{q_2} } \cos\left(-\frac{\varphi_{\text{BZ}}}{6}+\frac{2\pi C}{3}\right) \\
    &- 2e^{ i \frac{\varphi_\text{BZ}}{6} \frac{2 k}{q_2} }\cos(\frac{\varphi_\text{BZ}}{6}+\frac{2\pi}{3}C)+ 2\phi(k^2+\frac{q_2^2}{4})e^{ i \frac{\varphi_\text{BZ}}{6} \frac{2 k}{q_2} }\cos(\frac{\varphi_\text{BZ}}{6}+\frac{2\pi}{3}C)+2i\phi kq_2 e^{ i \frac{\varphi_\text{BZ}}{6} \frac{2 k}{q_2} }\sin(\frac{\varphi_\text{BZ}}{6}+\frac{2\pi}{3}C)\bigg]
    \\& -b\Omega_{tot}\int_{-\frac{q_2}{2}}^{\frac{q_2}{2}}dk' \bigg(2i \sin\left(\frac{\varphi_\text{BZ}}{6} (\frac{2k'}{q_2}- \frac{2k}{q_2})\right)  \\
    &-\phi b_1^2e^{i \frac{\varphi_\text{BZ}}{6} (\frac{2k'}{q_2}- \frac{2k}{q_2})} +\phi(k-k')^2
    e^{-i \frac{\varphi_\text{BZ}}{6} (\frac{2k'}{q_2}- \frac{2k}{q_2})} \bigg) g_{1,k'},
\end{eqnarray}
where we have made the notational replacement $\delta k_{1y}\rightarrow k$ for simplicity. We split the order parameter into (symmetric) real and (anti-symmetric) imaginary parts
\begin{eqnarray}
    g_{1,k}=R_{k}+iI_{k},
\end{eqnarray}
which satisfy
\begin{eqnarray}
    &\frac{\Omega_{tot}}{V_0} R_k=N_{1,2}\bigg[
    2 (1-\phi b_1^2)\cos(\frac{\varphi_\text{BZ}}{6}\frac{2k}{q_2})\cos\left(-\frac{\varphi_{\text{BZ}}}{6}+\frac{2\pi C}{3}\right) \\
    &- 2\cos(\frac{\varphi_\text{BZ}}{6}\frac{2k}{q_2})\cos(\frac{\varphi_\text{BZ}}{6}+\frac{2\pi}{3}C)+ 2\phi(k^2+\frac{q_2^2}{4})\cos(\frac{\varphi_\text{BZ}}{6}\frac{2k}{q_2})\cos(\frac{\varphi_\text{BZ}}{6}+\frac{2\pi}{3}C)-2\phi kq_2 \sin(\frac{\varphi_\text{BZ}}{6}\frac{2k}{q_2})\sin(\frac{\varphi_\text{BZ}}{6}+\frac{2\pi}{3}C)\bigg]
    \\& -b\Omega_{tot}\int_{-\frac{q_2}{2}}^{\frac{q_2}{2}}dk' \bigg(-2\cos(\frac{\varphi_\text{BZ}}{6}\frac{2k}{q_2})\sin(\frac{\varphi_\text{BZ}}{6}\frac{2k'}{q_2})I_{k'} \\
    &+\phi(k^2+k'^2-b_1^2)\cos(\frac{\varphi_\text{BZ}}{6}\frac{2k}{q_2})\cos(\frac{\varphi_\text{BZ}}{6}\frac{2k'}{q_2})R_{k'}+\phi(k^2+k'^2+b_1^2)\cos(\frac{\varphi_\text{BZ}}{6}\frac{2k}{q_2})\sin(\frac{\varphi_\text{BZ}}{6}\frac{2k'}{q_2})I_{k'}\\
    &-2\phi kk'\left(\sin(\frac{\varphi_\text{BZ}}{6}\frac{2k}{q_2})\sin(\frac{\varphi_\text{BZ}}{6}\frac{2k'}{q_2})R_{k'}-\sin(\frac{\varphi_\text{BZ}}{6}\frac{2k}{q_2})\cos(\frac{\varphi_\text{BZ}}{6}\frac{2k'}{q_2})I_{k'}\right)\bigg)
\end{eqnarray}
\begin{eqnarray}
    &\frac{\Omega_{tot}}{V_0} I_k=N_{1,2}\bigg[
    -2 (1-\phi b_1^2)\sin(\frac{\varphi_\text{BZ}}{6}\frac{2k}{q_2})\cos\left(-\frac{\varphi_{\text{BZ}}}{6}+\frac{2\pi C}{3}\right) \\
    &-2\sin(\frac{\varphi_\text{BZ}}{6}\frac{2k}{q_2})\cos(\frac{\varphi_\text{BZ}}{6}+\frac{2\pi}{3}C)+ 2\phi(k^2+\frac{q_2^2}{4})\sin(\frac{\varphi_\text{BZ}}{6}\frac{2k}{q_2})\cos(\frac{\varphi_\text{BZ}}{6}+\frac{2\pi}{3}C)+2\phi kq_2 \cos(\frac{\varphi_\text{BZ}}{6}\frac{2k}{q_2})\sin(\frac{\varphi_\text{BZ}}{6}+\frac{2\pi}{3}C)\bigg]
    \\& -b\Omega_{tot}\int_{-\frac{q_2}{2}}^{\frac{q_2}{2}}dk' \bigg(-2\sin(\frac{\varphi_\text{BZ}}{6}\frac{2k}{q_2})\cos(\frac{\varphi_\text{BZ}}{6}\frac{2k'}{q_2})R_{k'} \\
    &+\phi(k^2+k'^2-b_1^2)\sin(\frac{\varphi_\text{BZ}}{6}\frac{2k}{q_2})\sin(\frac{\varphi_\text{BZ}}{6}\frac{2k'}{q_2})I_{k'}+\phi(k^2+k'^2+b_1^2)\sin(\frac{\varphi_\text{BZ}}{6}\frac{2k}{q_2})\cos(\frac{\varphi_\text{BZ}}{6}\frac{2k'}{q_2})R_{k'}\\
    &-2\phi kk'\left(\cos(\frac{\varphi_\text{BZ}}{6}\frac{2k}{q_2})\cos(\frac{\varphi_\text{BZ}}{6}\frac{2k'}{q_2})I_{k'}-\cos(\frac{\varphi_\text{BZ}}{6}\frac{2k}{q_2})\sin(\frac{\varphi_\text{BZ}}{6}\frac{2k'}{q_2})R_{k'}\right)\bigg).
\end{eqnarray}
We parameterize the solution in powers of $\phi$. We recall that $b$ also depends on the order parameter and should similarly be expanded in powers of $\phi$
\begin{eqnarray}
    & R_k=\mathcal{R}_{0,k}+\phi\mathcal{R}_{1,k}\\
    & I_k=\mathcal{I}_{0,k}+\phi\mathcal{I}_{1,k}\\
    & b=B_0+\phi B_1.
\end{eqnarray}
We can extract $B_0$ and $B_1$ in the following way (recalling that the order parameter is purely real at $k=0$):
\begin{eqnarray}
    & b=  \frac{1}{ v'}  \frac{1}{ (2 \pi)^2 }  \log\frac{2 v' \Lambda}{|\mathcal{R}_{0,0}+\phi \mathcal{R}_{1,0}|}=\frac{1}{ v'}  \frac{1}{ (2 \pi)^2 }  \log\frac{2 v' \Lambda}{|\mathcal{R}_{0,0}|(1+\phi \mathcal{R}_{1,0}/\mathcal{R}_{0,0})}\approx \frac{1}{ v'}  \frac{1}{ (2 \pi)^2 } \left( \log\frac{2 v' \Lambda}{|\mathcal{R}_{0,0}|}-\phi \frac{\mathcal{R}_{1,0}}{\mathcal{R}_{0,0}}\right)
\end{eqnarray}
leading to
\begin{eqnarray}
    & B_0=\frac{1}{ v'}  \frac{1}{ (2 \pi)^2 } \log\frac{2 v' \Lambda}{|\mathcal{R}_{0,0}|},\quad B_1=-\frac{1}{ v'}  \frac{1}{ (2 \pi)^2 }\frac{\mathcal{R}_{1,0}}{\mathcal{R}_{0,0}}.
\end{eqnarray}

The zeroth-order part of the self-consistent equation is
\begin{eqnarray}
    &\frac{\Omega_{tot}}{2V_0} \mathcal{R}_{0,k}=2N_{1,2}\cos(\frac{\varphi_\text{BZ}}{6}\frac{2k}{q_2}) \sin(\frac{\varphi_\text{BZ}}{6}) \sin(\frac{2\pi}{3}C) +B_0\Omega_{tot}\cos(\frac{\varphi_\text{BZ}}{6}\frac{2k}{q_2})\int_{-\frac{q_2}{2}}^{\frac{q_2}{2}}dk' \sin(\frac{\varphi_\text{BZ}}{6}\frac{2k'}{q_2})\mathcal{I}_{0,k'}
\end{eqnarray}
\begin{eqnarray}
    &\frac{\Omega_{tot}}{2V_0} \mathcal{I}_{0,k}=-2N_{1,2}\sin(\frac{\varphi_\text{BZ}}{6}\frac{2k}{q_2}) \cos(\frac{\varphi_\text{BZ}}{6}) \cos(\frac{2\pi}{3}C) +B_0\Omega_{tot}\sin(\frac{\varphi_\text{BZ}}{6}\frac{2k}{q_2})\int_{-\frac{q_2}{2}}^{\frac{q_2}{2}}dk' \cos(\frac{\varphi_\text{BZ}}{6}\frac{2k'}{q_2})\mathcal{R}_{0,k'}
\end{eqnarray}
which recovers the $\phi=0$ limit obtained previously in App.~\ref{subsecapp:k1_k12_HF_phi0}. The solution can be parameterized as
\begin{eqnarray}
    &\mathcal{R}_{0,k}=2G_R\cos (\frac{\varphi_{\text{BZ}}}{6} \frac{2 k}{q_2}),\quad \mathcal{I}_{0,k}=2G_I \sin (\frac{\varphi_{\text{BZ}}}{6} \frac{2 k}{q_2}),
\end{eqnarray}
where the real parameters $G_R$ and $G_I$ satisfy
\begin{eqnarray}
    &\frac{\Omega_{tot}}{V_0} G_R=2N_{1,2} \sin(\frac{\varphi_\text{BZ}}{6}) \sin(\frac{2\pi}{3}C) +2B_0\Omega_{tot}G_I q_2\xi_-
\end{eqnarray}
\begin{eqnarray}
    &\frac{\Omega_{tot}}{V_0} G_I=-2N_{1,2}
    \cos(\frac{\varphi_\text{BZ}}{6}) \cos(\frac{2\pi}{3}C) +2B_0\Omega_{tot}G_R q_2\xi_+,
\end{eqnarray}
where we have defined $\xi_\pm=\frac{1}{2}\pm\frac{3 \sin \left(\frac{\varphi_\text{BZ}}{3}\right)}{2 \varphi_\text{BZ}}$. We recall the definitions of the momentum areas of the $k_1$ and $k_{12}$ regions are $A_{k_1}=q_2\Lambda$ and $A_{k_{12}}=N_{1,2}\frac{(2\pi)^2}{\Omega_{tot}}$.  We also define dimensionless variables $r=\frac{A_{k_{12}}}{A_{k_1}}$ and $\gamma=\frac{(2\pi)^2v'}{q_2V_0}$. In terms of the dimensionless variables $z$ and $z'$, we have
\begin{eqnarray}
    &z=\frac{G_R}{v'\Lambda}=\frac{2}{\gamma}\left[
    r\sin(\frac{\varphi_\text{BZ}}{6}) \sin(\frac{2\pi}{3}C)-\xi_- z'\ln|z|
    \right]\\
    &z'=\frac{G_I}{v'\Lambda}=\frac{2}{\gamma}\left[
    -r\cos(\frac{\varphi_\text{BZ}}{6}) \cos(\frac{2\pi}{3}C)-\xi_+z\ln|z|
    \right].
\end{eqnarray}
In certain cases, $z$ and $z'$ can be solved for analytically as shown in App.~\ref{subsecapp:k1_k12_HF_phi0}. Regardless, we will assume from now on that we have (numerically) solved for $\mathcal{R}_{0,k},\mathcal{I}_{0,k},B_0$ in terms of $z$ and $z'$ (or equivalently $G_R$ and $G_I$).

The first-order part of the self-consistent equation is
\begin{eqnarray}
    &\frac{\Omega_{tot}}{2V_0} \mathcal{R}_{1,k}=N_{1,2}\bigg[
    -b_1^2\cos(\frac{\varphi_\text{BZ}}{6}\frac{2k}{q_2})\cos\left(-\frac{\varphi_{\text{BZ}}}{6}+\frac{2\pi C}{3}\right) \\
    &(k^2+\frac{q_2^2}{4})\cos(\frac{\varphi_\text{BZ}}{6}\frac{2k}{q_2})\cos(\frac{\varphi_\text{BZ}}{6}+\frac{2\pi}{3}C)- kq_2 \sin(\frac{\varphi_\text{BZ}}{6}\frac{2k}{q_2})\sin(\frac{\varphi_\text{BZ}}{6}+\frac{2\pi}{3}C)\bigg]
    \\& -B_0\Omega_{tot}\int_{-\frac{q_2}{2}}^{\frac{q_2}{2}}dk' \bigg(-\cos(\frac{\varphi_\text{BZ}}{6}\frac{2k}{q_2})\sin(\frac{\varphi_\text{BZ}}{6}\frac{2k'}{q_2})\mathcal{I}_{1,k'} \\
    &+\frac{1}{2}(k^2+k'^2-b_1^2)\cos(\frac{\varphi_\text{BZ}}{6}\frac{2k}{q_2})\cos(\frac{\varphi_\text{BZ}}{6}\frac{2k'}{q_2})\mathcal{R}_{0,k'}+\frac{1}{2}(k^2+k'^2+b_1^2)\cos(\frac{\varphi_\text{BZ}}{6}\frac{2k}{q_2})\sin(\frac{\varphi_\text{BZ}}{6}\frac{2k'}{q_2})\mathcal{I}_{0,k'}\\
    &- kk'\left(\sin(\frac{\varphi_\text{BZ}}{6}\frac{2k}{q_2})\sin(\frac{\varphi_\text{BZ}}{6}\frac{2k'}{q_2})\mathcal{R}_{0,k'}-\sin(\frac{\varphi_\text{BZ}}{6}\frac{2k}{q_2})\cos(\frac{\varphi_\text{BZ}}{6}\frac{2k'}{q_2})\mathcal{I}_{0,k'}\right)\bigg) \\& +B_1\Omega_{tot}\int_{-\frac{q_2}{2}}^{\frac{q_2}{2}}dk' \cos(\frac{\varphi_\text{BZ}}{6}\frac{2k}{q_2})\sin(\frac{\varphi_\text{BZ}}{6}\frac{2k'}{q_2})\mathcal{I}_{0,k'} 
\end{eqnarray}
\begin{eqnarray}
    &\frac{\Omega_{tot}}{2V_0} \mathcal{I}_{1,k}=N_{1,2}\bigg[
    b_1^2\sin(\frac{\varphi_\text{BZ}}{6}\frac{2k}{q_2})\cos\left(-\frac{\varphi_{\text{BZ}}}{6}+\frac{2\pi C}{3}\right) \\
    &
    +(k^2+\frac{q_2^2}{4})\sin(\frac{\varphi_\text{BZ}}{6}\frac{2k}{q_2})\cos(\frac{\varphi_\text{BZ}}{6}+\frac{2\pi}{3}C)+ kq_2 \cos(\frac{\varphi_\text{BZ}}{6}\frac{2k}{q_2})\sin(\frac{\varphi_\text{BZ}}{6}+\frac{2\pi}{3}C)\bigg]
    \\& -B_0\Omega_{tot}\int_{-\frac{q_2}{2}}^{\frac{q_2}{2}}dk' \bigg(-\sin(\frac{\varphi_\text{BZ}}{6}\frac{2k}{q_2})\cos(\frac{\varphi_\text{BZ}}{6}\frac{2k'}{q_2})\mathcal{R}_{1,k'} \\
    &+\frac{1}{2}(k^2+k'^2-b_1^2)\sin(\frac{\varphi_\text{BZ}}{6}\frac{2k}{q_2})\sin(\frac{\varphi_\text{BZ}}{6}\frac{2k'}{q_2})\mathcal{I}_{0,k'}+\frac{1}{2}(k^2+k'^2+b_1^2)\sin(\frac{\varphi_\text{BZ}}{6}\frac{2k}{q_2})\cos(\frac{\varphi_\text{BZ}}{6}\frac{2k'}{q_2})\mathcal{R}_{0,k'}\\
    &-kk'\left(\cos(\frac{\varphi_\text{BZ}}{6}\frac{2k}{q_2})\cos(\frac{\varphi_\text{BZ}}{6}\frac{2k'}{q_2})\mathcal{I}_{0,k'}-\cos(\frac{\varphi_\text{BZ}}{6}\frac{2k}{q_2})\sin(\frac{\varphi_\text{BZ}}{6}\frac{2k'}{q_2})\mathcal{R}_{0,k'}\right)\bigg)\\& +B_1\Omega_{tot}\int_{-\frac{q_2}{2}}^{\frac{q_2}{2}}dk' \sin(\frac{\varphi_\text{BZ}}{6}\frac{2k}{q_2})\cos(\frac{\varphi_\text{BZ}}{6}\frac{2k'}{q_2})\mathcal{R}_{0,k'}.
\end{eqnarray}
Using the following integrals (which implicitly define the dimensionless variables $\xi_\pm,\chi_\pm,\zeta$ which are functions solely of $\varphi_\text{BZ}$)
\begin{eqnarray}
    &  \int   dk' \sin( \frac{\varphi_\text{BZ}}{6} \frac{2 k'}{q_2} )^2= \frac{q_2}{2}-\frac{3 q_2 \sin \left(\frac{\varphi_\text{BZ}}{3}\right)}{2 \varphi_\text{BZ}} = q_2\xi_-\approx q_2\times \frac{\varphi_{\text{BZ}}^2}{108} \nonumber \\  & \int   dk' \cos( \frac{\varphi_\text{BZ}}{6} \frac{2 k'}{q_2} )^2= \frac{q_2}{2}+ \frac{3 q_2 \sin \left(\frac{\varphi_\text{BZ}}{3}\right)}{2 \varphi_\text{BZ}}= q_2\xi_+ = q_2(1- \xi_-)\approx q_2 \nonumber \\ & \int   dk' k'^2 \sin( \frac{\varphi_\text{BZ}}{6} \frac{2 k'}{q_2} )^2 = \frac{q_2^3 \left(\varphi_\text{BZ}^3-9 \left(\varphi_\text{BZ}^2-18\right) \sin \left(\frac{\varphi_\text{BZ}}{3}\right)-54 \varphi_\text{BZ} \cos \left(\frac{\varphi_\text{BZ}}{3}\right)\right)}{24 \varphi_\text{BZ}^3} =q_2^3\chi_- \approx q_2^3\times \frac{\varphi_\text{BZ}^2}{720}\nonumber \\ & \int   dk' k'^2 \cos( \frac{\varphi_\text{BZ}}{6} \frac{2 k'}{q_2} )^2  = \frac{q_2^3 \left(\varphi_\text{BZ}^3+9 \left(\varphi_\text{BZ}^2-18\right) \sin \left(\frac{\varphi_\text{BZ}}{3}\right)+54 \varphi_\text{BZ} \cos \left(\frac{\varphi_\text{BZ}}{3}\right)\right)}{24 \varphi_\text{BZ}^3}= q_2^3\chi_+ = \frac{q_2^3}{12}- q_2^3\chi_- \approx \frac{q_2^3}{12} \nonumber \\ & \int   dk' k' \sin( \frac{\varphi_\text{BZ}}{6} \frac{2k'}{q_2} )  \cos( \frac{\varphi_\text{BZ}}{6} \frac{2 k'}{q_2} )= -\frac{3 q_2^2 \left(\varphi_\text{BZ} \cos \left(\frac{\varphi_\text{BZ}}{3}\right)-3 \sin \left(\frac{\varphi_\text{BZ}}{3}\right)\right)}{4 \varphi_\text{BZ}^2}= q_2^2\zeta\approx q_2^2\times \frac{\varphi_\text{BZ}}{36},
\end{eqnarray}
we can substitute in the zeroth order solutions and perform the corresponding integrals. Above, we have also displayed the lowest-order expansions for small $\varphi_\text{BZ}$. Separating out the equations into distinct functions of $k$ leads to
\begin{eqnarray}
    &\frac{\Omega_{tot}}{2V_0} \mathcal{R}_{1,k}=\bigg[-N_{1,2}
    b_1^2\cos\left(-\frac{\varphi_{\text{BZ}}}{6}+\frac{2\pi C}{3}\right)
+N_{1,2}\frac{q_2^2}{4}\cos(\frac{\varphi_\text{BZ}}{6}+\frac{2\pi}{3}C)\\
&B_0\Omega_{tot}\int_{-\frac{q_2}{2}}^{\frac{q_2}{2}}dk' \sin(\frac{\varphi_\text{BZ}}{6}\frac{2k'}{q_2})\mathcal{I}_{1,k'}
-q_2B_0\Omega_{tot}\left(G_R(q_2^2\chi_+-b_1^2\xi_+)+G_I(q^2_2\chi_-+b_1^2\xi_-)\right)+2q_2B_1\Omega_{tot}G_I\xi_-
    \bigg]\cos(\frac{\varphi_\text{BZ}}{6}\frac{2k}{q_2})\\
    &+\bigg[
    N_{1,2}\cos(\frac{\varphi_\text{BZ}}{6}+\frac{2\pi}{3}C)    
        -q_2B_0\Omega_{tot}(G_R\xi_++G_I\xi_-)
    \bigg]k^2\cos(\frac{\varphi_\text{BZ}}{6}\frac{2k}{q_2})\\
    &+\bigg[
    - N_{1,2}q_2 \sin(\frac{\varphi_\text{BZ}}{6}+\frac{2\pi}{3}C)
    +2q_2^2B_0\Omega_{tot}(G_R-G_I)\zeta
    \bigg]k\sin(\frac{\varphi_\text{BZ}}{6}\frac{2k}{q_2})
\end{eqnarray}
\begin{eqnarray}
    &\frac{\Omega_{tot}}{2V_0} \mathcal{I}_{1,k}=\bigg[N_{1,2}
    b_1^2\cos\left(-\frac{\varphi_{\text{BZ}}}{6}+\frac{2\pi C}{3}\right)
+N_{1,2}\frac{q_2^2}{4}\cos(\frac{\varphi_\text{BZ}}{6}+\frac{2\pi}{3}C)\\
&B_0\Omega_{tot}\int_{-\frac{q_2}{2}}^{\frac{q_2}{2}}dk' \cos(\frac{\varphi_\text{BZ}}{6}\frac{2k'}{q_2})\mathcal{R}_{1,k'}
-q_2B_0\Omega_{tot}\left(G_R(q_2^2\chi_++b_1^2\xi_+)+G_I(q_2^2\chi_--b_1^2\xi_-)\right)+2q_2B_1\Omega_{tot}G_R\xi_+
    \bigg]\sin(\frac{\varphi_\text{BZ}}{6}\frac{2k}{q_2})\\
    &+\bigg[
    N_{1,2}\cos(\frac{\varphi_\text{BZ}}{6}+\frac{2\pi}{3}C)    
        -q_2B_0\Omega_{tot}(G_R\xi_++G_I\xi_-)
    \bigg]k^2\sin(\frac{\varphi_\text{BZ}}{6}\frac{2k}{q_2})\\
    &+\bigg[
     N_{1,2}q_2 \sin(\frac{\varphi_\text{BZ}}{6}+\frac{2\pi}{3}C)
    -2q_2^2B_0\Omega_{tot}(G_R-G_I)\zeta
    \bigg]k\cos(\frac{\varphi_\text{BZ}}{6}\frac{2k}{q_2}).
\end{eqnarray}
It is clear that we can express the solutions of $\mathcal{R}_{1,k}$ and $\mathcal{I}_{1,k}$ as the following
\begin{eqnarray}
    & \mathcal{R}_{1,k}=\mathfrak{r}_0\cos( \frac{\varphi_\text{BZ}}{6}  \frac{2 k}{q_2})+\mathfrak{r}_1k^2\cos( \frac{\varphi_\text{BZ}}{6}  \frac{2 k}{q_2})+\mathfrak{r}_2k\sin(  \frac{\varphi_\text{BZ}}{6} \frac{2 k}{q_2} ) \\
    &\mathcal{I}_{1,k}=\mathfrak{i}_0\sin( \frac{\varphi_\text{BZ}}{6}  \frac{2 k}{q_2})+\mathfrak{i}_1k^2\sin( \frac{\varphi_\text{BZ}}{6}  \frac{2 k}{q_2})+\mathfrak{i}_2k\cos(  \frac{\varphi_\text{BZ}}{6} \frac{2 k}{q_2} ) .
\end{eqnarray}
Since $\mathcal{R}_{1,k}$ has dimensions of $[EL^{-2}]$, then $\mathfrak{r}_0,\mathfrak{r}_1,\mathfrak{r}_2$ has dimensions of $[EL^{-2}],[E],[EL^{-1}]$ respectively (and analogously for the imaginary components). 
In fact, four of these coefficients are determined already in terms of known quantities as
\begin{eqnarray}
    &\mathfrak{r}_1=\mathfrak{i}_1=\frac{2V_0}{\Omega_{tot}}\bigg[N_{1,2}\cos(\frac{\varphi_\text{BZ}}{6}+\frac{2\pi}{3}C)    
        -q_2B_0\Omega_{tot}(G_R\xi_++G_I\xi_-)\bigg]\\
     &\mathfrak{r}_2=-\mathfrak{i}_2=  \frac{2q_2V_0}{\Omega_{tot}}\bigg[- N_{1,2} \sin(\frac{\varphi_\text{BZ}}{6}+\frac{2\pi}{3}C)
    +2q_2B_0\Omega_{tot}(G_R-G_I)\zeta\bigg],
\end{eqnarray}
or in terms of dimensionless variables $w_1,w_2$
\begin{eqnarray}\label{eqapp:w1w2}
    &w_1=\frac{\mathfrak{r}_1}{v'\Lambda}=\frac{2}{\gamma}\left[
    r\cos(\frac{\varphi_\text{BZ}}{6}+\frac{2\pi}{3}C)+(z\xi_++z'\xi_-)\ln|z|
    \right]\\
     &w_2=\frac{\mathfrak{r}_2}{q_2v'\Lambda}=-\frac{2}{\gamma}\left[
    r\sin(\frac{\varphi_\text{BZ}}{6}+\frac{2\pi}{3}C)+2(z-z')\zeta\ln|z|
    \right],
\end{eqnarray}

There is no simple relation between $\mathfrak{r}_0$ and $\mathfrak{i}_0$. The equations for the latter quantities are coupled
\begin{eqnarray}
    &\frac{\Omega_{tot}}{2V_0} \mathfrak{r}_{0}=\bigg[-N_{1,2}
    b_1^2\cos\left(-\frac{\varphi_{\text{BZ}}}{6}+\frac{2\pi C}{3}\right)
+N_{1,2}\frac{q_2^2}{4}\cos(\frac{\varphi_\text{BZ}}{6}+\frac{2\pi}{3}C)\\
&q_2B_0\Omega_{tot}(\mathfrak{i}_0\xi_-+\mathfrak{i}_1q_2^2\chi_-+\mathfrak{i}_2q_2\zeta)
-q_2B_0\Omega_{tot}\left(G_R(q_2^2\chi_+-b_1^2\xi_+)+G_I(q_2^2\chi_-+b_1^2\xi_-)\right)+2q_2B_1\Omega_{tot}G_I\xi_-
    \bigg]
\end{eqnarray}
\begin{eqnarray}
    &\frac{\Omega_{tot}}{2V_0} \mathfrak{i}_{0}=\bigg[N_{1,2}
    b_1^2\cos\left(-\frac{\varphi_{\text{BZ}}}{6}+\frac{2\pi C}{3}\right)
+N_{1,2}\frac{q_2^2}{4}\cos(\frac{\varphi_\text{BZ}}{6}+\frac{2\pi}{3}C)\\
&q_2B_0\Omega_{tot}(\mathfrak{r}_0\xi_++\mathfrak{r}_1q_2^2\chi_++\mathfrak{r}_2q_2\zeta)
-q_2B_0\Omega_{tot}\left(G_R(q_2^2\chi_++b_1^2\xi_+)+G_I(q_2^2\chi_--b_1^2\xi_-)\right)+2q_2B_1\Omega_{tot}G_R\xi_+
    \bigg].
\end{eqnarray}
We also recall
\begin{eqnarray}
&B_1=-\frac{1}{ v'}  \frac{1}{ (2 \pi)^2 }\frac{\mathfrak{r}_0}{2G_R}.
\end{eqnarray}
Hence we have a linear system of equations
\begin{eqnarray}
    &\begin{pmatrix}
        M_{RR} & M_{RI} \\
        M_{IR} & M_{II}
    \end{pmatrix}
    \begin{pmatrix}
        \mathfrak{r}_0 \\ 
        \mathfrak{i}_0
    \end{pmatrix}
    =
    \begin{pmatrix}
        C_R\\C_I
    \end{pmatrix}\\
    &M_{RR}=\frac{\Omega_{tot}}{2V_0}+2\Omega_{tot}q_2G_I\xi_-\frac{1}{ v'}  \frac{1}{ (2 \pi)^2 }\frac{1}{2G_R}\\
    &M_{RI}=-q_2B_0\Omega_{tot}\xi_-\\
    &M_{IR}=-q_2B_0\Omega_{tot}\xi_+
    +2\Omega_{tot}q_2G_R\xi_+\frac{1}{ v'}  \frac{1}{ (2 \pi)^2 }\frac{1}{2G_R}\\
    &M_{II}=\frac{\Omega_{tot}}{2V_0}\\
    &C_R=\bigg[-N_{1,2}
    b_1^2\cos\left(-\frac{\varphi_{\text{BZ}}}{6}+\frac{2\pi C}{3}\right)
+N_{1,2}\frac{q_2^2}{4}\cos(\frac{\varphi_\text{BZ}}{6}+\frac{2\pi}{3}C)\\
&q_2B_0\Omega_{tot}(\mathfrak{i}_1q^2_2\chi_-+\mathfrak{i}_2q_2\zeta)
-q_2B_0\Omega_{tot}\left(G_R(q_2^2\chi_+-b_1^2\xi_+)+G_I(q_2^2\chi_-+b_1^2\xi_-)\right)
    \bigg]\\
    &C_I=\bigg[N_{1,2}
    b_1^2\cos\left(-\frac{\varphi_{\text{BZ}}}{6}+\frac{2\pi C}{3}\right)
+N_{1,2}\frac{q_2^2}{4}\cos(\frac{\varphi_\text{BZ}}{6}+\frac{2\pi}{3}C)\\
&q_2B_0\Omega_{tot}(\mathfrak{r}_1q_2^2\chi_++\mathfrak{r}_2q_2\zeta)
-q_2B_0\Omega_{tot}\left(G_R(q_2^2\chi_++b_1^2\xi_+)+G_I(q_2^2\chi_--b_1^2\xi_-)\right)  \bigg]
\end{eqnarray}
that can be inverted to obtain $\mathfrak{r}_0$ and $\mathfrak{i}_0$. Alternatively, consider the dimensionless variables
\begin{eqnarray}
    &w_0=\frac{\mathfrak{r}_0}{q_2^2v'\Lambda},\quad y_0=\frac{\mathfrak{i}_0}{q_2^2v'\Lambda}.
\end{eqnarray}
We recall that $b_1^2=3 q_2^2$. The equations for $w_0$ and $y_0$ read
\begin{eqnarray}\label{eqapp:w0y0}
    &\begin{pmatrix}
        m_{RR} & m_{RI} \\
        m_{IR} & m_{II}
    \end{pmatrix}
    \begin{pmatrix}
        w_0 \\ 
        y_0
    \end{pmatrix}
    =
    \begin{pmatrix}
        c_R\\c_I
    \end{pmatrix}\\
    &m_{RR}=1+\frac{2\xi_-z'}{\gamma z}\\
    &m_{RI}=\frac{2\xi_-\ln|z|}{\gamma}\\
    &m_{IR}=\frac{2\xi_+}{\gamma}(\ln|z|+1)\\
    &m_{II}=1\\
    &c_R=\frac{2r}
    {\gamma}\bigg[-3\cos\left(-\frac{\varphi_{\text{BZ}}}{6}+\frac{2\pi C}{3}\right)
+\frac{1}{4}\cos(\frac{\varphi_\text{BZ}}{6}+\frac{2\pi}{3}C)\bigg]\\
&+\frac{2\ln|z|}{\gamma}\bigg[
-w_1\chi_-+w_2\zeta+z(\chi_+-\xi_+)+z'(\chi_-+\xi_-)
\bigg]
\\
&c_I=\frac{2r}
    {\gamma}\bigg[3\cos\left(-\frac{\varphi_{\text{BZ}}}{6}+\frac{2\pi C}{3}\right)
+\frac{1}{4}\cos(\frac{\varphi_\text{BZ}}{6}+\frac{2\pi}{3}C)\bigg]\\
&+\frac{2\ln|z|}{\gamma}\bigg[
-w_1\chi_+-w_2\zeta+z(\chi_++\xi_+)+z'(\chi_-- \xi_-)
\bigg].
\end{eqnarray}

To summarize, the full parametrization of the order parameter is
\begin{eqnarray}\label{eqapp:k1_12_finitephi_finalparam}
    & g_{1,k}=R_k+iI_k\\
    &R_k=2G_R \cos(  \frac{\varphi_\text{BZ}}{6} \frac{2 k}{q_2} )+\phi\left( \mathfrak{r}_0\cos( \frac{\varphi_\text{BZ}}{6}  \frac{2 k}{q_2})+\mathfrak{r}_1k^2\cos( \frac{\varphi_\text{BZ}}{6}  \frac{2 k}{q_2})+\mathfrak{r}_2k\sin(  \frac{\varphi_\text{BZ}}{6} \frac{2 k}{q_2} )\right) \\
    &=v'\Lambda \bigg[
    2z\cos(  \frac{\varphi_\text{BZ}}{6} \frac{2 k}{q_2} )+\phi q_2^2\left( 
    w_0\cos( \frac{\varphi_\text{BZ}}{6}  \frac{2 k}{q_2})+w_1\frac{k^2}{q_2^2}\cos( \frac{\varphi_\text{BZ}}{6}  \frac{2 k}{q_2})+w_2\frac{k}{q_2}\sin(  \frac{\varphi_\text{BZ}}{6} \frac{2 k}{q_2} )
    \right)
    \bigg] \\
    &I_{k}=2G_I \sin(  \frac{\varphi_\text{BZ}}{6} \frac{2 k}{q_2} ) +\phi\left(\mathfrak{i}_0\sin( \frac{\varphi_\text{BZ}}{6}  \frac{2 k}{q_2})+\mathfrak{r}_1k^2\sin( \frac{\varphi_\text{BZ}}{6}  \frac{2 k}{q_2})-\mathfrak{r}_2k\cos(  \frac{\varphi_\text{BZ}}{6} \frac{2 k}{q_2} )\right)\\
    &=v'\Lambda \bigg[
    2z'\sin(  \frac{\varphi_\text{BZ}}{6} \frac{2 k}{q_2} )+\phi q_2^2\left( 
    y_0\sin( \frac{\varphi_\text{BZ}}{6}  \frac{2 k}{q_2})+w_1\frac{k^2}{q_2^2}\sin( \frac{\varphi_\text{BZ}}{6}  \frac{2 k}{q_2})-w_2\frac{k}{q_2}\cos(  \frac{\varphi_\text{BZ}}{6} \frac{2 k}{q_2} )
    \right)
    \bigg],
\end{eqnarray}
with dimensionless coefficients $z$ and $z'$ (Eq.~\ref{eqapp:k1_k12_phi0_z}), $w_0$ and $y_0$ (Eq.~\ref{eqapp:w0y0}), and $w_1$ and $w_2$ (Eq.~\ref{eqapp:w1w2}).

\begin{figure}
    \centering
    \includegraphics[width = 0.8\linewidth]{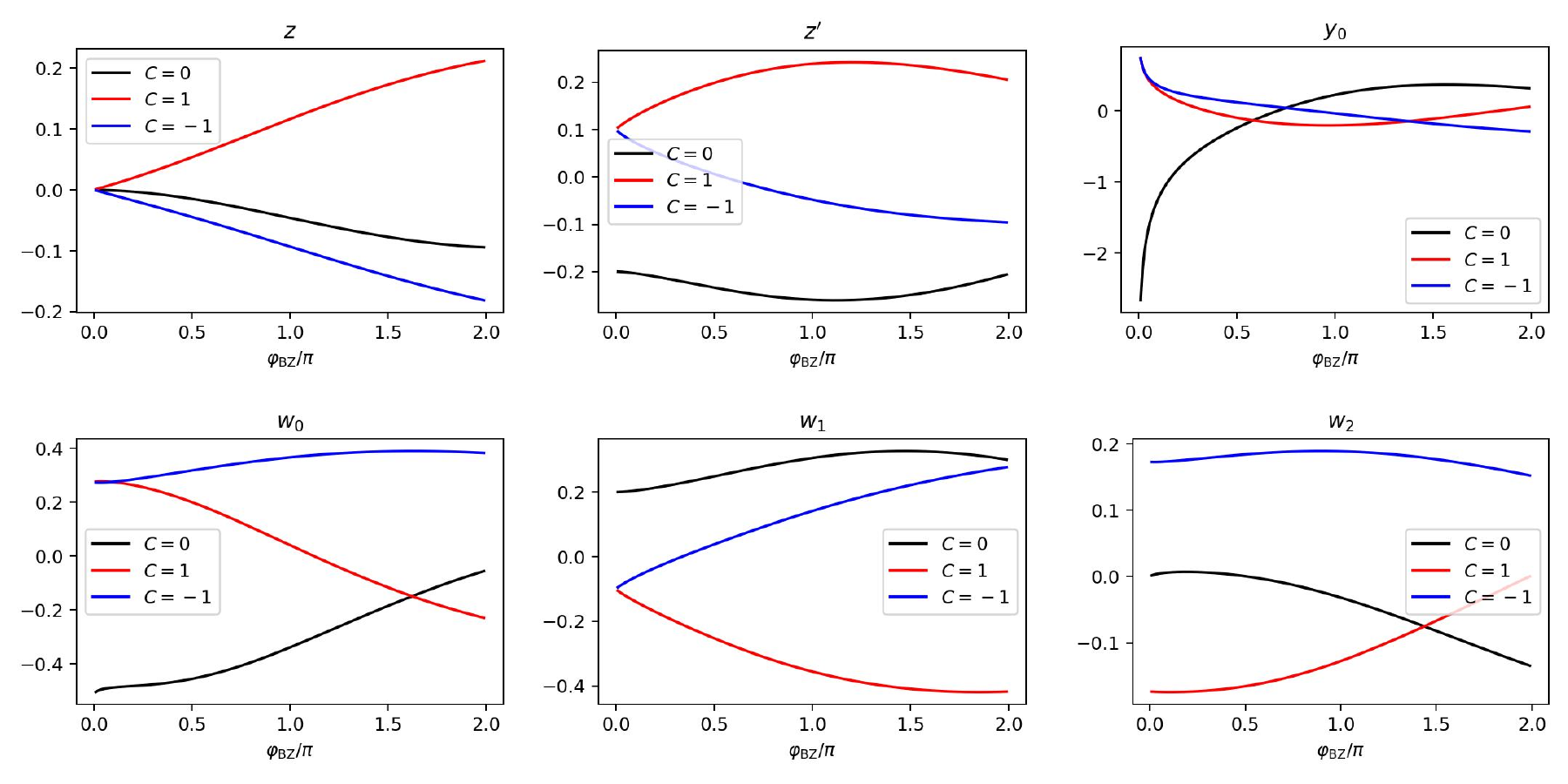}
    \caption{Plots of $z,z',y_0,w_0,w_1,w_2$ as function of $\varphi_{\text{BZ}}$ for $r=0.3$, $\gamma=3$, and different values of $C\mod 3$. $z,z',y_0,w_0,w_1,w_2$ are dimensionless coefficients that characterize the order parameter for the problem to linear order in $\phi$ (see Eq.~\ref{eqapp:k1_12_finitephi_finalparam}). }
    \label{figapp:k1_k12_finitephi_coeffs}
\end{figure}

Recall that in the $\phi=0$ limit in App.~\ref{subsecapp:k1_k12_HF_phi0}, there was no self-consistent gapped $C=0\mod 3$ solution for $\varphi_\text{BZ}=0$. We now check whether this holds for finite $\phi>0$. Setting $\varphi_\text{BZ}=0$ and $C=0\mod 3$, we find $z=0,w_1=\frac{2r}{\gamma},w_2=0,w_0=-\frac{11r}{2\gamma}$ ($z'$ and $y_0$ do not contribute to the order parameter for $\varphi_\text{BZ}=0$). Hence, $g_{1,k}$ is purely real and negative for $-\frac{q_2}{2}\leq k\leq \frac{q_2}{2}$. This means that $C=0\mod6$ is a self-consistent solution at $\varphi_\text{BZ}=0$ for finite $\phi>0$. In fact, by considering the analysis of App.~\ref{subsecapp:full_Chern}, we find that this solution has a Chern number $C=0$, because $g_{1,k}$ (which remains purely real and negative) does not induce any change of the phase of the order parameter as a function of $k$.
For $C=1\mod3$, we find $z=0,w_1=-\frac{r}{\gamma},w_2=-\frac{\sqrt{3}r}{\gamma},w_0=\frac{11r}{4\gamma}$. 

In Fig.~\ref{figapp:k1_k12_finitephi_coeffs}, we numerically compute the coefficients $z,z',y_0,w_0,w_1,w_2$ as a function of $\varphi_\text{BZ}$ for different $C\mod3$ for some representative parameters $r,\gamma$.

We now evaluate the part of the total energy that depends on the Chern number. We first evaluate the contribution $E_{\text{tot},12}$ to the total mean-field energy $E_{\text{tot}}=E_{\text{tot},1}+E_{\text{tot},12}$ from region $k_{12}$
\begin{eqnarray}
    &E_{\text{tot},12}=6N_{1,2}
    \text{Re}\left[g_{12}e^{-i\frac{2\pi}{3}C}\right]\\
    &=6N_{1,2}\bigg[\phi\mathfrak{r}_1\frac{q_2^2}{4}\cos(\frac{\varphi_{\text{BZ}}}{6}+\frac{2\pi}{3}C)+\phi\mathfrak{r}_2\frac{q_2}{2}\sin(\frac{\varphi_{\text{BZ}}}{6}+\frac{2\pi}{3}C)\\
    &+(2G_R+\phi\mathfrak{r}_0)\cos(\frac{\varphi_{\text{BZ}}}{6})\cos(\frac{2\pi}{3}C)-(2G_I+\phi\mathfrak{i}_0)\sin(\frac{\varphi_{\text{BZ}}}{6})\sin(\frac{2\pi}{3}C)
    \bigg]
    \\
    &=\frac{6\Omega_{tot}A^2_{k_1}\gamma V_0}{(2\pi)^4}r\bigg[\phi q_2^2\frac{w_1}{4}\cos(\frac{\varphi_{\text{BZ}}}{6}+\frac{2\pi}{3}C)+\phi q_2^2\frac{w_2}{2}\sin(\frac{\varphi_{\text{BZ}}}{6}+\frac{2\pi}{3}C)\\
    &+(2z+\phi q_2^2 w_0)\cos(\frac{\varphi_{\text{BZ}}}{6})\cos(\frac{2\pi}{3}C)-(2z'+\phi q_2^2 y_0)\sin(\frac{\varphi_{\text{BZ}}}{6})\sin(\frac{2\pi}{3}C)
    \bigg].
\end{eqnarray}

We now consider the contribution from the $k_1$ region, assuming that the renormalized velocity $v'$ is dominated by the kinetic velocity
\begin{eqnarray}
    E_\text{tot,1}&=-        \frac{3}{2} \frac{\Omega_{tot}}{ (2 \pi)^2v'}  \int_{- \frac{q_2}{2}}^{ \frac{q_2}{2}} dk    | g_{1,k}|^2.
\end{eqnarray}
To be consistent, we only retain terms up to linear order in $\phi$
\begin{eqnarray}
    E_\text{tot,1}&=-        \frac{6q_2\Omega_{tot}}{ (2 \pi)^2v'} \bigg[
    (G_R^2\xi_++G_I^2\xi_-)+\phi\left(
    G_R(\mathfrak{r}_0\xi_++\mathfrak{r}_1q_2^2\chi_++\mathfrak{r}_2q_2\zeta)+G_I(\mathfrak{i}_0\xi_-+\mathfrak{r}_1q_2^2\chi_--\mathfrak{r}_2q_2\zeta)
    \right)
    \bigg]\\
    &=-        \frac{6\Omega_{tot}A_{k_1}^2\gamma V_0}{ (2 \pi)^4} \bigg[
    (z^2\xi_++z'^2\xi_-)+\phi q_2^2\left(
    z(w_0\xi_++w_1\chi_++w_2\zeta)+z'(y_0\xi_-+w_1\chi_--w_2\zeta)
    \right)
    \bigg].
\end{eqnarray}

\begin{figure}
    \centering
    \includegraphics[width = 1.0\linewidth]{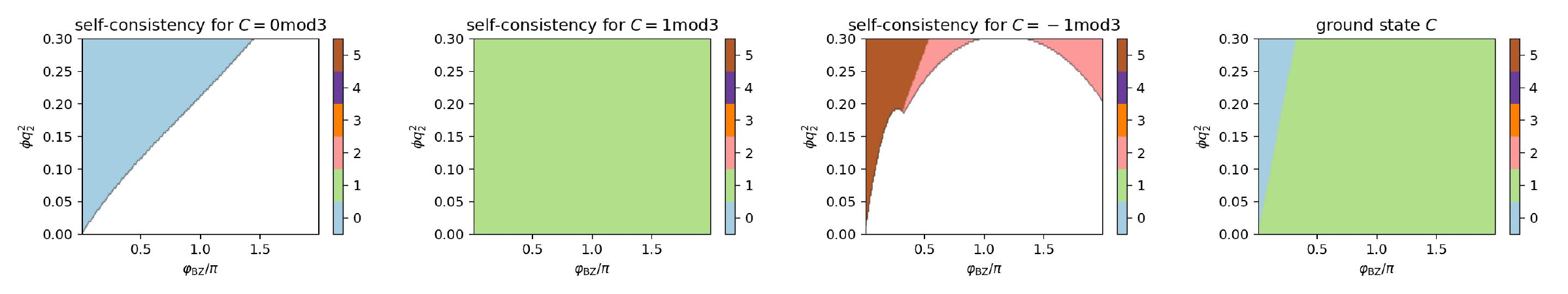}
    \caption{First three plots show self-consistent values of $C\mod 6$ for solutions corresponding to $C=0,1,-1\mod 3$ as a function of $\varphi_\text{BZ}$ and $\phi$ for $r=0.3,\gamma=3$. Last plot shows the Chern number $C\mod 6$ of the lowest energy self-consistent solution. Calculations are performed to linear order in $\phi$, including both Hartree and Fock terms (see App.~\ref{subsecapp:k1_k12_linearphi_HF}).}
    \label{figapp:k1_k12_finitephi_selfcons_gs}
\end{figure}

At $\varphi_\text{BZ}=0$, we find from the analytical expressions that the $C=0\mod 3$ solution has a lower energy than the $C=1\mod 3$ solution for finite $\phi>0$. This reflects the relative lowering of the Hartree penalty for the $C=0\mod 3$ solution.

In Fig.~\ref{figapp:k1_k12_finitephi_selfcons_gs}, we numerically check the self-consistency conditions and determine $C\mod 6$ of the lowest energy solution for the interval $0<\varphi_\text{BZ}<2\pi$. For $\phi=0$, only the $C=1\mod6$ state is self-consistent within this interval, as demonstrated previously in App.~\ref{subsecapp:k1_k12_HF_phi0}. For finite $\phi$, the $C=0\mod6$ becomes the (self-consistent) ground state for small values of $\varphi_\text{BZ}$, and the phase boundary between $C=0\mod 6$ and $C=1\mod6$ moves to larger $\varphi_\text{BZ}$ for increasing $\phi$. Hence, the suppression of the Hartree penalty for finite $\phi>0$ enables the $C=0\mod6$ solution to become the ground state for small values of $\varphi_\text{BZ}$.
In fact, by continuity, we can conclude for Fig.~\ref{figapp:k1_k12_finitephi_selfcons_gs} that the $C=0\mod 6$ solution has $C=0$ (since we have shown this above for $\phi>0$ and $\varphi_\text{BZ}=0$) and the $C=1\mod 6$ solution has $C=1$ (since we have shown this for $\phi=0$ in App.~\ref{subsecapp:k1_k12_HF_phi0}).

\subsubsection{Fock only, $\phi=0$}\label{subsecapp:k1_k12_Fockonly_phi0}

We now consider the limit where we only keep Fock terms. This is equivalent to setting $V_{b_1}=0$, which removes any non-trivial Hartree terms. This limit is motivated by the fact that a physical interaction $V_q$ (such as a gate-screened Coulomb interaction) is generally expected to decay with $q$ and satisfy $V_{b_1}<V_{q_2}$. Combined with the Gaussian decay of the GMP form factors, this suppresses the contributions from $q=b_1$. In this section, we consider the case $\phi=0$ (so $V_q$ actually increases from $q=0$, but then vanishes at $q=b_1$), where the equations are easier to handle analytically. 

We consider removing the $e^{-\phi b_1^2}$ terms and setting $\phi=0$ in the other terms in Eq.~\ref{eq:g1_GMP_general}
\begin{eqnarray}\label{eqapp:Fockonly_phi0_selfconseqn}
    &\frac{\Omega_{tot}}{V_0} g_{1,\delta k_{1y} }=e^{ i \frac{\varphi_\text{BZ}}{6} \frac{2 \delta k_{1y}}{q_2}}\left[-2N_{1,2}\cos\left(\frac{\varphi_{\text{BZ}}}{6}+\frac{2\pi C}{3}\right)+b\Omega_{tot}\int_{-\frac{q_2}{2}}^{\frac{q_2}{2}}d\delta k'_{1y} 
    e^{-i \frac{\varphi_\text{BZ}}{6} \frac{2\delta k_{1y}'}{q_2}} g_{1,\delta k'_{1y}}\right].
\end{eqnarray}
The hybridization field thus has constant magnitude (parameterized by $2\tilde{G}$) and winds its phase around the BZ boundary
\begin{eqnarray}
    &g_{1,\delta k_{1y} }=2\tilde{G}e^{ i \frac{\varphi_\text{BZ}}{6} \frac{2 \delta k_{1y}}{q_2}}\\
    &\frac{\Omega_{tot}}{V_0} \tilde{G}=-N_{1,2}\cos\left(\frac{\varphi_{\text{BZ}}}{6}+\frac{2\pi C}{3}\right)+bq_2\Omega_{tot}\tilde{G}\\
    &b= \frac{1}{(2\pi)^2 v'}\log\left(\frac{v'\Lambda}{|\tilde{G}|}\right).
\end{eqnarray}
Note that self-consistently, the parameter $b_{k_{1y}}$ (see Eq.~\ref{eqapp:bk1y}) does not depend on $k_{1y}$. We define the momentum areas of the $k_1$ and $k_{12}$ regions as $A_{k_1}=q_2\Lambda$ and $A_{k_{12}}=N_{1,2}\frac{(2\pi)^2}{\Omega_{tot}}$. We also recall $\Omega_{tot}=\frac{(2\pi)^2 N_{\text{UC}}}{A_{\text{BZ}}}$, where $N_{\text{UC}}$ is the number of Wigner unit cells in the system. We also define dimensionless variables $\kappa=\frac{A_{k_{12}}}{A_{k_1}}\cos(\frac{\varphi_{\text{BZ}}}{6}+\frac{2\pi C}{3})$ and $\gamma=\frac{(2\pi)^2v'}{q_2V_0}$. In terms of $z=\frac{\tilde{G}}{v'\Lambda}$, whose magnitude should be much less than 1 (see Eq.~\ref{eqapp:O01_log}), we have 
\begin{eqnarray}
    & z+\frac{\kappa}
{\gamma+\ln |z|}=0.
\end{eqnarray}
For $\kappa=0$, we find
\begin{eqnarray}
    |z|=e^{-\gamma},
\end{eqnarray}
where either sign of $z$ is a valid solution. This limit motivates considering large $\gamma$ to ensure $|z|\ll 1$. For general $\kappa$, we find
\begin{eqnarray}\label{eqapp:zsoln}
    z=-\frac{\kappa}{W_0(|\kappa|e^{\gamma})}.
\end{eqnarray}
in terms of the principal branch $W_0$ of the Lambert $W$ function, which is positive for positive arguments.

Having obtained $g_{1,\delta k_{1y}}$ (and hence $g_{12}$ as well), we discuss the self-consistency conditions for obtaining a ground state with $C\mod3$. We first consider the condition at the $M_M$ points. Recall that even (odd) $C$ requires negative (positive) $g_{1,\delta k_{1y}=0}$. Hence, an even (odd) $C$ requires negative (positive) $\tilde{G}$. Note that from Eq.~\ref{eqapp:zsoln}, the sign of $\tilde{G}$ is opposite to the sign of $\cos(\frac{\varphi_{\text{BZ}}}{6}+\frac{2\pi C}{3})$.

We now consider the $K_M$-point condition, where the hybridization field is $g_{1,\delta k_{1y}=-q_2/2}=g_{12}$. This satisfies $\text{arg}\,g_{12}=-\varphi_{\text{BZ}}/6+\frac{\text{sgn}\,\tilde{G}-1}{2}\pi$. We define $\text{arg}\,g_{12}$ to lie in the range $-\pi$ to $\pi$. Recall that the lowest energy solution of the mean-field Hamiltonian at $\bm{k}=\bm{q}_2$ corresponds to $C=m\mod 3$ with $m$ satisfying $\text{arg}(g_{12}e^{-i\frac{2(m+1)\pi}{3}})\in[0,\frac{2\pi}{3}]$. 
\begin{itemize}
\item For $C=0\mod 3$, we require $\frac{2\pi}{3}<\text{arg}\,g_{12}$ or $\text{arg}\,g_{12}<-\frac{2\pi}{3}$. If $\tilde{G}$ is positive (which further constrains $C=3\mod 6$), this means $\varphi_{\text{BZ}}\in [4\pi,8\pi] \mod 12 \pi$. If $\tilde{G}$ is negative (which further constrains $C=0\mod 6$), this means $\varphi_{\text{BZ}}\in [-2\pi,2\pi] \mod 12 \pi$.
\item For $C=1\mod 3$, we require $-\frac{2\pi}{3}<\text{arg}\,g_{12}<0$. If $\tilde{G}$ is positive (which further constrains $C=1\mod 6$), this means $\varphi_{\text{BZ}}\in [0,4\pi] \mod 12 \pi$. If $\tilde{G}$ is negative (which further constrains $C=4\mod 6$), this means $\varphi_{\text{BZ}}\in [6\pi,10\pi] \mod 12 \pi$.
\item For $C=-1\mod 3$, we require $0<\text{arg}\,g_{12}<\frac{2\pi}{3}$. If $\tilde{G}$ is positive (which further constrains $C=-1\mod 6$), this means $\varphi_{\text{BZ}}\in [8\pi,12\pi] \mod 12 \pi$. If $\tilde{G}$ is negative (which further constrains $C=2\mod 6$), this means $\varphi_{\text{BZ}}\in [2\pi,6\pi] \mod 12 \pi$.
\end{itemize}

We now evaluate the part of the total mean-field energy that depends on the Chern number (see Eq.~\ref{eqapp:Etot_distinguishC}). We begin with the $\bm{k}_{12}$ contribution from Eq.~\ref{eqapp:Etot12_N12}
\begin{eqnarray}
    &E_{\text{tot},12}=12N_{1,2}\tilde{G}\cos(\frac{\varphi_{\text{BZ}}}{6}+\frac{2\pi}{3}C).
\end{eqnarray}
For the $\bm{k}_1$ contribution, we have
\begin{align}
    E_\text{tot,1}&=\text{const.}-    \frac{3}{2} \frac{\Omega_{tot}}{ (2 \pi)^2} q_2v'     \left[ \left(1+\frac{v}{2v'}\right)\Lambda_x^2 \sqrt{\frac{ 4\tilde{G}^2}{v'^2\Lambda_x^2}+1}+\left(1-\frac{v}{2v'}\right)\frac{ 4\tilde{G}^2}{v'^2} \tanh ^{-1}\left(\frac{1}{\sqrt{\frac{4\tilde{G}^2}{v'^2\Lambda_x^2}+1}}\right) \right].
\end{align}
Note that for $\phi=0$, we have that $v'=\frac{v}{2}$ is entirely set by the kinetic velocity (see App.~\ref{subsubsecapp:f_GMP_exp}), so  Eq.~\ref{eqapp:Etot1_kineticv} holds.  Considering the regime $|z|\ll 1$, we obtain for the total mean-field energy
\begin{align}
    E_\text{tot}&\approx\text{const.}+12N_{1,2}\tilde{G}\cos(\frac{\varphi_{\text{BZ}}}{6}+\frac{2\pi}{3}C)
    -\frac{6\Omega_{tot}q_2\tilde{G}^2}{(2\pi)^2v'}\\
    &=\text{const.}+\frac{12\Omega_{tot}A_{k_1}A_{k_{12}}v'}{(2\pi)^2q_2}z\cos(\frac{\varphi_{\text{BZ}}}{6}+\frac{2\pi}{3}C)-\frac{6\Omega_{tot}A_{k_1}^2v'}{(2\pi)^2q_2}z^2\\
    &=\text{const.}+\frac{6A_{k_1}^2\Omega_{tot}\gamma V_0}{(2\pi)^4}z(2\kappa - z)\\
    &=\text{const.}-\frac{6A_{k_1}^2\Omega_{tot}\gamma V_0}{(2\pi)^4}\frac{\kappa^2}{W_0(|\kappa|e^\gamma)}\left(2+\frac{1}{W_0(|\kappa|e^\gamma)}\right).
\end{align}
Note that all the dependence on the Chern number is encoded in $\kappa$. Since $W_0(x)$ is an increasing and concave function for all $x>0$, then $E_{\text{tot}}$ is a monotonically decreasing function of $|\kappa|$. Hence, the ground state corresponds to the value of $C$ that maximizes $|\kappa|$. We conclude that $\varphi_{\text{BZ}}\in[-\pi,\pi]$ leads to a $C=0\mod6$ ground state, $\varphi_{\text{BZ}}\in[\pi,3\pi]$ leads to a $C=1\mod6$ ground state, $\varphi_{\text{BZ}}\in[3\pi,5\pi]$ leads to a $C=2\mod6$ ground state, etc. The ground state is always self-consistent.

Using the analysis of App.~\ref{subsecapp:full_Chern}, we can further refine the Chern number of our analytic solution. Consider the relative phase $\beta_{\delta k_{1y}}$ between the coefficients of the Bloch state at $\bm{k}=(-b_1/2,\delta k_{1y})$ and $(b_1/2,\delta k_{1y})$ in the HF wavefunction for the $\bm{k}_1$ region.
Even values of $C$ lead to negative $\tilde{G}$, so that $\beta_{\delta k_{1y}}$ varies from $-\varphi_\text{BZ}/6$ to $\varphi_\text{BZ}/6$ as $\delta k_{1y}$ goes from $-q_2/2$ to $q_2/2$ in the $\bm{k}_1$ region. If $\varphi_{\text{BZ}}\in[-\pi,\pi]$, then the nearest multiple of $2\pi/3$ to $-\varphi_\text{BZ}/6$ and $\varphi_\text{BZ}/6$ is $0$ in both cases. Hence, App.~\ref{subsecapp:full_Chern} shows that $\beta_{\delta k_{1y}}$ varies from $0$ to $0$ as $\delta k_{1y}$ traverses vertically the entire left edge of the BZ (accounting now also for the $\bm{k}_{12}$ regions), leading to $C=0$. If on the other hand $\varphi\in[3\pi,5\pi]$, then App.~\ref{subsecapp:full_Chern} finds that $\beta_{\delta k_{1y}}$ varies from $-2\pi/3$ to $2\pi/3$ as $\delta k_{1y}$ traverses vertically the entire left edge of the BZ, leading to $C=2$. Analogous arguments can be made for other intervals of $\varphi_\text{BZ}$, as well as odd values of $C$ (where $\tilde{G}$ is positive). Thus, we conclude that $\varphi_{\text{BZ}}\in [2\pi(m-\frac{1}{2}),2\pi(m+\frac{1}{2})]$ leads to a $C=m$ ground state for integer $m$.
In other words, the Chern number of the ground state is the one whose integrated Berry curvature is closest to $\varphi_\text{BZ}$. Without reference to a specific microscopic model, Ref.~\cite{dong2024stability} has previously obtained a similar `rounding' (up to $C\mod 3$) of $\varphi_\text{BZ}$ due to the Fock term, in a small-$q$ analysis of the properties of the Bloch functions on the entire BZ boundary.

While we have neglected the Hartree term in the self-consistent equation for the order parameter in Eq.~\ref{eqapp:Fockonly_phi0_selfconseqn}, we can ask what the Hartree energy (at finite momentum) of the solutions obtained here would be assuming some interaction potential $V_{b_1}$ and keeping the solutions unchanged. The expression for the Hartree energy is
\begin{eqnarray}
    &E_{\text{Hartree}}=\frac{3}{\Omega_{\text{tot}}}V_{b_1}\left|\sum_{\bm{k}}M_{\bm{k},-\bm{b}_1}O_{\bm{k},\bm{k}+\bm{b}_1}\right|^2=\frac{3}{\Omega_{\text{tot}}}V_{b_1}\left|\rho_{\bm{b}_1}\right|^2
\end{eqnarray}
which is valid with $C_3$ symmetry. $\rho_{\bm{b}_1}$ is simply the amplitude of the total charge density at momentum $\bm{b}_1$, i.e.~the spatially inhomogeneous component. Due to $C_6$ symmetry, $\rho_{\bm{b}_1}$ alone can be used to reconstruct the real-space charge profile. Straightforward manipulations lead to the result
\begin{eqnarray}\label{eqapp:Fockonly_phi0_chargedens}
    &\rho_{\bm{b}_1}=2\Omega_\text{tot}\frac{A_{k_{1}}}{(2\pi)^2}e^{-\frac{\sqrt{3}\varphi_{\text{BZ}}}{6}}\bigg[r\cos\left(-\frac{\varphi_\text{BZ}}{6}+\frac{2\pi}{3}C\right)-z\log(1/|z|)
    \frac{\sin\left(\frac{\varphi_\text{BZ}}{3}\right)}{\frac{\varphi_\text{BZ}}{3}}\bigg]
\end{eqnarray}
where we have defined $r=\frac{A_{k_{12}}}{A_{k_1}}$. The first term describes the contribution from the $\bm{k}_{12}$ region while the second term describes the contribution from the $\bm{k}_{1}$ region. The $\bm{k}_0$ region does not contribute to the spatially-modulated part of the charge density, since these momenta cannot hybridize with other momenta. Consider the interval $0\leq \varphi_{\text{BZ}}<\pi$ where the Fock term favors $C=0$ over $C=1$. For $C=0$, both contributions are positive, while for $C=1$, both contributions are negative. Furthermore, the magnitude of each term is larger for $C=0$ than for $C=1$:
\begin{itemize}
\item For the $\bm{k}_{12}$ region, this arises from the interference of the charge densities from $K_M$ and $K_M'$ points~\cite{crepel2024efficientpredictionsuperlatticeanomalous, soejima2024AHC2,dong2024stability}. This is simplest to visualize in the limit of vanishing Berry curvature $\varphi_\text{BZ}=0$, where the underlying wavefunctions of the model can be considered as trivial plane waves. For a $C=0\mod 3$ solution, the plane wave coefficients of the HF Bloch function at both the $K_M$ and $K_M'$ points are $\sim (1,1,1)$. Therefore, the HF coefficients do not introduce additional phases, and the HF wavefunctions at $K_M$ and $K_M'$ both have a peak at $\bm{r}=0$. For a $C=1\mod 3$ solution, the plane wave coefficients of the HF Bloch function are instead $\sim (1,\omega,\omega^2)$. Therefore, the coefficient of say $e^{i\bm{b}_1\cdot\bm{r}}$ in the HF wavefunction has a relative phase of $\omega$ between $K_M$ and $K_M'$. The corresponding charge densities are therefore shifted relative to each other, which suppresses the total magnitude of the charge modulation.
\item For the $\bm{k}_1$ region, this arises from the larger order parameter for $C=0$ (since the phase winding, apart from an overall sign, is identical for both $C=0,1$ in this $\phi=0$ limit). 
\end{itemize}
Hence, the $C=0$ state has a larger modulation of the charge density, and is relatively disfavored by the Hartree energy, which would shift the phase boundary between these two states to lower values $\varphi_\text{BZ}<\pi$.

\subsubsection{Fock only, linear order in $\phi$}\label{secapp:fockonly_linearphi}

In this section, we consider the limit where we only keep Fock terms, but unlike in  App.~\ref{subsecapp:k1_k12_Fockonly_phi0}, we retain up to linear order in $\phi$. We therefore remove the $e^{-\phi b_1^2}$ terms in Eq.~\ref{eq:g1_GMP_general}
\begin{eqnarray}\label{eqapp:fockonly_finitephi_startingpoint}
    &\frac{\Omega_{tot}}{V_0} g_{1,k }=N_{1,2}\bigg[
    -  e^{-\phi(k+\frac{q_2}{2})^2}  e^{ i \frac{\varphi_\text{BZ}}{6}(1+ \frac{2 k}{q_2} )} e^{i\frac{2\pi}{3}C}- e^{-\phi(-k+\frac{q_2}{2})^2}  e^{ -i \frac{\varphi_\text{BZ}}{6}(1-\frac{2 k}{q_2} )} e^{-i\frac{2\pi}{3}C}
    \bigg]
    \\& +b\Omega_{tot}\int_{-\frac{q_2}{2}}^{\frac{q_2}{2}}dk'  e^{-\phi(k-k')^2}
    e^{-i \frac{\varphi_\text{BZ}}{6} (\frac{2k'}{q_2}- \frac{2k}{q_2})} g_{1,k'},
\end{eqnarray}
where we have relabeled $\delta k_{1y}\rightarrow k$ and $\delta k_{1y}'\rightarrow k' $ for notational clarity. 

Before proceeding, we comment that the solutions of the order parameter for certain different $C\mod 3$ are related at certain values of $\varphi_{\text{BZ}}$. To see this, we let $g_{1,k}=h_{k}e^{ i \frac{\varphi_\text{BZ}}{6}\frac{2 k}{q_2} }$, leading to
\begin{eqnarray}\label{eqapp:fockonly_hk}
    &\frac{\Omega_{tot}}{V_0} h_k=N_{1,2}\bigg[
    -  e^{-\phi(k+\frac{q_2}{2})^2}  e^{ i \left(\frac{\varphi_\text{BZ}}{6}+\frac{2\pi}{3}C\right)}- e^{-\phi(-k+\frac{q_2}{2})^2}  e^{ -i \left(\frac{\varphi_\text{BZ}}{6}+\frac{2\pi}{3}C\right)}
    \bigg]
    \\& +b\Omega_{tot}\int_{-\frac{q_2}{2}}^{\frac{q_2}{2}}dk'  e^{-\phi(k-k')^2} h_{k'}.
\end{eqnarray}
Consider $\varphi_\text{BZ}=\pi$ as an example. The term proportional to $N_{1,2}$ differs between the $C=0$ and $C=1$ cases by a negative complex conjugation. Hence, we have that $h^{C=0}_{k}=-\left(h^{C=1}_k\right)^*$. Note that this is true even if the $k$ dependence of $b_k$ is restored, instead of taking the approximation $b_k\approx b_{k=0}=b$, since $b_k$ only depends on the modulus of $h_k$. Similarly for $\varphi_{\text{BZ}}=0$, we have $h^{C=-1}_{k}=\left(h^{C=1}_k\right)^*$. Generally for $\varphi_{\text{BZ}}=m\pi$ for integer $m$, we have a relation between the order parameters for $C=(-m+1)\mod3$ and $C=(-m+2)\mod 3$. This remains true for arbitrary interaction potentials, since changing $V(q)$ would only affect the real coefficients of each term in Eq.~\ref{eqapp:fockonly_hk} in a way that is independent of $\varphi_{\text{BZ}}$ and $C$.

We return to our original goal of working to linear order in $\phi$. Expanding the exponentials in Eq.~\ref{eqapp:fockonly_finitephi_startingpoint} up to linear order in $\phi$ leads to
\begin{eqnarray}
    &\frac{\Omega_{tot}}{V_0} g_{1,k }=2N_{1,2}e^{ i \frac{\varphi_\text{BZ}}{6}\frac{2 k}{q_2} } \bigg[
    -  \cos(\frac{\varphi_{\text{BZ}}}{6}+\frac{2\pi}{3}C) \\
    & +\phi(k^2+\frac{q_2^2}{4})\cos(\frac{\varphi_{\text{BZ}}}{6}+\frac{2\pi}{3}C) +i\phi kq_2\sin(\frac{\varphi_{\text{BZ}}}{6}+\frac{2\pi}{3}C)
    \bigg]
    \\& +b\Omega_{tot}e^{ i \frac{\varphi_\text{BZ}}{6}\frac{2 k}{q_2} }\int_{-\frac{q_2}{2}}^{\frac{q_2}{2}}dk'
    e^{-i \frac{\varphi_\text{BZ}}{6} \frac{2k'}{q_2}}
    \left(
    1 + \phi(-k^2-k'^2+2kk')
    \right)
    g_{1,k'}.
\end{eqnarray}
 We parameterize the solution in powers of $\phi$. We recall that $b$ also depends on the order parameter and should similarly be expanded in powers of $\phi$
\begin{eqnarray}
    & g_{1,k}=\mathcal{G}_{0,k}+\phi\mathcal{G}_{1,k}\\
    & b=B_0+\phi B_1.
\end{eqnarray}
We can extract $B_0$ and $B_1$ in the following way (recalling that the order parameter is purely real at $k=0$):
\begin{eqnarray}
    & b=  \frac{1}{ v'}  \frac{1}{ (2 \pi)^2 }  \log\frac{2 v' \Lambda}{|\mathcal{G}_{0,0}+\phi \mathcal{G}_{1,0}|}=\frac{1}{ v'}  \frac{1}{ (2 \pi)^2 }  \log\frac{2 v' \Lambda}{|\mathcal{G}_{0,0}|(1+\phi \mathcal{G}_{1,0}/\mathcal{G}_{0,0})}\approx \frac{1}{ v'}  \frac{1}{ (2 \pi)^2 } \left( \log\frac{2 v' \Lambda}{|\mathcal{G}_{0,0}|}-\phi \frac{\mathcal{G}_{1,0}}{\mathcal{G}_{0,0}}\right)
\end{eqnarray}
leading to
\begin{eqnarray}
    & B_0=\frac{1}{ v'}  \frac{1}{ (2 \pi)^2 } \log\frac{2 v' \Lambda}{|\mathcal{G}_{0,0}|},\quad B_1=-\frac{1}{ v'}  \frac{1}{ (2 \pi)^2 }\frac{\mathcal{G}_{1,0}}{\mathcal{G}_{0,0}}.
\end{eqnarray}

The zeroth-order part of the self-consistent equation is
\begin{eqnarray}
    &\frac{\Omega_{tot}}{V_0} \mathcal{G}_{0,k}=-2N_{1,2}e^{ i \frac{\varphi_\text{BZ}}{6}\frac{2 k}{q_2} } \cos(\frac{\varphi_{\text{BZ}}}{6}+\frac{2\pi}{3}C)
     +B_0\Omega_{tot}e^{ i \frac{\varphi_\text{BZ}}{6}\frac{2 k}{q_2} }\int_{-\frac{q_2}{2}}^{\frac{q_2}{2}}dk'
    e^{-i \frac{\varphi_\text{BZ}}{6} \frac{2k'}{q_2}}
    \mathcal{G}_{0,k'}.
\end{eqnarray}
As expected, this reduces to the self-consistent equation for the $\phi=0$ limit in App.~\ref{subsecapp:k1_k12_Fockonly_phi0}. The solution can be parametrized as
\begin{eqnarray}
    \mathcal{G}_{0,k}=2\tilde{G}e^{ i \frac{\varphi_\text{BZ}}{6}\frac{2 k}{q_2} } 
\end{eqnarray}
where $\tilde{G}$ is a real quantity with dimensions of energy. We recall the definitions of the momentum areas of the $k_1$ and $k_{12}$ regions are $A_{k_1}=q_2\Lambda$ and $A_{k_{12}}=N_{1,2}\frac{(2\pi)^2}{\Omega_{tot}}$.  We also define dimensionless variables $r=\frac{A_{k_{12}}}{A_{k_1}}$ (note that this differs from $\kappa$ defined previously by the absence of the cosine) and $\gamma=\frac{(2\pi)^2v'}{q_2V_0}$. In terms of $z=\frac{\tilde{G}}{v'\Lambda}$, we have
\begin{eqnarray}
    &z+\frac{r\cos(\frac{\varphi_{\text{BZ}}}{6}+\frac{2\pi}{3}C)}{\gamma+\ln|z|}=0.
\end{eqnarray}
This can be solved for analytically as in Eq.~\ref{eqapp:zsoln}. Hence, $\mathcal{G}_{0,k}$ and $B_0$ are known quantities in terms of $\tilde{G}$.

The first-order part of the self-consistent equation is
\begin{eqnarray}
    &\frac{\Omega_{tot}}{V_0} \mathcal{G}_{1,k }=2N_{1,2}e^{ i \frac{\varphi_\text{BZ}}{6}\frac{2 k}{q_2} } \bigg[
    (k^2+\frac{q_2^2}{4})\cos(\frac{\varphi_{\text{BZ}}}{6}+\frac{2\pi}{3}C) +ikq_2\sin(\frac{\varphi_{\text{BZ}}}{6}+\frac{2\pi}{3}C)
    \bigg]
    \\& +B_0\Omega_{tot}e^{ i \frac{\varphi_\text{BZ}}{6}\frac{2 k}{q_2} }\int_{-\frac{q_2}{2}}^{\frac{q_2}{2}}dk'
    e^{-i \frac{\varphi_\text{BZ}}{6} \frac{2k'}{q_2}}
    \left(
    \mathcal{G}_{1,k'} + (-k^2-k'^2+2kk') \mathcal{G}_{0,k'}
    \right)\\
    &+B_1\Omega_{tot}e^{ i \frac{\varphi_\text{BZ}}{6}\frac{2 k}{q_2} }\int_{-\frac{q_2}{2}}^{\frac{q_2}{2}}dk'
    e^{-i \frac{\varphi_\text{BZ}}{6} \frac{2k'}{q_2}}
   \mathcal{G}_{0,k'}.
\end{eqnarray}
Inserting the expression for $\mathcal{G}_{0,k}$ and integrating leads to
\begin{eqnarray}
    &\frac{\Omega_{tot}}{V_0} \mathcal{G}_{1,k }=2N_{1,2}e^{ i \frac{\varphi_\text{BZ}}{6}\frac{2 k}{q_2} } \bigg[
    (k^2+\frac{q_2^2}{4})\cos(\frac{\varphi_{\text{BZ}}}{6}+\frac{2\pi}{3}C) +ikq_2\sin(\frac{\varphi_{\text{BZ}}}{6}+\frac{2\pi}{3}C)
    \bigg]
    \\& +B_0\Omega_{tot}e^{ i \frac{\varphi_\text{BZ}}{6}\frac{2 k}{q_2} }\int_{-\frac{q_2}{2}}^{\frac{q_2}{2}}dk'
    e^{-i \frac{\varphi_\text{BZ}}{6} \frac{2k'}{q_2}}
    \mathcal{G}_{1,k'}-2\tilde{G}q_2B_0\Omega_{tot}e^{ i \frac{\varphi_\text{BZ}}{6}\frac{2 k}{q_2} }(k^2+\frac{q_2^2}{12})
\\
    &+2q_2\tilde{G}B_1\Omega_{tot}e^{ i \frac{\varphi_\text{BZ}}{6}\frac{2 k}{q_2} }.
\end{eqnarray}
It is clear that we can express the solution of the above equation as follows
\begin{eqnarray}
    \mathcal{G}_{1,k}=e^{ i \frac{\varphi_\text{BZ}}{6}\frac{2 k}{q_2} } \left[\mathfrak{g}_0+k^2\mathfrak{g}_1-ik\mathfrak{g}_2\right]
\end{eqnarray}
with real coefficients $\mathfrak{g}_0,\mathfrak{g}_1,\mathfrak{g}_2$. Since $\mathcal{G}_{1,k}$ has dimensions of $[EL^{-2}]$, then $\mathfrak{g}_0,\mathfrak{g}_1,\mathfrak{g}_2$ has dimensions of $[EL^{-2}],[E],[EL^{-1}]$ respectively.  We can read off
\begin{eqnarray}
    &\mathfrak{g}_1=\frac{V_0}{\Omega_{tot}}\left[2N_{1,2}\cos(\frac{\varphi_{\text{BZ}}}{6}+\frac{2\pi}{3}C)-2\tilde{G}q_2B_0\Omega_{tot} \right]\\
    &\mathfrak{g}_2=\frac{V_0}{\Omega_{tot}}\left[-2N_{1,2}q_2\sin(\frac{\varphi_{\text{BZ}}}{6}+\frac{2\pi}{3}C)\right]
\end{eqnarray}
in terms of known quantities. In terms of dimensionless variables $w_1,w_2$, we have
\begin{eqnarray}
    &w_1=\frac{\mathfrak{g}_1}{v'\Lambda}=\frac{2}{\gamma}\left[r\cos(\frac{\varphi_{\text{BZ}}}{6}+\frac{2\pi}{3}C)+z\ln|z|\right]\\
    &w_2=\frac{\mathfrak{g}_2}{q_2v'\Lambda}=-\frac{2r}{\gamma}\sin(\frac{\varphi_{\text{BZ}}}{6}+\frac{2\pi}{3}C).
\end{eqnarray}

The equation for $\mathfrak{g}_0$ reads
\begin{eqnarray}
    &\frac{\Omega_{tot}}{V_0} \mathfrak{g}_0=\frac{1}{2}N_{1,2}q_2^2\cos(\frac{\varphi_{\text{BZ}}}{6}+\frac{2\pi}{3}C)
 +q_2B_0\Omega_{tot}(\mathfrak{g}_0+\frac{q_2^2}{12}\mathfrak{g}_1)-\frac{1}{6}\tilde{G}q_2^3B_0\Omega_{tot}
+2q_2\tilde{G}B_1\Omega_{tot}.
\end{eqnarray}
Using that $B_1=-\frac{1}{ v'}  \frac{1}{ (2 \pi)^2 }\frac{\mathcal{G}_{1,0}}{\mathcal{G}_{0,0}}=-\frac{1}{ v'}  \frac{1}{ (2 \pi)^2 }\frac{\mathfrak{g}_0}{2\tilde{G}}$, we can solve for $\mathfrak{g}_0$
\begin{eqnarray}
    & \mathfrak{g}_0=\frac{V_0\left(\frac{1}{2}\frac{N_{1,2}}{\Omega_{tot}}q_2^2\cos( \frac{\varphi_\text{BZ}}{6}+  \frac{2\pi}{3} C)-\frac{1}{6}\tilde{G}B_0q_2^3+B_0\mathfrak{g}_1\frac{q_2^3}{12}\right)}{1-B_0 V_0 q_2+q_2V_0\frac{1}{ v'}  \frac{1}{ (2 \pi)^2 }}.
\end{eqnarray}
In terms of the dimensionless variable $w_0$, we have
\begin{eqnarray}
    &w_0=\frac{\mathfrak{g}_0}{q_2^2v'\Lambda}=\frac{\frac{r}{2}\cos( \frac{\varphi_\text{BZ}}{6}+  \frac{2\pi}{3} C)+\frac{1}{12}(2z-w_1)\ln|z|}{1+\gamma+\ln|z|}.
\end{eqnarray}
Hence, all quantities depend only on the dimensionless parameters $\gamma,r,C,\varphi_{\text{BZ}},\phi q_2^2$. To summarize, the order parameter is
\begin{eqnarray}\label{eqapp:Fockonly_finitephi_orderparameter_dimensionless}
    g_{1,k}=v'
    \Lambda e^{ i \frac{\varphi_\text{BZ}}{6}\frac{2 k}{q_2} }\left[2z+\phi q_2^2\left(w_0+\frac{k^2}{q_2^2}w_1-i\frac{k}{q_2}w_2\right)\right].
\end{eqnarray}

\begin{figure}
    \centering
    \includegraphics[width = 1.0\linewidth]{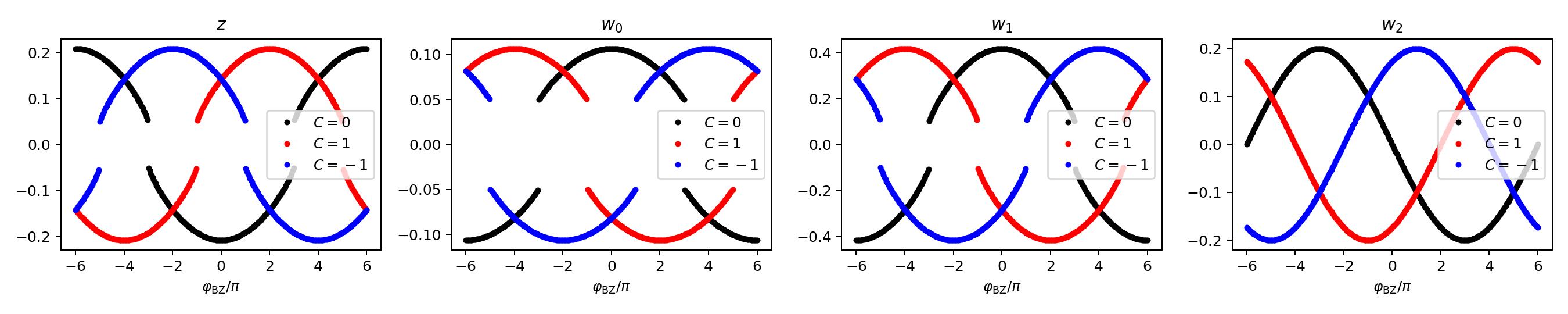}
    \caption{Plots of $z,w_0,w_1,w_2$ as function of $\varphi_{\text{BZ}}$ for $r=0.3$, $\gamma=3$, and different values of $C\mod 3$. $z,w_0,w_1,w_2$ are dimensionless coefficients that characterize the order parameter for the Fock-only problem to linear order in $\phi$ (see Eq.~\ref{eqapp:Fockonly_finitephi_orderparameter_dimensionless}). }
    \label{figapp:Fockonly_finitephi_zw}
\end{figure}

\begin{figure}
    \centering
    \includegraphics[width = 1.0\linewidth]{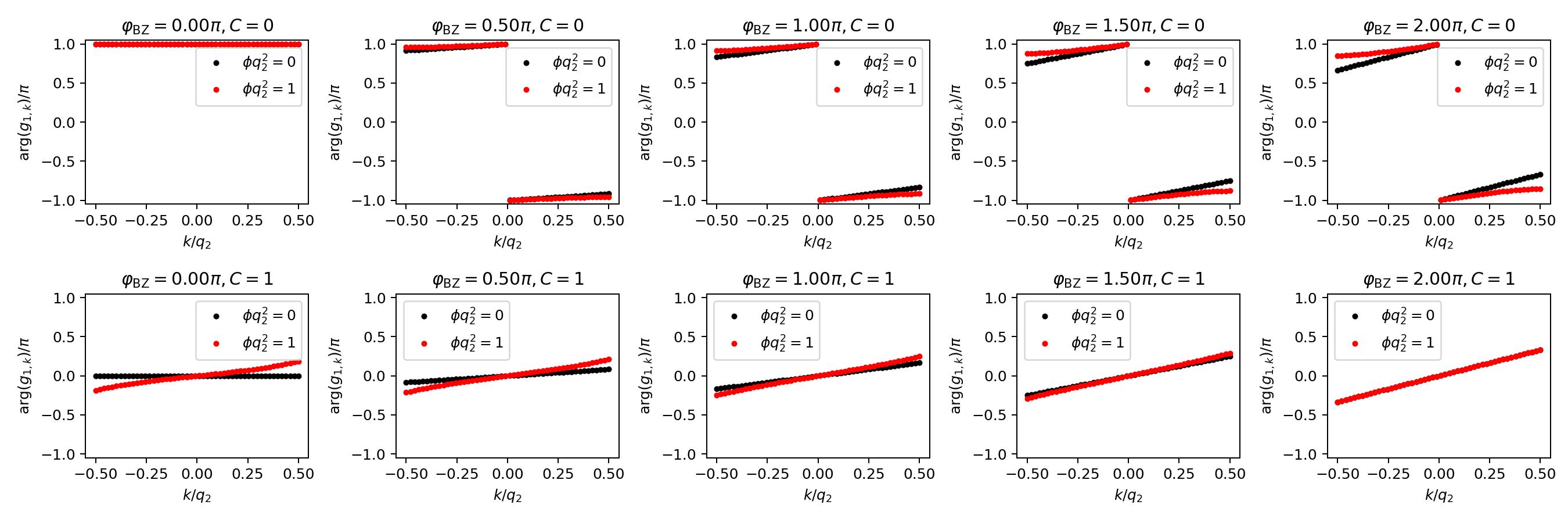}
    \caption{Argument of $g_{1,k}$ (see Eq.~\ref{eqapp:Fockonly_finitephi_orderparameter_dimensionless}) as a function of $k$ for $r=0.3$, $\gamma=3$, and different values of $\varphi_\text{BZ}$ and $C\mod 3$. For illustration, we plot the results for $\phi q_2^2=0,1$. }
    \label{figapp:Fockonly_finitephi_g_winding}
\end{figure}

\begin{figure}
    \centering
    \includegraphics[width = 0.4\linewidth]{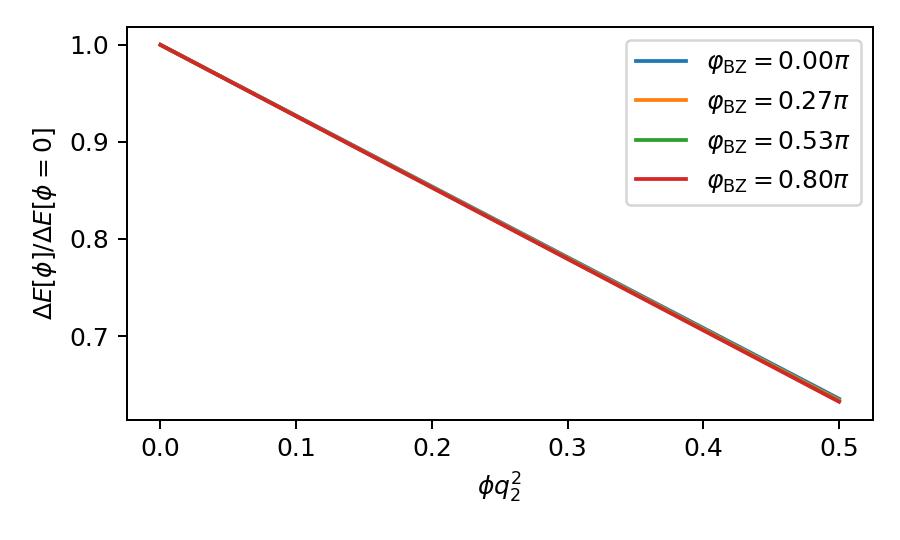}
    \caption{Difference of total energies $\Delta E$ (see Eq.~\ref{eqapp:Fock_finitetheta_Etot12final} and \ref{eqapp:Fock_finitetheta_Etot1final}) of the $C=0$ and $C=1$ solutions for $r=0.3$, $\gamma=3$, and different values of $\varphi_\text{BZ}$, as a function of $\phi q_2^2$. The energy difference is normalized to its value at $\phi=0$. For these values of $\varphi_{\mathrm{BZ}}$, the $C=0$ solution has lower energy.}
    \label{figapp:Fockonly_finitephi_dEphi}
\end{figure}

In Fig.~\ref{figapp:Fockonly_finitephi_zw}, we plot the dimensionless coefficients $z,w_0,w_1,w_2$ as a function of $\varphi_{\text{BZ}}$ for different $C\mod3$ for some representative parameters $r,\gamma$. In Fig.~\ref{figapp:Fockonly_finitephi_g_winding}, we plot the argument of $g_{1,k}$ as a function of $k$ for $\varphi_{\text{BZ}}$ different $C\mod3$. We focus on $0\leq \varphi_{\text{BZ}}\leq 2\pi$ where only the $C=0,1\mod 3$ solutions are relevant (see later discussion on the ground state phase diagram). For $\phi q_2^2=0$, the order parameter winds identically for $C=0,1\mod3$, apart from an overall sign. In this interval of $\varphi_{\text{BZ}}$, a finite $\phi q_2^2$ tends to steepen the winding of $g_{1,k}$ for $C=1\mod 3$, while it does the opposite for $C=0\mod 3$. 

For $\varphi_{\text{BZ}}=0$, the winding for $C=0$ is unchanged to order $\phi q_2^2$, while for $\varphi_{\text{BZ}}=2\pi$, the winding is unchanged for $C=1$ to order $\phi q_2^2$. This can be deduced by recognizing that the equation for $h_k$ in Eq.~\ref{eqapp:fockonly_hk} becomes purely real for these conditions, so that the phase of $g_{1,k}=h_{k}e^{ i \frac{\varphi_\text{BZ}}{6}\frac{2 k}{q_2} }$ is determined.

We now evaluate the part of the total energy that depends on the Chern number. We first evaluate the contribution $E_{\text{tot},12}$ to the total mean-field energy $E_{\text{tot}}=E_{\text{tot},1}+E_{\text{tot},12}$ from region $k_{12}$
\begin{eqnarray}\label{eqapp:Fock_finitetheta_Etot12final}
    &E_{\text{tot},12}=6N_{1,2}
    \text{Re}\left[g_{12}e^{-i\frac{2\pi}{3}C}\right]=6N_{1,2}\left[\left(
    2\tilde{G}
    +\phi(\mathfrak{g}_0+\mathfrak{g}_1\frac{q_2^2}{4})\right)\cos(\frac{\varphi_{\text{BZ}}}{6}+\frac{2\pi}{3}C)+\phi\mathfrak{g}_2\frac{q_2}{2}\sin(\frac{\varphi_{\text{BZ}}}{6}+\frac{2\pi}{3}C)
    \right]\\
    &=\frac{6\Omega_{tot}A^2_{k_1}\gamma V_0}{(2\pi)^4}r\left[\left(2z+\phi q_2^2\left(w_0+\frac{w_1}{4}\right)\right)\cos(\frac{\varphi_{\text{BZ}}}{6}+\frac{2\pi}{3}C)+\phi q_2^2\frac{w_2}{2}\sin(\frac{\varphi_{\text{BZ}}}{6}+\frac{2\pi}{3}C)\right].
\end{eqnarray}

We now consider the contribution from the $\bm{k}_1$ region, assuming that the renormalized velocity $v'$ is dominated by the bare velocity of the trashcan dispersion
\begin{eqnarray}
    E_\text{tot,1}&=-        \frac{3}{2} \frac{\Omega_{tot}}{ (2 \pi)^2v'}  \int_{- \frac{q_2}{2}}^{ \frac{q_2}{2}} dk  | g_{1,k}|^2.
\end{eqnarray}
Fortunately, the trigonometric functions simplify considerably, and we find
\begin{eqnarray}
    E_\text{tot,1}&=-        \frac{3}{2} \frac{\Omega_{tot}}{ (2 \pi)^2v'}  \int_{- \frac{q_2}{2}}^{ \frac{q_2}{2}} dk\left[
    \left(2\tilde{G}+\phi(\mathfrak{g}_0+\mathfrak{g}_1k^2)\right)^2+\phi^2\mathfrak{g}_2^2k^2
    \right].
\end{eqnarray}
To be consistent, we only retain terms up to linear order in $\phi$
\begin{eqnarray}\label{eqapp:Fock_finitetheta_Etot1final}
    E_\text{tot,1}&=-        6\tilde{G}\frac{\Omega_{tot}}{ (2 \pi)^2v'}  \int_{- \frac{q_2}{2}}^{ \frac{q_2}{2}} dk \left[\tilde{G}+\phi(\mathfrak{g}_0+\mathfrak{g}_1k^2)\right]=
    -        6q_2\tilde{G}\frac{\Omega_{tot}}{ (2 \pi)^2v'}  \left[\tilde{G}+\phi(\mathfrak{g}_0+\mathfrak{g}_1\frac{q^2}{12})\right]\\
    &=-\frac{6\Omega_{tot}A_{k_1}^2\gamma V_0}{(2\pi)^4}z\left[z+\phi q_2^2\left(w_0+\frac{w_1}{12}\right)\right].
\end{eqnarray}

We note that for certain values of $\varphi_{\text{BZ}}$, the total mean-field energy is identical for certain different values of $C\mod 3$. This continues to hold when keeping the full dependence on $\phi$, and using arbitrary interaction potentials. This stems from the relation between the order parameters for different $C\mod 3$ at such values of $\varphi_{\text{BZ}}$, as discussed below Eq.~\ref{eqapp:fockonly_hk}. Consider for example $\varphi_{\text{BZ}}=\pi$, where we can write the self-consistent order parameters $g^{C=0}_{1,k}=h_{k}e^{ i \frac{\varphi_\text{BZ}}{6}\frac{2 k}{q_2} }$ and $g^{C=1}_{1,k}=-h_k^*e^{ i \frac{\varphi_\text{BZ}}{6}\frac{2 k}{q_2} }$ for some $h_k$. It is straightforward to verify that $E_{\text{tot},1}$ and $E_{\text{tot},12}$ are the same for both values of $C$. Generally for $\varphi_{\text{BZ}}=m\pi$ for integer $m$, the total energies are the same when using the order parameters for $C=(-m+1)\mod3$ and $C=(-m+2)\mod 3$. 

Hence for small $\phi$, the ground state phase boundaries are unaffected compared to the $\phi=0$ analysis. However, increasing $\phi q_2^2$ reduces the energy difference between the solutions (see Fig.~\ref{figapp:Fockonly_finitephi_dEphi}) because the combination of the form factor and the interaction potential gets suppressed as a function of the momentum transfer. Note that by using the analysis of App.~\ref{subsecapp:full_Chern} (and using continuity from the $\phi=0$ results of App.~\ref{subsecapp:k1_k12_Fockonly_phi0}), we can further refine the Chern number, and conclude that $\varphi_{\text{BZ}}\in [2\pi(m-\frac{1}{2}),2\pi(m+\frac{1}{2})]$ leads to a $C=m$ ground state for integer $m$.

\subsection{Analysis of the full Chern number}
\label{subsecapp:full_Chern}

The actual Chern number $C$ (without modding by $2,3$ or $6$) of the gapped HF insulator can be determined by taking the line integral of the Berry connection $\bm{A}^\text{HF}(\bm{k})$ of the occupied HF wavefunction around the BZ boundary. Computing $C$ in this manner requires using a smooth gauge for the HF Bloch function $\ket{\psi^\text{HF}_{\bm{k}}}$, which cannot also be periodic in $\bm{k}$ in the case of a non-zero Chern number.
Generally, the HF Bloch function can be expanded as 
\begin{equation}
    \ket{\psi^\text{HF}_{\bm{k}}}=\sum_{\bm{G}}v_{\bm{k},\bm{G}}\ket{\psi_{\bm{k}+\bm{G}}},
\end{equation}
where $\ket{\psi_{\bm{k}+\bm{G}}}$ is the wavefunction of the underlying continuum band for momentum $\bm{k}+\bm{G}$. Equivalently, we have in second quantization for the creation operator $b^\dagger_{\bm{k}}$ for the lowest HF band
\begin{equation}
    b^\dagger_{\bm{k}}=\sum_{\bm{G}}v_{\bm{k},\bm{G}}\gamma^\dagger_{\bm{k}+\bm{G}}.
\end{equation}

For all $\bm{k}$ in BZ 0 (including its boundary), $v_{\bm{k},\bm{0}}$ is non-vanishing for the gapped HF states of interest here. Hence for such values of $\bm{k}$, we enforce a smooth gauge by taking $v_{\bm{k},\bm{0}}$ to be real and positive. The expression for the cell-periodic part of the HF Bloch function reads
\begin{equation}
    \ket{u^\text{HF}_{\bm{k}}}=\sum_{\bm{G}}v_{\bm{k},\bm{G}}\ket{u_{\bm{k}+\bm{G}}}.
\end{equation}
The Berry connection of the HF band is
\begin{equation}
    \bm{A}^\text{HF}(\bm{k})=i\braket{u^\text{HF}_{\bm{k}}}{\partial_{\bm{k}}u^\text{HF}_{\bm{k}}}=\sum_{\bm{G}}\left[iv^*_{\bm{k},\bm{G}}\partial_{\bm{k}}v_{\bm{k},\bm{G}}+|v_{\bm{k},\bm{G}}|^2 \bm{a}_{\bm{k}+\bm{G}}\right],
\end{equation}
where $\bm{a}_{\bm{k}}=i\braket{u_{\bm{k}}}{\partial_{\bm{k}}u_{\bm{k}}}$ is the Berry connection  of the underlying continuum band.

The Chern number is expressed as the line integral (taken counterclockwise)
\begin{equation}
    C=\frac{1}{2\pi}\oint_\text{BZ}d\bm{k}\cdot \bm{A}^\text{HF}(\bm{k}).
\end{equation}
Consider the segments of the integral along the $\pm k_y$ directions
\begin{align}
    \phantom{=}&\frac{1}{2\pi}\int_{-q_2/2}^{q_2/2}dk\, \left(A_y^\text{HF}(b_1/2,k)-A_y^\text{HF}(-b_1/2,k)\right)\\
    =&\frac{1}{2\pi}\int_{-q_2/2}^{q_2/2}dk\, \sum_{\bm{G}}\bigg[iv^*_{\frac{b_1}{2}\hat{x}+k\hat{y},\bm{G}}\partial_{k}v_{\frac{b_1}{2}\hat{x}+k\hat{y},\bm{G}}+|v_{\frac{b_1}{2}\hat{x}+k\hat{y},\bm{G}}|^2 a^y_{\frac{b_1}{2}\hat{x}+k\hat{y}+\bm{G}}\\
    &-iv^*_{-\frac{b_1}{2}\hat{x}+k\hat{y},\bm{G}}\partial_{k}v_{-\frac{b_1}{2}\hat{x}+k\hat{y},\bm{G}}-|v_{-\frac{b_1}{2}\hat{x}+k\hat{y},\bm{G}}|^2 a^y_{-\frac{b_1}{2}\hat{x}+k\hat{y}+\bm{G}}\bigg]\\
    =&\frac{1}{2\pi}\int_{-q_2/2}^{q_2/2}dk\, \sum_{\bm{G}}\bigg[iv^*_{\frac{b_1}{2}\hat{x}+k\hat{y},\bm{G}}\partial_{k}v_{\frac{b_1}{2}\hat{x}+k\hat{y},\bm{G}}-iv^*_{-\frac{b_1}{2}\hat{x}+k\hat{y},\bm{G}}\partial_{k}v_{-\frac{b_1}{2}\hat{x}+k\hat{y},\bm{G}}\bigg].
\end{align}
The $a^y_{\bm{k},\bm{G}}$ terms cancel above by taking $\bm{G}\rightarrow \bm{G}+b_1\hat{x}$ in the last term, and using the fact that $|v_{\frac{b_1}{2}\hat{x}+k\hat{y},\bm{G}}|=|v_{-\frac{b_1}{2}\hat{x}+k\hat{y},\bm{G}+b_1\hat{x}}|$. Hence, the Berry connection of the underlying continuum band does not explicitly enter the expression for $C$ (the connection and form factors for the continuum band of course still impact the mean-field equations and the energetic competition between different solutions). The Chern number can be written
\begin{gather}
    \tilde{\bm{A}}(\bm{k})=\sum_{\bm{G}}iv^*_{\bm{k},\bm{G}}\partial_{\bm{k}}v_{\bm{k},\bm{G}}\\
    C=\frac{1}{2\pi}\oint_\text{BZ}d\bm{k}\cdot \tilde{\bm{A}}(\bm{k})=-6\times \frac{1}{2\pi}\int_{-q_2/2}^{q_2/2}dk\, \tilde{A}_y(-b_1/2,k),
\end{gather}
where we have used $C_6$ symmetry $v_{C_6\bm{k},C_6\bm{G}}=v_{\bm{k},\bm{G}}$  of the gauge-fixed HF coefficients in the last equality, so that the integration can be taken over just a single line segment . We can thus focus on momenta $\bm{k}=(-b_1/2,k)$ that lie on the left vertical edge of the BZ. From now on, we use the simplified notation
\begin{gather}
    v_{k,\bm{G}}
    \equiv v_{-\frac{b_1}{2}\hat{x}+k\hat{y},\bm{G}}\\
    A(k)\equiv \tilde{A}_y(-b_1/2,k)=\sum_{\bm{G}}iv^*_{k,\bm{G}}\partial_{k}v_{k,\bm{G}}\\
    C=- \frac{3}{\pi}\int_{-q_2/2}^{q_2/2}dk\, A(k),
\end{gather}
where $k\in[-q_2/2,q_2/2]$ runs over the left vertical edge of the BZ (see Fig.~\ref{figapp:fullChernzones}).

\begin{figure}
    \centering
    \includegraphics[width = 0.6\linewidth]{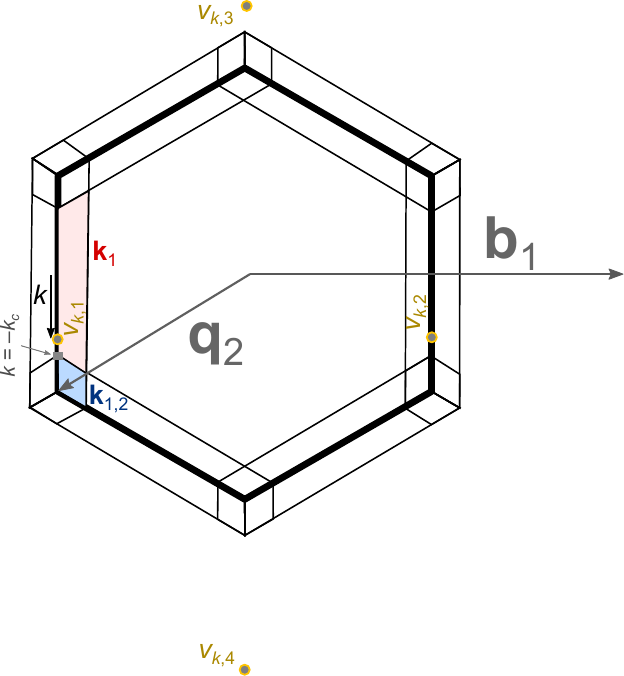}
    \caption{Setup for the computation of the full Chern number. The momentum variable $k\in[-q_2/2,q_2/2]$ runs over the left vertical edge of the BZ boundary (thick hexagon). For $|k|<k_c$, the corresponding momentum lies in the $\bm{k}_1$ region, while for $-q_2/2<k<-k_c$, the corresponding momentum lies in the $\bm{k}_{12}$ region. The wavefunction of the occupied HF band for $k$ can have up to four non-zero coefficients $v_{k,1},v_{k,2},v_{k,3},v_{k,4}$. Note that for the specific value of $k$ illustrated here, both $v_{k,3}$ and $v_{k,4}$ would vanish in our treatment of the Berry trashcan model, since their corresponding momenta lie outside the cutoff.}
    \label{figapp:fullChernzones}
\end{figure}

For our analytical mean-field analysis of the Berry trashcan model, the HF coefficients for $\bm{k}$ along the left vertical edge of the BZ can only take non-zero values for up to four possible $\bm{G}$'s
\begin{equation}
    v_{k,1}\equiv v_{k,\bm{0}},\quad v_{k,2}\equiv v_{k,\bm{b}_1},\quad v_{k,3}\equiv v_{k,C_6\bm{b}_1},\quad v_{k,4}\equiv v_{k,C^{-1}_6\bm{b}_1},
\end{equation}
which we collect into the (normalized) four-component vector $\Psi_k=[v_{k,1},v_{k,2},v_{k,3},v_{k,4}]^T$. See Fig.~\ref{figapp:fullChernzones} for an illustration of these coefficients.
We first parameterize the general form of $\Psi_k$ as 
\begin{equation}
    \Psi_k=\begin{pmatrix}
    a_k\\b_k e^{i\beta_k}\\
    c_k e^{i\gamma_k}\\
    d_k e^{i\delta_k}
    \end{pmatrix},
\end{equation}
where $a_k,b_k,c_k,d_k$ are non-negative, and we have fixed the gauge of $v_{k,1}$ to be real and positive. The combination of $C_6$ and $M_1\mathcal{T}$ symmetries, which are obeyed by the HF solution, strongly constrains the form of $\Psi_k$:
\begin{itemize}
    \item The action of $C_2$ symmetry (which relates $\Psi_{-k}$ to $\Psi_{k}$) leads to
    \begin{gather}
        \Psi_{-k}\propto \begin{pmatrix}
            0 & 1 & 0 & 0 \\
            1 & 0 & 0 & 0 \\
            0 & 0 & 0 & 1 \\
            0 & 0 & 1 & 0
        \end{pmatrix}\Psi_k\\
        \rightarrow b_{-k}=a_{k},\quad \beta_{k}=-\beta_{-k}\mod 2\pi,\quad d_{k}=c_{-k},\quad \delta_{-k}=\gamma_k-\beta_k\mod 2\pi.
    \end{gather}
    
        \item The action of $M_1\mathcal{T}$ symmetry (which relates $\Psi_{-k}$ to $\Psi^*_{k}$) leads to
    \begin{gather}
        \Psi_{-k}\propto \begin{pmatrix}
            1 & 0 & 0 & 0 \\
            0 & 1 & 0 & 0 \\
            0 & 0 & 0 & 1 \\
            0 & 0 & 1 & 0
        \end{pmatrix}\Psi^*_k\\
        \rightarrow b_k=b_{-k},\quad \delta_{-k}=-\gamma_k\mod 2\pi,
    \end{gather}
    where we have only listed the additional constraints beyond those imposed by $C_2$. 
    \item Combining the above constraints leads to the form 
    \begin{gather}
\Psi_k=\begin{pmatrix}
    b_k\\b_k e^{i\beta_k}\\
    c_k(-1)^{m_3} e^{i\beta_k/2}\\
    c_{-k}(-1)^{m_3}e^{-i\beta_{-k}/2}
\end{pmatrix}\\
\beta_k=-\beta_{-k}\mod 2\pi,\quad b_k=b_{-k},\quad  2b_k^2+c_k^2+c_{-k}^2=1,
\end{gather}
where $b_k>0$, $c_k\geq 0$, and $m_3$ is an integer. Since we are working with a smooth gauge, $\beta_k$ is smooth and $m_3$ does not depend on $k$.
\end{itemize}
$A(k)$ can then be evaluated as
\begin{align}
    A(k)&=i(2b_k\partial_kb_k+c_k\partial_kc_k+c_{-k}\partial_k c_{-k})-b_k^2\partial_k\beta_k-c_k^2\partial_k \beta_k/2 +c_{-k}^2\partial_k \beta_{-k}/2
    \\
    &=-b_k^2\partial_k\beta_k-c_k^2\partial_k \beta_k/2 +c_{-k}^2\partial_k \beta_{-k}/2=-\frac{1}{2}\partial_k\beta_k,
\end{align}
where we have used the normalization condition $2b_k^2+c_k^2+c_{-k}^2=1$.
The Chern number is then
\begin{equation}\label{eqapp:full_C_beta}
    C=\frac{3}{2\pi}(\beta_{q_2/2}-\beta_{-q_2/2}).
\end{equation}
Hence, $C$ is determined solely by the phase winding of the $v_{k,2}$ component (in the smooth gauge where $v_{k,1}$ is real and positive) as $k$ runs along the left vertical edge of the BZ. 

Our mean-field analysis of the Berry trashcan model is naturally framed in terms of the hybridization field $g_{1,k}$, which evolves smoothly with $k$. Within our decomposition of the BZ into $\bm{k}_1$ and $\bm{k}_{12}$ regions (see Fig.~\ref{figapp:fullChernzones}), the thin sliver approximation allows us to approximate the hybridization field to be constant in the $\bm{k}_{12}$ region, i.e.~we consider a uniform $g_{12}\simeq g_{1,-q_2/2}$. In deriving the mean-field equations, we have also replaced the order parameter in the $\bm{k}_{12}$ region by its value at $\bm{k}_{12}=\bm{q}_2$, since a simple analytical solution to the $3\times3$ Hamiltonian in this region is not forthcoming. Here, $\beta_{-q_2}$ can only take values $0,2\pi/3,4\pi/3\mod 2\pi$ owing to $C_3$ symmetry.

For the purposes of computing the full Chern number though, we need to revisit the order parameter in the $\bm{k}_{12}$ region, since Eq.~\ref{eqapp:full_C_beta} requires a smooth solution throughout $k\in[-q_2/2,q_2/2]$. As illustrated in Fig.~\ref{figapp:fullChernzones}, the $\bm{k}_{1}$ and $\bm{k}_{12}$ regions meet at $k=-k_c$, and we need to ensure that the HF wavefunction on the edge of the BZ smoothly evolves across this point. Therefore, we need to consider $\beta_k$ for $-q_2/2<k<-k_c$ in more detail (the problem for $k_c<k<q_2/2$ is related by $M_1\mathcal{T}$ symmetry). For such values of $k$, the mean-field Hamiltonian only acts on the single-particle states associated with the HF coefficients $v_{k,1},v_{k,2},v_{k,3}$. Using the thin sliver approximation for the hybridization field $g_{12}=g_{1,-q_2/2}$, the $3\times 3$ Hamiltonian reads in this basis
\begin{equation}
    H_k=g\begin{pmatrix}
        0 & e^{-i\theta} & e^{i\theta} \\
        e^{i\theta} & 0 & e^{-i\theta} \\
        e^{-i\theta} & e^{i\theta} & w_k 
    \end{pmatrix},\label{eqapp:Hk_theta}
\end{equation}
where we have parameterized $g_{12}=ge^{i\theta}$ with $g>0$. $w_k=v'(k+q_2/2)/g$ is positive semi-definite, and captures the sharp kinetic dispersion experienced at $\bm{q}_2+C_6\bm{b}_1+k\hat{y}$. From our symmetry analysis above, the eigenfunctions of $H_k$ can be parameterized as 
\begin{equation}
    \begin{pmatrix}
        b_k\\
        b_ke^{i\beta_k}\\
        c_k(-1)^{m_3}e^{i\beta_k/2}
    \end{pmatrix},
\end{equation}
where $b_k,c_k>0$.

Recall that the interface between the $\bm{k}_1$ and $\bm{k}_{12}$ regions is at $-k_c$, which is close to $-q_2/2$ in the thin sliver limit. Hence, as we take $k\rightarrow -k_c$ in the $\bm{k}_1$ region, the order parameter phase $\beta_k$ becomes $\theta-\pi$, which arises from solving the $2\times 2$ Hamiltonian acting on just the $v_{k,1}$ and $v_{k,2}$ coefficients, using $g_{1,-k_c}\approx g_{12}=ge^{i\theta}$. The task is now to determine $\beta_k$ for $k$ in the $\bm{k}_{12}$ region. We focus on $2\pi/3 <\theta < 4\pi/3$ (modulo $2\pi$), since the case with $\theta+2\pi/3$ can be obtained by multiplying the components of the eigenfunction according to $v_{k,n}\rightarrow v_{k,n} \omega^{n-1}$. We review the ground state solution in the following limiting cases:
\begin{itemize}
   \item At the BZ corner $\bm{k}=\bm{q}_2$, we have $w_k=0$ and the eigenfunctions of $H_k$ are just the $C_3$ eigenvectors. For $2\pi/3 <\theta < 4\pi/3$, the ground state of $H_{-q_2/2}$ is $\frac{1}{\sqrt{3}}[1,1,1]^T$ and corresponds to $C=0\mod 3$. Hence, we can take $\beta_{-q_2/2}=0$ and $m_3=0$. Since $H_k$ changes continuously with $k$ and we are interested in smooth $\beta_k$, we take $m_3=0$ for $k>-q_2/2$ as well. 
   \item For large enough $k$ such that $w_k\gg 1$, the ground state of $H_{k}$ is approximately $\frac{1}{\sqrt{2}}[1,-e^{i\theta},0]$, so that $\beta_k=\theta-\pi$. Note that this recovers the solution at the lower boundary $k\rightarrow -k_c$ of the $\bm{k}_1$ region as discussed above. 
\end{itemize}

\begin{figure}
    \centering
    \includegraphics[width = 0.4\linewidth]{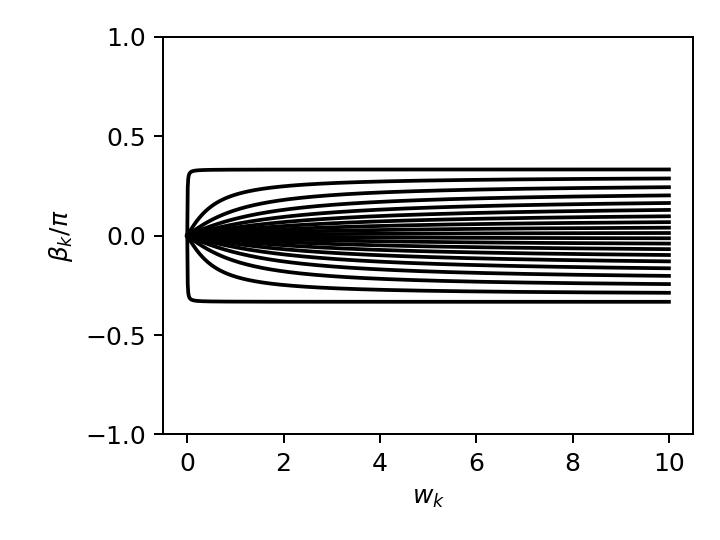}
    \caption{Relative phase $\beta_k$ of the first two components of the lowest eigenvector of Eq.~\ref{eqapp:Hk_theta}, as a function of $w_k$. Different lines correspond to different choices of $2\pi/3<\theta<4\pi/3$. For such values of $\theta$, the phase $\beta_k$ always remains within the interval $[-\pi/3,\pi/3]$, and reaches $0$ for $w_k\rightarrow0$.}
    \label{figapp:betak}
\end{figure}

Between these limiting cases $\beta_{-q_2/2}=0$ and $\beta_{-k_c}=\theta-\pi$, $H_k$ does not admit a tractable analytic solution for general $k$, but the behavior of $\beta_k$ can be easily computed numerically as shown in Fig.~\ref{figapp:betak}. We find that the ground state HF solution always satisfies $-\pi/3 < \beta_k < \pi/3$ for the case $2\pi/3 <\theta < 4\pi/3$ considered here. Note that we have $\beta_{-k_c}=\theta-\pi$ which also lies in the interval $-\pi/3<\beta_k<\pi/3$. Hence, the order parameter does not encounter any additional `unnecessary' windings in the $\bm{k}_{12}$ region. As mentioned above, the solution for $\theta+2\pi/3$ can be obtained by multiplying the components of the eigenfunction according to $v_{k,n}\rightarrow v_{k,n} \omega^{n-1}$, which takes $\beta_k\rightarrow \beta_k+2\pi/3$. 

We can summarize the key result of our analysis: Whatever value the phase $\beta_k$ takes at $k=-k_c$, it will simply move to the nearest multiple of $2\pi/3$ as $k\rightarrow-q_2/2$, without going through any additional windings. The behavior of $\beta_k$ for $k>k_c$ can be obtained using $M_1\mathcal{T}$ symmetry. Hence, given the solution of the order parameter in the $\bm{k}_1$ region, the full Chern number can be extracted by determining the evolution of $\beta_k$ for $-q_2/2\leq k\leq q_2/2$ as described above, and using Eq.~\ref{eqapp:full_C_beta}.

\clearpage

\end{document}

%% file: main.bbl
%